\documentclass[12pt,a4paper,twoside]{book}

\usepackage[ngerman,english]{babel} 
\usepackage[english]{babel}                                                 
\usepackage[latin1,utf8]{inputenc}                                                  
\usepackage{fancyhdr}                                                       
\usepackage{amsmath}
\usepackage{graphicx}
\usepackage{amssymb}
\usepackage{xcolor}
\usepackage{float}
\usepackage{appendix} 
\usepackage{setspace}
\usepackage{enumitem}
\usepackage{subcaption}
\usepackage{cite}
\usepackage{epsfig}
\usepackage{graphicx}
\usepackage{color}
\usepackage{amsmath}
\usepackage{bbold}
\usepackage{float}
\usepackage{longtable}
\usepackage{etaremune}
\usepackage{tabularx}
\usepackage{adjustbox}
\usepackage{xifthen}
\usepackage{amsbsy}
\usepackage{dsfont}
\usepackage{mathrsfs}
\usepackage{siunitx}
\allowdisplaybreaks
\usepackage[
    colorlinks,
    linkcolor={blue!80!black},
    citecolor={blue!80!black},
    urlcolor={blue!80!black}
]{hyperref}

\setlist[itemize]{leftmargin=4.0mm}

\DeclareMathOperator{\tr}{Tr}

\makeatletter
\def\@makechapterhead#1{%
  \vspace*{25\p@} 
  {\parindent \z@ \raggedright \normalfont
    \interlinepenalty\@M
    \Huge \bfseries
    \thechapter\hspace{20pt}#1\par\nobreak
    \vskip 40\p@
  }}
  
\def\@makeschapterhead#1{             
  \vspace*{25\p@}
  {\parindent \z@ \raggedright \normalfont
    \interlinepenalty\@M
    \Huge \bfseries
    #1\par\nobreak
    \vskip 40\p@
  }}
\makeatother

\makeatletter
\g@addto@macro\appendices{%
  \def\@makechapterhead#1{%
    \vspace*{0\p@}
    { \parindent \z@ \raggedright \normalfont
      \Huge \bfseries 
      \thechapter 
      \hspace{10pt}
      \interlinepenalty\@M
      \Huge \bfseries
      \parbox[t]{410pt}{#1} \par\nobreak
      \vskip 40\p@
    }}
}
\makeatother

\usepackage{afterpage}

\begin{document}     

\def\bcA{{\bf \mathcal{A}}}
\def\bS{{\bf S}}
\def\bB{{\bf B}}
\def\bp{{\bf p}}
\def\bR{{\bf R}}
\def\br{{\bf r}}
\def\bQ{{\bf Q}}
\def\bA{{\bf A}}
\def\bj{{\bf j}}
\def\bq{{\bf q}}
\def\bG{{\bf G}}
\def\bV{{\bf V}}
\def\bd{{\bf d}}
\def\bL{{\bf L}}
\def\bU{{\bf U}^{}}
\def\bE{{\bf E}}
\def\bv{{\bf v}}
\def\brho{\boldsymbol\rho}
\def\bUdag{{\bf U}^\dagger}
\def\Udag{U^\dagger}
\def\lam{\lambda}
\def\blam{\boldsymbol{\lambda}}
\def\blamt{\boldsymbol{\tilde\lambda}}
\def\bsigma{\boldsymbol{\sigma}}
\def\bOmega{\mathbf{\Omega}}
\def\alf{\alpha}
\def\eps{\epsilon}
\def\c{c^{}}
\def\U{U^{}}
\def\cdag{c^\dag}
\def\Tr{\text{Tr}}
\def\tr{\text{tr}}
\def\re{\text{Re}}
\def\im{\text{Im}}

\def\cD{\mathcal{D}}
\def\cS{\mathcal{S}}
\def\cE{\mathcal{E}}
\def\cA{\mathcal{A}}
\def\cF{\mathcal{F}}
\def\cG{\mathcal{G}}
\def\cX{\mathcal{X}}
\def\cY{\mathcal{Y}}
\def\cZ{\mathcal{Z}}
\def\cC{\mathcal{C}}
\def\cI{\mathcal{I}}
\def\cL{\mathcal{L}}
\def\cT{\mathcal{T}}
\def\cg{\mathcal{g}}
\def\cR{\mathcal{R}}
\def\cN{\mathcal{N}}
\def\cM{\mathcal{M}}
\def\cP{\mathcal{P}}
\def\cH{\mathcal{H}}

\def\dX{\mathcal{X}}
\def\dY{\mathcal{Y}}
\def\dM{\mathcal{M}}
\def\dF{\mathcal{F}}

\def\sA{\mathscr{A}}
\def\sP{\mathscr{P}}
\def\sG{\mathscr{G}}
\def\sV{\mathscr{V}}

\def\PIabgd{\Pi^{H,\alf\beta\gam\delta}_{iq_0}}
\def\PIabdg{\Pi^{H,\alf\beta\delta\gam}_{iq_0}}
\def\PIbagd{\Pi^{H,\beta\alf\gam\delta}_{iq_0}}
\def\PIbadg{\Pi^{H,\beta\alf\delta\gam}_{iq_0}}

\def\TrH{{\rm Tr}_\text{\tiny EB}}
\def\wideTrH{{\rm Tr}_\text{\tiny H}}

\def\ic{\nu}
\def\pauma{\tau}
\def\bpauma{\boldsymbol{\tau}}
\def\g{\textsl{g}}


\begin{titlepage}

{\setstretch{1.4}
%
%
%
\vspace{10mm}
\begin{center}
\textcolor{black}{
{\Huge{\bf Electrical Conductivity in \\
\vspace{3mm}
Quantum Materials}}\\
}
\end{center}

\vspace{15mm} \par \noindent

\begin{center}
Von der Fakult\"at Mathematik und Physik der Universit\"at
Stuttgart \\ zur Erlangung der W\"urde eines Doktors der Naturwissenschaften \\
(Dr. rer. nat.) genehmigte Abhandlung
\end{center}

} 

\vspace{15mm} 
\begin{center}
vorgelegt von
\end{center}

\begin{center}\textcolor{black}{
{\LARGE{\bf Johannes Mitscherling}}\\
}\end{center}

\begin{center}
aus Aachen, Deutschland
\end{center}

\vspace{17mm}
\begin{center}
\begin{tabular}{ll}
Hauptberichter: & Prof. Dr. Walter Metzner\\[0mm]
Mitberichter: & Prof. Dr. Maria Daghofer \\[0mm]
& Prof. Dr. Oleg Sushkov \\[0mm]
Pr\"ufungsvorsitzender: & Prof. Dr. Tilman Pfau \\[4mm]
Tag der Einreichung: & 27. Januar 2021\\
Tag der m\"undlichen Pr\"ufung: & 25. M\"arz 2021
\end{tabular}
\end{center}

\vspace{17mm}
\begin{center}
{\large{\bf
Max-Planck-Institut f\"ur Festk\"orperforschung \\
Universit\"at Stuttgart
\vspace{2mm}\\ \vspace{2mm} 2021
}}
\end{center}

\end{titlepage}

\clearpage{\pagestyle{empty}\cleardoublepage}


\begin{titlepage}     

\pagenumbering{arabic} 
\addtocounter{page}{2} 
                                                                            
\thispagestyle{empty}                                                       
\topmargin=6.5cm                                                            
\raggedleft                                                                 
\emph{
To my family, friends and colleagues.
}
\vspace{1cm}

\end{titlepage}
\clearpage{\pagestyle{empty}\cleardoublepage}                               




\cfoot[\thepage]{\thepage}

\setcounter{tocdepth}{2}

\rhead[\fancyplain{}{\bfseries                                             
Contents}]{\fancyplain{}{
}}                          
\lhead[\fancyplain{}{
}]{\fancyplain{}{\bfseries            
Contents}}
{\hypersetup{hidelinks} 
\renewcommand{\contentsname}{\vspace{-2.5cm} \raggedright \bfseries\Huge Contents \vspace{-0.5cm}}
    \tableofcontents  
}
\cleardoublepage

%







\rhead[\fancyplain{}{\bfseries                                             
Abstract}]{\fancyplain{}{
}}                          
\lhead[\fancyplain{}{
}]{\fancyplain{}{\bfseries            
Abstract}}
\chapter*{Abstract}
\addcontentsline{toc}{chapter}{Abstract} 

Not only in the past decades, material science and other research fields of condensed matter physics revealed fascinating new phenomena, which had or may have a huge impact on our daily life. The manifestation of quantum phenomena arising from the interplay of many particles still results in unexpected effects that have yet to be understood. The electrical conductivity is one of the fundamental properties of solids and, thus, may lead to a better understanding of quantum materials that host such phenomena.  In order to describe transport properties in materials theoretically, the concept of bands has proven to be very powerful. In recent years, there is an increasing interest in transport phenomena that are fundamentally linked to the presence of multiple bands. So-called interband contributions to the conductivity formulas are, for instance, necessary to capture pure interband phenomena, which are not describable by a single-band model. They are also relevant to provide a well understood connection between multiband models and their measurable consequences.

In this thesis, we develop, discuss, and apply a theory of the electrical conductivity that includes interband contributions within a microscopic approach. We focus our study on the conductivity for a general momentum-block-diagonal two-band model as a minimal model that is able to show interband effects. This model captures a broad variety of very different physical phenomena. For instance, it describes systems with magnetic order like N\'eel antiferromagnetism and spiral spin density waves as well as topological systems like Chern insulators. We derive formulas for the conductivity tensor $\sigma^{\alf\beta}$ and the Hall conductivity tensor $\sigma^{\alf\beta\eta}_\text{H}$, which describe the current in the presence of an external electric field and the Hall current in the presence of both an external electric and an external magnetic field, respectively. We identify two criteria that allow for a unique and physically motivated decomposition of the conductivity tensors. On the one hand, we distinguish {\it intraband} and {\it interband} contributions that are defined by the involved quasiparticle spectral functions of one or both bands. On the other hand, we distinguish {\it symmetric} and {\it antisymmetric} contributions that are defined by the symmetry under the exchange of the current and the electric field directions. The (symmetric) intraband contribution generalizes the formula of standard Boltzmann transport theory, whereas the interband contributions capture effects exclusively linked to the presence of multiple bands.  In order to obtain a non-diverging conductivity, a finite momentum relaxation is required. We include a phenomenological relaxation rate $\Gamma$ of arbitrary size. This allows us to generalize previous results and study the relevance of the interband contributions systematically. 

We apply the microscopic theory to models and experiments of recent interest. The antisymmetric interband contributions of the conductivity tensor $\sigma^{\alf\beta}$ describe the so-called intrinsic anomalous Hall effect, a transverse current without any external magnetic field that is not caused by (skew) scattering. Its deep connection to the Berry curvature and, thus, to the Chern number can lead to the quantization of the intrinsic anomalous Hall conductivity. We study the impact of a nonzero relaxation rate $\Gamma$ on this quantization. The scaling behavior with respect to $\Gamma$ is crucial for disentangling different extrinsic, that is, based on scattering off impurities, or intrinsic origins of the anomalous Hall effect. The validity of the conductivity formulas for $\Gamma$ of arbitrary size allows us to identify parameter regimes with typical scaling behavior and crossover regimes of the intrinsic anomalous Hall conductivity, which are consistent with experimental results. Recent experiments on hole-doped cuprates under very high magnetic fields, which are needed to suppress the superconductivity to sufficiently low temperature, show a drastic change of the Hall number when entering the pseudogap regime. This indicates a Fermi surface reconstruction at that doping. The onset of spiral antiferromagnetic order, which is closely related to N\'eel antiferromagnetic order with a slightly modified ordering wave vector, is consistent with the experimental findings. We discuss spiral magnetic order as an example of an order that can be incommensurate with the underlying lattice but that is still captured by a two-band model. We clarify the range of validity of simplified Boltzmann-like formulas for the longitudinal and the Hall conductivity, which do not include interband contributions. Those were used previously in the original theoretical proposal to describe the Fermi surface reconstruction. We show that these simplified formulas are valid for the experiments on cuprates not due to a general argument comparing energy scales but due to the small numerical size of the previously neglected contributions.

\clearpage{\pagestyle{empty}\cleardoublepage}


\begin{otherlanguage}{ngerman} 
\rhead[\fancyplain{}{\bfseries                                             
Zusammenfassung}]{\fancyplain{}{
}}                          
\lhead[\fancyplain{}{
}]{\fancyplain{}{\bfseries            
Zusammenfassung}}
\thispagestyle{empty}
\chapter*{Zusammenfassung}
\addcontentsline{toc}{chapter}{Zusammenfassung} 
\label{Zusammenfassung}

Die Forschung in den Materialwissenschaften und weiteren Feldern der Physik der kondensierten Materie hat in ihrer Geschichte immer wieder faszinierende Ph\"anomene aufgedeckt, die zum Teil mittlerweile einen gro{\ss}en Einfluss auf unser t\"agliches Leben haben. Besonders Quantenph\"anomene, die durch das Zusammenspiel vieler Teilchen entstehen, f\"uhren weiterhin zu unvorhergesehenen Ph\"anomenen, die im Fokus aktueller Forschung stehen. Als eine der fundamentalen Eigenschaften von Festk\"orpern kann die elektrische Leitf\"ahigkeit dazu beitragen, Quantenmaterialien, die solche Vielteilchenph\"anomene zeigen, besser zu verstehen. Bei der theoretischen Beschreibung der Leitf\"ahigkeit hat sich das Konzept von B\"andern als sehr n\"utzlich erwiesen. Besonders bekannt ist dabei die Transporttheorie nach Boltzmann, die in einem semiklassischen Ansatz die Impulsableitung der B\"ander mit der Leitf\"ahigkeit in Verbindung setzt. In den letzten Jahren hat sich ein wachsendes Interesse an Transportph\"anomenen entwickelt, die ma{\ss}geblich mit der Pr\"asenz und dem Zusammenspiel mehrerer B\"ander verbunden sind. Um nun die theoretische Vorhersagbarkeit von messbaren Eigenschaften aus Mehrbandmodellen zu verbessern, aber auch um Ph\"anomene zu verstehen, f\"ur deren Beschreibung ein einzelnes Band beziehungsweise mehrere unabh\"angige B\"ander nicht mehr ausreichen, wird ein besseres theoretisches Verst\"andnis der sogenannten Interbandbeitr\"age zur Leitf\"ahigkeit ben\"otigt. Deren theoretische Beschreibung geht \"uber die Standardversion der Boltzmannschen Transporttheorie hinaus und kann zum Beispiel mithilfe eines mikroskopischen Zugangs systematisch erforscht werden.

Im Fokus dieser Doktorarbeit steht die Herleitung, Analyse und Anwendung einer mikroskopische Theorie der elektrischen Leitf\"ahigkeit, die Interbandbeitr\"age mitber\"ucksichtigt. Wir leiten die Formeln der Leitf\"ahigkeit f\"ur ein allgemeines Zweibandmodell her, das blockdiagonal im Impuls ist. Dieses Minimalmodell f\"ur ein Mehrbandsystem beschreibt eine gro{\ss}e Bandbreite verschiedener physikalischer Systeme. Die Bandbreite reicht von Systemen mit magnetischer Ordnung wie N\'eel-Antiferromagnetismus und spiraler Spindichtewelle bis hin zu topologischen Systemen wie Chern-Isolatoren. Unsere Herleitung umfasst die longitudinale Leitf\"ahigkeit, die sogenannte intrinsische anomale Hall-Leitf\"ahigkeit sowie die gew\"ohnliche Hall-Leitf\"ahigkeit. Der intrinsische anomale Hall-Effekt beschreibt dabei einen transversalen Strom ohne externes magnetisches Feld, der jedoch nicht auf einer (extrinsischen) asymmetrischen Streuung an St\"orstellen basiert. Neben der Herleitung von einfach anzuwendenden Formeln liegt ein Fokus dieser Arbeit auf der Identifizierung und der Anwendung zweier Kriterien, die es uns erlauben, die Leitf\"ahigkeit in eindeutige Beitr\"age mit einer sinnvollen physikalischen Interpretation zu zerlegen. Als erstes Kriterium unterscheiden wir Intra- und Interbandbeitr\"age, die dar\"uber definiert sind, ob die Quasiteilchenspektralfunktionen von einem oder beiden B\"andern Bestandteil dieses Beitrages sind. Als zweites Kriterium unterscheiden wir symmetrische und antisymmetrische Beitr\"age. Diese definieren wir \"uber die Symmetrie unter Vertauschung der Strom- und der elektrischen Feldrichtung. Die (symmetrischen) Intrabandbeitr\"age verallgemeinern die Formel der Boltzmannschen Transporttheorie, wohingegen die Interbandbeitr\"age die Effekte beschreiben, die ausschlie{\ss}lich auf der Pr\"asenz und dem Zusammenspiel von mehreren B\"andern beruhen. Ohne die Verletzung der Impulserhaltung divergiert die Leitf\"ahigkeit und erm\"oglicht damit keine sinnvolle theoretische Beschreibung. Aus diesem Grund f\"uhren wir eine ph\"anomenologischen Relaxierungsrate $\Gamma$ in unsere Theorie ein. Im Gegensatz zu vorherigen Arbeiten erlauben wir in unserer Herleitung ein $\Gamma$ von beliebiger Gr\"o{\ss}e, was es uns erm\"oglicht, vorherige Formeln zu verallgemeinern und die Relevanz der Interbandbeitr\"age systematisch zu analysieren.

Als Beispiel der allgemeinen Eigenschaften der neu hergeleitetenden Formeln sowie um ein tieferes Verst\"andnis zu gewinnen, wenden wir unsere mikroskopische Theorie auf mehrere konkrete Modelle und Experimente von aktuellem Interesse an. Die antisymmetrischen Interbandbeitr\"age der normalen Leitf\"ahigkeit beschreiben den intrinsischen anomalen Hall-Effekt, der durch seine Verbindung zur Berry-Kr\"ummung und damit zur Chernzahl quantisiert sein kann. Wir untersuchen den Einfluss der Relaxierungssrate $\Gamma$ auf diese Quantisierung. Wir nutzen die G\"ultigkeit der neu hergeleitetenden Leitf\"ahigkeitsformeln f\"ur $\Gamma$ von beliebiger Gr\"o{\ss}e, um Parameterbereiche mit typischem Skalierungsverhalten in Bezug auf $\Gamma$ beziehungsweise in Bezug auf die longitudinale Leitf\"ahigkeit zu identifizieren. Wir zeigen, dass das Skalierungsverhalten sowohl qualitativ als auch quantitativ konsistent mit experimentellen Messungen der anomalen Hall-Leitf\"ahigkeit als Funktion der longitudinalen Leitf\"ahigkeit sind. 

Kuprate z\"ahlen zu den Hochtemperatursupraleitern, deren Phasendiagramm weiterhin im Fokus aktueller Forschung steht. Mit Hilfe starker magnetischer Felder l\"asst sich die Supraleitung  auch im Bereich der maximalen Sprungtemperatur bei optimaler Lochdotierung soweit unterdr\"ucken, dass die Messung der Hall-Leitf\"ahigkeit des normalleitenden Kuprats bei ausreichend niedrigen Temperaturen m\"oglich ist. Aktuelle Experimente zeigen in den lochdotierten Kupraten unter diesen sehr starken magnetischen Feldern eine drastische \"Anderung der Hall-Zahl beim \"Ubergang in den Bereich des Phasendiagramms, der als Pseudogap Phase bezeichnet wird. Die \"Anderung der Hall-Zahl l\"asst sich mit einer Restrukturierung der Fermifl\"achen bei dieser Dotierung erkl\"aren. Eine solche Restrukturierung der Fermifl\"ache kann durch das Einsetzen einer spiral-antiferromagnetische Ordnung bei dieser Dotierung hervorgerufen werden und liefert eine Dotierungsabh\"angigkeit der Hall-Zahl, die konsistent mit den experimentellen Beobachtungen ist. Eine besondere Eigenschaft der spiralmagnetischen Ordnung ist, dass der Ordnungsvektor des Spiralmagnetismus nicht kommensurabel mit dem zugrunde liegenden Gitter sein muss, um mithilfe eines Zweibandmodells beschreibbar zu sein. Die theoretische Beschreibung der Spiralordnung schlie\ss t dabei Ferromagnetismus und N\'eel-Antiferromagnetismus als Spezialf\"alle mit ein. Bei der Formel, die in der urspr\"unglichen theoretischen Arbeit zur Erkl\"arung der Fermifl\"achenrestrukturierung genutzt wurden, wurden die Interbandbeitr\"age in Anwesenheit der spiralmagnetischen Ordnung nicht ber\"ucksichtigt. Unsere Theorie erlaubt es uns nun, den G\"ultigkeitsbereich der vereinfachten Formeln ohne Interbandbeitr\"age zu kl\"aren. Wir zeigen explizit durch einen Vergleich der Intra- und Interbandbeitr\"age, dass die vereinfachten Formeln f\"ur diese Experimente mit Kupraten und daher auch die damit verbundenen Schlussfolgerungen g\"ultig sind. Die Begr\"undung l\"asst sich jedoch nicht auf einen einfachen Vergleich von Energieskalen reduzieren, sondern liegt in der kleinen numerischen Gr\"o{\ss}e der Interbandbeitr\"age.

\newpage\leavevmode\thispagestyle{empty}\newpage
\end{otherlanguage}




\rhead[\fancyplain{}{\bfseries                                               
Introduction}]{\fancyplain{}{
}}
\lhead[\fancyplain{}{
}]{\fancyplain{}{\bfseries             
Introduction}}
\chapter*{Introduction}
\addcontentsline{toc}{chapter}{Introduction} 

The electrical conductivity is one of the fundamental properties of solids. As such, measurements of the electrical conductivity are used to explore new physical phenomena, to characterize new materials, and to find evidence for theoretical predictions, which, finally, might lead to new technologies and impact our daily life. Thus, an improved understanding of the electrical conductivity itself is of ongoing interest in physics for both theory and experiment. In recent years, advances in experimental techniques revealed the need of reconsidering theoretical descriptions of the conductivity in multiband systems. An improved theoretical description is required in order to expand our knowledge of those phenomena that are rooted in the presence of multiple bands.

In this thesis, we will derive and study formulas of the electrical conductivity for a very general two-band model as the simplest example of a multiband system. The model under consideration captures a broad spectrum of physically very different systems. This spectrum includes models with N\'eel antiferromagnetic or spiral magnetic order as well as models that involve spin-orbit interaction and are known to show topological properties. The focus of our discussion will lie on a systematic decomposition of the conductivity formulas into contributions of distinct physical interpretation and properties. We will generalize different well-known results and clarify different aspects. We will be able to give physical and analytical insights and strategies, which may provide a better intuitive understanding of effects that are specific to multiband systems. In a second part, we will apply the general formalism to various examples. We will discuss how an improved understanding of the conductivity formulas affects the interpretation of recent experimental results. 

In this introduction, we summarize some basic aspects of the electrical conductivity in multiband systems. We give a glimpse which type of multiband effects will be captured by our general formalism and how they are important in two topics of recent interest, that is, in experiments on cuprates and topologically nontrivial materials. We close by a short sketch of two criteria for a useful decomposition of the conductivity formulas. Those criteria will be introduced in Chapter~\ref{sec:theory} and will guide us in the derivation. We will come back to the applications in Chapter~\ref{sec:application}. Large parts of the presented results are already published in 
\begin{itemize}
 \item J.~Mitscherling and W.~Metzner, \\ 
 Longitudinal conductivity and Hall coefficient in two-dimensional metals with spiral magnetic order, \\
 \href{https://doi.org/10.1103/PhysRevB.98.195126}{Phys. Rev. B {\bf 98}, 195126 (2018)}.
 \item P.~M.~Bonetti${}^*$, J.~Mitscherling${}^*$, D.~Vilardi, and W.~Metzner,	\\
 Charge carrier drop at the onset of pseudogap behavior in the two-dimensional Hubbard model, \\
 \href{https://doi.org/10.1103/PhysRevB.101.165142}{Phys. Rev. B {\bf 101}, 165142 (2020)}.
 \let\thefootnote\relax\footnotetext{*: equal contribution}
 \item J.~Mitscherling, \\
 Longitudinal and anomalous Hall conductivity of a general two-band model, \\
 \href{https://doi.org/10.1103/PhysRevB.102.165151}{Phys. Rev. B {\bf 102}, 165151 (2020)}.
\end{itemize}
and re-arranged, combined and expanded for a consistent presentation throughout this thesis \cite{Mitscherling2018, Bonetti2020, Mitscherling2020, Bonetti2020Authors}.

%
%

\subsection*{Electrical conductivity in multiband systems}

Applying external electric and magnetic fields to a material may induce a current. The current density $j^\alf$ in spatial direction $\alf=x,y,z$ that is induced by an electric field $E^\beta$ in $\beta=x,y,z$ direction and a magnetic field $B^\eta$ in $\eta=x,y,z$ direction can be expanded in the form
\begin{align}
 j^\alf=\sigma^{\alf\beta} E^\beta+\sigma^{\alf\beta\eta} E^\beta B^\eta + ... \, ,
\end{align}
by which we introduce the conductivity tensors $\sigma^{\alf\beta}$ and $\sigma^{\alf\beta\eta}$. These tensors capture a broad spectrum of transport phenomena. For instance, the diagonal elements $\sigma^{xx}$, $\sigma^{yy}$ and $\sigma^{zz}$ are the longitudinal conductivities. The Hall current in the $x$-$y$ plane due to a perpendicular magnetic field in $z$ direction is described by the Hall conductivity $\sigma^{xyz}_\text{H}=-\sigma^{yxz}_\text{H}$, which is a component of the antisymmetric contribution of $\sigma^{\alf\beta\eta}$ with respect to $\alf\leftrightarrow\beta$. A transverse current in the absence of any external magnetic field, which is known as anomalous Hall current, is captured by the antisymmetric part of $\sigma^{\alf\beta}$. Thus, a key task is the calculation of the conductivity tensors $\sigma^{\alf\beta}$ and $\sigma^{\alf\beta\eta}$ for a given model. 

Semiclassical approaches like (standard) Boltzmann transport theory \cite{Ashcroft1976, Rickayzen1980, Mahan2000} leads to the following formula of the conductivity 
\begin{align}
 \label{eqn:condBoltzmann}
 \sigma^{\alf\beta}=-e^2\tau\int\hspace{-2mm}\frac{d^d\bp}{(2\pi)^d}\, f'(\eps_\bp-\mu)\,\frac{\partial \eps_\bp}{\partial p^\alf}\,\frac{\partial \eps_\bp}{\partial p^\beta}=e^2\tau\int\hspace{-2mm}\frac{d^d\bp}{(2\pi)^d}\, f(\eps_\bp-\mu)\, \frac{\partial^2 \eps_\bp}{\partial p^\alf\partial p^\beta} 
\end{align}
in the relaxation time approximation for one band with dispersion $\eps_\bp$. We do not consider spin for simplicity. The electric charge is denoted by $e$. We use the convention $e>0$ throughout this thesis. A relaxation time $\tau$ is required for momentum relaxation in order to get a finite, non-diverging conductivity. We introduce the Fermi function $f(\omega)=(e^{\omega/T}+1)^{-1}$ and its derivative $f'(\omega)$ at temperature $T$ with the Boltzmann constant set to unity, $k_B=1$. The chemical potential is denoted by $\mu$. The expression is integrated over momentum in $d$ dimensions. In the second step in \eqref{eqn:condBoltzmann}, we performed a partial integration in momentum so that the conductivity involves the Fermi function and the second derivative of the dispersion. Thus for a quadratic dispersion $\eps_\bp=\bp^2/2m$ with mass $m$, we immediately get Drude's formula $\sigma^{xx}=\sigma^{yy}=e^2\tau n/m$ with the carrier density $n$. 

We derive the formula of the Hall conductivity $\sigma^{xyz}_\text{H}$ as the low-field limit of the semiclassical result in a uniform magnetic field \cite{Ashcroft1976}. We show in Appendix~\ref{appendix:lowfield} by an expansion up to linear order in the magnetic field that the formula for the Hall conductivity reads
\begin{align}
 \label{eqn:HallBoltzmann}
 \sigma^{xyz}_\text{H}&=-e^3\tau^2\int\hspace{-2mm}\frac{d^d\bp}{(2\pi)^2}\, f(\eps_\bp-\mu)\,\bigg[\frac{\partial^2 \eps_\bp}{\partial p^x\partial p^x}\,\frac{\partial^2\eps_\bp}{\partial p^y\partial p^y}-\Big(\frac{\partial^2\eps_\bp}{\partial p^x\partial p^y}\Big)^2\bigg]
 \\[2mm]
 &=\frac{e^3\tau^2}{2}\hspace{-2mm}\int\hspace{-2mm}\frac{d^d\bp}{(2\pi)^2}\, f'(\eps_\bp-\mu)\,\bigg[\Big(\frac{\partial \eps_\bp}{\partial p^x}\Big)^2\,\frac{\partial^2\eps_\bp}{\partial p^y\partial p^y}\hspace{-0mm}+\hspace{-0mm}\Big(\frac{\partial \eps_\bp}{\partial p^y}\Big)^2\,\frac{\partial^2\eps_\bp}{\partial p^x\partial p^x}-2\frac{\partial \eps_\bp}{\partial p^x}\,\frac{\partial \eps_\bp}{\partial p^y}\,\frac{\partial^2\eps_\bp}{\partial p^x\partial p^y}\bigg]
\end{align}
for $\omega_c\tau\ll 1$, where $\omega_c$ is the cyclotron frequency, which is proportional to the magnetic field. In order to obtain the formula in the second line, we once more performed a partial integration in momentum. For a quadratic dispersion, we obtain $\sigma^{xyz}_\text{H}=-e^3\tau^2 n/m^2$ and, thus, have the useful property of the Hall coefficient $R_\text{H}=\sigma^{xyz}_\text{H}/\sigma^{xx}\sigma^{yy}$ to depend only on the carrier density, $R_\text{H}=-1/en$. This suggests the definition of the Hall number $n_\text{H}$ as $R_\text{H}\equiv 1/e n_\text{H}$. The property of the Hall number to be equal to the carrier density, that is, $n_\text{H}=-n$ for electron-like contributions, is strictly valid only for a quadratic dispersion or in the high-field limit in the external magnetic field, that is, $\omega_c\tau\gg 1$ \cite{Ashcroft1976}. 

The two formulas in \eqref{eqn:condBoltzmann} and \eqref{eqn:HallBoltzmann} are valid for arbitrary dispersions $\eps_\bp$ and make them very useful in a broad range of applications, for instance, for lattice models. However, a natural question arises. What is the correct generalization if more than one band is present, that is, if we do not have only one dispersion $\eps_\bp$ but a multiband system with several dispersions $\eps^{(n)}_\bp$,
\begin{align}
 \eps_\bp \rightarrow \{\eps^{(1)}_\bp, \eps^{(2)}_\bp,... \} \, ?
\end{align}
It seems a reasonable, simple and useful approximation to assume that the conductivity of the multiband system is simply the sum of the single-band formulas with the bare dispersion replaced by the dispersion of the respective band,
\begin{align}
 \label{eqn:indQPP}
 \sigma^{\alf\beta}_\text{ind}= \sigma^{\alf\beta}\big[\eps_\bp\rightarrow \eps^{(1)}_\bp\big]+\sigma^{\alf\beta}\big[\eps_\bp\rightarrow \eps^{(2)}_\bp\big]+... =\sum_{n\in\text{bands}} \sigma^{\alf\beta}\big[\eps_\bp\rightarrow \eps^{(n)}_\bp\big]\, .
\end{align}
However, the precise condition on the range of validity of this {\it independent-band approximation} remains unclear without further considerations. Furthermore, we might miss phenomena that rely on the interplay between several bands. We will come back to both aspects in this thesis.

\begin{figure}[t!]
\centering
\includegraphics[width=0.49\textwidth]{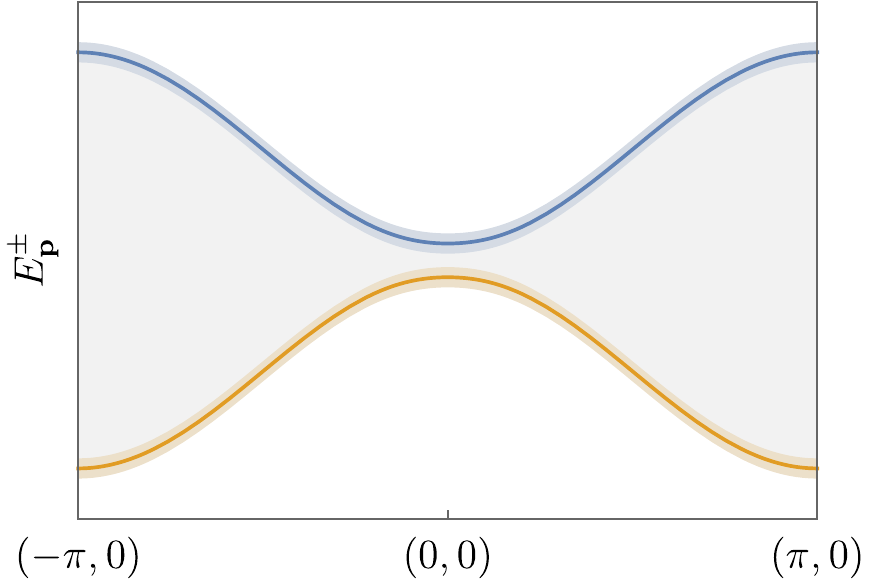}
\includegraphics[width=0.49\textwidth]{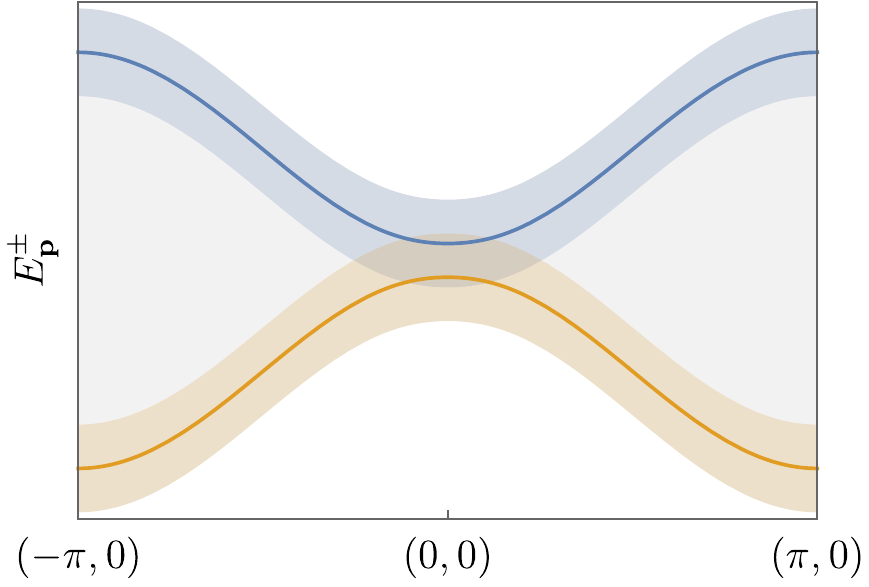}
\caption{An upper band $E^+_\bp$ (blue) and a lower band $E^-_\bp$ (orange) as a function of momentum $\bp=(p^x,0)$ over the Brillouin zone. We indicate the characteristic scale of the band broadening by a constant $\Gamma$ as colored area around the bands with $\Gamma$ of the left figure to be smaller than the $\Gamma$ of the right figure.  For sufficiently large $\Gamma$ (right figure) the two broadened bands overlap significantly. A certain type of interband contribution to the conductivities is present in these regions of overlap. Another type is present in the full area between the two bands (gray). The band structure corresponds to our example of a Chern insulator, which we will discuss in Sec.~\ref{sec:application:anomalousHallEffect:Wilson} in more detail.
\label{fig:Intro}}
\end{figure}

In order to get a better intuitive picture whether or not the approximation in \eqref{eqn:indQPP} might be correct, we consider a simple two-dimensional two-band model, which we will discuss in more detail as an example in Sec.~\ref{sec:application:anomalousHallEffect:Wilson}. We sketch the two bands $E^\pm_\bp$ as a function of momentum $\bp=(p^x,0)$ within the Brillouin zone in Fig.~\ref{fig:Intro}. The upper band (blue) and the lower band (orange) have no band crossings, but the two bands are close at zero momentum. They are far apart at the Brillouin zone boundary. Due to, for instance, impurities, electron-electron interaction or other phenomena, the bands (or, more precise, the spectral functions corresponding to the two bands) might be smeared out (or broadened) with a characteristic scale $\Gamma$, which is referred to as (momentum) relaxation rate with $\Gamma=1/2\tau$. Other synonyms for relaxation rate as scattering rate or decay rate are often used in the literature. The relaxation rate is, in general, frequency, momentum and band dependent. For simplicity, we assume a constant broadening. In the left and right panel of Fig.~\ref{fig:Intro}, we show the bands with $\Gamma$ of different size. We identify different regimes: For a broadening much smaller than the direct band gap (left panel), that is, $2\Gamma\ll E^+_\bp-E^-_\bp$ for all momenta, we have two bands that do not overlap.  In the limit $2\Gamma\gg E^+_\bp-E^-_\bp$ for some momenta, the two bands are no longer distinguishable over this momentum range (right panel). For very large $\Gamma$, the bands are indistinguishable for all momenta.

On the one hand, it is quite natural to expect the approximation in \eqref{eqn:indQPP} to hold if $\Gamma$ is chosen to be sufficiently small compared to the direct band gaps. On the other hand, we expect the interplay between the two bands to be important in regions of large overlap. As we will see throughout this thesis, this is indeed the case for one type of interband contributions to the electrical conductivity. Whether or not this type is relevant in a concrete model, depends on further details like the chemical potential or the size of the gap. We will discuss in the next section how the question of interband contributions and its relevance for the longitudinal conductivity and the Hall number arise in the context of recent Hall experiments on cuprates. 

There is another type of interband contribution that is not restricted to the regions of overlap of the two bands. This type of interband contribution is, for instance, responsible for the intrinsic anomalous Hall effect, that is, a transverse current in the absence of any applied magnetic field that is not caused by (skew) scattering. Under certain conditions, it is responsible for a quantized current known as the quantum anomalous Hall effect. A nonzero effect requires a chemical potential in between both bands (gray area in Fig.~\ref{fig:Intro}). We will give an introduction to this type of phenomena after the following section. There are further phenomena that may be crucial for transport like, for instance, weak and strong localization, especially, in low dimensions \cite{Akkermans2007}. These phenomena are, however, beyond the scope of this thesis.

%
%

\subsection*{Hall experiments on cuprates}

Understanding the ground state in the absence of superconductivity is the key to understanding the fluctuations that govern the anomalous behavior of cuprate superconductors above the critical temperature $T_c$, at which the superconductivity sets in \cite{Broun2008}. Superconductivity can be suppressed by applying a magnetic field, but very high fields are required for a complete elimination in those high-temperature superconductors. In the past years, magnetic fields up to almost 100 Tesla were achieved, such that the critical temperature of $\rm YBa_2Cu_3O_y$ (YBCO) and other cuprate compounds could be substantially suppressed even at optimal hole doping, at which the critical temperature is maximal. Charge transport measurements in such high magnetic fields indicate a drastic reduction of the charge-carrier density in a narrow doping range upon entering the pseudogap regime \cite{Badoux2016, Laliberte2016, Collignon2017, Proust2019}, whose origin is still debated. In particular, the Hall number $n_\text{H}$ drops from $1+p$ to $p$ in a relatively narrow range of hole doping $p$ below the critical doping $p^*$ at the edge of the pseudogap regime.

The observed drop in the charge carrier density below $p^*$ indicates a phase transition associated with a Fermi-surface reconstruction. The Hall number drop is qualitatively consistent with the formation of a N\'eel state \cite{Storey2016, Storey2017, Verret2017}, spiral magnetic order \cite{Eberlein2016, Chatterjee2017, Verret2017}, charge order \cite{Caprara2017, Sharma2018}, and nematic order \cite{Maharaj2017}. Alternatively, it may be explained by strongly fluctuating states without long-range order such as fluctuating antiferromagnets \cite{Qi2010, Chatterjee2016, Morice2017, Scheurer2018} and the Yang-Rice-Zhang (YRZ) state \cite{Yang2006a, Storey2016, Verret2017}, while it appears difficult to relate the experimental data to incommensurate collinear magnetic order \cite{Charlebois2017}. As long as no direct spectroscopic measurements are possible in high magnetic fields, it is hard to confirm or rule out any of these candidates experimentally.
Most recently, magnetic scenarios received considerable support 
from nuclear magnetic resonance (NMR) and ultrasound experiments in high magnetic fields by Frachet {\it et al.} \cite{Frachet2020}. They observed glassy antiferromagnetic order in $\rm La_{2-x} Sr_x Cu O_4$ (LSCO) at low temperatures up to the critical doping $p^*$ for pseudogap behavior. By contrast, in the superconducting state forming in the absence of a strong external magnetic field, magnetic order exists only in the low doping regime \cite{Keimer2015}. For strongly underdoped cuprates with respect to their optimal doping, where superconductivity is found to be absent or very weak, neutron scattering probes show that the N\'eel state is quickly destroyed upon doping, in agreement with theoretical findings \cite{Shraiman1989, Kotov2004, Sushkov2004, Sushkov2006, Luscher2007, Sushkov2009}. For underdoped YBCO, incommensurate antiferromagnetic order has been observed \cite{Yamada1998, Fujita2002,Haug2009, Haug2010}.

In this thesis, we will have a closer look on the theoretical proposal that the onset of planar spiral magnetic state may explain the observed drop in the Hall number \cite{Eberlein2016}. The spiral magnetic state is characterized by two parameters, the magnetic gap $\Delta$ and the ordering wave vector $\bQ$. It includes the two special cases of ferromagnetic order with $\bQ=(0,0)$ and N\'eel antiferromagnetic order $\bQ=(\pi,\pi)$ for a two-dimensional system. In Sec.~\ref{sec:application:spiral}, we will give a general introduction of spiral magnetic order, which involves its proper definition, several fundamental properties, and the appearance of the spiral magnetic state in the two-dimensional Hubbard model, which seems to well capture the competition between antiferromagnetism and superconductivity in cuprates \cite{Scalapino2012}. In a spiral magnet, the electron band is split into only two quasiparticle bands in spite of an arbitrary ordering wave vector, which might be incommensurate with the underlying lattice, so that the order is never repeated over the full lattice. In this respect, the spiral state is as simple as the N\'eel state despite the broken translation invariance. By contrast, all other magnetically ordered states entail a fractionalization in many subbands, actually infinitely many in the case of incommensurate order. Hence, only the spiral magnet forms a metal with a simple Fermi surface topology for arbitrary wave vectors with only a small number of electron and hole pockets. For a sufficiently large magnetic gap $\Delta$ there are only hole pockets in the hole-doped system. The spectral weight for single-electron excitations is strongly anisotropic, so that the spectral function resembles Fermi arcs \cite{Eberlein2016}, which are a characteristic feature of the pseudogap phase in high-$T_c$ cuprates \cite{Damascelli2003}.

\begin{figure}[b!]
\centering
\includegraphics[width=0.9\textwidth]{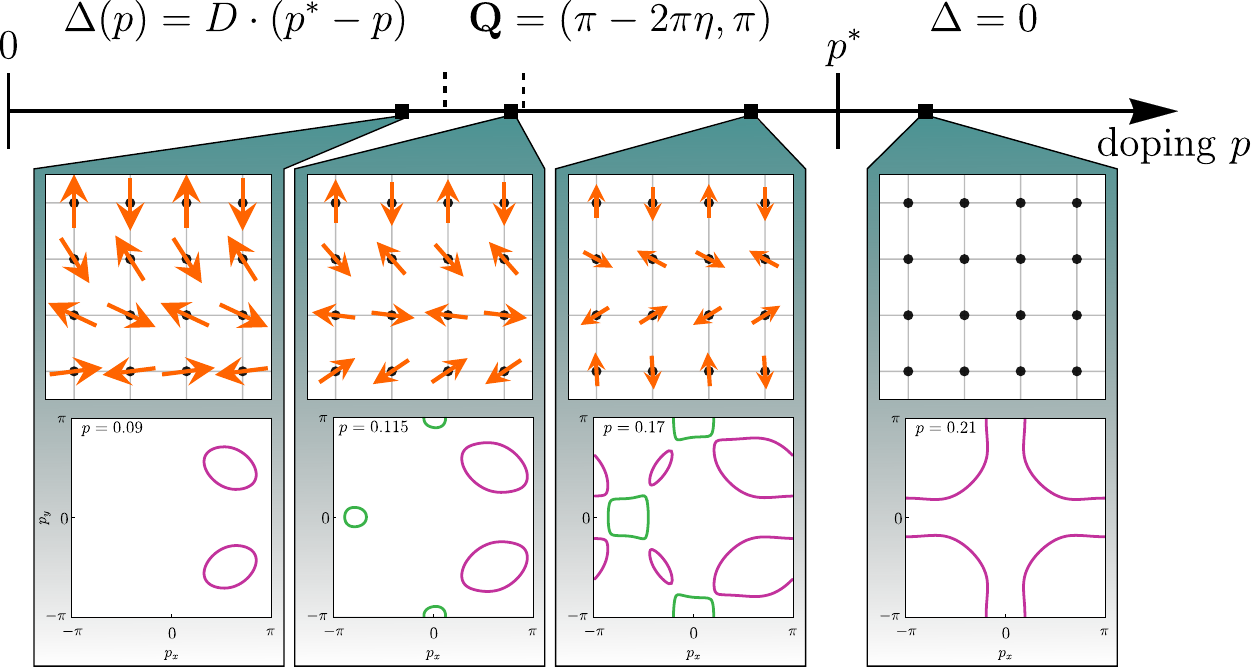}
\caption{Assuming a phenomenological form of the spiral magnetic gap $\Delta$ and the ordering wave vector $\bQ$ as a function of hole doping $p$ results in a Fermi surface reconstruction from a large hole-like Fermi surface (purple) above $p^*$ to small hole pockets far below $p^*$. We have additional electron pockets (green) in the intermediate regime. The observed drop of the Hall number from $1+p$ to $p$ \cite{Badoux2016, Laliberte2016, Collignon2017} is reproduced by this model \cite{Eberlein2016}. We plot the corresponding local magnetic moments $\langle \bS_i\rangle$ of the spiral magnetic order on a square lattice.
\label{fig:modelSpiral}}
\end{figure}

The electromagnetic response of spiral magnetic states has already been analyzed by Voruganti {\it et al.} \cite{Voruganti1992}. They derived formulas for the zero-frequency (DC) longitudinal and Hall conductivity in the low-field limit $\omega_c\tau\ll 1$ up to linear order in the magnetic field. The spiral states were treated in a mean-field approximation. The resulting expressions have the same form as in Eqns.~\eqref{eqn:condBoltzmann} and \eqref{eqn:HallBoltzmann}, with the bare dispersion relation replaced by the two quasiparticle bands. Assuming a simple phenomenological form of the spiral order parameter as a function of doping, Eberlein {\it et al.} \cite{Eberlein2016} showed that the Hall conductivity computed with the formula derived by Voruganti {\it et al.} indeed exhibits a drop of the Hall number consistent with the recent experiments \cite{Badoux2016, Laliberte2016, Collignon2017} on cuprates in high magnetic fields. In Fig.~\ref{fig:modelSpiral}, we sketch the evolution of the Fermi surface reconstruction and the corresponding local magnetic moments of the spiral magnetic order. Recently, an expression for the thermal conductivity in a spiral state has been derived along the same lines, and a similar drop in the carrier density has been found \cite{Chatterjee2017}.

The expressions for the electrical and Hall conductivities derived by Voruganti {\it et al.} \cite{Voruganti1992} have been obtained for small relaxation rates $\Gamma=1/2\tau$. However, the relaxation rate in the cuprate samples studied experimentally is sizable. For example, in spite of the high magnetic fields, the product of cyclotron frequency and relaxation time $\omega_c \tau$ extracted from the experiments on $\rm La_{1.6-x} Nd_{0.4} Sr_x Cu O_4$ (Nd-LSCO) samples is as low as $0.075$ \cite{Collignon2017}, which is clearly in the low-field limit $\omega_c\tau\ll 1$. Since $\omega_c\propto B$, the relaxation time $\tau$ has to be reduced. Moreover, from the derivation of Voruganti {\it et al.}, the precise criterion for a ``small'' relaxation rate is not clear. The direct band gap $E^+_\bp-E^-_\bp$ is of the order of the magnetic gap $\Delta$ for several momenta. Thus, our general argument, which we discussed previously and sketched in Fig.~\ref{fig:Intro}, suggests that interband contributions might be crucial at the onset of order at $p^*$, where $\Gamma\sim\Delta$, and might have an impact on the previous conclusions.  Therefore, it is worth to reconsider the transport properties of spiral magnetic states for arbitrarily large relaxation rate $\Gamma$ in order to clarify the discrepancy in the arguments and the impact on the interpretation of the experimental results using the model of spiral magnetic order. 

%
%

\subsection*{Anomalous Hall effect}

The existence of a transverse current perpendicular to an applied electric field in the absence of any external magnetic field is known as anomalous Hall effect. In contrast, the (ordinary) Hall effect, which we have just considered in the context of cuprates, relies on the presence of an external magnetic field. It is important to distinguish the extrinsic and the intrinsic anomalous Hall effects \cite{Nagaosa2010, Xiao2010}, which differ by their physical origin. The extrinsic mechanisms are based on the scattering off impurities, which can be further specified as skew-scattering \cite{Smit1955, Smit1958} and side-jump \cite{Berger1970} mechanisms. The intrinsic mechanism was first introduced by an additional contribution to the group velocity known as anomalous velocity in a semiclassical theory \cite{Karplus1954}. This anomalous velocity is nowadays understood in terms of the Berry curvature of the underlying band structure and the corresponding (Bloch) eigenstates \cite{Berry1984, Zak1989} and is, as such, intrinsic. In recent years, there is an increasing interest in transport properties of systems with topological properties in material science \cite{Culcer2020, Xu2020, Sun2020}, including Heusler compounds \cite{Kubler2014, Manna2018, Noky2020}, Weyl semimetals \cite{Destraz2020, Li2020, Nagaosa2020}, and graphene \cite{Sharpe2019, McIver2020}, and in other physical systems like in microcavities \cite{Gianfrate2020} and cold atoms \cite{Cooper2019}. The established connection between the intrinsic anomalous Hall conductivity and the Berry curvature \cite{Thouless1982, Niu1985, Kohmoto1985, Onoda2002, Jungwirth2002} combined with {\it ab initio} band structure calculations \cite{Fang2003,Yao2004} has become a powerful tool for combining theoretical and experimental results and is state-of-the-art in recent studies \cite{Culcer2020, Xu2020, Sun2020, Kubler2014, Manna2018, Noky2020, Destraz2020, Li2020, Nagaosa2020, Sharpe2019, McIver2020, Gianfrate2020, Cooper2019}. 

The formula in \eqref{eqn:condBoltzmann} derived in the standard semiclassical transport theory within a relaxation time approximation does neither capture the intrinsic nor the extrinsic anomalous Hall effect. A modern semiclassical description of the anomalous Hall effect can be found in the review of Sinitsyn \cite{Sinitsyn2008}. Microscopic approaches beyond semiclassical assumptions provide a systematic framework but are usually physically less transparent \cite{Nagaosa2010}. The combination of both approaches, that is, a systematic microscopic derivation combined with a Boltzmann-like physical interpretation, in order to find further relevant mechanisms, is still subject of recent research \cite{Sinitsyn2007}. 

In this thesis, we will consider a very general two-band model, which may be topological and may have a nonzero Berry curvature. Our microscopic approach to calculate the conductivity of this model should capture the intrinsic anomalous Hall effect if we are taking interband contributions into account. Indeed, we will see that the anomalous Hall conductivity arises as one type of interband contribution within our calculation.  We do not assume any restriction on the size of the relaxation rate $\Gamma$, which will allow us to perform a detailed analysis of the limiting behavior of the conductivities with respect to $\Gamma$. We find that the intrinsic anomalous Hall effect is relaxation-rate independent for sufficiently small $\Gamma$ consistent with previous results \cite{Nagaosa2010}. The anomalous Hall conductivity can be quantized \cite{Thouless1982, Niu1985, Kohmoto1985}. Within a simple model of a Chern insulator, we will discuss the modification of the quantized anomalous Hall effect due to a finite relaxation rate $\Gamma$. We will consider a ferromagnetic multi-d-orbital model by Kontani {\it et al.} \cite{Kontani2007} to discuss the scaling behavior of the anomalous Hall conductivity as a function of the relaxation rate $\Gamma$ as well as the longitudinal conductivity. We will see that the result is both qualitatively and quantitatively in good agreement with experimental results for ferromagnets (see Ref.~\cite{Onoda2008} and references therein). 

The microscopic derivation of conductivities in multiband systems is a longstanding question. We have already mentioned above that the longitudinal and the Hall conductivity of a system with spiral magnetic order were discussed by Voruganti {\it et al.} \cite{Voruganti1992}, but both a complete formula including the interband contributions and a detailed discussions of their importance were still missing. Common microscopic approaches to the anomalous Hall conductivity, which already include interband contributions in order to capture the intrinsic anomalous Hall effect, are based on the work of Bastin {\it et al.} and St\v{r}eda \cite{Bastin1971, Streda1982, Crepieux2001}. Starting from Kubo's linear response theory \cite{Mahan2000} in a Matsubara Green's function formalism, Bastin {\it et al.} \cite{Bastin1971} expanded in the frequency $\omega$ of the external electric field $\bE(\omega)$ after analytic continuation and obtained the DC conductivity $\sigma^{\alf\beta}$, where $\alf,\beta=x,y,z$ are the directions of the induced current and the electric field, respectively. St\v{r}eda \cite{Streda1982} further decomposed this result into so-called Fermi-surface and Fermi-sea contributions $\sigma^{\alf\beta,I}$ and $\sigma^{\alf\beta,II}$ that are defined by involving the derivative of the Fermi function or the Fermi function itself, respectively. This or similar decompositions are common starting points for further theoretical investigations \cite{Nagaosa2010, Sinitsyn2007, Crepieux2001, Dugaev2005, Onoda2006, Yang2006, Kontani2007, Nunner2008, Onoda2008, Tanaka2008, Kovalev2009, Streda2010, Pandey2012, Burkov2014, Chadova2015, Kodderitzsch2015, Mizoguchi2016}. However, the decomposition by St\v{r}eda and similar ones are usually not unique, which can be directly seen by a simple partial integration in the internal frequency, and are {\it a priori} not motivated by stringent mathematical or physical reasons.

%
%

\subsection*{Criteria for a useful decomposition}

Throughout this thesis, we will motivate, analyze and apply two criteria for a unique decomposition of the conductivity formulas with, furthermore, clear physical meaning. Using these criteria will not only provide a deeper physical insight due to unique properties of the individual contributions but will also help to reduce the technical complexity of the derivation. Thus, it may also pave the way for future studies of multiband systems beyond the scope of this thesis. For a first idea, we give a short preliminary definition of the criteria in the following. We will discuss them in detail in Chapter~\ref{sec:theory}.

The Hamiltonian that we will assume in this thesis is block-diagonal in momentum. The matrix form of the Hamiltonian, which is a consequence of the two-band structure, is captured by a momentum-dependent matrix $\lam_\bp$, which is called the Bloch Hamiltonian. We will show in detail that a key ingredient of the conductivity is this matrix $\lam_\bp$ since both the Green's functions and the electromagnetic vertices involve $\lam_\bp$ in a particular form, for instance, its derivative. The final conductivity formulas are eventually obtained by tracing a particular combination of Green's function and vertex matrices. Changing to the eigenbasis of $\lam_\bp$ at fixed momentum will separate the momentum derivative of $\lam_\bp$ into a diagonal quasivelocity matrix, which consists of the momentum derivative of the eigenvalues, and an off-diagonal matrix, which involves the derivative of the eigenbasis in a particular form, that is, the Berry connection. The former one leads to the {\it intraband contribution} $\sigma^{\alf\beta}_\text{intra}$ that involves only quasiparticle spectral functions of one band in each term. The latter one mixes the quasiparticle spectral functions of both bands and leads to the {\it interband contribution} $\sigma^{\alf\beta}_\text{inter}$. It is this interband contribution that captures the interplay between the two bands and, thus, the phenomena that are missing in the semiclassical approach. 

Besides the separation into intra- and interband contributions, we will identify a second criterion. The conductivity depends on the direction of the current and the external electric field. We can uniquely decompose the conductivity into a {\it symmetric}, $\sigma^{\alf\beta,s}=\sigma^{\beta\alf,s}$, and an {\it antisymmetric}, $\sigma^{\alf\beta,a}=-\sigma^{\beta\alf,a}$, contribution. We will see that the intraband contribution is already symmetric, but the interband contribution consists of both a symmetric and an antisymmetric part. Thus, we will obtain a decomposed conductivity tensor of the form
\begin{align}
 \label{eqn:introDecomp}
 \sigma^{\alf\beta}=\sigma^{\alf\beta}_\text{intra}+\sigma^{\alf\beta,s}_\text{inter}+\sigma^{\alf\beta,a}_\text{inter} \, .
\end{align}
The result of Boltzmann transport theory in \eqref{eqn:condBoltzmann} will arise from the intraband contribution in the limit of a small relaxation rate $\Gamma$ and is shown to be precisely the result of independent quasiparticle bands, which we motivated in \eqref{eqn:indQPP}. We will show that the symmetric interband contribution is a correction only present for a finite relaxation rate $\Gamma$ and is controlled by the quantum metric. This is the first type of interband contribution, which we discussed above. We will show that its mayor contributions are from regions of band overlap sketched in Fig.~\ref{fig:Intro}. The antisymmetric interband contribution involves the Berry curvature and generalizes previous formulas for the anomalous Hall conductivity \cite{Thouless1982, Niu1985, Kohmoto1985, Onoda2002, Jungwirth2002} to a finite relaxation rate $\Gamma$. This interband contribution can essentially be only nonzero for a chemical potential in between both bands, which we indicated as a gray area in Fig.~\ref{fig:Intro} and referred to as the second type of interband contribution. Similar to the decomposition of the conductivity tensor $\sigma^{\alf\beta}$ in \eqref{eqn:introDecomp}, we decompose the Hall conductivity $\sigma^{\alf\beta\eta}_\text{H}$ in order to generalize the semiclassical result in \eqref{eqn:HallBoltzmann}. 
We will motivate, discuss and apply all these aspects in more detail as part of the derivation in the following Chapter~\ref{sec:theory}.

\newpage


\rhead[\fancyplain{}
{\bfseries Theory of electrical conductivity}
]{\fancyplain{}{
}}
\lhead[\fancyplain{}{
}]{\fancyplain{}
{\bfseries Theory of electrical conductivity}
}

\chapter{Theory of electrical conductivity}
\label{sec:theory} %

In this chapter, we derive and discuss the electrical conductivity of a general momentum-block-diagonal two-band model including interband contributions within a microscopic approach. We start by specifying our model under consideration (Sec.~\ref{sec:theory:twobandsystem}). We describe the coupling to the electromagnetic fields, calculate the related currents and identify the conductivity tensors, which act as the starting point for the subsequent derivations. Several fundamental concepts that will be useful in the derivation in order to obtain the conductivity formulas in a transparent and simple form are introduced (Sec.~\ref{sec:theory:fundamentalconcepts}). We derive and discuss the longitudinal and the anomalous Hall conductivity of the general two-band model (Sec.~\ref{sec:theory:conductivity}). The goal of this derivation will be a unique decomposition for disentangling conductivity contributions of different physical origin and unique properties. We close this chapter by deriving and discussing the Hall conductivity for the general two-band model with momentum-independent coupling between the two subsystems of the two-band model (Sec.~\ref{sec:theory:Hall}).

%
%

\section{General two-band model}
\label{sec:theory:twobandsystem}

The model that we consider in this thesis is a general momentum-block-diagonal two-band model, a minimal model for a multiband system. It captures a broad spectrum of physically very different systems including models with N\'eel antiferromagnetic order and spiral spin density waves as well as models that involve spin-orbit interaction and are known to show topological properties. In this section, we define and introduce the model (Sec.~\ref{sec:theory:twobandsystem:model}) and present the coupling to an external electric and magnetic field (Sec.~\ref{sec:theory:twobandsystem:coupling}). We present the induced currents and give the general expressions of the conductivities (Sec.~\ref{sec:theory:twobandsystem:currentandconductivity}), which will be the starting point for the subsequent sections. More details of the derivations can be found in the Appendices~\ref{appendix:peierls},~\ref{appendix:current} and \ref{appendix:Mass}. 

%
%

\subsection{Definition of the model}
\label{sec:theory:twobandsystem:model}

We assume the quadratic momentum-block-diagonal tight-binding Hamiltonian
\begin{align}
 \label{eqn:H}
 H=\sum_\bp \Psi^\dagger_\bp \lam^{}_\bp \Psi^{}_\bp \, ,
\end{align}
where $\lam_\bp$ is a hermitian $2\times2$ matrix, $\Psi_\bp$ is a fermionic spinor and $\Psi^\dag_\bp$ is its hermitian conjugate. Without loss of generality, we parameterize $\lam_\bp$ as
\begin{align}
 \label{eqn:lam}
  \lam_\bp=\begin{pmatrix} \eps_{\bp,A} && \Delta_\bp \\[3mm] \Delta^*_\bp && \eps_{\bp,B}\end{pmatrix} \,,
\end{align}
where $\eps_{\bp,\ic}$ are two arbitrary (real) bands of the subsystems $\ic=A,B$. The complex function $\Delta_\bp$ describes the coupling between $A$ and $B$. The spinor $\Psi_\bp$ consists of the annihilation operator $c_{\bp,\ic}$ of the subsystems, 
\begin{align}
\label{eqn:spinor}
 \Psi^{}_\bp=\begin{pmatrix} \c_{\bp+\bQ_A,A} \\[1mm] \c_{\bp+\bQ_B,B} \end{pmatrix} \, ,
\end{align}
where $\bQ_\ic$ are arbitrary but fixed offsets of the momentum. The subsystems $A$ and $B$ can be further specified by a set of spatial and/or non-spatial quantum numbers like spin or two atoms in one unit cell. We label the positions of the unit cells via the Bravais lattice vector $\bR_i$. If needed, we denote the spatial position of subsystem $\ic$ within a unit cell as $\brho_\ic$. The Fourier transformation of the annihilation operator from momentum to real space and vice versa are given by
\begin{align}
\label{eqn:FourierC}
 &\c_{j,\ic}=\frac{1}{\sqrt{L}}\sum_\bp\,\c_{\bp,\ic}\,e^{i\bp\cdot(\bR_j+\brho_\ic)} \, , \\
 \label{eqn:FourierCInv}
 &\c_{\bp,\ic}=\frac{1}{\sqrt{L}}\sum_j\,\c_{j,\ic}\,e^{-i\bp\cdot(\bR_j+\brho_\ic)} \, ,
\end{align}
where $L$ is the number of unit cells. By choosing a unit of length so that a single unit cell has volume 1, $L$ is the volume of the system. Note that we included the precise position $\bR_i+\brho_\ic$ of the subsystem $\ic$ in the Fourier transformation \cite{Tomczak2009, Nourafkan2018}. 

The considered momentum-block-diagonal Hamiltonian \eqref{eqn:H} is not necessarily (lattice) translational invariant due to the $\bQ_\ic$ in \eqref{eqn:spinor}. The translational invariance is present only for $\bQ_A=\bQ_B$, that is, if there is no relative momentum difference between the spinor components. In the case $\bQ_A\neq\bQ_B$, the Hamiltonian is invariant under combined translation and rotation in spinor space. This difference can be explicitly seen in the real space hoppings presented in Appendix~\ref{appendix:peierls}. Using the corresponding symmetry operator one can map a spatially motivated model to \eqref{eqn:H} and, thus, obtain a physical interpretation of the parameters \cite{Sandratskii1998}.

%
%

\subsection{Coupling to electric and magnetic fields}
\label{sec:theory:twobandsystem:coupling}

We couple the Hamiltonian \eqref{eqn:H} to external electric and magnetic fields $\bE(\br,t)$ and $\bB(\br,t)$ via the Peierls substitution, that is, a phase factor gained by spatial motion, and neglect further couplings. The derivation in this and the following subsection generalizes the derivation performed in the context of spiral spin density waves \cite{Voruganti1992}. We Fourier transform the Hamiltonian \eqref{eqn:H} via \eqref{eqn:FourierC} defining
\begin{align}
 \label{eqn:FourierH}
 \sum_\bp\Psi^\dagger_\bp\lam^{}_\bp\Psi^{}_\bp=\sum_{j,\,j'}\Psi^\dagger_j\lam^{}_{jj'}\Psi^{}_{j'} \, ,
\end{align}
where the indices $j$ and $j'$ indicate the unit cell coordinates $\bR_j$ and $\bR_{j'}$, respectively. We modify the entries of the real space hopping matrix $\lam_{jj'}=(t_{jj',\,\ic\ic'})$ by
\begin{align}
\label{eqn:Peierls}
 t^{}_{jj',\,\ic\ic'}\rightarrow t^{}_{jj',\,\ic\ic'}\,e^{-ie\int^{\bR_j+\brho_\ic}_{\bR_{j'}+\brho_{\ic'}}\bA(\br,t)\cdot d\br} \, .
\end{align}
$\bA(\br,t)$ is the vector potential. We have set the speed of light $c=1$ and the reduced Planck's constant $\hbar=1$ to unity. We have chosen the coupling charge to be the electron charge $q=-e$ with our convention $e>0$. Note that we have included hopping inside the unit cell by using the precise position $\bR_j+\brho_\ic$ of the subsystems $\ic$ \cite{Tomczak2009, Nourafkan2018}. Neglecting $\brho_\ic$ would lead to unphysical results \cite{Tomczak2009, Nourafkan2018, Mitscherling2020}. The coupling \eqref{eqn:Peierls} does not include local processes induced by the vector potential, for instance, via $\cdag_{j,A}c^{\phantom{\dag}}_{j,B}$ if the two subsystems are at the same position within the unit cell, $\brho_A=\brho_B$. Using the (incomplete) Weyl gauge such that the scalar potential is chosen to vanish, the electric and magnetic fields are entirely described by the vector potential via $\bE(\br,t)=-\partial_t \bA(\br,t)$ and $\bB(\br,t)=\nabla\times\bA(\br,t)$, where $\partial_t=\frac{\partial}{\partial t}$ is the time derivative and $\nabla=(\frac{\partial}{\partial x},\frac{\partial}{\partial y},\frac{\partial}{\partial z})$ is the spacial derivative.

We are interested in the long-wavelength regime of the external fields and, in particular, in the zero-frequency (DC) conductivity. Assuming that the vector potential $\bA(\br,t)$ varies only slowly over the hopping ranges defined by nonzero $t_{jj',\,\ic\ic'}$ allows us to approximate the integral inside the exponential in \eqref{eqn:Peierls}. As shown in Appendix~\ref{appendix:peierls}, we get
\begin{align}
\label{eqn:HA}
 H[\bA]=\sum_\bp\Psi^\dagger_\bp \lam^{}_\bp \Psi^{}_\bp+\sum_{\bp,\bp'}\Psi^\dagger_{\bp}\sV^{}_{\bp\bp'}\Psi^{}_{\bp'} \, .
\end{align}
The first term is the unperturbed Hamiltonian \eqref{eqn:H}. The second term involves the electromagnetic vertex $\sV_{\bp\bp'}$ that captures the effect of the vector potential and vanishes for zero potential, that is, $\sV_{\bp\bp'}[\bA=0]=0$. The Hamiltonian is no longer block-diagonal in momentum $\bp$ due to the spatial modulation of the vector potential. The vertex is given by
\begin{align}
\label{eqn:Vpp'}
 \sV_{\bp\bp'}=\sum^\infty_{n=1} \frac{e^n}{n!}&\sum_{\substack{\bq_1,...,\bq_n \\ \alf_1,...,\alf_n}}
 \lam^{\alf_1...\alf_n}_{\frac{\bp+\bp'}{2}}\,A^{\alf_1}_{\bq_1}(t)\,...\,A^{\alf_n}_{\bq_n}(t)\,\delta_{\sum_m \bq_m,\bp-\bp'} \,.
\end{align}
The n-th order of the vertex is proportional to the product of $n$ modes $\bA_\bq(t)$ of the vector potential given by 
\begin{align}
\label{eqn:FourierAq}
 \bA(\br,t)=\sum_\bq \bA_\bq(t)e^{i\bq\cdot\br} \, .
\end{align}
$A^\alf_\bq$ denotes the $\alf=x,y,z$ component of the mode. The Dirac delta function assures momentum conservation. Each order of the vertex is weighted by the n-th derivative of the bare Bloch Hamiltonian in \eqref{eqn:lam}, that is,
\begin{align}
 \label{eqn:DlamDef}
 \lam^{\alf_1...\alf_n}_\bp=\partial_{\alf_1}...\partial_{\alf_n}\lam^{}_\bp \, ,
\end{align}
at momentum $(\bp+\bp')/2$, where $\partial_\alf=\frac{\partial}{\partial p^\alf}$ is the derivative of momentum $\bp=(p^x,p^y,p^z)$ in $\alf$ direction. As can be seen in the derivation in Appendix~\ref{appendix:peierls}, both the use of the precise position of the subsystems in the Fourier transformation \cite{Tomczak2009, Nourafkan2018} as well as the momentum-block-diagonal Hamiltonian are crucial for this result. Note that $\lam^{\alf_1...\alf_n}_\bp$ is symmetric in the indices $\alf_1$ to $\alf_n$.

%
%

\subsection{Current and conductivity}
\label{sec:theory:twobandsystem:currentandconductivity}

We derive the current of Hamiltonian \eqref{eqn:HA} induced by the vector potential within an imaginary-time path-integral formalism in order to assure consistency and establish notation. The matrices, which are present due to the two subsystems, do not commute in general and, thus, the order of the Green's function and the vertex matrices are crucial. We sketch the steps in the following. Details of the derivation are given in Appendix~\ref{appendix:current}. We have set $k_B=1$ and $\hbar=1$.

We rotate the vector potential modes $\bA_\bq(t)$ in the vertex \eqref{eqn:Vpp'} to imaginary time $\tau=i t$ and Fourier transform $\bA_\bq(\tau)$ via
\begin{align}
 \bA_\bq(\tau)=\sum_{q_0}\bA_q\,e^{-iq_0\tau} \, ,
\end{align}
where $q_0=2n\pi T$ are bosonic Matsubara frequencies for $n\in \mathds{Z}$ and temperature $T$. We combine these frequencies $q_0$ and the momenta $\bq$ in four-vectors for shorter notation, $q=(iq_0,\bq)$. The real frequency result will be recovered by analytic continuation, $iq_0\rightarrow \omega+i0^+$, at the end of the calculations. After the steps above, the electromagnetic vertex $\sV_{pp'}$ involving Matsubara frequencies is of the same form as \eqref{eqn:Vpp'} with the momenta replaced by the four-vector $p$ and $p'$. The Dirac delta function assures both frequency and momentum conservation. The (euclidean) action of \eqref{eqn:HA} reads
\begin{align}
\label{eqn:S}
 S[\Psi,\Psi^*]=-\sum_p \Psi^*_p \sG^{-1}_p \Psi^{}_p+\sum_{p,p'} \Psi^*_p \sV^{}_{pp'} \Psi^{}_{p'} \,
\end{align}
where $\Psi_p$ and $\Psi_p^*$ are (complex) Grassmann fields. The inverse (bare) Green's function is given by
\begin{align}
\label{eqn:Green}
 \sG^{-1}_p=ip_0+\mu-\lam_\bp + i\Gamma\,\text{sign}(p_0) \, .
\end{align}
We include the chemical potential $\mu$. $p_0=(2n+1)\pi T$ are fermionic Matsubara frequencies for $n\in \mathds{Z}$ and temperature $T$. We assume the simplest possible momentum-relaxation rate $\Gamma>0$ as a constant imaginary part proportional to the identity matrix. $\text{sign}(p_0)=\pm 1$ is the sign function for positive and negative $p_0$, respectively. 

The assumed phenomenological relaxation rate $\Gamma$ is momentum  and frequency independent as well as diagonal and equal for both subsystems $\ic=A,B$. Such approximations on $\Gamma$ are common in the literature of multiband systems \cite{Verret2017, Eberlein2016, Mizoguchi2016, Tanaka2008, Kontani2007}. A microscopically derived relaxation rate $\Gamma$, for instance, due to interaction or impurity scattering depends on details of the models, which we do not further specify in our general two-band system. A microscopic derivation can, for instance, be performed within a Born approximation \cite{Rickayzen1980}, which then can be used to concretize the range of validity. We are aware that each generalization of $\Gamma$ may effect parts of the following derivations and conclusions. We do not assume that $\Gamma$ is small and derive the current for $\Gamma$ of arbitrary size. 

The current density $j^\alf_q$ in $\alf=x,y,z$ direction that is induced by the external electric and magnetic fields is given by the functional derivative of the grand canonical potential $\Omega[\bA]$ with respect to the vector potential \cite{Tremblay2020},
\begin{align}
 j^\alf_q=-\frac{1}{L}\frac{\delta \Omega[\bA]}{\delta A^\alf_{-q}} \,.
\end{align}
The explicit form of $\Omega[\bA]$ is given in Appendix~\ref{appendix:current:grandcanonicalpotential}. We are interested in the current up to second order in the vector potential. We define 
\begin{align}
 \label{eqn:jexpDef}
 j^\alf_q=j^\alf_0-\sum_\beta\Pi^{\alf\beta}_qA^\beta_q-\sum_{\beta\gamma}\sum_{q'}\Pi^{\alf\beta\gamma}_{q,q'}A^\beta_{q'}A^\gamma_{q-q'}+\mathcal{O}(A^3) \, .
\end{align}
In Appendix~\ref{appendix:current:expansion}, we present the derivation of the current and the expansion up to second order, so that we can identify the zeroth-, first- and second-order contributions $j^\alf_0$, $\Pi^{\alf\beta}_q$ and $\Pi^{\alf\beta\gamma}_{q,q'}$ explicitly. An explicit formula of the current $j^\alf_0$, that is, a current without any external fields, is given in Appendix~\ref{appendix:current:paramagneticcurrent}.  It vanishes for all cases that we will discuss. Thus, we will omit it in the following.  The expansion of the current in \eqref{eqn:jexpDef} describes a broad variety of different physical transport phenomena. We specify two different cases in the following. 

The current $j^\alf_\text{\tiny E}(\omega)$ induced by only a uniform electric field $E^\beta(\omega)$ and no magnetic field can be described by a conductivity tensor $\sigma^{\alf\beta}_\text{\tiny E}(\omega)$. The current $j^\alf_\text{\tiny EB}(\omega)$ induced by a uniform electric field $E^\beta(\omega)$ and a static magnetic field $B^\eta$ can be described by a conductivity tensor $\sigma^{\alf\beta\eta}_\text{\tiny EB}(\omega)$. The indices $\alf,\,\beta,\,\eta=x,y,z$ indicate the component of the respective spatial direction. We have 
\begin{alignat}{2}
 &\label{eqn:jE}
 j^\alf_\text{\tiny E}(\omega)&&=\sum_\beta\sigma^{\alf\beta}_\text{\tiny E}(\omega)E^{\beta}(\omega) \, ,\\
 &\label{eqn:jEB}
 j^\alf_\text{\tiny EB}(\omega)&&=\sum_{\beta,\eta}\sigma^{\alf\beta\eta}_\text{\tiny EB}(\omega)E^\beta(\omega)B^\eta\, .
\end{alignat}
We perform the calculations at finite frequency, but we will mainly focus on the zero frequency (DC) limit. Note that the order of momentum and frequency limit is crucial in order to obtain the correct DC conductivities. First, we perform the limit $\bq\rightarrow 0$ for uniform external fields, and secondly, we perform the limit $\omega\rightarrow 0$ for static external fields. The reversed order would describe properties of the system at equilibrium instead of transport properties \cite{Tremblay2020}. For a transparent notation, we will label the component index of the current as $\alf$, the component index of the electric field as $\beta$ and the component index of the magnetic field as $\eta$. 

We identify the relation between the conductivity tensors in \eqref{eqn:jE} and \eqref{eqn:jEB} and the corresponding components of the polarization tensors $\Pi^{\alf\beta}$ and $\Pi^{\alf\beta\gamma}_{q,q'}$ in \eqref{eqn:jexpDef} by using the identities $E^\beta(\omega,\bq)=i\omega A^\beta(\omega,\bq)$ and $B^\eta(\omega,\bq)=\big(i\bq\times \bA(\omega,\bq)\big)^\eta=i\sum_{\delta \gamma} q^\delta \eps^{\eta\delta\gamma} A^\gamma(\omega,\bq)$, where $\eps^{\eta\delta\gamma}$ is the Levi-Civita symbol. We get 
\begin{alignat}{4}
 \label{eqn:sE}\sigma^{\alf\beta}_\text{\tiny E}(\omega)=&& -\frac{1}{i\omega}\Pi^{\alf\beta}_\text{\tiny E}(\omega)\equiv&& -\frac{1}{i\omega}\left.\Pi^{\alf\beta}_{\text{\tiny E},iq_0}\right|_{iq_0\,\rightarrow\,\omega+i0^+}&&&\, ,
 \\[2mm]
 \label{eqn:sH}\sum_\eta \eps^{\eta\gamma\delta}\sigma^{\alf\beta\eta}_\text{\tiny EB}(\omega)=&& -\frac{1}{\omega}\Pi^{\alf\beta\gamma\delta}_\text{\tiny EB}(\omega)\equiv&& -\frac{1}{\omega}\left.\Pi^{\alf\beta\gamma\delta}_{\text{\tiny EB},iq_0}\right|_{iq_0\,\rightarrow\, \omega+i0^+}&&& \, ,
\end{alignat}
where $iq_0\rightarrow \omega+i0^+$ indicates the analytic continuation from Matsubara to real frequencies. It is intuitive to specify several components. For instance, we have $\sigma^{xyz}_\text{\tiny EB}(\omega)=-\Pi^{xyxy}_\text{\tiny EB}(\omega)/\omega$ and $\sigma^{xxz}_\text{\tiny EB}(\omega)=-\Pi^{xxxy}_\text{\tiny EB}(\omega)/\omega$. 

We present the derivation of $\Pi^{\alf\beta}_{\text{\tiny E},iq_0}$ in Appendix~\ref{appendix:current:piE}. We obtain 
\begin{align}
 \label{eqn:PiE}\Pi^{\alf\beta}_{\text{\tiny E},iq_0}=e^2\frac{T}{L}\sum_{ip_0,\bp}\tr\big[\sG^{}_{ip_0+iq_0,\bp}\,\lam^\beta_\bp\,\sG^{}_{ip_0,\bp}\,\lam^\alf_\bp-(iq_0=0)\big]\,.
\end{align}
The polarization tensor involves a product of Green's function and (first-order) vertex matrices defined in \eqref{eqn:Green} and \eqref{eqn:DlamDef}, which may not commute. It vanishes at zero external Matsubara frequency since the $iq_0=0$ contribution of the first term is subtracted. The final result is obtained by a trace over the matrix, the Matsubara frequency and the momentum summation. We present the derivation of $\Pi^{\alf\beta\gamma\delta}_{\text{\tiny EB},iq_0}$ in Appendix~\ref{appendix:current:piH}. We obtain
\begin{align}
 \label{eqn:PiH}\Pi^{\alf\beta\gamma\delta}_{\text{\tiny EB},iq_0}=\big(\Pi^{\alf\beta\gamma\delta}_{\text{\tiny EB},iq_0}\big)^{(\text{tri})}+\big(\Pi^{\alf\beta\gamma\delta}_{\text{\tiny EB},iq_0}\big)^{(\text{rec})} \, .
\end{align}
We separated the contribution that involve three vertices and four vertices into a {\it triangular} and a {\it rectangular} contribution, respectively, following the terminology introduced by Nourafkan and Tremblay \cite{Nourafkan2018}. The triangular contribution with three vertices reads
\begin{align}
  \label{eqn:PiHTri}\big(\Pi^{\alf\beta\gamma\delta}_{\text{\tiny EB},iq_0}\big)^{(\text{tri})} =\frac{1}{4}\TrH\big[&\sG^{}_{ip_0+iq_0,\bp}\lam^{\beta}_{\bp}\sG^{}_{ip_0,\bp}\lam^\delta_{\bp}\sG^{}_{ip_0,\bp}\lam^{\alf\gamma}_{\bp}
 -\sG^{}_{ip_0+iq_0,\bp}\lam^{\beta\gamma}_{\bp}\sG^{}_{ip_0,\bp}\lam^\delta_\bp\sG^{}_{ip_0,\bp}\lam^\alf_{\bp}
 \nonumber\\
  +&\sG^{}_{ip_0-iq_0,\bp}\lam^\alf_{\bp}\sG^{}_{ip_0,\bp}\lam^\delta_\bp\sG^{}_{ip_0,\bp}\lam^{\beta\gamma}_{\bp} 
  -\sG^{}_{ip_0-iq_0,\bp}\lam^{\alf\gamma}_{\bp}\sG^{}_{ip_0,\bp}\lam^\delta_\bp\sG^{}_{ip_0,\bp}\lam^\beta_{\bp}\big] \, .
\end{align}
We introduced the compact notation $\TrH[\,\cdot\,]\hspace{-0.8mm}=\hspace{-0.8mm}e^3TL^{-1}\sum_p\tr[\,\cdot\,-(iq_0\hspace{-0.5mm}=\hspace{-0.5mm}0)-(\gamma\hspace{-0.5mm}\leftrightarrow\hspace{-0.5mm}\delta)]$, where the dot $\cdot$ indicates the argument over which the trace is performed. The compact notation includes the prefactors, both summations over Matsubara frequencies and momenta, the matrix trace as well as the subtraction of the zero Matsubara frequency contribution of the argument and the subtraction of the previous terms with $\gamma\leftrightarrow\delta$ exchanged. Thus when writing all terms explicitly, each product in \eqref{eqn:PiHTri} gives four terms. The rectangular contribution with four vertices reads
\begin{align}
  \label{eqn:PiHRec}\big(\Pi^{\alf\beta\gamma\delta}_{\text{\tiny EB},iq_0}\big)^{(\text{rec})} =\frac{1}{2}\TrH\big[&\sG^{}_{ip_0+iq_0,\bp}\lam^{\beta}_{\bp}\sG^{}_{ip_0,\bp}\lam^\delta_{\bp}\sG^{}_{ip_0,\bp}\lam^\gamma_\bp\sG^{}_{ip_0,\bp}\lam^\alf_{\bp}
 \nonumber\\
 +&\sG^{}_{ip_0-iq_0,\bp}\lam^\alf_{\bp}\sG^{}_{ip_0,\bp}\lam^\delta_\bp\sG^{}_{ip_0,\bp}\lam^\gamma_{\bp}\sG^{}_{ip_0,\bp}\lam^\beta_{\bp}\big] \, .
\end{align}
We see that both contributions involve different combinations of Green's function and vertex matrices. All contributions involve the first-order vertex. The second-order vertex is also present in the triangular contribution. Note that the decomposition of the form in \eqref{eqn:PiHTri} and \eqref{eqn:PiHRec} is not unique due to the possibility of partial integration in momentum. The presented form turns out to be a convenient decomposition for further calculations. Both contributions and, thus, the polarization tensor $\Pi^{\alf\beta\gamma\delta}_{\text{\tiny EB},iq_0}$ in \eqref{eqn:PiH} are antisymmetric with respect to $\gamma\leftrightarrow\delta$ as required by the relation \eqref{eqn:sH} and vanishes at zero $iq_0$. Note that the Green's function matrices and the vertex matrices do not commute in general. We respected this issue in the derivation.

%
%

\section{Fundamental concepts}
\label{sec:theory:fundamentalconcepts}

Our goal is to derive conductivity formulas, which decompose into contributions with unique properties and involve quantities of clear physical interpretation. For given Bloch Hamiltonian $\lam_\bp$, chemical potential $\mu$, temperature $T$ and relaxation rate $\Gamma$, the polarization tensors in \eqref{eqn:PiE} and \eqref{eqn:PiH} can be evaluated directly by performing the Matsubara summation explicitly. However, we are interested in an analytic result and are faced with two key steps in the following derivation: performing the trace over the two subsystems and the analytic continuation from Matsubara frequency to real frequency. Both steps are tedious and not physically transparent without any further strategy. Before performing the actual derivation in the next Sec. \ref{sec:theory:conductivity} and \ref{sec:theory:Hall}, we present different fundamental concepts that will eventually guide us to a deeper understanding of the underlying structure and, thus, not only to a transparent calculation but also to a physically motivated decomposition of the conductivities. 

We structure this section as follows: We introduce a different representation of the Bloch Hamiltonian $\lam_\bp$ in \eqref{eqn:lam}. The Bloch Hamiltonian $\lam_\bp$ is the crucial quantity for both the Green's function and the vertex matrices. The eigenvalues and eigenbasis of $\lam_\bp$ are particular simple in this different representation (Sec.~\ref{sec:theory:fundamentalconcepts:sphericalrepresentation}). We express the Green's function and the vertex matrices in the eigenbasis of $\lam_\bp$. Interband contributions for the first-order vertex, that is, for the momentum derivative of the Bloch Hamiltonian $\partial_\alf\lam_\bp$, after the basis change are found and analyzed (Sec.~\ref{sec:theory:fundamentalconcepts:coherence}). We have a more general view on interband contributions for more than two bands and draw connection to concepts of quantum geometry, in particular, to the quantum metric and the Berry curvature (Sec.~\ref{sec:theory:fundamentalconcepts:quantumgeometry}). Those concepts are then applied and specified for our two-band model (Sec.~\ref{sec:theory:fundamentalconcepts:quantumgeometrytwoband}). We close this section by performing similar steps to the second derivative of the Bloch Hamiltonian $\partial_\alf\partial_\beta\lam_\bp$ in a general multiband system and for our two-band model (Sec.~\ref{sec:theory:fundamentalconcepts:effectivemass}).

%
%

\subsection{Spherical representation}
\label{sec:theory:fundamentalconcepts:sphericalrepresentation}

\begin{figure}[t!]
\centering
\includegraphics[width=10cm]{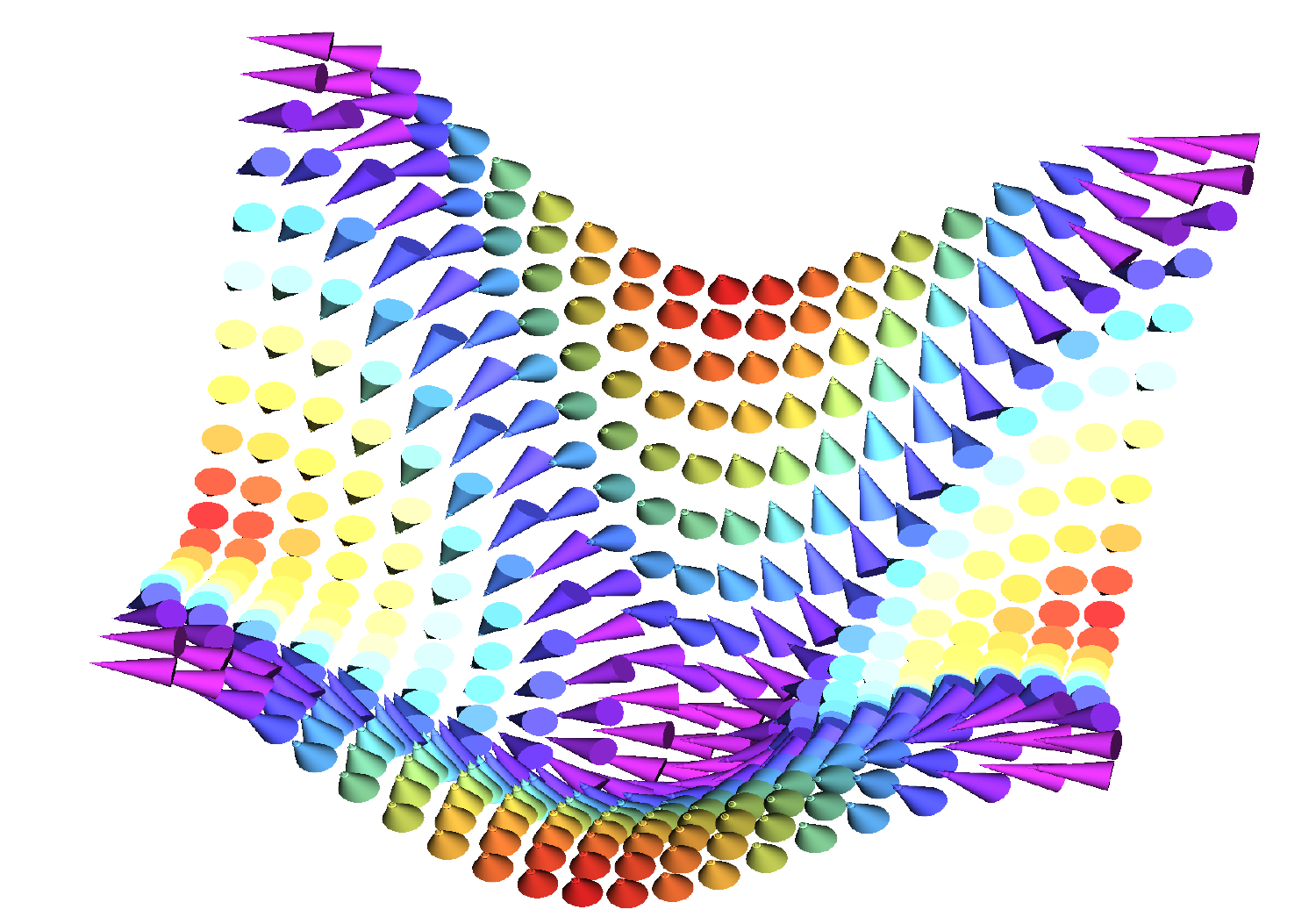}
\caption{We can represent the Bloch Hamiltonian $\lam_\bp$ by a function $\g_\bp$ and a vector-valued function $\bd_\bp$. For a two-dimensional system, we can visualize $\g_\bp$ as a surface on top of which we indicate the vector $\bd_\bp$ by its length $r_\bp$ (color from purple to red) and direction given by the angles $\theta_\bp$ and $\varphi_\bp$. The conductivity is given by the modulation of these functions. 
Here, we show $\lam_\bp$ of the example in Sec.~\ref{sec:application:anomalousHallEffect:scaling}.  
\label{fig:plotLam}}
\end{figure}

The crucial quantity to evaluate \eqref{eqn:PiE} and \eqref{eqn:PiH} is the Bloch Hamiltonian matrix $\lam_\bp$, which is present in the Green's function $\sG_{ip_0,\bp}$ and the first- and second-order vertices $\lambda^\alf_\bp$ and $\lam^{\alf\beta}_\bp$. The basic property of the $2\times2$ matrix $\lam_\bp$ is its hermiticity, which allows us to expand it in the identity matrix $\mathds{1}$ and the three Pauli matrices 
\begin{align}
 \pauma_x=\begin{pmatrix} 0 && 1 \\ 1 && 0 \end{pmatrix}, \,
 \pauma_y=\begin{pmatrix} 0 && -i \\ i && 0 \end{pmatrix}, \,
 \pauma_z=\begin{pmatrix} 1 && 0 \\ 0 && -1 \end{pmatrix} \,,
\end{align}
which we combine to the Pauli vector $\bpauma=(\pauma_x,\pauma_y,\pauma_z)$. The indexing $x,y,z$ must not be confused with the spatial directions. We get the compact notation
\begin{align}
 \lam_\bp&=\g_\bp\,\mathds{1}+\bd_\bp\cdot\bpauma 
\end{align}
with a function $\g_\bp$ and a vector-valued function $\bd_\bp$ \cite{Gianfrate2020, Volovik1988, Dugaev2008, Asboth2016, Bleu2018}. It is very useful to represent the vector $\bd_\bp$ via its length $r_\bp$ and the two angles $\theta_\bp$ and $\varphi_\bp$ in spherical coordinates \cite{Gianfrate2020, Asboth2016, Bleu2018}, that is,
\begin{align}
 \bd_\bp=r_\bp\begin{pmatrix}
          \cos\varphi_\bp\,\sin\theta_\bp\\[1mm]
          \sin\varphi_\bp\,\sin\theta_\bp\\[1mm]
          \cos\theta_\bp
          \end{pmatrix} \,.
\end{align}
The Bloch Hamiltonian matrix $\lam_\bp$ in spherical coordinates reads
\begin{align}
\label{eqn:lamPolar}
 \lam_\bp= \begin{pmatrix} \g_\bp+r_\bp\cos\theta_\bp && r_\bp\sin\theta_\bp \,e^{-i\varphi_\bp} \\[2mm] r_\bp\sin\theta_\bp\, e^{i\varphi_\bp} && \g_\bp-r_\bp\cos\theta_\bp\end{pmatrix} \, .
\end{align}
The momentum derivative of $\lam_\bp$, in particular, the first- and second-order vertices $\lam^\alf_\bp$ and $\lam^{\alf\beta}_\bp$, can be expressed in the derivatives of these functions $\g_\bp$, $r_\bp$, $\theta_\bp$, and $\varphi_\bp$. In a two-dimensional system, we can visualize $\g_\bp$ as a surface on top of which we indicate the vector $\bd_\bp$ by its length and direction. We give an example in Fig.~\ref{fig:plotLam}. 

Both \eqref{eqn:lam} and \eqref{eqn:lamPolar} are equivalent and impose no restriction on the Hamiltonian other than hermiticity. In the following, we exclusively use $\lam_\bp$ in spherical coordinates. For given $\eps_{\bp,A},\,\eps_{\bp,B}$ and $\Delta_\bp$ in \eqref{eqn:lam} the construction of \eqref{eqn:lamPolar} is straightforward. We give the relations explicitly since they may provide a better intuitive understanding of the involved quantities. We have
 \begin{align}
 \label{eqn:gh}
  &\g_\bp=\frac{1}{2}(\eps_{\bp,A}+\eps_{\bp,B}) \, , \\[1mm]
  &h_\bp=\frac{1}{2}(\eps_{\bp,A}-\eps_{\bp,B}) \, , \\[1mm]
  &r_\bp=\sqrt{h^2_\bp+|\Delta_\bp|^2} \, ,
\end{align}
where we defined the function $h_\bp$. The radius $r_\bp$ involves $h_\bp$ and the absolute value of $\Delta_\bp$. The angle $\theta_\bp$ describes the ratio between $h_\bp$ and $|\Delta_\bp|$. The angle $\varphi_\bp$ is equal to the negative phase of $\Delta_\bp$. They are given by 
\begin{align}
 \label{eqn:Theta}
  &\cos\theta_\bp = \frac{h_\bp}{r_\bp} &&\sin\theta_\bp = \frac{|\Delta_\bp|}{r_\bp} \, , \\
  \label{eqn:Phi}
  &\cos\varphi_\bp=\re\frac{\Delta_\bp}{|\Delta_\bp|}&&\sin\varphi_\bp=-\im\frac{\Delta_\bp}{|\Delta_\bp|} \, ,
 \end{align}
where the real and imaginary part are denoted by $\re$ and $\im$, respectively. The advantage of the spherical form \eqref{eqn:lamPolar} is its simplicity of the eigenvalues and eigenvectors. We denote the eigensystem at momentum $\bp$ as $\pm_\bp$. The eigenenergies are 
\begin{align}
 E^\pm_\bp=\g^{}_\bp\pm r^{}_\bp
\end{align}
with corresponding eigenvectors
\begin{align}
 &\label{eqn:+}|+_\bp\rangle=e^{i\phi^+_\bp}\begin{pmatrix} \cos \frac{1}{2}\theta_\bp \\[2mm] e^{i\varphi_\bp}\,\sin \frac{1}{2}\theta_\bp \end{pmatrix}\,,
 \\[5mm]
 &\label{eqn:-}|-_\bp\rangle=e^{i\phi^-_\bp}\begin{pmatrix} -e^{-i\varphi_\bp}\,\sin \frac{1}{2}\theta_\bp \\[2mm]\cos \frac{1}{2}\theta_\bp \end{pmatrix}\,.
\end{align}
These eigenvectors are normalized and orthogonal, that is, $\langle +_\bp|+_\bp\rangle=\langle -_\bp|-_\bp\rangle=1$ and $\langle +_\bp|-_\bp\rangle=\langle -_\bp|+_\bp\rangle=0$. The two phases $\phi^\pm_\bp$ reflect the freedom to choose a phase of the normalized eigenvectors when diagonalizing at fixed momentum $\bp$, that is, a ``local'' $U(1)$ gauge symmetry in momentum space. We include it explicitly for an easier comparison with other gauge choices and to make gauge-dependent quantities more obvious in the following calculations. 

%
%

\subsection{First-order vertex}
\label{sec:theory:fundamentalconcepts:coherence}

The polarization tensors in \eqref{eqn:PiE} and \eqref{eqn:PiH} are the trace of the product of Green's function matrices and vertex matrices. A trace is invariant under unitary transformations (or, in general, similarity transformations) due to its cyclic property. We transform all matrices by the $2\times 2$ unitary transformation $U_\bp=\begin{pmatrix}|+_\bp\rangle & |-_\bp\rangle\end{pmatrix}$, whose columns are composed of the eigenvectors $|\pm_\bp\rangle$. The matrix $U_\bp$ diagonalizes the Bloch Hamiltonian matrix
\begin{align}
 \label{eqn:Ep}
 \cE_\bp=\Udag_\bp\lam^{}_\bp \U_\bp=\begin{pmatrix} E^+_\bp && 0 \\[1mm] 0 && E^-_\bp \end{pmatrix} \, ,
\end{align}
where we defined the quasiparticle band matrix $\cE_\bp$. We transform the Green's function matrix in \eqref{eqn:Green} and get the diagonal Green's function
\begin{align}
\label{eqn:Gdiag}
 \cG^{}_{ip_0,\bp}&=\Udag_\bp \sG^{}_{ip_0,\bp}\U_\bp= \big[ip_0+\mu-\cE_\bp+i\Gamma \,\text{sign}(p_0)\big]^{-1} \, .
\end{align}
Note that the assumptions of $\Gamma$ to be proportional to the identity matrix is crucial to obtain a diagonal Green's function matrix by this transformation. 

In general, the vertex matrices $\lam^\alf_\bp$ and $\lam^{\alf\beta}_\bp$ will not be diagonal after unitary transformation with $U_\bp$, since they involve the momentum derivative $\lam^\alf_\bp=\partial^{}_\alf \lam^{}_\bp$ and $\lam^{\alf\beta}_\bp=\partial^{}_\alf\partial^{}_\beta\lam^{}_\bp$. The derivatives do not commute with the momentum-dependent $U_\bp$. In a first step, we focus on $\lam^\alf_\bp$. Expressing $\lam_\bp$ in terms of $\cE_\bp$ we get
\begin{align}
 \label{eqn:UdagLamUDeriv}
 \Udag_\bp\lam^\alf_\bp \U_\bp=\Udag_\bp\big[\partial^{}_\alf\lam^{}_\bp\big] \U_\bp=\Udag_\bp\big[\partial^{}_\alf\big(\U_\bp\cE^{}_\bp\Udag_\bp\big)\big]\U_\bp \,.
\end{align}
The derivative of $\cE_\bp$ leads to the eigenvelocities $\cE^\alf_\bp=\partial^{}_\alf \cE^{}_\bp$. The two other terms from the derivative contain the momentum derivative of $U_\bp$. Using the identity 
$\big(\partial^{}_\alf\Udag_\bp\big)\U_\bp=-\Udag_\bp\big(\partial^{}_\alf \U_\bp\big)$ of unitary matrices we end up with 
\begin{align}
\label{eqn:UdagLamU}
 \Udag_\bp\lam^\alf_\bp \U_\bp=\cE_\bp^\alf+\cF^\alf_\bp\, ,
\end{align}
where we defined $\cF^\alf_\bp=i\big[\cE^{}_\bp,\cA_\bp^\alf\big]$ with
\begin{align}
\label{eqn:BerryConnection}
 \cA_\bp^\alf=i\Udag_\bp\big(\partial^{}_\alf \U_\bp\big) \,.
\end{align}
Since $\cF^\alf_\bp$ involves the commutator with the diagonal matrix $\cE_\bp$, it is an off-diagonal matrix. Thus, we see already at this stage that $\cF^\alf_\bp$ causes the mixing of the two quasiparticle bands and, thus, captures the interband effects induced by the vertex $\lam^\alf_\bp$. We refer to $\cF^\alf_\bp$ as ``(first-order) interband matrix''.

Let us have a closer look at $\cA^\alf_\bp$ defined in \eqref{eqn:BerryConnection}. The matrix $U_\bp$ consists of the eigenvectors $|\pm_\bp\rangle$. Its complex conjugation $\Udag_\bp$ consists of the corresponding $\langle\pm_\bp|$. Thus, we can identify the diagonal elements of $\cA_\bp^\alf$ as the Berry connection of the eigenstates $|\pm_\bp\rangle$, that is $\cA^{\alf,\pm}_\bp=i\langle\pm_\bp|\partial_\alf\pm_\bp\rangle$, where $|\partial_\alf \pm_\bp\rangle=\partial_\alf |\pm_\bp\rangle$ is the momentum derivative of the eigenstate \cite{Berry1984,Zak1989}. $\cA_\bp^\alf$ is hermitian due to the unitarity of $U_\bp$. This allows us to express it in terms of the identity and the Pauli matrices, $\cA_\bp^\alf=\cI^\alf_\bp+\cX_\bp^\alf+\cY_\bp^\alf+\cZ_\bp^\alf$, where 
\begin{align}
 \label{eqn:I}
 \cI^\alf_\bp&= -\frac{1}{2}\big[\phi^{+,\alf}_\bp+\phi^{-,\alf}_\bp\big]\,\mathds{1} \, ,\\
 \label{eqn:X}
 \cX^\alf_\bp&=-\frac{1}{2}\big[\varphi^\alf_\bp\sin\theta_\bp\cos\tilde\varphi_\bp+\theta^\alf_\bp\sin\tilde\varphi_\bp\big]\,\pauma_x \, , \\
 \label{eqn:Y}
 \cY^\alf_\bp&=-\frac{1}{2}\big[\varphi^\alf_\bp\sin\theta_\bp\sin\tilde\varphi_\bp-\theta_\bp^\alf\cos\tilde\varphi_\bp\big]\,\pauma_y  \, ,\\
 \label{eqn:Z}
 \cZ^\alf_\bp&=-\frac{1}{2}\big[\phi^{+,\alf}_\bp-\phi^{-,\alf}_\bp+\varphi^\alf_\bp\big(1-\cos\theta_\bp\big)\big]\,\pauma_z \, ,
\end{align}
and $\tilde\varphi_\bp=\varphi_\bp+\phi^+_\bp-\phi^-_\bp$. We calculated the prefactors by using \eqref{eqn:+} and \eqref{eqn:-} and used the short notation $\theta^\alf_\bp=\partial_\alf\theta_\bp$ and $\varphi^\alf_\bp=\partial_\alf\varphi_\bp$ for the momentum derivative in $\alf$ direction. Each component of $\cA_\bp$ is gauge dependent by involving $\tilde\varphi^{}_\bp$ or $\phi^{\pm,\alf}_\bp=\partial^{}_\alf \phi^\pm_\bp$. The interband matrix $\cF^\alf_\bp$ involves only the off-diagonal matrices $\cX^\alf_\bp$ and $\cY^\alf_\bp$ since the diagonal contributions $\cI_\bp$ and $\cZ_\bp$ vanish by the commutator with the diagonal matrix $\cE_\bp$. We see that the interband matrix $\cF^\alf_\bp$ is gauge dependent due to $\tilde\varphi_\bp$.


%
%

\subsection{Quantum geometric tensor}
\label{sec:theory:fundamentalconcepts:quantumgeometry}

The identification of a matrix $\cA^\alf_\bp$ in \eqref{eqn:UdagLamU}, which involves the Berry connection, suggests a deeper and more general connection to concepts of quantum geometry. 
Expressing the momentum derivative $\partial_\alf \hat \lam_\bp$ of a general multiband (and not necessarily two-band) Bloch Hamiltonian $\hat \lam_\bp$ in its orthonormal and complete eigenbasis $|n_\bp\rangle$ with eigenvalues $E^n_\bp$ naturally leads to intraband and interband contributions via 
\begin{align}
 \label{eqn:DlamGeneral}
 \langle n_\bp|(\partial_\alf \hat \lam_\bp)|m_\bp\rangle = \delta_{nm}\,E^{n,\alf}_\bp+i(E^n_\bp-E^m_\bp)\cA^{\alf,nm}_\bp 
\end{align}
after treating the momentum derivative and the momentum dependence of the eigenbasis carefully. The first term on the right-hand side involves the quasiparticle velocities $E^{n,\alf}_\bp=\partial_\alf E^n_\bp$ and is only present for $n=m$. The second term involves the Berry connection $\cA^{\alf,nm}_\bp=i\langle n_\bp|\partial_\alf m_\bp\rangle$, where $|\partial_\alf m_\bp\rangle$ is the momentum derivative of the eigenstate $|m_\bp\rangle$ \cite{Berry1984,Zak1989}, and is only present for $n\neq m$. 

The Berry connection $\cA^{\alf,nm}_\bp$ is not invariant under the ``local`` $U(1)$ gauge transformation $|n_\bp\rangle\rightarrow e^{i\phi^n_\bp}|n_\bp\rangle$ in momentum space and, thus, should not show up in physical quantities like the conductivity. In other words, not the Hilbert space but the projective Hilbert space is physically relevant \cite{Provost1980,Anandan1991,Anandan1990,Cheng2013}. In general, the transformation of the Berry connection with respect to the gauge transformation above reads
\begin{align}
 \label{eqn:AnmGauge}\cA^{\alf,nm}_\bp&\rightarrow \cA^{\alf,nm}_\bp e^{-i(\phi^n_\bp-\phi^m_\bp)}-\delta_{nm}\phi^{n,\alf}_\bp 
\end{align}
with $\phi^{n,\alf}_\bp=\partial^{}_\alf \phi^n_\bp$ only present for $n=m$. Obviously, the combination 
\begin{align}
\label{eqn:T}
 \cT^{\alf\beta,n}_\bp=\sum_{m\neq n}\cA^{\alf,nm}_\bp\cA^{\beta,mn}_\bp
\end{align}
is gauge independent. The quantity $\cT^{\alf\beta,n}_\bp$ is a momentum-dependent tensor for each band $n$ with components $\alf$ and $\beta$. The tensor of band $n$ involves the summation over all other bands. We rewrite \eqref{eqn:T} by using the orthogonality and the completeness of the eigenbasis, $\langle n_\bp|\partial_\alf m_\bp\rangle=-\langle \partial_\alf n_\bp|m_\bp\rangle$ and $\sum_{m\neq n} |m_\bp\rangle\langle m_\bp|=1-|n_\bp\rangle\langle n_\bp|$, and obtain
\begin{align}
 \cT^{\alf\beta,n}_\bp=\langle\partial_\alf n_\bp|\partial_\beta n_\bp\rangle-\langle \partial_\alf n_\bp|n_\bp\rangle\langle n_\bp|\partial_\beta n_\bp\rangle \, .
\end{align}
We have recovered the quantum geometric tensor, which is the Fubini-Study metric of the projective Hilbert space \cite{Provost1980,Anandan1991, Anandan1990, Cheng2013, Bleu2018}.

It turns out to be convenient to decompose the quantum geometric tensor $\cT^{\alf\beta,n}_\bp$ into its symmetric and antisymmetric part with respect to $\alf\leftrightarrow\beta$. This decomposition is unique. Using the property of the Berry connection under complex conjugation in \eqref{eqn:T}, we see that the symmetric part is the real part and the antisymmetric part is the imaginary part of $\cT^{\alf\beta,n}_\bp$, respectively. We define
\begin{align}
 \cT^{\alf\beta,n}_\bp=\frac{1}{2}\big(C^{\alf\beta,n}_\bp-i\,\Omega^{\alf\beta,n}_\bp\big)
\end{align}
with the symmetric real-valued function $\cC^{\alf\beta,n}_\bp=\cC^{\beta\alf,n}_\bp$ and the antisymmetric real-valued function $\Omega^{\alf\beta,n}_\bp=-\Omega^{\beta\alf,n}_\bp$. 

Both $\cC^{\alf\beta,n}_\bp$ and $\Omega^{\alf\beta,n}_\bp$ have a clear physical interpretation. By the latter one, we have recovered the Berry curvature
\begin{align}
\label{eqn:OmegaRot}
 \Omega^{\alf\beta,n}_\bp&=-2\,\im\cT^{\alf\beta,n}_\bp=\partial^{}_\alf\,\cA^{\beta,n}_\bp-\partial^{}_\beta\,\cA^{\alf,n}_\bp \, .
\end{align}
The Berry curvature is the curl of the Berry connection. Using \eqref{eqn:T}, one can show that $\sum_n \Omega^{\alf\beta,n}_\bp=0$, that is, the Berry curvature summed over all bands vanishes. In order to understand the meaning of the symmetric part $C^{\alf\beta,n}_\bp$ we consider the squared distance function
\begin{align}
\label{eqn:QuantumDistance}
 D\big(|n_\bp\rangle,|n_{\bp'}\rangle\big)^2&=1-|\langle n_\bp|n_{\bp'}\rangle|^2 \, ,
\end{align}
where $|n_\bp\rangle$ and $|n_{\bp'}\rangle$ are two normalized eigenstates of the same band $E^n_\bp$ at different momenta \cite{Provost1980, Anandan1991,Anandan1990, Cheng2013, Bleu2018}. The distance function is invariant under the gauge transformations $|n_\bp\rangle\rightarrow e^{i\phi^n_\bp}|n_\bp\rangle$. It is maximal, if the two states are orthogonal, and zero, if they differ only by a phase. We can understand the function in \eqref{eqn:QuantumDistance} as the distance of the projective Hilbert space in the same manner as $||n_\bp\rangle - |n_{\bp'}\rangle|$ is the natural distance in the Hilbert space, which is, in contrast, not invariant under the upper gauge transformation \cite{Provost1980}. If we expand the distance between the two eigenstates $|n_\bp\rangle$ and $|n_{\bp+d\bp}\rangle$, whose momenta differ only by an infinitesimal momentum $d\bp$, up to second order, we find a metric tensor $g^{\alf\beta,n}_\bp$ that is given by the real part of the quantum geometric tensor. We see that 
\begin{align}
\label{eqn:Cbasic}
 C^{\alf\beta,n}_\bp=2\,g^{\alf\beta,n}_\bp=2\,\re\cT^{\alf\beta,n}_\bp \, .
\end{align}
We summarize that the momentum derivative of the Bloch Hamiltonian, or first-order vertex, expressed in the eigenbasis of the Bloch Hamiltonian in \eqref{eqn:DlamGeneral} naturally leads to a quasiparticle velocity and a Berry-connection term. A gauge-independent combination of the latter one defines the quantum geometric tensor, which decomposes into the quantum metric and the Berry curvature of the corresponding band.

%
%

\subsection[Quantum metric factor and Berry curvature of the two-band model]{Quantum metric factor and Berry curvature of the\\ two-band model}
\label{sec:theory:fundamentalconcepts:quantumgeometrytwoband}

After the general considerations in the previous section, we apply those concepts explicitly to our two-band model. In the decomposition of the first-order vertex in \eqref{eqn:DlamGeneral}, the first term corresponds to $\cE^\alf_\bp$ in \eqref{eqn:Ep}, the second term to $\cF^\alf_\bp$ in \eqref{eqn:UdagLamU} and the $\cA^{\alf,nm}_\bp$ are the elements of the matrix $\cA_\bp$ in \eqref{eqn:BerryConnection} with indices $n,m$ for both bands $\pm$, that is
\begin{align}
 \cA_\bp=i\Udag_\bp\partial^{}_\alf U^{}_\bp=\begin{pmatrix}\cA^{\alf,+}_\bp & \cA^{\alf,+-}_\bp \\[2mm] \cA^{\alf,-+}_\bp & \cA^{\alf,-}_\bp
\end{pmatrix}\, .
\end{align}
We used the short notation $\cA^{\alf,+}_\bp\equiv\cA^{\alf,++}_\bp$ and $\cA^{\alf,-}_\bp\equiv\cA^{\alf,--}_\bp$ for the diagonal elements. The diagonal elements $\cA^{\alf,+}_\bp$ and $\cA^{\alf,-}_\bp$ correspond to $\cI^\alf_\bp+\cZ^\alf_\bp$ in \eqref{eqn:I} and \eqref{eqn:Z}. The off-diagonal elements $\cA^{\alf,+-}_\bp$ and $\cA^{\alf,-+}_\bp$  correspond to $\cX^\alf_\bp+\cY^\alf_\bp$ in \eqref{eqn:X} and \eqref{eqn:Y}. We consider the gauge dependence of $\cA^{\alf,nm}_\bp$ in \eqref{eqn:AnmGauge} by allowing the phases $\phi^\pm_\bp$ in \eqref{eqn:+} and \eqref{eqn:-} explicitly. The quasiparticle velocity $\cE^\alf_\bp$ is gauge independent, whereas the interband matrix $\cF^\alf_\bp$ is gauge dependent. In analogy to \eqref{eqn:T}, the product $\cF^\alf_\bp\cF^\beta_\bp$ is gauge independent, which can be see by
\begin{align}
\label{eqn:FF}
 \cF^\alf_\bp\cF^\beta_\bp\propto\big(\cX^\alf_\bp+\cY^\alf_\bp\big)\big(\cX^\beta_\bp+\cY^\beta_\bp\big)\propto \begin{pmatrix} 0 & e^{-i\tilde\varphi_\bp} \\ e^{i\tilde\varphi_\bp} & 0\end{pmatrix}
 \begin{pmatrix} 0 & e^{-i\tilde\varphi_\bp} \\ e^{i\tilde\varphi_\bp} & 0 \end{pmatrix}\propto \mathds{1} \, ,
\end{align}
where we dropped gauge-independent quantities in each step. The (diagonal) elements of the product $\cF^\alf_\bp\cF^\beta_\bp$ are proportional to the quantum geometric tensor $\cT^{\alf\beta,\pm}_\bp$.

The product $\cF^\alf_\bp\cF^\beta_\bp$ is neither symmetric nor antisymmetric with respect to $\alf\leftrightarrow\beta$. Up to a prefactor, its symmetric and antisymmetric parts read
\begin{align}
 &\cF^\alf_\bp\cF^\beta_\bp+\cF^\beta_\bp\cF^\alf_\bp\propto\{\cX^\alf_\bp,\cX^\beta_\bp\}+\{\cY^\alf_\bp,\cY^\beta_\bp\}=\cC^{\alf\beta}_\bp\,,\\[2mm]
 &\cF^\alf_\bp\cF^\beta_\bp-\cF^\beta_\bp\cF^\alf_\bp\propto [\cX^\alf_\bp,\cY^\beta_\bp]+[\cY^\alf_\bp,\cX^\beta_\bp]=-i\,\Omega^{\alf\beta}_\bp\,,
 \end{align}
which defines the symmetric function $\cC^{\alf\beta}_\bp$ and antisymmetric function $\Omega^{\alf\beta}_\bp$, which are both real-valued diagonal matrices. Using \eqref{eqn:X} and \eqref{eqn:Y} we get
\begin{align}
\label{eqn:CM}
 \cC^{\alf\beta}_\bp&=\frac{1}{2}\big(\theta^\alf_\bp\theta^\beta_\bp+\varphi^\alf_\bp\varphi^\beta_\bp\sin^2\theta_\bp\big)\mathds{1}\,,\\
 \label{eqn:OmegaM}
 \Omega^{\alf\beta}_\bp&=\frac{1}{2}\big(\varphi^\alf_\bp\theta^\beta_\bp-\varphi^\beta_\bp\theta^\alf_\bp\big)\sin\theta_\bp\,\pauma_z\,.
 \end{align}
We see explicitly that $\cC^{\alf\beta}_\bp$ and $\Omega^{\alf\beta}_\bp$ are gauge independent. Note that $\cC^{\alf\beta}_\bp$ involves equal contributions for both quasiparticle bands, whereas $\Omega^{\alf\beta}_\bp$ involves contributions of opposite sign for the two quasiparticle bands. We have $\Omega^{\alf\beta}_\bp=\partial_\alf\cZ^\beta_\bp-\partial_\beta\cZ^\alf_\bp$ since $\Omega^{\alf\beta}_\bp$ is the Berry curvature of the eigenbasis $|\pm_\bp\rangle$. The two definitions \eqref{eqn:CM} and \eqref{eqn:OmegaM} are the matrix versions of the general expressions in \eqref{eqn:OmegaRot} and \eqref{eqn:Cbasic}. The matrix $\cC^{\alf\beta}_\bp$ involves the quantum metric of the two quasiparticle bands, which are equal in this case. Thus, we refer to $\cC^{\alf\beta}_\bp$ as ''quantum metric factor`` in the following, which is more precise than the previous used ''coherence factor`` \cite{Voruganti1992}. Note that both $\cC^{\alf\beta}_\bp$ and $\Omega^{\alf\beta}_\bp$ only involve the angles $\theta_\bp$ and $\varphi_\bp$. 

%
%

\subsection{Second-order vertex}
\label{sec:theory:fundamentalconcepts:effectivemass}

Whereas the considerations above are sufficient to study the polarization tensors $\Pi^{\alf\beta}_{\text{\tiny E},iq_0}$ in \eqref{eqn:PiE}, the polarization tensor $\Pi^{\alf\beta\gamma\delta}_{\text{\tiny EB},iq_0}$ in \eqref{eqn:PiH} does also involve the second derivative of the Bloch Hamiltonian $\lam^{\alf\beta}_\bp=\partial^{}_\alf\partial^{}_\beta\lam^{}_\bp$, the second-order vertex. In order to identify the interband contributions due to $\lam^{\alf\beta}_\bp$, we can perform similar steps as for the first-order vertex in \eqref{eqn:UdagLamUDeriv} by calculating 
\begin{align}
 \Udag_\bp\lam^{\alf\beta}_\bp\U_\bp=\Udag_\bp\big[\partial^{}_\alf\partial^{}_\beta\lam_\bp\big]\U_\bp=\Udag_\bp\big[\partial^{}_\alf\partial^{}_\beta\big(\U_\bp\cE^{}_\bp\Udag_\bp\big)\big]\U_\bp \, .
\end{align}
We see that we obtain nine contributions after evaluating the derivatives: The second derivative of $\cE^{\alf\beta}_\bp=\partial^{}_\alf\partial^{}_\beta\cE^{}_\bp$ gives the inverse quasiparticle effective mass. We have four contributions involving the quasiparticle velocity $\cE^\nu_\bp=\partial^{}_\nu \cE^{}_\bp$ and the Berry connection matrix $\cA^\nu_\bp=i\Udag_\bp\big(\partial^{}_\nu\U_\bp\big)$ with $\nu=\alf,\beta$. We have two contributions involving the product of $\cA^\alf_\bp$ and $\cA^\beta_\bp$. The remaining two contributions are involving $\Udag_\bp\big(\partial^{}_\alf\partial^{}_\beta \U_\bp\big)$ and $\big(\partial^{}_\alf\partial^{}_\beta\Udag_\bp\big)\U_\bp$. The final result is symmetric in $\alf\leftrightarrow\beta$.

The derivation above is straightforward and gives an idea about the contributions that are present. However, the final result is not transparent with respect to the gauge transformation $|n_\bp\rangle=e^{i\phi^n_\bp}|n_\bp\rangle$ of the eigenbasis and, thus, is not sufficient for the identification of different terms that are physical individually. Thus, before deriving our final result of $\lam^{\alf\beta}_\bp$ in our two-band model, we consider again a general (not necessarily two-band) Bloch Hamiltonian $\hat \lam_\bp$ with orthonormal eigenbasis $|n_\bp\rangle$ in a first step and specify the result for our two-band model in a second step. 

We express $\partial_\alf \partial_\beta\hat \lam_\bp$ in the eigenbasis. After carefully taking the momentum dependence of the eigenbasis into account, we can express the final result in only the eigenenergies $E^n_\bp$, the Berry connection $\cA^{\alf,nm}_\bp$ and their derivatives. We present the derivation in Appendix~\ref{appendix:Mass}. The diagonal elements read
\begin{align}
 \label{eqn:Mn}
 \langle n_\bp|\big(\partial_\alf\partial_\beta\hat\lam_\bp\big)|n_\bp\rangle=E^{n,\alf\beta}_\bp+\sum_l\big(E^l_\bp-E^n_\bp\big)\big(A^{\alf,nl}_\bp A^{\beta,ln}_\bp+A^{\beta,nl}_\bp A^{\alf,ln}_\bp\big)
\end{align}
with the inverse quasiparticle effective mass $E^{n,\alf\beta}_\bp=\partial^{}_\alf\partial^{}_\beta E^n_\bp$. We sum over all bands in the second term on the right-hand side, which we indicate by the summation over $l$. This term vanishes in a one-band system and is, thus, a pure interband contribution. Note that the second term is close to the combination $\cT^{\alf\beta,n}_\bp+\cT^{\beta\alf,n}_\bp$ of the quantum geometric tensor in \eqref{eqn:T} but not equal due to the prefactor $E^l_\bp$. Both terms in \eqref{eqn:Mn} are gauge invariant. The off-diagonal elements for $n\neq m$ read
\begin{align}
 \label{eqn:Mnm}
 \langle n_\bp|\big(\partial_\alf\partial_\beta\hat\lam_\bp\big)|m_\bp\rangle&=i\big(E^{n,\alf}_\bp-E^{m,\alf}_\bp\big)A^{\beta,nm}_\bp
 \\[2mm]
 &\label{eqn:Mnm2}+\frac{i}{2}\big(E^n_\bp-E^m_\bp\big)\big[\partial_\alf-i\big(A^{\alf,n}_\bp-A^{\alf,m}_\bp\big)\big]A^{\beta,nm}_\bp
 \\[1mm]
 &\label{eqn:Mnm3}+\sum_{l\neq n,m}\big[E^l_\bp-\frac{1}{2}\big(E^n_\bp+E^m_\bp\big)\big]A^{\alf,nl}_\bp A^{\beta,lm}_\bp
 \\[-1mm]
 &\label{eqn:Mnm4}+(\alf\leftrightarrow\beta)\, .
\end{align}
The third term \eqref{eqn:Mnm3} is only present for more than two bands and captures the interband effects due to all other bands than the considered $n$  and $m$. The first term \eqref{eqn:Mnm} has a form very similar to the off-diagonal component of the first-order vertex in \eqref{eqn:DlamGeneral} but involves the quasiparticle velocities instead of the eigenenergies. The second term \eqref{eqn:Mnm2} involves the derivative of the Berry connection in a gauge covariant form, $\partial_\alf-i\big(A^{\alf,n}_\bp-A^{\alf,m}_\bp\big)$. We find that each line transforms individually with a phase factor $e^{-i(\phi^n_\bp-\phi^m_\bp)}$ under the ''local`` gauge transformation $|n_\bp\rangle\rightarrow e^{i\phi^n_\bp}|n_\bp\rangle$ in momentum space.

We return to our two-band model in the following. By considering the diagonal and off-diagonal components in \eqref{eqn:Mn} and \eqref{eqn:Mnm}, respectively, for the two eigenstates $|+_\bp\rangle$ and $|-_\bp\rangle$, we identify a diagonal and an off-diagonal part of $\Udag_\bp\lam^{\alf\beta}_\bp\U_\bp$, which we label as
\begin{align}
 \label{eqn:DecompLamAB}
 \Udag_\bp\lam^{\alf\beta}_\bp\U_\bp=(\dM^{-1})^{\alf\beta}_\bp+\dF^{\alf\beta}_\bp\,.
\end{align}
We use the upper notation to indicate the interpretation of $(\dM^{-1})^{\alf\beta}_\bp$ as ''inverse generalized effective mass`` and to symbolize the analogy of the ''(second-order) interband matrix`` $\dF^{\alf\beta}_\bp$ with respect to the first-order interband matrix $\cF^\alf_\bp$ in \eqref{eqn:UdagLamU}. The inverse generalized effective mass $(\dM^{-1})^{\alf\beta}_\bp$ reads
\begin{align}
 (\dM^{-1})^{\alf\beta}_\bp=\cE^{\alf\beta}_\bp-2r_\bp\pauma_z\,\cC^{\alf\beta}_\bp \,,
\end{align}
where $\cE^{\alf\beta}_\bp=\partial^{}_\alf\partial^{}_\beta\cE_\bp$ is the inverse quasiparticle effective mass, that is, the second derivative of the quasiparticle dispersion. The second term on the right-hand side involves the quantum metric factor $\cC^{\alf\beta}_\bp$ with different sign for the upper and the lower band due to the Pauli matrix $\pauma_z$ \cite{Iskin2019}. 

We calculate the second-order interband matrix $\dF^{\alf\beta}_\bp$ for our two-band model in the following. We combine the four contributions in \eqref{eqn:Mnm} that involve $\cE^\nu_\bp$ and $\cA^\nu_\bp$ to two commutators and expand $\cA^\nu_\bp$ in the identity and the Pauli matrices. Due to the commutator, only $\cX^\nu_\bp$ and $\cY^\nu_\bp$ contribute, which are given in \eqref{eqn:X} and \eqref{eqn:Y}, respectively. We calculate the second factor in \eqref{eqn:Mnm2} explicitly in a first step and decompose it into two terms proportional to $\pauma_x$ and $\pauma_y$ in a second step. We obtain the off-diagonal matrix entries with $n\neq m$
\begin{align}
\big[\partial_\alf-i\big(A^{\alf,n}_\bp-A^{\beta,m}_\bp\big)\big]A^{\beta,nm}_\bp \equiv \big(\dX^{\alf\beta}_\bp+\dY^{\alf\beta}_\bp\big)^{nm}
\end{align}
of the matrices
\begin{alignat}{2}
 \label{eqn:dX}
 \dX^{\alf\beta}_\bp&=-\frac{1}{2}\big[\big(\varphi^{\alf\beta}_\bp\sin\theta_\bp+\big(\theta^\alf_\bp\varphi^\beta_\bp+\varphi^\alf_\bp\theta^\beta_\bp\big)\cos\theta_\bp\big)&&\cos\tilde\varphi_\bp\nonumber\\&\hspace{2.75cm} +\big(\theta^{\alf\beta}_\bp-\varphi^\alf_\bp\varphi^\beta_\bp\cos\theta_\bp\sin\theta_\bp\big)&&\sin\tilde\varphi_\bp\big]\,\pauma_x \, ,\\
 \label{eqn:dY}
 \dY^{\alf\beta}_\bp&=-\frac{1}{2}\big[\big(\varphi^{\alf\beta}_\bp\sin\theta_\bp+\big(\theta^\alf_\bp\varphi^\beta_\bp+\varphi^\alf_\bp\theta^\beta_\bp\big)\cos\theta_\bp\big)&&\sin\tilde\varphi_\bp\nonumber\\&\hspace{2.75cm} -\big(\theta^{\alf\beta}_\bp-\varphi^\alf_\bp\varphi^\beta_\bp\cos\theta_\bp\sin\theta_\bp\big)&&\cos\tilde\varphi_\bp\big]\,\pauma_y \, .
\end{alignat}
We defined the short notation for the first derivative $\varphi^\nu_\bp=\partial^{}_\nu\varphi^{}_\bp$, the second derivative $\varphi^{\alf\beta}_\bp=\partial^{}_\alf\partial^{}_\beta\varphi^{}_\bp$ and equivalently for $\theta_\bp$. Note that $\dX^{\alf\beta}_\bp$ and $\dY^{\alf\beta}_\bp$ are symmetric in $\alf\leftrightarrow \beta$. Combining all results, the second-order interband matrix $\dF^{\alf\beta}_\bp$ reads
\begin{align}
 \dF^{\alf\beta}_\bp=i\big[\cE^\alf_\bp,\cX^\beta_\bp\big]+i\big[\cE^\beta_\bp,\cX^\alf_\bp\big]+i\big[\cE^{}_\bp,\dX^{\alf\beta}_\bp\big]+i\big[\cE^\alf_\bp,\cY^\beta_\bp\big]+i\big[\cE^\beta_\bp,\cY^\alf_\bp\big]+i\big[\cE^{}_\bp,\dY^{\alf\beta}_\bp\big] \, .
\end{align}
%
It is symmetric in $\alf\leftrightarrow\beta$ as expected. Note that $\dF^{\alf\beta}_\bp$ is gauge dependent due to $\tilde\varphi_\bp=\varphi+\phi^+_\bp-\phi^-_\bp$ in $\cX^\nu_\bp$, $\cY^\nu_\bp$, $\dX^{\alf\beta}_\bp$ and $\dY^{\alf\beta}_\bp$. Thus, $\dF^{\alf\beta}_\bp$ is not a good physical quantity by itself. The combination $\cF^\alf_\bp\dF^{\beta\gamma}_\bp$ is gauge independent for arbitrary indices $\alf$, $\beta$ and $\gamma$, which can be seen in analogy to \eqref{eqn:FF}. We can identify a particular combination as derivative of the quantum metric factor and the Berry curvature
\begin{align}
 &\frac{1}{2}\big[\big(\cX^\alf_\bp\,\dX^{\beta\gamma}_\bp+\cY^\alf_\bp\,\dY^{\beta\gamma}_\bp\big)+\big(\alf\leftrightarrow\beta\big)\big]= \frac{1}{4}\big(\partial^{}_\gamma \,\cC^{\alf\beta}_\bp\big) \, ,\\
 &\frac{1}{2}\big[\big(\cX^\alf_\bp\,\dY^{\beta\gamma}_\bp+\cY^\alf_\bp\,\dX^{\beta\gamma}_\bp\big)-\big(\alf\leftrightarrow\beta\big)\big]=\frac{i}{4}\big(\partial^{}_\gamma\,\Omega^{\alf\beta}_\bp\big) \, .
\end{align}
Note that the results, which we derived above for the general Bloch Hamiltonian in \eqref{eqn:lamPolar}, drastically simplify for a constant angle $\varphi_\bp=\varphi$, that is, for a momentum-independent phase of the coupling between the two subsystems of the two-band model $\Delta_\bp$.

%
%

\section{Longitudinal and anomalous Hall conductivity}
\label{sec:theory:conductivity}

After the introduction of our general two-band model, the conductivity formulas in imaginary-time formalism and several fundamental concepts in the last two sections, we continue by deriving the longitudinal and the anomalous Hall conductivity. Therefore, we assume only a uniform electric field and no magnetic field in the following. An electric field with frequency $\omega$ in direction $\beta=x,y,z$ induces a current in direction $\alf=x,y,z$. The proportionality is described by the conductivity tensor $\sigma^{\alf\beta}(\omega)\equiv\sigma^{\alf\beta}_\text{\tiny E}(\omega)$ in \eqref{eqn:sE}, where we omit the lower index for shorter notation in the following. The tensor describes two conceptional different phenomena: the longitudinal conductivity and the anomalous Hall conductivity. In a simplified picture, the difference between those conductivities is whether the current is induced parallel or transverse to the applied electric field. A more precise definition is given in the derivation. In the following presentation, we have a special focus on a unique and physically meaningful decomposition of the conductivity formulas. 

This section is structured as follows: In the first three subsections, we present the derivation of the conductivity formulas, which we analyze in the last four subsections. We use the matrix structure and properties of the matrix trace to decompose the polarization tensor (Sec.~\ref{sec:theory:conductivity:decomposition}). After having performed the Matsubara summation (Sec.~\ref{sec:theory:conductivity:matsubara}), we identify the distinct form of the different contributions, which results in our main formulas for the DC conductivity (Sec.~\ref{sec:theory:conductivity:formulas}). We relate our result to the commonly used approach by Bastin and St\v{r}eda (Sec.~\ref{sec:theory:conductivity:BastinStreda}), which can be seen as a different derivation. We discuss the reduction of the computational effort due to the used decomposition (Sec.~\ref{sec:theory:conductivity:basis}) and, in detail, the limits of small and large relaxation rate, and the low temperature limit (Sec.~\ref{sec:theory:conductivity:limits}). Finally, we relate our result to the anomalous Hall effect, discuss the possibility of anisotropic longitudinal conductivity and quantization (Sec.~\ref{sec:theory:conductivity:anomalousHall}).

%
%

\subsection{Decomposition}
\label{sec:theory:conductivity:decomposition}

With the general concepts that we derived in the previous Sec.~\ref{sec:theory:fundamentalconcepts}, we evaluate the polarization tensor $\Pi^{\alf\beta}_{iq_0}\equiv \Pi^{\alf\beta}_{\text{\tiny E},iq_0}$ given in \eqref{eqn:PiE}, where we omit the lower index for shorter notation. Using the invariance under unitary transformations of the matrix trace as well as the Green's function \eqref{eqn:Gdiag} and first-order vertex matrices \eqref{eqn:UdagLamU} expressed in the eigenbasis, we obtain
\begin{align}
 \Pi^{\alf\beta}_{iq_0}=\Tr\big[\cG^{}_{ip_0+iq_0,\bp}\big(\cE^\beta_\bp+\cF^\beta_\bp\big)\cG^{}_{ip_0,\bp}\big(\cE^\alf_\bp+\cF^\alf_\bp\big)\big] \, .
\end{align}
We have introduced the compact notation $\Tr[\,\cdot\,]=e^2TL^{-1}\sum_p\tr[\,\cdot\,-(iq_0=0)]$, where the dot $\cdot$ indicates the argument over which the trace is performed. The compact notation involves the prefactors, the summation over Matsubara frequencies and momenta, the matrix trace as well as the subtraction of the argument at $iq_0=0$. The Green's function matrices \eqref{eqn:Gdiag} are diagonal, whereas the vertices \eqref{eqn:UdagLamU} contain the diagonal matrix $\cE^\alf_\bp$ and the off-diagonal matrix $\cF^\alf_\bp$. The matrix trace only gives a nonzero contribution if the product of the four matrices involves an even number of off-diagonal matrices, that is, zero or two in this case. Thus, the mixed terms involving both $\cE^\alf_\bp$ and $\cF^\alf_\bp$ vanish. This leads to the decomposition of $\Pi^{\alf\beta}_{iq_0}$ into an {\it intraband} and an {\it interband contribution}:
\begin{align}
 \label{eqn:DecompIntraInter}
 \Pi^{\alf\beta}_{iq_0}=\Pi^{\alf\beta}_{iq_0,\text{intra}}+\Pi^{\alf\beta}_{iq_0,\text{inter}} \,.
\end{align}
In the intraband contribution, the two eigensystems $\pm_\bp$ are not mixed, whereas they mix in the interband contribution due to the interband matrix $\cF^\alf_\bp$. The individual contributions in \eqref{eqn:DecompIntraInter} are gauge independent due to \eqref{eqn:FF}. 

The matrix trace is invariant under transposition of the matrix, of which the trace is performed. For the product of several symmetric and antisymmetric (or skew-symmetric) matrices $A,\,B,\,C,\,D$ this leads to $\tr\big(ABCD\big)\hspace{-0.7mm}=\hspace{-0.6mm}\tr\big(D^\text{T}C^\text{T}B^\text{T}A^\text{T}\big)\hspace{-0.7mm}=\hspace{-0.7mm}(-1)^{n}\,\hspace{-0.1mm}\tr\big(DCBA\big)$
with $A^\text{T}$ being the transposed matrix of $A$, and so on, and $n$ the number of antisymmetric matrices involved. We call the procedure 
\begin{align}
 \label{eqn:TraceTrans}
 \tr\big[M_1...M_n\big]\rightarrow\tr\big[M_n...M_1\big]
\end{align}
with arbitrary square matrices $M_i$ ``trace transposition'' or ``reversing the matrix order under the trace'' in the following \cite{Mitscherling2018}. We call the trace that remains equal with a positive overall sign after trace transposition {\it symmetric} and a trace that remains equal up to a negative overall sign after trace transposition {\it antisymmetric}. Every trace of arbitrary square matrices can be uniquely decomposed in this way. We analyze the intra- and interband contribution in \eqref{eqn:DecompIntraInter} with respect to their behavior under trace transposition. The intraband contribution involves the quasiparticle velocities $\cE^\alf_\bp$ and the Green's functions, that is
\begin{align}
\label{eqn:intra}
 \Pi^{\alf\beta}_{iq_0,\text{intra}}=\Tr\big[\cG^{}_{ip_0+iq_0,\bp}\cE^\beta_\bp \cG^{}_{ip_0,\bp}\cE^\alf_\bp\big] \, .
\end{align}
All matrices are diagonal and, thus, symmetric. We see that the intraband contribution is symmetric under trace transposition. 
The interband contribution involves diagonal Green's functions and $\cF^\alf_\bp$, which is neither symmetric nor antisymmetric. We decompose it into its symmetric and antisymmetric part
\begin{align}
 &\cF^{\alf,s}_\bp=\frac{1}{2}\big(\cF^\alf_\bp+(\cF^\alf_\bp)^\text{T}\big)=i\big[\cE_\bp,\cY^\alf_\bp\big] \, ,\\
 &\cF^{\alf,a}_\bp=\frac{1}{2}\big(\cF^\alf_\bp-(\cF^\alf_\bp)^\text{T}\big)=i\big[\cE_\bp,\cX^\alf_\bp\big] \, .
\end{align}
%
By this, the interband contribution decomposes into a symmetric and antisymmetric contribution under trace transposition,
\begin{align}
 \Pi^{\alf\beta}_{iq_0,\,\text{inter}}=\Pi^{\alf\beta,\text{s}}_{iq_0,\,\text{inter}}+\Pi^{\alf\beta,\text{a}}_{iq_0,\,\text{inter}} \,,
\end{align}
where
\begin{align}
\label{eqn:inter_sym}
 \Pi^{\alf\beta,\text{s}}_{iq_0,\,\text{inter}}&=\Tr\big[4r_\bp^2\cG^{}_{ip_0+iq_0,\bp}\cX^\beta_\bp \cG^{}_{ip_0,\bp}\cX^\alf_\bp\big]+\Tr\big[4r_\bp^2\cG^{}_{ip_0+iq_0,\bp}\cY^\beta_\bp \cG^{}_{ip_0,\bp}\cY^\alf_\bp\big]\,,\\[1mm]
\label{eqn:inter_antisym}
 \Pi^{\alf\beta,\text{a}}_{iq_0,\,\text{inter}}&=\Tr\big[4r_\bp^2\cG^{}_{ip_0+iq_0,\bp}\cX^\beta_\bp \cG^{}_{ip_0,\bp}\cY^\alf_\bp\big]+\Tr\big[4r_\bp^2\cG^{}_{ip_0+iq_0,\bp}\cY^\beta_\bp \cG^{}_{ip_0,\bp}\cX^\alf_\bp\big]\,.
\end{align} 
We used $\cE_\bp=\g_\bp+r_\bp\pauma_z$ and performed the commutator explicitly. Interestingly, the symmetry under trace transposition, which is due to the multiband character, is connected to the symmetry of the polarization tensor or, equivalently, of the conductivity tensor $\sigma=(\sigma^{\alf\beta})$ itself: Trace transposition of \eqref{eqn:intra}, \eqref{eqn:inter_sym} and \eqref{eqn:inter_antisym} is equal to the exchange of $\alf\leftrightarrow\beta$, the directions of the current and the external electric field.


%
%

\subsection{Matsubara summation}
\label{sec:theory:conductivity:matsubara}

We continue by performing the Matsubara summations and the analytic continuation. The sum over the (fermionic) Matsubara frequency $p_0$ in \eqref{eqn:intra}, \eqref{eqn:inter_sym} and \eqref{eqn:inter_antisym} is of the form
\begin{align}
\label{eqn:MatsumE}
 I_{iq_0}\equiv T\sum_{p_0}\tr\big[(\cG_{iq_0}-\cG)M_1\cG M_2\big]
\end{align}
with two matrices $M_1$ and $M_2$ that are symmetric and/or antisymmetric. $T$ is the temperature. We omit the momentum dependence for simplicity in this subsection. We further shorten the notation of the Green's functions $\cG \equiv \cG_{ip_0}$ and $\cG_{\pm iq_0}\equiv \cG_{ip_0\pm iq_0}$. If $I_{iq_0}$ is symmetric under trace transposition, that is, for the intraband and the symmetric interband contribution, we split \eqref{eqn:MatsumE} into two equal parts. In the second part, we reverse the matrix order under the trace and shift the Matsubara summation $ip_0\rightarrow ip_0-iq_0$. We get
\begin{align}
\label{eqn:Isq0}
 &I^\text{s}_{iq_0}=\frac{T}{2}\sum_{p_0}\tr\big[\big((\cG_{iq_0}-\cG)+(\cG_{-iq_0}-\cG)\big)M_1\cG M_2\big] \, .
\end{align}
If $I_{iq_0}$ is antisymmetric, that is, for the antisymmetric interband contribution, we obtain
\begin{align}
 \label{eqn:Iaq0}
 &I^{\text{a}}_{iq_0}=\frac{T}{2}\sum_{p_0}\tr\big[(\cG_{iq_0}-\cG_{-iq_0})M_1\cG M_2\big] 
\end{align}
after the same steps. We perform the Matsubara summation and analytic continuation $iq_0\rightarrow \omega+i0^+$ of the external frequency leading to $I^\text{s}_\omega$ and $I^\text{a}_\omega$. We are interested in the zero-frequency (DC) limit. The detailed Matsubara summation and the zero frequency limit are performed in Appendix~\ref{appendix:Matsubara}. We end up with
\begin{align}
\label{eqn:Isw}
 &\lim_{\omega\rightarrow 0}\frac{I^\text{s}_\omega}{i\omega}=\frac{\pi}{2}\hspace{-1mm}\int\hspace{-1mm}d\eps\, f_\epsilon'\,\hspace{0.5mm}\tr\big[A^{}_\epsilon \,M^{}_1 \,A^{}_\epsilon\, M^{}_2+A^{}_\epsilon \,M^{}_2 \,A^{}_\epsilon \,M^{}_1\big] \, ,\\
 \label{eqn:Iaw}
 &\lim_{\omega\rightarrow 0}\frac{I^\text{a}_\omega}{i\omega}=-i\hspace{-1mm}\int\hspace{-1.5mm} d\eps \,f_\epsilon\,\hspace{0.2mm} \tr\big[P'_\epsilon\, M^{}_1\,A^{}_\epsilon\, M^{}_2-P'_\epsilon \,M^{}_2\,A^{}_\epsilon\, M^{}_1\big],
\end{align}
where $f_\epsilon = (e^{\epsilon/T}+1)^{-1}$ is the Fermi function and $f'_\epsilon$ its derivative. Furthermore, it involves the spectral function matrix $A_\epsilon=-(\cG^R_\epsilon-\cG^A_\epsilon)/2\pi i$ and the derivative of the principle-value function matrix $P'_\epsilon=\partial_\eps(\cG^R_\epsilon+\cG^A_\epsilon)/2$, where $\cG^R_\eps$ and $\cG^A_\eps$ are the retarded and advanced Green's function matrices, respectively.

In \eqref{eqn:Isw} and \eqref{eqn:Iaw}, we exclusively used the spectral function $A_\eps$ and the principle-value function $P_\eps$, which are both real-valued functions, and avoided the complex-valued retarded or advanced Green's functions. As we have a real-valued DC conductivity, the combination of $M_1$ and $M_2$ has to be purely real in \eqref{eqn:Isw} and purely complex in \eqref{eqn:Iaw}. The symmetric part \eqref{eqn:Isw} involves the derivative of the Fermi function $f'_\epsilon$, whereas the antisymmetric part \eqref{eqn:Iaw} involves the Fermi function $f_\epsilon$. This suggests to call the latter one the {\it Fermi-surface contribution} and the former one the {\it Fermi-sea contribution}. However, this distinction is not unique, since we can perform partial integration in the internal frequency $\epsilon$. For instance, the decomposition proposed by St\v{r}eda \cite{Streda1982} is different. 

Using the explicit form of the Green's function in \eqref{eqn:Gdiag}, the spectral function matrix reads 
\begin{align}
\label{eqn:AM}
 A_\epsilon = \begin{pmatrix} A^+_\epsilon && 0 \\[1mm] 0 && A^-_\epsilon \end{pmatrix}
\end{align}
with the spectral functions of the two quasiparticle bands
\begin{align}
\label{eqn:Apm}
 A^\pm_\epsilon=\frac{\Gamma/\pi}{(\epsilon+\mu-E^\pm_\bp)^2+\Gamma^2} \, .
\end{align}
For our specific choice of $\Gamma$, the spectral function is a Lorentzian function, which peaks at $E^\pm_\bp-\mu$ for small $\Gamma$. Using \eqref{eqn:Apm}, the derivative of the principle-value function $P'_\epsilon$ can be rewritten in terms of the spectral function as
\begin{align}
\label{eqn:Pprime}
 P'_\epsilon=2\pi^2 A^2_\epsilon-\frac{\pi}{\Gamma}A_\epsilon \,.
\end{align}
When inserting this into \eqref{eqn:Iaw} the second, linear term drops out. We see that \eqref{eqn:Isw} and \eqref{eqn:Iaw} can be completely expressed by combinations of quasiparticle spectral functions. Note that \eqref{eqn:Pprime} is valid only for a relaxation rate $\Gamma$ that is frequency-independent as well as proportional to the identity matrix.

We apply the result of the Matsubara summation \eqref{eqn:Isw} and \eqref{eqn:Iaw} to the symmetric and antisymmetric interband contributions \eqref{eqn:inter_sym} and \eqref{eqn:inter_antisym}. Since $M_1$ and $M_2$ are off-diagonal matrices in both cases, the commutation with the diagonal spectral function matrix $A_\eps$ simply flips its diagonal entries, that is $M_iA_\eps=\overline{A}_\eps M_i$ where $\overline{A}_\eps$ is given by \eqref{eqn:AM} with $A^+_\eps\leftrightarrow A^-_\eps$ exchanged. We collect the product of involved matrices and identify
\begin{align}
 &A_\eps\,\big(\cX^\beta\cX^\alf+\cX^\alf\cX^\beta+\cY^\beta\cY^\alf+\cY^\alf\cY^\beta\big)\,\overline{A}_\eps
 =\hspace{2mm}A_\eps\,C^{\alf\beta}\,\overline{A}_\eps \, ,\\[2mm]
 &A^2_\eps\,\big(\cX^\beta\cY^\alf-\cY^\alf\cX^\beta+\cY^\beta\cX^\alf-\cX^\alf\cY^\beta\big)\,\overline{A}_\eps
 =i\,A^2_\eps\,\Omega^{\alf\beta}\,\overline{A}_\eps \, ,
\end{align}
where $\cC^{\alf\beta}_\bp$ and $\Omega^{\alf\beta}$ were defined in \eqref{eqn:CM} and \eqref{eqn:OmegaM}.

%
%

\subsection{Formulas of the conductivity tensor}
\label{sec:theory:conductivity:formulas}

As the final step of the derivation, we combine all our results. The conductivity and the polarization tensor are related via \eqref{eqn:sE}. We write out the trace over the two quasiparticle bands explicitly. The zero-frequency (DC) conductivity $\sigma^{\alf\beta}\equiv\sigma^{\alf\beta}_\text{\tiny E}$ decomposes into five different contributions: 
\begin{align}
\label{eqn:DecompSigma}
 \sigma^{\alf\beta}&=\sigma^{\alf\beta}_{\text{intra},+}+\sigma^{\alf\beta}_{\text{intra},-}+\sigma^{\alf\beta,\text{s}}_\text{inter}+\sigma^{\alf\beta,\text{a}}_{\text{inter},+}+\sigma^{\alf\beta,\text{a}}_{\text{inter},-} \, .
\end{align}
These contributions are distinct by three categories: (a) intra- and interband, (b) symmetric and antisymmetric with respect to $\alf\leftrightarrow\beta$ (or, equivalently, with respect to trace transposition) and (c) quasiparticle band $\pm$. As the symmetric interband contribution $\sigma^{\alf\beta,s}_\text{inter}$ is shown to be symmetric in $+\leftrightarrow -$ for our two-band model, we dropped the band index for simplicity. Each contribution consists of three essential parts: i) the Fermi function $f(\epsilon)$ or its derivative $f'(\epsilon)$, ii) a spectral weighting factor involving a specific combination of the quasiparticle spectral functions $A^n_\bp(\epsilon)$ of quasiparticle bands $n=\pm$, that is,
\begin{align}
 \label{eqn:Wintra}&w^n_{\bp,\text{intra}}(\epsilon)=\pi\big(A^n_\bp(\epsilon)\big)^2 \, ,\\[2mm]
 \label{eqn:Wsinter}&w^s_{\bp,\text{inter}}(\epsilon)=4\pi r^2_\bp A^+_\bp(\epsilon)A^-_\bp(\epsilon) \, ,\\[2mm]
 \label{eqn:Wainter}&w^{a,n}_{\bp,\text{inter}}(\epsilon)=8\pi^2r^2_\bp \big(A^n_\bp(\epsilon)\big)^2A^{-n}_\bp(\epsilon)\, ,
\end{align}
with $-n$ denoting the opposite band, and iii) a momentum-dependent weighting factor involving the quasiparticle velocities $E^{\pm,\alf}_\bp$, the quantum metric factor $C^{\alf\beta}_\bp$ or the Berry curvatures $\Omega^{\alf\beta,\pm}_\bp$ given as 
\begin{align}
 &E^{\pm,\alf}_\bp=\g^\alf_\bp\pm r^\alf_\bp \, ,\\[2mm]
 &C^{\alf\beta}_\bp=\frac{1}{2}\big(\theta^\alf_\bp\theta^\beta_\bp+\varphi^\alf_\bp\varphi^\beta_\bp\sin^2\theta_\bp\big) \, ,\\[1mm]
 \label{eqn:Omega}&\Omega^{\alf\beta,\pm}_\bp=\pm \frac{1}{2}\big(\varphi^\alf_\bp\theta^\beta_\bp-\varphi^\beta_\bp\theta^\alf_\bp\big)\sin\theta_\bp \, ,
\end{align}
where $\g^\alf_\bp=\partial_\alf \g_\bp$, $r^\alf_\bp=\partial_\alf r_\bp$, $\theta^\alf_\bp=\partial_\alf \theta_\bp$ and $\varphi^\alf_\bp=\partial_\alf \varphi_\bp$ with the momentum derivative in $\alf$ direction $\partial_\alf=\frac{\partial}{\partial p^\alf}$. The conductivity is expressed in units of the conductance quantum $2\pi\sigma_0=e^2/\hbar$ by writing $\hbar$ explicitly, which is set to unity in the rest of this thesis. We perform the thermodynamic limit by replacing $L^{-1}\sum_\bp\rightarrow \int \frac{d^d\bp}{(2\pi)^d}$, where $d$ is the dimension of the system. We end up with  
\begin{align}
 &\sigma^{\alf\beta}_{\text{intra},n}\hspace{-1.0mm}=\hspace{-0mm}-\frac{e^2}{\hbar}\hspace{-1.5mm}\int\hspace{-1.9mm}\frac{d^d\bp}{(2\pi)^d}\hspace{-1.7mm}\int\hspace{-1.5mm}d\epsilon \,f'(\epsilon) w^n_{\bp,\text{intra}}(\epsilon) E^{n,\alf}_\bp E^{n,\beta}_\bp\hspace{-1mm}, \label{eqn:SintraN}
 \\[2mm]
 &\sigma^{\alf\beta,\text{s}}_\text{inter}\hspace{1.5mm}=\hspace{-0.2mm}-\frac{e^2}{\hbar}\hspace{-1.5mm}\int\hspace{-1.9mm}\frac{d^d\bp}{(2\pi)^d}\hspace{-1.7mm}\int\hspace{-1.5mm}d\epsilon\,f'(\epsilon)w^s_{\bp,\text{inter}}(\epsilon) \,C^{\alf\beta}_\bp\,, \label{eqn:SinterS}
 \\[2mm]
 &\sigma^{\alf\beta,\text{a}}_{\text{inter},n}\hspace{-1mm}=\hspace{-0.2mm}-\frac{e^2}{\hbar}\hspace{-1.5mm}\int\hspace{-1.9mm}\frac{d^d\bp}{(2\pi)^d}\hspace{-1.7mm}\int\hspace{-1.5mm}d\epsilon\,\,f(\epsilon)\, w^{a,n}_{\bp,\text{inter}}(\epsilon)\,\Omega^{\alf\beta,n}_\bp\,. \label{eqn:SinterAN}
\end{align}
If we restore SI units, the conductivity has units $1/\Omega\,\text{m}^{d-2}$ for dimension $d$. Note that we have $\sigma^{\alf\beta}\propto e^2/h$ in a two-dimensional system and $\sigma^{\alf\beta}\propto e^2/ha$ in a stacked quasi-two-dimensional system, where $a$ is the interlayer distance. For given $\lam_\bp$, $\mu$, $T$ and $\Gamma$ the evaluation of \eqref{eqn:SintraN}, \eqref{eqn:SinterS} and \eqref{eqn:SinterAN} is straightforward. The mapping of $\lam_\bp$ to spherical coordinates is given in \eqref{eqn:gh}-\eqref{eqn:Phi}. The spectral function $A^\pm_\bp(\epsilon)$ is defined in \eqref{eqn:Apm}.

%
%

\subsection{Relation to the Bastin and the St\v{r}eda formula}
\label{sec:theory:conductivity:BastinStreda}

Microscopic approaches to the anomalous Hall conductivity are frequently based on the formulas of Bastin {\it et al.} \cite{Bastin1971} and St\v{r}eda \cite{Streda1982}. A modern derivation is given by Cr\'epieux {\it et al.}  \cite{Crepieux2001}. We present a different derivation that follows the steps of Bastin {\it et al.} \cite{Bastin1971} in our notation and discuss the relation to our results. We omit the momentum dependence for a simpler notation in this section. 

We start with the polarization tensor $\Pi^{\alf\beta}_{iq_0}$ in \eqref{eqn:PiE} before analytic continuation. In contrast to our discussion above, we perform the Matsubara sum and the analytic continuation in \eqref{eqn:sE} immediately and get
\begin{align}
 \sigma^{\alf\beta}_\omega=-\frac{1}{i\omega}\Tr_{\eps,\bp}\big[f_\eps\, \big(\sA^{}_\eps\lam^\beta \sG^A_{\eps-\omega}\lam^\alf+\sG^R_{\eps+\omega}\lam^\beta \sA^{}_\eps \lam^\alf -\sA_\eps \lam^\beta \sP_\eps \lam^\alf-\sP_\eps\lam^\beta \sA_\epsilon \lam^\alf\big)\big] \, .
\end{align}
We combined the prefactors, the summation over momenta and the frequency integration as well as the matrix trace in the short notation $\Tr_{\eps,\bp}\big[\cdot\big]=e^2L^{-1}\sum_\bp\int d\eps\,\tr\big[\cdot\big]$, where the dot $\cdot$ indicates the argument. The first two and the last two terms are obtained by the argument explicitly given in \eqref{eqn:PiE} and its $(iq_0=0)$ contribution, respectively. Details of the Matsubara summation and the analytic continuation are given in Appendix \ref{appendix:Matsubara}. $\sG^R_\eps$ and $\sG^A_\eps$ are the retarded and advanced Green's function of \eqref{eqn:Green}, respectively.  $\sA_\epsilon=-(\sG^R_\eps-\sG^A_\eps)/2\pi i$ is the spectral function matrix and $\sP_\eps=(\sG^R_\eps+\sG^A_\eps)/2$ is the principle-value function matrix. $f_\eps$ is the Fermi function. 

We derive the zero-frequency (DC) limit by expanding $\sigma^{\alf\beta}_\omega$ in the frequency $\omega$ of the external electric field $\bE(\omega)$. The diverging term $\propto 1/\omega$ vanishes, which can be checked by using $\sG^R_\epsilon=\sP_\epsilon-i\pi \sA_\epsilon$ and $\sG^A_\epsilon=\sP_\epsilon+i\pi \sA_\epsilon$. The constant term is
\begin{align}
\label{eqn:Bastin}
 \sigma^{\alf\beta}_\text{Bastin}=i\,\Tr_{\eps,\bp}\big[f_\epsilon&\, \big(-\sA_\epsilon\lam^\beta (\sG^A_\epsilon)'\lam^\alf+(\sG^R_\epsilon)'\lam^\beta \sA_\epsilon \lam^\alf\big)\big]\, ,
\end{align}
which was derived by Bastin {\it et al.} \cite{Bastin1971}. The derivative with respect to the internal frequency $\epsilon$ of the retarded and advanced Green's function matrices is denoted by $\big(\sG^{R/A}_\epsilon\big)'=\partial_\eps \sG^{R/A}_\epsilon$, respectively. The expression in \eqref{eqn:Bastin} is written in the subsystem basis, in which we expressed the Bloch Hamiltonian $\lam_\bp$ in \eqref{eqn:lam}. Due to the matrix trace, we can change to the diagonal basis via \eqref{eqn:Gdiag} and \eqref{eqn:UdagLamU}.

In Sec.~\ref{sec:theory:conductivity:decomposition}, we identified the symmetry under exchange of $\alf\leftrightarrow\beta$ as a good criterion for a decomposition. The Bastin formula is neither symmetric nor antisymmetric in $\alf\leftrightarrow\beta$. When we decompose $\sigma^{\alf\beta}_\text{Bastin}$ into its symmetric and antisymmetric part, we can easily identify our result \eqref{eqn:DecompSigma}, that is,
\begin{align}
 &\frac{1}{2}\big(\sigma^{\alf\beta}_\text{Bastin}+\sigma^{\beta\alf}_\text{Bastin}\big)=\sigma^{\alf\beta}_{\text{intra},+}+\sigma^{\alf\beta}_{\text{intra},-}+\sigma^{\alf\beta,s}_\text{inter} \, , \label{eqn:BastinSym}\\
 &\frac{1}{2}\big(\sigma^{\alf\beta}_\text{Bastin}-\sigma^{\beta\alf}_\text{Bastin}\big)=\sigma^{\alf\beta,a}_{\text{inter},+}+\sigma^{\alf\beta,a}_{\text{inter},-} \label{eqn:BastinAntiSym} \, .
\end{align}
This identification is expected as the decomposition into the symmetric and antisymmetric part is unique. We note that this separation naturally leads to a Fermi-surface \eqref{eqn:BastinSym} and a Fermi-sea contribution \eqref{eqn:BastinAntiSym} of the same form that we defined in Sec.~\ref{sec:theory:conductivity}. Based on our derivation, we argue that we should see the symmetry under $\alf\leftrightarrow\beta$ as the fundamental difference between \eqref{eqn:BastinSym} and \eqref{eqn:BastinAntiSym} instead of the property involving $f_\eps$ or $f'_\eps$. 

The Bastin formula \eqref{eqn:Bastin} is the starting point for the derivation of the St\v{r}eda formula \cite{Streda1982,Crepieux2001}. We split $\sigma^{\alf\beta}_\text{Bastin}$ into two equal parts and perform partial integration in the internal frequency $\eps$ on the latter one. We obtain
\begingroup
 \allowdisplaybreaks[0]
\begin{align}
 \sigma^{\alf\beta}_\text{Bastin}=\,\,&\frac{i}{2}\Tr_{\eps,\bp}\big[ \, f_\epsilon\, \big(-\sA_\epsilon\lam^\beta (\sG^A_\epsilon)'\lam^\alf+(\sG^R_\epsilon)'\lam^\beta \sA_\epsilon \lam^\alf\big)\big]  \\
 -&\frac{i}{2}\Tr_{\eps,\bp}\big[ \, f'_\epsilon\, \big(-\sA_\epsilon\lam^\beta \sG^A_\epsilon\lam^\alf+\sG^R_\epsilon\lam^\beta \sA_\epsilon \lam^\alf\big)\big] \\
 -&\frac{i}{2}\Tr_{\eps,\bp}\big[ \, f_\epsilon\, \big(-\sA'_\epsilon\lam^\beta \sG^A_\epsilon\lam^\alf+\sG^R_\epsilon\lam^\beta \sA'_\epsilon \lam^\alf\big)\big].
\end{align}
\endgroup
We replace the spectral function by its definition $\sA_\epsilon=-(\sG^R_\epsilon-\sG^A_\epsilon)/2\pi i$ and sort by $f_\eps$ and $f'_\eps$. By doing so, the St\v{r}eda formula decomposes into two contributions, historically labeled as
\begin{align}
\label{eqn:DecompStreda}
 \sigma^{\alf\beta}_\text{Streda}=\sigma^{\alf\beta,I}_\text{Streda}+\sigma^{\alf\beta,II}_\text{Streda}
\end{align}
with the ``Fermi-surface contribution''
\begin{align}
 &\sigma^{\alf\beta,I}_\text{Streda}=\frac{1}{4\pi}\Tr_{\eps,\bp}\big[ \, f'_\epsilon\, \big(-(\sG^R_\epsilon-\sG^A_\epsilon)\lam^\beta \sG^A_\epsilon \lam^\alf+G^R_\epsilon\lam^\beta (\sG^R_\epsilon-\sG^A_\epsilon) \lam^\alf\big)\big] \, ,\label{eqn:StredaI}
\end{align}
and the ``Fermi-sea contribution''
\begin{align}
 \sigma^{\alf\beta,II}_\text{Streda}=-\frac{1}{4\pi}\Tr_{\eps,\bp}\big[ \, f_\epsilon\, \big(&\sG^A_\epsilon\lam^\beta (\sG^A_\epsilon)'\lam^\alf-(\sG^A_\epsilon)'\lam^\beta \sG^A_\epsilon \lam^\alf\nonumber\\+&(\sG^R_\epsilon)'\lam^\beta \sG^R_\epsilon\lam^\alf-\sG^R_\epsilon\lam^\beta (\sG^R_\epsilon)' \lam^\alf\big)\big] \, . \label{eqn:StredaII}
\end{align}
The decomposition \eqref{eqn:DecompStreda} explicitly shows the ambiguity in the definition of Fermi-sea and Fermi-surface contributions due to the possibility of partial integration in the internal frequency $\eps$. Following our distinction by the symmetry with respect to $\alf\leftrightarrow\beta$, we notice that the second contribution \eqref{eqn:StredaII} is antisymmetric, whereas the first contribution \eqref{eqn:StredaI} is neither symmetric nor antisymmetric. If we decompose \eqref{eqn:StredaI} into its symmetric and antisymmetric part and combine the latter one with \eqref{eqn:StredaII}, we recover our findings
\begin{align}
 &\frac{1}{2}\big(\sigma^{\alf\beta,I}_\text{Streda}+\sigma^{\beta\alf,I}_\text{Streda}\big)=\sigma^{\alf\beta}_{\text{intra},+}+\sigma^{\alf\beta}_{\text{intra},-}+\sigma^{\alf\beta,s}_\text{inter} \,, \\
 \label{eqn:StredaAntiSym}
 &\frac{1}{2}\big(\sigma^{\alf\beta,I}_\text{Streda}-\sigma^{\beta\alf,I}_\text{Streda}\big)+\sigma^{\alf\beta,II}_\text{Streda}=\sigma^{\alf\beta,a}_{\text{inter},+}\hspace{-0.3mm}+\hspace{-0.3mm}\sigma^{\alf\beta,a}_{\text{inter},-}\, ,
\end{align}
as expected by the uniqueness of this decomposition. We see that the antisymmetric interband contribution, which will be shown to be responsible for the anomalous Hall effect, is given by parts of St\v{r}eda's Fermi-surface and Fermi-sea contributions combined \cite{Kodderitzsch2015}. In the literature different parts of \eqref{eqn:StredaI} and \eqref{eqn:StredaII} are identified to be relevant when treating disorder effects via quasiparticle lifetime broadening or beyond \cite{Nagaosa2010, Sinitsyn2007, Crepieux2001, Dugaev2005, Onoda2006, Yang2006, Kontani2007, Nunner2008, Onoda2008, Tanaka2008, Kovalev2009, Streda2010, Pandey2012, Burkov2014, Chadova2015, Kodderitzsch2015, Mizoguchi2016}. Due to the mathematical uniqueness and the clear physical interpretation we propose \eqref{eqn:BastinAntiSym} or, equivalently, \eqref{eqn:StredaAntiSym} as a good starting point for further studies on the anomalous Hall conductivity.

%
%

\subsection{Basis choice and subsystem basis}
\label{sec:theory:conductivity:basis}

The polarization tensor $\Pi^{\alf\beta}_{iq_0}$ in \eqref{eqn:PiE} is the trace of a matrix and is, thus, invariant under unitary (or, more general, similarity) transformations of this matrix. In other words, the conductivities can be expressed within a different basis, for instance, the eigenbasis, which we used for the final formulas in \eqref{eqn:SintraN}-\eqref{eqn:SinterAN} in Sec.~\ref{sec:theory:conductivity:formulas}. The obvious advantage of the eigenbasis is that we can easily identify terms with clear physical interpretation like the quasiparticle spectral functions $A^\pm_\bp(\eps)$, the quasiparticle velocities $E^{\pm,\alf}_\bp$, the quantum metric factor $C^{\alf\beta}_\bp$, and the Berry curvature $\Omega^{\alf\beta,\pm}_\bp$. 

In general, we can use any invertible matrix $U_\bp$ and perform similar steps as we did in our derivation: In analogy to \eqref{eqn:Ep} and \eqref{eqn:Gdiag}, we obtain a transformed Bloch Hamiltonian matrix $\tilde \lam^{}_\bp=U^{-1}_\bp\lam^{}_\bp U^{}_\bp$ and a corresponding Green's function matrix. Reconsidering the steps in \eqref{eqn:UdagLamUDeriv}, we obtain a new decomposition \eqref{eqn:UdagLamU} of the velocity matrix with an analogue of the Berry-connection-like matrix in \eqref{eqn:BerryConnection}. We see that the following steps of decomposing the Berry-connection-like matrix, separating the involved matrices of the polarization tensor into their diagonal and off-diagonal parts and splitting the off-diagonal matrices into their symmetric and antisymmetric components under transposition are possible but lengthy. 

A special case is $U_\bp=\mathds{1}$, by which we express the conductivity in the subsystem basis, in which we defined the Bloch Hamiltonian $\lam_\bp$ in \eqref{eqn:lam}. Following the derivation in Sec.~\ref{sec:theory:conductivity:BastinStreda}, we obtain \eqref{eqn:Bastin}, which we further decompose into the symmetric and antisymmetric part with respect to $\alf\leftrightarrow\beta$, $\sigma^{\alf\beta}=\sigma^{\alf\beta,s}+\sigma^{\alf\beta,a}$. We obtain
\begingroup
 \allowdisplaybreaks[0]
\begin{align}
\label{eqn:SigmaSAB}
 &\sigma^{\alf\beta,s}=\,-\pi\,\Tr_{\eps,\bp}\big[f'_\eps \,\sA_\eps \,\lam^\beta \,\sA_\eps \,\lam^\alf\big]\,,\\[2mm]
 \label{eqn:SigmaAAB}
 &\sigma^{\alf\beta,a}=2\pi^2\, \Tr_{\eps,\bp}\big[f_\eps\,\big(\sA^2_\eps\, \lam^\beta \,\sA_\eps \,\lam^\alf\,-\,\sA_\eps\, \lam^\beta\, \sA^2_\eps\, \lam^\alf\big)\big]\, .
\end{align}
\endgroup
We replaced $\sP_\eps'$ by using \eqref{eqn:Pprime}. These expressions still involve the matrix trace. Obviously, an immediate evaluation of this trace without any further simplifications would produce very lengthy expressions. 

A major reduction of the effort to perform the matrix trace is the decomposition into symmetric and antisymmetric parts with respect to the trace transposition, which was defined in \eqref{eqn:TraceTrans}. We expand $\sA_\eps$, $\lam^\alf$ and $\lam^\beta$ into their diagonal and off-diagonal components, which we further decompose into parts proportional to $\pauma_x$ and $\pauma_y$. For instance in \eqref{eqn:SigmaSAB}, we obtain 81 combinations, where several combinations vanish by tracing an off-diagonal matrix. We get symmetric as well as antisymmetric contributions under trace transposition. However, the latter ones will eventually vanish due to the antisymmetry in $\alf\leftrightarrow\beta$. Similarly, the symmetric contributions under trace transposition will drop out in \eqref{eqn:SigmaAAB}.

%
%

\subsection{Small and large relaxation rate $\Gamma$ and low temperature}
\label{sec:theory:conductivity:limits}

In our derivation in Sec.~\ref{sec:theory:conductivity:decomposition} to Sec.~\ref{sec:theory:conductivity:formulas}, we did not assume any restrictions on the size of the relaxation rate $\Gamma$. Thus, the formulas \eqref{eqn:SintraN}-\eqref{eqn:SinterAN} are valid for a relaxation rate $\Gamma$ of arbitrary size. In the following, we discuss both the clean limit (small $\Gamma$) and the dirty limit (large $\Gamma$) analytically. We are not only interested in the limiting behavior of the full conductivity $\sigma^{\alf\beta}$ in \eqref{eqn:DecompSigma}, but also in the behavior of the individual contributions \eqref{eqn:SintraN}-\eqref{eqn:SinterAN}. The dependence on $\Gamma$ is completely captured by the three different spectral weighting factors $w^n_{\bp,\text{intra}}$, $w^s_{\bp,\text{inter}}$ and $w^{a,n}_{\bp,\text{inter}}$, which involve a specific product of quasiparticle spectral functions and are defined in \eqref{eqn:Wintra}-\eqref{eqn:Wainter}.

The spectral weighting factor of the intraband conductivities $w^n_{\bp,\text{intra}}$ in \eqref{eqn:Wintra} involves the square of the spectral function of the same band, $\big(A^n_\bp(\eps)\big)^2$, and, thus, peaks at the corresponding quasiparticle Fermi surface, which is defined by $E^n_\bp-\mu=0$, for small $\Gamma$. If $\Gamma$ is so small that the quasiparticle velocities $E^{\pm,\alf}_\bp$ are almost constant in a momentum range in which the variation of $E^\pm_\bp$ is of order $\Gamma$, we can approximate 
\begin{align}
 w^n_{\bp,\text{intra}}(\epsilon)\approx \frac{1}{2\Gamma}\delta(\epsilon+\mu-E^n_\bp)\sim \mathcal{O}(\Gamma^{-1})\,. \label{eqn:winG0}
\end{align}
Thus, the intraband conductivities $\sigma^{\alf\beta}_{\text{intra},\pm}$ diverge as $1/\Gamma$, consistent with Boltzmann transport theory \cite{Mahan2000}. 

\begin{figure}[t!]
\centering
\includegraphics[width=0.6\textwidth]{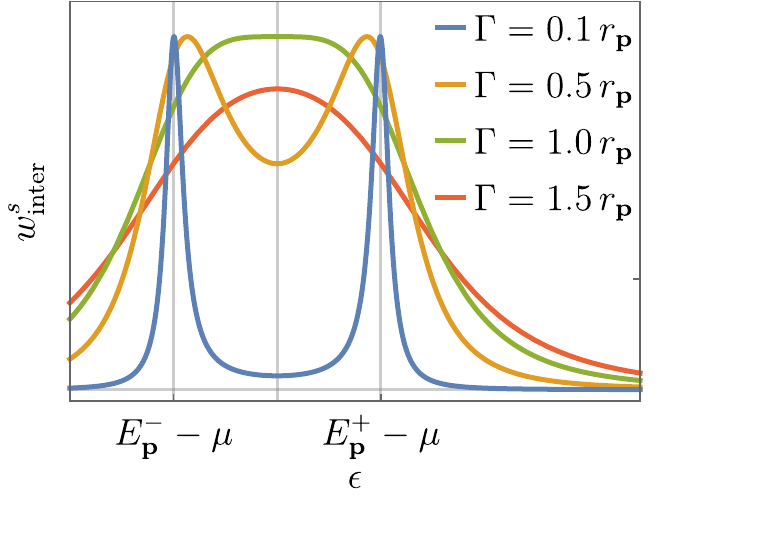}\\
\includegraphics[width=0.6\textwidth]{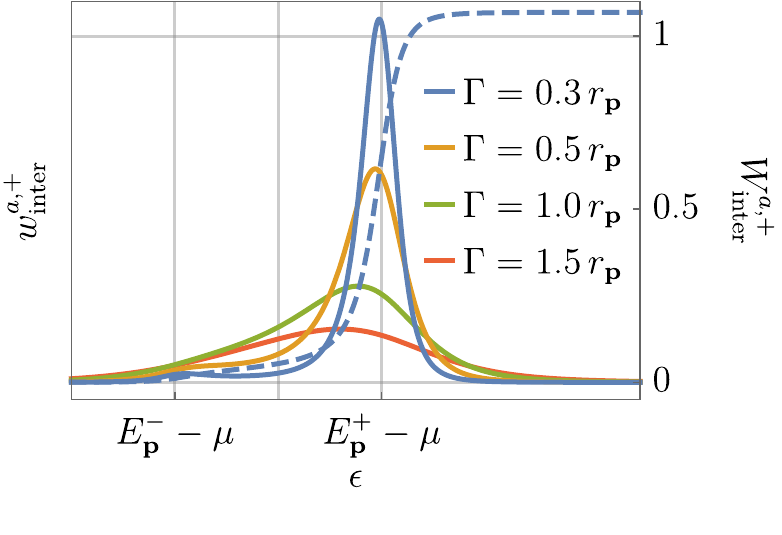}
\caption{The spectral weighting factors $w^{s}_{\bp,\text{inter}}$ (top) and $w^{a,+}_{\bp,\text{inter}}$ (bottom, solid), and its primitive $W^{a,+}_{\bp,\text{inter}}$ (bottom, dashed) for different choices of $\Gamma$. \label{fig:WInter}}
\end{figure}

The spectral weighting factor of the symmetric interband conductivity $w^s_{\bp,\text{inter}}$ in \eqref{eqn:Wsinter} is the product of the spectral functions of the two bands, $A^+_\bp(\eps)A^-_\bp(\eps)$. For small $\Gamma$, $w^s_{\bp,\text{inter}}$ peaks equally at the Fermi surface of both bands. For increasing $\Gamma$, the gap starts to fill up until the peaks merge and form one broad peak at $(E^+_\bp+E^-_\bp)/2-\mu=\g_\bp-\mu$. It decreases further for even larger $\Gamma$. Since each spectral function $A^n_\bp(\eps)$ has half width of $\Gamma$ at half the maximum value, the relevant scale for the crossover is $2\Gamma=E^+_\bp-E^-_\bp=2r_\bp$. We sketch $w^s_{\bp,\text{inter}}$ in Fig.~\ref{fig:WInter} for several choices of $\Gamma$. If the quantum metric factor $C^{\alf\beta}_\bp$ is almost constant in a momentum range in which the variation of $E^\pm_\bp$ is of order $\Gamma$ and, furthermore, if $\Gamma\ll r_\bp$, we can approximate  
\begin{align}
 w^s_{\bp,\text{inter}}(\epsilon)\approx\Gamma\sum_{n=\pm}\delta(\epsilon+\mu-E^n_\bp)\sim\mathcal{O}(\Gamma^1)\,. \label{eqn:wsG0}
\end{align}
We see that the symmetric interband conductivity $\sigma^{\alf\beta,s}_\text{inter}$ scales linearly in $\Gamma$ and is suppressed by a factor $\Gamma^2$ in the clean limit compared to the intraband conductivities. 

The spectral weighting factor of the antisymmetric interband conductivities $w^{a,n}_{\bp,\text{inter}}$ in \eqref{eqn:Wainter} is the square of the spectral function of one band multiplied by the spectral function of the other band, $\big(A^n_\bp(\eps)\big)^2A^{-n}_\bp(\eps)$. In the clean limit, it is dominated by a peak at $E^n_\bp-\mu$. For increasing $\Gamma$, the peak becomes asymmetric due to the contribution of the spectral function of the other band at $E^{-n}_\bp-\mu$ and develops a shoulder. For $2\Gamma\gg E^+_\bp-E^-_\bp=2r_\bp$, it eventually becomes one broad peak close to $(E^+_\bp+E^-_\bp)/2-\mu=\g_\bp-\mu$. We sketch $w^{a,+}_\text{inter}$ in Fig.~\ref{fig:WInter} for several choices of $\Gamma$. If the Berry curvature $\Omega^{\alf\beta,n}_\bp$ is almost constant in a momentum range in which the variation of $E^n_\bp$ is of order $\Gamma$ and, furthermore, if $\Gamma\ll r_\bp$, we can approximate 
\begin{align}
 w^{n,a}_{\bp,\text{inter}}(\epsilon)\approx \delta(\epsilon+\mu-E^n_\bp) \sim \mathcal{O}(\Gamma^0)\, . \label{eqn:wanG0}
\end{align}
Thus, the antisymmetric interband conductivities $\sigma^{\alf\beta,a}_{\text{inter},\pm}$ become $\Gamma$ independent, or ``dissipationless'' \cite{Nagaosa2010}. The symmetric interband conductivity is suppressed by a factor $\Gamma$ compared to the antisymmetric interband conductivities. The antisymmetric interband conductivities are suppressed by a factor $\Gamma$ compared to the intraband conductivities. However, note that the leading order might vanish, for instance, when integrating over momenta or due to zero Berry curvature.  

Using \eqref{eqn:winG0}, \eqref{eqn:wsG0} and \eqref{eqn:wanG0} we see that the intraband conductivities and the symmetric interband conductivity are proportional to $-f'(E^\pm_\bp-\mu)$ whereas the antisymmetric interband conductivities involve the Fermi function $f(E^\pm_\bp-\mu)$ in the clean limit. Thus, the former ones are restricted to the vicinity of the Fermi surface at low temperature $k_BT\ll 1$. In contrast, all occupied states contribute to the antisymmetric interband conductivities. The consistency with the Landau Fermi liquid picture was discussed by Haldane \cite{Haldane2004}.

The Fermi function $f(\eps)$ and its derivative $f'(\eps)$ capture the temperature broadening effect in the different contributions \eqref{eqn:SintraN}-\eqref{eqn:SinterAN} of the conductivity. In the following, we have a closer look at the low temperature limit. Since $f'(\eps)\rightarrow -\delta(\eps)$ for $k_BT\ll 1$ the spectral weighting factors of the intraband and the symmetric interband conductivity read $-w^{n}_{\bp,\text{intra}}(0)$ and $-w^{s}_{\bp,\text{inter}}(0)$, respectively, after the frequency integration over $\eps$. The antisymmetric interband conductivities involve the Fermi function, which results in the Heaviside step function for $k_BT\ll 1$, that is $f(\eps)\rightarrow \Theta(-\eps)$. Thus, the frequency integration has still to be performed from $-\infty$ to $0$. In order to circumvent this complication, we define the primitive $(W^{n,a}_{\bp,\text{inter}}(\eps))'=w^{n,a}_{\bp,\text{inter}}(\eps)$ with the boundary condition $W^{n,a}_{\bp,\text{inter}}(-\infty)=0$. The zero temperature limit is then performed after partial integration in $\eps$ by 
\begin{align}
 \int\hspace{-1mm} d\eps f(\eps)\,w^{n,a}_{\bp,\text{inter}}(\eps)&=-\int\hspace{-1mm} d\eps f'(\eps)\,W^{n,a}_{\bp,\text{inter}}(\eps)\approx W^{n,a}_{\bp,\text{inter}}(0)\,. 
\end{align}
In Fig.~\ref{fig:WInter}, we sketch $W^{n,a}_{\bp,\text{inter}}(\eps)$ for $\Gamma=0.3\,r_\bp$. At finite $\Gamma$, it is a crossover from zero to approximately one, which eventually approaches a step function at $E^n_\bp-\mu$ for small $\Gamma$. At low temperature $k_BT\ll 1$, the occupied states with $E^n_\bp-\mu<0$ contribute significantly to the antisymmetric interband conductivities as expected. Note that $\int\hspace{-1mm}d\eps\, w^{n,a}_{\bp,\text{inter}}(\eps)=r^2_\bp(r^2_\bp+3\Gamma^2)/(r^2_\bp+\Gamma^2)^2\approx 1+\Gamma^2/r_\bp^2$, so that a step function of height 1 is only approached in the limit $\Gamma\rightarrow 0$.

In the following, we discuss the limiting cases of the spectral weighting factors $w^n_{\bp,\text{intra}}(0)$, $w^s_{\bp,\text{inter}}(0)$, and $W^{n,a}_{\bp,\text{inter}}(0)$, that is, in the low or zero temperature limit. We start with the case of a band insulator in the clean limit and assume a chemical potential below, above or in between the two quasiparticle bands as well as a relaxation rate much smaller than the gap, $\Gamma\ll|E^n_\bp-\mu|$. Within this limit, we find very distinct behavior of the spectral weighting factors of the intraband conductivities and of the symmetric interband conductivity on the one hand and the spectral weighting factor of the antisymmetric interband conductivities on the other hand. The former ones scale as 
\begin{align}
 &w^n_{\bp,\text{intra}}(0)\approx \frac{\Gamma^2}{\pi(\mu-E^n_\bp)^4}\sim \mathcal{O}(\Gamma^2) \, ,\\
 &w^s_{\bp,\text{inter}}(0)\approx \frac{4r^2_\bp\Gamma^2}{\pi(\mu-E^+_\bp)^2(\mu-E^-_\bp)^2}\sim\mathcal{O}(\Gamma^2) \, .
\end{align}
We see that the intraband and the symmetric interband conductivity for filled or empty bands are only present due to a finite relaxation rate. The spectral weighting factor of the antisymmetric interband conductivities has a different behavior whether the bands are all empty, all filled or the chemical potential is in between both bands. By expanding $W^{n,a}_{\bp,\text{inter}}(0)$ we get 
\begin{align}
 W^{n,a}_{\bp,\text{inter}}(0)&= \frac{1}{2}\big[1+\text{sign}(\mu-E^n_\bp)\big]+\big[2+\sum_{\nu=\pm}\text{sign}(\mu-E^\nu_\bp)\big]\frac{\Gamma^2}{4r^2_\bp}+\mathcal{O}(\Gamma^3) \, .
\end{align}
Note that a direct expansion of $w^{n,a}_{\bp,\text{inter}}(\eps)$ followed by the integration over $\eps$ from $-\infty$ to $0$ is not capable to capture the case of fully occupied bands, which shows that the regularization by a finite $\Gamma$ is crucial to avoid divergent integrals in the low temperature limit. For completely filled bands $\mu>E^+_\bp,E^-_\bp$ we have $W^{n,a}_{\bp,\text{inter}}(0)\approx 1+\Gamma^2/r^2_\bp$ in agreement with the discussion above. For completely empty bands $\mu<E^+_\bp,E^-_\bp$ we have $W^{n,a}_{\bp,\text{inter}}(0)\propto \Gamma^3$. If the chemical potential lies in between both bands $E^-_\bp<\mu<E^+_\bp$ we have $W^{-,a}_{\bp,\text{inter}}(0)=1+\Gamma^2/2r^2_\bp$ and $W^{+,a}_{\bp,\text{inter}}(0)=\Gamma^2/2r^2_\bp$. The antisymmetric interband conductivities involve the Berry curvature, which is equal for both bands up to a different sign, $\Omega^{\alf\beta,+}=-\Omega^{\alf\beta,-}$. Thus, the antisymmetric interband conductivity summed over both bands involves
\begin{align}
 &W^{+,a}_{\bp,\text{inter}}(0)-W^{-,a}_{\bp,\text{inter}}(0)\nonumber\\[1mm]&= \frac{1}{2}\big[\text{sign}(\mu-E^+_\bp)-\text{sign}(\mu-E^-_\bp)\big]-\frac{16r^3_\bp\Gamma^3}{3\pi(\mu-E^+_\bp)^3(\mu-E^-_\bp)^3}+\mathcal{O}(\Gamma^5) \, .
\end{align}
We see that a scattering-independent or ``dissipationless'' term is only present for a chemical potential in between the two bands. The next order in $\Gamma$ is at least cubic. Note that different orders can vanish in the conductivities after the integration over momenta.

Our formulas \eqref{eqn:SintraN}-\eqref{eqn:SinterAN} are valid for an arbitrarily large relaxation rate $\Gamma$. We study the dirty limit (large $\Gamma$) in the following. In contrast to the clean limit, it is crucial to distinct the two following cases: fixed chemical potential and fixed particle number (per unit cell) $\rho_N=N/L$. We use the notation $\rho_N$ here to avoid confusion with the band index, but we will use the standard notation $n\equiv\rho_N$ elsewhere throughout this thesis. The condition of fixed particle number density leads to a scattering-dependent chemical potential $\mu(\Gamma)$, which modifies the scaling of the spectral weighting factors. To see this, we calculate the total particle number per unit cell at small temperature and get
\begin{align}
 \rho_N&=\sum_{\nu=\pm}\int\hspace{-1.5mm} d\eps \int\hspace{-1.5mm} \frac{d^d\bp}{(2\pi)^d} A^\nu_\bp(\eps) f(\eps)\\[1mm]
 &\approx1-\sum_{\nu=\pm}\frac{1}{\pi}\int\hspace{-1.5mm} \frac{d^d\bp}{(2\pi)^d}\arctan\frac{E^\nu_\bp-\mu}{\Gamma}\\[2mm]
 &\approx 1-\frac{2}{\pi}\arctan\frac{c-\mu}{\Gamma} \,.
 \label{eqn:muGamma}
\end{align}
\\
In the last step, we assumed that $\Gamma$ is much larger than the band width, that is, $(E^+_\text{max}-E^-_\text{min})/2=W\ll \Gamma$, where $E^+_\text{max}$ is the maximum of the upper band and $E^-_\text{min}$ is the minimum of the lower band. We denote the center of the bands as $c=(E^+_\text{max}+E^-_\text{min})/2$. 
Solving for the chemical potential gives the linear dependence on $\Gamma$, $\mu(\Gamma)=c+\mu_\infty\Gamma$ with
\begin{align}
 \mu_\infty=-\tan \frac{(1-\rho_N)\pi}{2} \,.
\end{align}
Note that at half filling, $\rho_N=1$, the chemical potential becomes scattering independent, $\mu_\infty=0$. At $\rho_N=0,2$ we have $\mu_\infty=\mp\infty$. We assume a relaxation rate much larger than the bandwidth $W\ll \Gamma$ in the following.

In a first step, we consider the case of fixed particle number density. We discuss the limiting cases of the spectral weighting factors $w^n_{\bp,\text{intra}}(0)$, $w^s_{\bp,\text{inter}}(0)$ and $W^{n,a}_{\bp,\text{inter}}(0)$ by expanding up to several orders in $1/\Gamma$. If needed, the expansion to even higher orders is straightforward. The expansion of the spectral weighting factor of the intraband conductivities $w^n_{\bp,\text{intra}}(0)$ in \eqref{eqn:Wintra} reads
\begin{align}
 &w^n_\text{intra}(0)\approx \frac{1}{(1+\mu^2_\infty)^2}\frac{1}{\pi \Gamma^2}+\frac{4\mu_\infty}{(1+\mu_\infty^2)^3}\frac{E^n_\bp-c}{\pi \Gamma^3}-\frac{2(1-5\mu_\infty^2)}{(1+\mu_\infty^2)^4}\frac{(E^n_\bp-c)^2}{\pi\Gamma^4} \, . \label{eqn:WnInfty}
\end{align}
The prefactors involve $\mu_\infty$ at each order and an additional momentum-dependent prefactor at cubic and quartic order. The expansion of the spectral weighting factor of the symmetric interband conductivity $w^s_{\bp,\text{inter}}(0)$ in \eqref{eqn:Wsinter} reads
\begin{align}
 &w^s_\text{inter}(0)\approx \frac{4}{(1+\mu^2_\infty)^2}\frac{r^2_\bp}{\pi \Gamma^2}+\frac{16\mu_\infty}{(1+\mu_\infty^2)^3}\frac{r^2_\bp(\g_\bp-c)}{\pi\Gamma^3}\nonumber\\[3mm]&\hspace{3cm}-\bigg[\frac{8(1-5\mu_\infty^2)}{(1+\mu^2_\infty)^4}\frac{r^2_\bp (\g_\bp-c)^2}{\pi\Gamma^4}+\frac{8(1-\mu_\infty^2)}{(1+\mu^2_\infty)^4}\frac{r^4_\bp}{\pi\Gamma^4}\bigg] \, . \label{eqn:WsInfty}
\end{align}
Note that all orders involve a momentum-dependent prefactor. In both $w^n_\text{intra}(0)$ and $w^s_\text{inter}(0)$ the cubic order vanishes at half filling by $\mu_\infty=0$. The expansion of the spectral weighting factor of the antisymmetric interband conductivities $W^{n,a}_{\bp,\text{inter}}(0)$ in \eqref{eqn:Wainter} reads
\begin{align}
 &W^{a,\pm}_\text{inter}(0)\approx \bigg[\frac{3\pi}{2}\hspace{-0.5mm}+\hspace{-0.5mm}3\arctan\mu_\infty\hspace{-0.5mm}+\hspace{-0.5mm}\frac{\mu_\infty(5+3\mu_\infty^2)}{(1+\mu_\infty^2)^2}\bigg]\frac{r^2_\bp}{\pi\Gamma^2}-\frac{8}{3(1+\mu_\infty^2)^3}\frac{3r^2_\bp (\g_\bp-c)\pm r^3_\bp}{\pi\Gamma^3} \, . \label{eqn:WaInfty}
\end{align}
Note that the expansion of $w^{a,\pm}_\text{inter}(\eps)$ with subsequent frequency integration from $-\infty$ to $0$ leads to divergences and predicts a wrong lowest order behavior. Due to the property of the Berry curvature, $\Omega^{\alf\beta,+}_\bp=-\Omega^{\alf\beta,-}_\bp$, the quadratic order drops out of the antisymmetric interband conductivity summed over the two bands, leading to
\begin{align}
 &W^{a,+}_\text{inter}(0)-W^{a,-}_\text{inter}(0)\approx -\frac{16}{3(1+\mu_\infty^2)^3}\frac{r^3_\bp}{\pi\Gamma^3}-\frac{32\mu_\infty}{(1+\mu^2_\infty)^4}\frac{r^3_\bp(\g_\bp-c)}{\pi\Gamma^4}\nonumber\\[3mm]&\hspace{5cm}+\bigg[\frac{16(1-7\mu^2_\infty)}{(1+\mu^2_\infty)^5}\frac{r^3_\bp (\g_\bp-c)^2}{\pi\Gamma^5}+\frac{16(3-5\mu^2_\infty)}{5(1+\mu^2_\infty)^5}\frac{r^5_\bp}{\pi\Gamma^5}\bigg] \, . \label{eqn:WaDiffInfty}
\end{align}
The antisymmetric interband conductivity summed over the two bands is at least cubic in $1/\Gamma$ in contrast to the intraband and the symmetric interband conductivity, which are at least quadratic. The integration over momenta in the conductivities can cause the cancellation of some orders or can reduce the numerical prefactor drastically, so that the crossover to lower orders take place far beyond the scale that is numerically or physically approachable. By giving the exact prefactors above, this can be checked not only qualitatively but also quantitatively for a given model. 

The dirty limit for fixed chemical potential does not involve orders due to the scattering dependence of $\mu(\Gamma)$, however modifies the prefactor due to a constant $\mu$. The corresponding expansion of the different spectral weighting factors can be obtained simply by setting $\mu_\infty=0$ and $c=\mu$ in \eqref{eqn:WnInfty} - \eqref{eqn:WaDiffInfty}.

The scaling behavior $\sigma^{xx}\sim \Gamma^{-2}$ of the longitudinal conductivity and $\sigma^{xy}\sim \Gamma^{-3}$ of the anomalous Hall conductivity (for zero $\sigma^{xy}_{\text{intra},\pm}$) is consistent with Kontani {\it et al.} \cite{Kontani2007} and Tanaka {\it et. al.} \cite{Tanaka2008}. We emphasize, however, that a scattering dependence of $\mu$ and the integration over momenta may modify the upper scalings. Thus, the scaling relation $\sigma^{xy}\propto (\sigma^{xx})^\nu$ useful in the analysis of experimental results (see, for instance, Ref.~\cite{Onoda2008}) is not necessarily $\nu=1.5$ in the limit $W\ll\Gamma$ \cite{Tanaka2008}. We will give an example in Sec.~\ref{sec:application:anomalousHallEffect:scaling}.   

%
%

\subsection[Anomalous Hall effect, anisotropic longitudinal conductivity and quantization]{Anomalous Hall effect, anisotropic longitudinal \\ conductivity and quantization}
\label{sec:theory:conductivity:anomalousHall}

The Berry curvature tensor $\Omega^{\alf\beta,n}_\bp$ is antisymmetric in $\alf\leftrightarrow\beta$ and, thus, has three independent components in a 3-dimensional system, which can be mapped to a Berry curvature vector $\bOmega^n_\bp=\begin{pmatrix}\Omega^{yz,n}_\bp, & -\Omega^{xz,n}_\bp, & \Omega^{xy,n}_\bp\end{pmatrix}$. In order to use the same notation in a 2-dimensional system we set the corresponding elements in $\bOmega^n_\bp$ to zero, for instance, $\Omega^{yz,n}_\bp=\Omega^{xz,n}_\bp=0$ for a system in the $x$-$y$ plane. By using the definition of the conductivity and our result \eqref{eqn:SinterAN} of the antisymmetric interband contribution we can write the current density vector $\bj^a_n$ of band $n=\pm$ induced by $\bOmega^n_\bp$ as 
\begin{align}
 &\bj^a_n\hspace{-0.5mm}=\hspace{-0.5mm}-\frac{e^2}{\hbar}\hspace{-1mm}\int\hspace{-1.5mm}\frac{d^d\bp}{(2\pi)^d}\hspace{-1mm}\int\hspace{-1.5mm}d\epsilon\,\,f(\epsilon)\, w^{a,n}_{\bp,\text{inter}}(\epsilon)\,\bE\times\bOmega^n_\bp\, 
\end{align}
The Berry curvature vector $\bOmega^n_\bp$ acts like an effective magnetic field \cite{Nagaosa2010,Xiao2010} in analogy to the Hall effect induced by an external magnetic field $\bB$. We see that the antisymmetric interband contribution of the conductivity in \eqref{eqn:SinterAN} is responsible for the intrinsic anomalous Hall effect, that is, a Hall current without an external magnetic field that is not caused by (skew) scattering.

In a $d$-dimensional system, the conductivity tensor is a $d\times d$ matrix $\sigma=(\sigma^{\alf\beta})$. Besides its antisymmetric part, which describes the anomalous Hall effect, it does also involve a symmetric part $\sigma_\text{sym}$ due to the intraband and the symmetric interband contributions \eqref{eqn:SintraN} and \eqref{eqn:SinterS}. We can diagonalize the, in general, non-diagonal matrix $\sigma_\text{sym}$ by a rotation $\cR$ of the coordinate system, which we fixed to an orthogonal basis $\mathbf{e}_x,\,\mathbf{e}_y,\,\mathbf{e}_z$ when labeling $\alf$ and $\beta$ in \eqref{eqn:jexpDef}. If the rotation $\cR$ is chosen such that $\mathcal{R}^T\sigma_\text{sym}\mathcal{R}$ is diagonal, the antisymmetric part in the rotated basis is described by the the rotated Berry curvature vector $\cR^T\bOmega^n_\bp$. We see that a rotation within the plane of a two-dimensional system does not affect $\bOmega^n_\bp$, which highlights the expected isotropy of the anomalous Hall effect consistent with the interpretation of $\bOmega^n_\bp$ as an effective magnetic field perpendicular to the plane. The possibility to diagonalize the symmetric part $\sigma_\text{sym}$ shows that the diagonal and off-diagonal intraband and symmetric interband contributions in \eqref{eqn:SintraN} and \eqref{eqn:SinterS} are part of the (anisotropic) longitudinal conductivity in a rotated coordinate system. 

Finally, we discuss the possibility of quantization of the anomalous Hall conductivity. Let us assume a two-dimensional system that is lying in the $x$-$y$ plane without loss of generality. The Chern number of band $n$ is calculated by the momentum integral of the Berry curvature over the full Brillouin zone (BZ), that is,
\begin{align}
\label{eqn:Chern}
 C_n=-\frac{1}{2\pi}\int_\text{BZ}\bOmega^n_\bp\cdot d\bS=-2\pi \int \frac{d^2 \bp}{(2\pi)^2} \,\Omega^{xy,n}_\bp 
\end{align}
and is quantized to integer numbers \cite{Thouless1982, Xiao2010, Nagaosa2010}. We can define a generalized Chern number dependent on the temperature, the relaxation rate and the chemical potential as
\begin{align}
 & C_n(T,\Gamma,\mu) \hspace{-0.5mm}=\hspace{-0.5mm}-2\pi\hspace{-1mm}\int\hspace{-1.5mm}\frac{d^2\bp}{(2\pi)^2}\hspace{-1mm}\int\hspace{-1.5mm}d\epsilon\,\,f(\epsilon)\, w^{a,n}_{\bp,\text{inter}}(\epsilon)\,\Omega^{xy,n}_\bp\,  ,
\end{align}
which is weighted by the Fermi function as well as by the spectral weighting factor $w^{a,n}_{\bp,\text{inter}}(\epsilon)$ defined in \eqref{eqn:Wainter}. Thus, we include the effect of band occupation, temperature and finite relaxation rate. The antisymmetric interband conductivity, that is the anomalous Hall conductivity, then reads
\begin{align}
 \sigma^{xy,a}_{\text{inter},n}=\frac{e^2}{h} C_n(T,\Gamma,\mu) \, .
\end{align}
In the clean limit $\Gamma\ll 1$, we recover the broadly used result of Onoda {\it et al.} \cite{Onoda2002} and  Jungwirth {\it et al.} \cite{Jungwirth2002}. If we further assume zero temperature $k_B T\ll 1$ and a completely filled band $n$, we recover the famous Thouless-Kohmoto-Nightingale-Nijs (TKNN) formula for the quantized anomalous Hall effect \cite{Thouless1982}, where the anomalous Hall conductivity is quantized to $\frac{e^2}{h}C_n$ due to the quantized integer Chern number $C_n$. Note that finite temperature, finite relaxation rate $\Gamma$ and partially filled bands break the quantization. 

Furthermore, we may be able to relate the antisymmetric interband conductivity to topological charges and, by this, obtain a quantized anomalous Hall conductivity. The Berry curvature $\bOmega^n_\bp$ is the curl of the Berry connection $\bcA^n_\bp=\begin{pmatrix}\cA^{x,n}_\bp, & \cA^{y,n}_\bp, & \cA^{z,n}_\bp\end{pmatrix}$, see \eqref{eqn:OmegaRot}. Via Stokes' theorem, the integral over a two-dimensional surface within the Brillouin zone can be related to a closed line integral. This line integral may define a quantized topological charge, which leads to a quantized value of $\sigma^{\alf\beta,a}_{\text{inter},n}$ integrated over this surface. For instance, this causes a quantized radial component of the current in a $\mathcal{P}\mathcal{T}$-symmetric Dirac nodal-line semimetal
\cite{Rui2018}.

%
%

\section{Hall conductivity}
\label{sec:theory:Hall}

After having derived and analyzed the formulas for the longitudinal and the anomalous Hall conductivity in the last section, we continue with the derivation of the formulas for the (ordinary) Hall conductivity. In the following, we consider the presence of both an electric and a magnetic field. The conductivity tensor $\sigma^{\alf\beta\eta}_\text{\tiny EB}(\omega)$ in \eqref{eqn:sH}, which describes the induced current in $\alf$ direction due to an electric field in $\beta$ and a magnetic field in $\eta$ direction, does not only capture the Hall conductivity. For instance, $\sigma^{xxz}_\text{\tiny EB}(\omega)$ describes the modification of the longitudinal conductivity $\sigma^{xx}$ due to a perpendicular magnetic field, that is, the effect of (linear) magnetoresistance. In this thesis, we are mainly interested in the antisymmetric contribution of $\sigma^{\alf\beta\eta}_\text{\tiny EB}(\omega)$ with respect to $\alf\leftrightarrow\beta$, which describes the Hall conductivity. Later on, we will denote this contribution as Hall conductivity tensor $\sigma^{\alf\beta\eta}_\text{H}(\omega)$, which involves the particular component $\sigma^{xyz}_\text{H}(\omega)=-\sigma^{yxz}_\text{H}(\omega)$ of interest for a two-dimensional system in the $x$-$y$ plane with an perpendicular magnetic field in $z$ direction. However, the majority of the derivations are performed for general indices, which may, thus, serve as a starting point for generalizations. In this section, we continue the evaluation of $\sigma^{\alf\beta\eta}_\text{\tiny EB}(\omega)$ in \eqref{eqn:sH} by simplifying the polarization tensor $\Pi^{\alf\beta\gamma\delta}_{\text{\tiny EB},iq_0}$ in \eqref{eqn:PiH} with a subsequent analytic continuation from Matsubara to real frequencies. The two additional indices $\gamma$ and $\delta$ are present since we have expressed the magnetic field as the curl of the corresponding vector potential.

We structure this section as follows: We use the symmetry under trace transposition to decompose the polarization tensor. Like for the longitudinal and the anomalous Hall conductivity, we will show that this symmetry is equivalent to the symmetry of the conductivities in the current and electric field directions. By this, we are able to disentangle the contributions that describe the effect of linear magnetoresistance and the (ordinary) Hall effect (Sec.~\ref{sec:theory:Hall:symmetryindirections}). We assume a Bloch Hamiltonian with momentum-independent coupling between the two subsystems, which is, for instance, sufficient for the planar spiral magnetic state, and show that this drastically simplifies the involved quantities (Sec.~\ref{sec:theory:Hall:simplifications}). These simplifications eventually lead to a compact form of the polarization tensor (Sec.~\ref{sec:theory:Hall:decomposition}), for which we can perform the Matsubara summation and the analytic continuation analytically (Sec.~\ref{sec:theory:Hall:matsubara}). We summarize our final results for the Hall conductivity (Sec.~\ref{sec:theory:Hall:formulas}), specify the components for a two-dimensional system, and discuss the limit of a small relaxation rate (Sec.~\ref{sec:theory:Hall:coefficient}). 

%
%

\subsection{Symmetry in current and electric field directions}
\label{sec:theory:Hall:symmetryindirections}

In Sec.~\ref{sec:theory:conductivity} in the context of the longitudinal and the anomalous Hall conductivity, we found the equivalence between the symmetry in trace transposition, which was defined in \eqref{eqn:TraceTrans}, and the symmetry in the indices $\alf\leftrightarrow\beta$, that is, the current and electric field directions. We can decompose any product of $n$ square matrices $M_i$ into its symmetric and antisymmetric part under trace transposition by
\begin{align}
 \label{eqn:DecompSymAntisymGeneral}
 \tr \big[M_1... M_n\big]=\frac{1}{2}\tr \big[M_1... M_n+M_n...M_1\big]+\frac{1}{2}\tr \big[M_1... M_n-M_n... M_1\big]\,.
\end{align}
Note that the involved matrices may not commute in general, so that the precise order is crucial. Let us consider the triangular and rectangular part of $\Pi^{\alf\beta\gamma\delta}_{\text{\tiny EB},iq_0}$ in \eqref{eqn:PiHTri} and \eqref{eqn:PiHRec}. We apply \eqref{eqn:DecompSymAntisymGeneral} to the triangular contribution and define
\begin{align}
  \label{eqn:PiHTriDecomp}
  \big(\Pi^{\alf\beta\gamma\delta}_{\text{\tiny EB},iq_0}\big)^{(\text{tri})} =\big(\Pi^{\alf\beta\gamma\delta,s}_{\text{\tiny EB},iq_0}\big)^{(\text{tri})}+\big(\Pi^{\alf\beta\gamma\delta,a}_{\text{\tiny EB},iq_0}\big)^{(\text{tri})}
\end{align}
with the symmetric and antisymmetric parts
\begingroup
 \allowdisplaybreaks[0]
\begin{alignat}{2}
  \label{eqn:PiHTriS}&\big(\Pi^{\alf\beta\gamma\delta,s}_{\text{\tiny EB},iq_0}\big)^{(\text{tri})} =\frac{1}{8}\TrH\big[&&\big(\sG^{}_{ip_0+iq_0,\bp}-\sG^{}_{ip_0-iq_0,\bp}\big)\lam^{\beta}_{\bp}\sG^{}_{ip_0,\bp}\lam^\delta_{\bp}\sG^{}_{ip_0,\bp}\lam^{\alf\gamma}_{\bp}\nonumber\\
 &\hspace{3.75cm}+&&\big(\sG^{}_{ip_0+iq_0,\bp}-\sG^{}_{ip_0-iq_0,\bp}\big)\lam^{\alf\gamma}_{\bp}\sG^{}_{ip_0,\bp}\lam^\delta_\bp\sG^{}_{ip_0,\bp}\lam^\beta_{\bp}\big]-(\alf\leftrightarrow\beta) \, ,
 \\[3mm]
 \label{eqn:PiHTriA}&\big(\Pi^{\alf\beta\gamma\delta,a}_{\text{\tiny EB},iq_0}\big)^{(\text{tri})} =\frac{1}{8}\TrH\big[&&\big(\sG^{}_{ip_0+iq_0,\bp}+\sG^{}_{ip_0-iq_0,\bp}\big)\lam^{\beta}_{\bp}\sG^{}_{ip_0,\bp}\lam^\delta_{\bp}\sG^{}_{ip_0,\bp}\lam^{\alf\gamma}_{\bp}\nonumber\\
 &\hspace{3.75cm}-&&\big(\sG^{}_{ip_0+iq_0,\bp}+\sG^{}_{ip_0-iq_0,\bp}\big)\lam^{\alf\gamma}_{\bp}\sG^{}_{ip_0,\bp}\lam^\delta_\bp\sG^{}_{ip_0,\bp}\lam^\beta_{\bp}\big]+(\alf\leftrightarrow\beta) \, .
\end{alignat}
\endgroup
Equivalently, we decompose the rectangular contribution into
\begin{align}
  \label{eqn:PiHRecDecomp}
  \big(\Pi^{\alf\beta\gamma\delta}_{\text{\tiny EB},iq_0}\big)^{(\text{rec})} = \big(\Pi^{\alf\beta\gamma\delta,s}_{\text{\tiny EB},iq_0}\big)^{(\text{rec})}+\big(\Pi^{\alf\beta\gamma\delta,a}_{\text{\tiny EB},iq_0}\big)^{(\text{rec})}
\end{align}
with the symmetric and antisymmetric parts
%
\begin{alignat}{2}
  \label{eqn:PiHRecS}&\big(\Pi^{\alf\beta\gamma\delta,s}_{\text{\tiny EB},iq_0}\big)^{(\text{rec})}\hspace{-0.15cm} =\hspace{-0.05cm}\frac{1}{8}\TrH\big[&&\big(\sG^{}_{ip_0+iq_0,\bp}-\sG^{}_{ip_0-iq_0,\bp}\big)\lam^{\beta}_{\bp}\sG^{}_{ip_0,\bp}\lam^\delta_{\bp}\sG^{}_{ip_0,\bp}\lam^\gamma_\bp\sG^{}_{ip_0,\bp}\lam^\alf_{\bp}
 \nonumber\\
 &\hspace{3.7cm}+&&\big(\sG^{}_{ip_0+iq_0,\bp}-\sG^{}_{ip_0-iq_0,\bp}\big)\lam^\alf_{\bp}\sG^{}_{ip_0,\bp}\lam^\gamma_\bp\sG^{}_{ip_0,\bp}\lam^\delta_{\bp}\sG^{}_{ip_0,\bp}\lam^\beta_{\bp}\big]\hspace{-0.05cm}-\hspace{-0.05cm}(\alf\leftrightarrow\beta)\, ,
 \\[3mm]
 \label{eqn:PiHRecA}&\big(\Pi^{\alf\beta\gamma\delta,a}_{\text{\tiny EB},iq_0}\big)^{(\text{rec})}\hspace{-0.15cm}=\hspace{-0.05cm}\frac{1}{8}\TrH\big[&&\big(\sG^{}_{ip_0+iq_0,\bp}+\sG^{}_{ip_0-iq_0,\bp}\big)\lam^{\beta}_{\bp}\sG^{}_{ip_0,\bp}\lam^\delta_{\bp}\sG^{}_{ip_0,\bp}\lam^\gamma_\bp\sG^{}_{ip_0,\bp}\lam^\alf_{\bp}
 \nonumber\\
 &\hspace{3.7cm}-&&\big(\sG^{}_{ip_0+iq_0,\bp}+\sG^{}_{ip_0-iq_0,\bp}\big)\lam^\alf_{\bp}\sG^{}_{ip_0,\bp}\lam^\gamma_\bp\sG^{}_{ip_0,\bp}\lam^\delta_{\bp}\sG^{}_{ip_0,\bp}\lam^\beta_{\bp}\big]\hspace{-0.05cm}+\hspace{-0.05cm}(\alf\leftrightarrow\beta) \, .
\end{alignat}
%
We recapitulate the short notation $\TrH[\,\cdot\,]=e^3TL^{-1}\sum_p\tr[\,\cdot\,-(iq_0\hspace{-0.5mm}=\hspace{-0.5mm}0)-(\gamma\hspace{-0.5mm}\leftrightarrow\hspace{-0.5mm}\delta)]$, which was introduced below \eqref{eqn:PiHTri}. We used the sign change due to the antisymmetry in $\gamma\leftrightarrow\delta$ in order to bring all expressions in a similar, comparable structure for further analysis. The obtained structure of the formulas in \eqref{eqn:PiHTriS}, \eqref{eqn:PiHTriA}, \eqref{eqn:PiHRecS} and \eqref{eqn:PiHRecA} emphasizes several aspects: We see that the symmetric part under trace transposition is antisymmetric in the indices $\alf\leftrightarrow\beta$. The antisymmetric part under trace transposition is symmetric in $\alf\leftrightarrow\beta$. Thus, the contributions that describe the Hall effect and the effect of linear magnetoresistance are clearly separated by their symmetry under trace transposition. For instance, we find the (symmetric) tensor component $\sigma^{xxz}_\text{\tiny EB}(\omega)$, which describes the linear magnetoresistance in $x$ direction due to a magnetic field in $z$ direction, to be antisymmetric under trace transposition, whereas the antisymmetric part in $x\leftrightarrow y$ of the tensor component $\sigma^{xyz}_\text{\tiny EB}(\omega)$ is symmetric under trace transposition and captures the (ordinary) Hall effect. Note that tensor element $\sigma^{xyz}_\text{\tiny EB}(\omega)$ may, in general, involve a symmetric contribution in $x\leftrightarrow y$ as part of the antisymmetric contribution under trace transposition. Such a term leads to an anisotropy of the longitudinal conductivity as described in Sec.~\ref{sec:theory:conductivity:anomalousHall}. A nonzero symmetric or antisymmetric contribution under trace transposition is clearly connected to the symmetry of the involved matrices under transposition, which we will use in the following. We find a characteristic dependence on the external Matsubara frequency $iq_0$. The zeroth component with $iq_0=0$ vanishes for \eqref{eqn:PiHTriS} and \eqref{eqn:PiHRecS}, whereas it is, in general, nonzero for \eqref{eqn:PiHTriA} and \eqref{eqn:PiHRecA}. 

%
%

\subsection{Simplifications for a momentum-independent gap}
\label{sec:theory:Hall:simplifications}

In analogy to the derivation of the formulas for the longitudinal and the anomalous Hall conductivity in Sec.~\ref{sec:theory:conductivity:decomposition}, we will use the invariance of the trace under unitary transformations of the involved matrices. We express all Green's functions and vertices in \eqref{eqn:PiHTriS}, \eqref{eqn:PiHTriA}, \eqref{eqn:PiHRecS} and \eqref{eqn:PiHRecA} in the eigenbasis $|\pm_\bp\rangle$ of the Bloch Hamiltonian $\lam_\bp$. The Green's function matrices are diagonal in this basis and are given in \eqref{eqn:Gdiag}. By decomposing the vertices into diagonal and off-diagonal parts, which are given in \eqref{eqn:UdagLamU} and \eqref{eqn:DecompLamAB} for the first-order and second-order vertex, respectively, we will be able to separate intra- and interband contributions. In a second step, we will use the symmetry under transposition of the individual matrices in order to take advantage of our decomposition with respect to the symmetry under trace transposition. Whereas the Green's function matrices are diagonal and, thus, symmetric under transposition, the off-diagonal matrices $\cF^\nu_\bp$ and $\dF^{\nu\mu}_\bp$ have to be decomposed into their parts proportional to the Pauli matrices $\pauma_x$ and $\pauma_y$, which are symmetric and antisymmetric under transposition, respectively. 

In this thesis, we will not perform this procedure for the general Hamiltonian in \eqref{eqn:lam} but for a special case, which is, for instance, sufficient to capture the model of planar spiral magnetic order, which we will discuss in the context of recent Hall experiments in cuprates in Chapter~\ref{sec:application}. We assume a constant and real coupling between the two subsystems $A$ and $B$ of the two-band system $\Delta_\bp=-\Delta$ so that the simplified Bloch Hamiltonian is given by
\begin{align}
 \label{eqn:lamSpecial}
 \lam_\bp=\begin{pmatrix} \eps_{\bp,A} && -\Delta \\[2mm] -\Delta && \eps_{\bp,B} \end{pmatrix} \, .
\end{align}
We will refer to $\Delta$ as gap in the following. The angle $\varphi_\bp$, which was shown to capture the negative phase of the gap $\Delta$, is thus constant and given by
\begin{align}
 \varphi_\bp=-\pi \, .
\end{align}
Eventually, the decomposition into diagonal and off-diagonal components does only lead to gauge-independent quantities under the ``local'' $U(1)$ gauge transformation in momentum space, $|\pm_\bp\rangle\rightarrow e^{i\phi^\pm_\bp}|\pm_\bp\rangle$. For simplicity, we fix the gauge to $\phi^+_\bp=\phi^-_\bp=0$ in the following. Note that the choice of the interband coupling to be real (and positive) is without loss of generality for the following reason: All gauge-dependent quantities involve the modified phase $\tilde\varphi=\varphi+\phi^+_\bp-\phi^-_\bp$, where $\varphi$ is the negative phase of $\Delta$. Thus, we can always choose an adequate momentum-independent gauge for a momentum-independent $\Delta$. For a complex gap $\Delta=|\Delta|e^{-i\varphi}$, we can choose $\phi^+_\bp=-\varphi$ and $\phi^-_\bp=0$. The constant angle $\varphi_\bp=-\pi$ drastically simplifies the different contributions of the interband matrices $\cF^\alf_\bp$ and $\cF^{\alf\beta}_\bp$ in \eqref{eqn:X}, \eqref{eqn:Y}, \eqref{eqn:dX} and \eqref{eqn:dY}. We have
\begin{alignat}{3}
 &\cX^\alf_\bp&&=0\, , \hspace{3.5cm}\cY^\alf_\bp&&=-\frac{1}{2}\theta^\alf_\bp\,\pauma_y\, ,\\
 &\dX^{\alf\beta}_\bp&&=0\, , \hspace{3.5cm} \dY^{\alf\beta}_\bp&&=-\frac{1}{2}\theta^{\alf\beta}_\bp\,\pauma_y \, .
\end{alignat}
Thus, the quantum metric factor and the Berry curvature are
\begin{align}
 \cC^{\alf\beta}_\bp=\frac{1}{2}\theta^\alf_\bp\theta^\beta_\bp\,\mathds{1}\, , \hspace{3.5cm} \Omega^{\alf\beta}_\bp=0 \, .
\end{align}
We see that a non-vanishing Berry curvature is connected to a momentum-dependent angle $\varphi_\bp$. The eigenenergies read
\begin{align}
 \label{eqn:ESpecial}
 E^\pm_\bp=\g_\bp\pm\sqrt{h^2_\bp+\Delta^2}
\end{align}
with $\g_\bp=\big(\eps_{\bp,A}+\eps_{\bp,B}\big)/2$ and $h_\bp=\big(\eps_{\bp,A}-\eps_{\bp,B}\big)/2$. We can express the derivative of the angle $\theta_\bp$ by the derivatives of the function $h_\bp$ via
\begin{align}
 &\theta^\alf_\bp=-\frac{\Delta\,h^\alf_\bp}{h^2_\bp+\Delta^2}\, , 
 \label{eqn:ThetaSpecial}
 &\theta^{\alf\beta}_\bp=-\frac{\Delta\,h^{\alf\beta}_\bp}{h^2_\bp+\Delta^2}+\frac{2h^{}_\bp h^\alf_\bp h^\beta_\bp}{\big(h^2_\bp+\Delta^2\big)^2}\,.
\end{align}
Using this and the derivative of $r_\bp=\sqrt{h^2_\bp+\Delta^2}=\big(E^+_\bp-E^-_\bp\big)/2$, we end up with the simplified form of the interband contributions of the first- and second-order vertices
\begingroup
 \allowdisplaybreaks[0]
\begin{align}
 \label{eqn:cFSpecial}
 &\cF^\alf_\bp=-r^{}_\bp\theta^\alf_\bp\,\pauma_x=\frac{2\Delta\,h^\alf_\bp}{E^+_\bp-E^-_\bp} \,\pauma_x\, ,
 \\[2mm]
 \label{eqn:dFSpecial}
 &\dF^{\alf\beta}_\bp=-\big(r^\alf_\bp\theta^\beta_\bp+r^\beta_\bp\theta^\alf_\bp+r^{}_\bp\theta^{\alf\beta}_\bp\big)\pauma_x=\frac{2\Delta\,h^{\alf\beta}_\bp}{E^+_\bp-E^-_\bp}\,\pauma_x\, .
\end{align}
\endgroup
For the last step in \eqref{eqn:dFSpecial} note that the second term of $\theta^{\alf\beta}_\bp$ in \eqref{eqn:ThetaSpecial} multiplied with $r_\bp$ is canceled by $r^\alf_\bp\theta^\beta_\bp+r^\beta_\bp\theta^\alf_\bp$. In order to summarize, we see that the interband matrices \eqref{eqn:cFSpecial} and \eqref{eqn:dFSpecial} have a very similar structure. They involve the first or second momentum derivative of $h_\bp$ and the ratio between the interband coupling $\Delta$ and the direct gap $E^+_\bp-E^-_\bp$. They are proportional to $\pauma_x$ due to the specific gauge choice that we have chosen. For the following calculations, we define the momentum-dependent prefactors $F^\alf_\bp=2\Delta h^\alf_\bp/\big(E^+_\bp-E^-_\bp\big)$ and $F^{\alf\beta}_\bp=2\Delta h^{\alf\beta}_\bp/\big(E^+_\bp-E^-_\bp\big)$ via $\cF^{\alf}_\bp=F^\alf_\bp\,\tau_x$ and $\dF^{\alf\beta}_\bp=F^{\alf\beta}_\bp\tau_x$, respectively.

As a consequence of the momentum-independent gap and an appropriate gauge choice, both the Green's function and the vertex matrices in the eigenbasis are symmetric under transposition. Thus, the antisymmetric contributions in \eqref{eqn:PiHTriA} and \eqref{eqn:PiHRecA} vanish, $\big(\Pi^{\alf\beta\gamma\delta,a}_{\text{\tiny EB},iq_0}\big)^{(\text{tri})}=\big(\Pi^{\alf\beta\gamma\delta,a}_{\text{\tiny EB},iq_0}\big)^{(\text{rec})}=0$. The polarization tensors $\big(\Pi^{\alf\beta\gamma\delta}_{\text{\tiny EB},iq_0}\big)^{(\text{tri})}$ and $\big(\Pi^{\alf\beta\gamma\delta}_{\text{\tiny EB},iq_0}\big)^{(\text{rec})}$ in \eqref{eqn:PiHTriDecomp} and \eqref{eqn:PiHRecDecomp} are entirely given by the remaining symmetric contributions in \eqref{eqn:PiHTriS} and \eqref{eqn:PiHRecS} and, thus, antisymmetric in $\alf\leftrightarrow\beta$. Using the trace transposition, we obtain
\begingroup
 \allowdisplaybreaks[0]
\begin{alignat}{2}
  \label{eqn:PiHTriSpecial}&\big(\Pi^{H,\alf\beta\gamma\delta}_{iq_0}\big)^{(\text{tri})}\equiv\big(\Pi^{\alf\beta\gamma\delta,s}_{\text{\tiny EB},iq_0}\big)^{(\text{tri})} =\frac{1}{4}\wideTrH\Big[&&\sG^{}_{ip_0+iq_0,\bp}\,\lam^\alf_\bp\,\sG^{}_{ip_0,\bp}\,\lam^\gamma_\bp\,\sG^{}_{ip_0,\bp}\,\lam^{\beta\delta}_\bp\Big] \, ,\\[2mm]
  \label{eqn:PiHRecSpecial}&\big(\Pi^{H,\alf\beta\gamma\delta}_{iq_0}\big)^{(\text{rec})}\equiv\big(\Pi^{\alf\beta\gamma\delta,s}_{\text{\tiny EB},iq_0}\big)^{(\text{rec})}\hspace{-0.1cm} =\frac{1}{4}\wideTrH\Big[&&\sG^{}_{ip_0+iq_0,\bp}\,\lam^\alf_\bp\,\sG^{}_{ip_0,\bp}\,\lam^\gamma_\bp\,\sG^{}_{ip_0,\bp}\,\lam^\delta_\bp\,\sG^{}_{ip_0,\bp}\,\lam^\beta_{\bp}\Big]\, .
\end{alignat}
\endgroup
We redefined the notation of the polarization tensor in order to shorten the notation and to highlight the connection to the Hall conductivity. We introduced the compact notation $\wideTrH[\,\cdot\,]=e^3TL^{-1}\sum_{ip_0,\bp}\tr[\,\cdot\,-(iq_0\rightarrow -iq_0)-(\gamma\leftrightarrow\delta)-(\alf\leftrightarrow\beta)]$, where the dot $\cdot$ denotes the argument of the trace. The compact notation captures the prefactors, the summation over frequency and momentum as well as the antisymmetry in $\gamma\leftrightarrow\delta$ and $\alf\leftrightarrow\beta$. We used the fact that the $iq_0=0$ contribution vanishes and introduced the notation $(iq_0\rightarrow -iq_0$) for the argument with replaced external Matsubara frequency. The subtraction of the corresponding contributions with replaced Matsubara frequency and exchanged indices apply to all previous terms, such that \eqref{eqn:PiHTriSpecial} and \eqref{eqn:PiHRecSpecial} consist of eight terms each when writing them explicitly. Note that we changed simultaneously both indices $\alf\leftrightarrow\beta$ and $\gamma\leftrightarrow\delta$, which does not change the overall sign.

%
%

\subsection{Decomposition and recombination}
\label{sec:theory:Hall:decomposition}

We further simplify the triangular and rectangular contributions in \eqref{eqn:PiHTriSpecial} and \eqref{eqn:PiHRecSpecial}, so that we will eventually be able to perform the Matsubara summation and the analytic continuation to real frequency analytically. In the following, we extensively use the simplifications for a momentum-independent gap, which we derived in the previous section. The detailed calculation is presented in Appendix~\ref{appendix:DecompRecomp}. We summarize the different steps and required identities for the calculation in the following.

We start by expressing both contributions in \eqref{eqn:PiHTriSpecial} and \eqref{eqn:PiHRecSpecial} in the eigenbasis $|\pm_\bp\rangle$. The Green's function matrices $\cG^{}_{ip_0,\bp}=\Udag_\bp \sG^{}_{ip_0,\bp} \U_\bp$ are diagonal. We apply the transformation of the first- and second-order vertices $\Udag_\bp\lam^\alf_\bp\U_\bp=\cE^\alf_\bp+\cF^\alf_\bp$ and $\Udag_\bp\lam^{\alf\beta}_\bp\U_\bp=(\dM^{-1})^{\alf\beta}_\bp+\dF^{\alf\beta}_\bp$, which were defined in \eqref{eqn:UdagLamU} and \eqref{eqn:DecompLamAB}, respectively. We split the expressions by using these decompositions into diagonal and off-diagonal components.  All terms that involve an odd number of off-diagonal matrices vanish by the matrix trace. We show that several terms vanish by the antisymmetry in the indices $\gamma\leftrightarrow\delta$ when using the simplified form of $\cF^\nu_\bp\propto \pauma_x$ explicitly. The antisymmetry in the indices $\alf\leftrightarrow\beta$ allow for further recombinations. 

The rectangular contribution involves four Green's function matrices, whereas the triangular contribution only involves three Green's function matrices. In order to provide a path for combining the rectangular and the triangular contribution, we reduce the number of Green's function matrices in the rectangular contribution by using the identity
\begin{align}
 \label{eqn:RedGreenFunc}
 \cG^{}_{ip_0,\bp}\,\cF^\delta_\bp\,\cG^{}_{ip_0,\bp}=\cF^\delta_\bp\,\cG^{}_{ip_0,\bp}\,\cS^{}_\bp+\cS^{}_\bp\,\cG^{}_{ip_0,\bp}\,\cF^\delta_\bp \, .
\end{align}
The identity can be verified by purely algebraic steps using the explicit form of the Green's function in \eqref{eqn:Green} and $\cF^\delta_\bp\propto\pauma_x$. We defined the short notation $\cS^{}_\bp=1/(E^+_\bp-E^-_\bp)\,\pauma_z$. A second useful identity is based on the explicit form of $F^\alf_\bp$ in \eqref{eqn:cFSpecial}. We have $E^{+,\nu}_\bp-E^{-,\nu}_\bp=2r^\nu_\bp=\frac{2h_\bp}{\Delta}F^\nu_\bp$ and, thus, the identity 
\begin{align} \label{eqn:changingindices}
  F^\delta_\bp \big(E^{+,\beta}_{\bp}-E^{-,\beta}_\bp\big)=F^\beta_\bp \big(E^{+,\delta}_{\bp}-E^{-,\delta}_\bp\big) \,.
\end{align}
Note that both identities \eqref{eqn:RedGreenFunc} and \eqref{eqn:changingindices} rely on the simplifications that were derived by considering a momentum-independent gap.

Using those two identities and the explicit expressions provided in the previous section as well as performing partial integration in momentum, we can combine all terms of the rectangular contribution with the terms of the triangular contribution. The result of this decomposition and recombination reads
\begin{align}
  \label{eqn:PiHDecomp}
  \Pi^{H,\alf\beta\gamma\delta}_{iq_0} = \Pi^{H,\alf\beta\gamma\delta}_{iq_0,\text{intra}}+\Pi^{H,\alf\beta\gamma\delta}_{iq_0,\text{inter}} \, ,
\end{align}
where we defined the intraband and interband contributions
\begin{alignat}{8}
  \label{eqn:PiHIntra}\Pi^{H,\alf\beta\gamma\delta}_{iq_0,\text{intra}}=&-\frac{1}{4}\wideTrH\Big[\,&&\cG^{}_{ip_0+iq_0,\bp}\,&&\cE^\alf_\bp\,&&\cG^{}_{ip_0,\bp}\,&&\cE^\delta_\bp\,&&\cG^{}_{ip_0,\bp}\,&&\cE^{\beta\gamma}_\bp\,&&\Big] \, ,  \\[3mm]
  \label{eqn:PiHInter}\Pi^{H,\alf\beta\gamma\delta}_{iq_0,\text{inter}}=&-\frac{1}{4}\wideTrH\Big[\,&&\cG^{}_{ip_0+iq_0,\bp}\,&&\cF^\alf_\bp\,&&\cG^{}_{ip_0,\bp}\,&&\cE^\delta_\bp\,&&\cG^{}_{ip_0,\bp}\,&&\dF^{\beta\gamma}_\bp\,&&\Big]  \\[1mm]
  \label{eqn:PiHInter2}&-\frac{1}{2}\wideTrH\Big[\,&&\cG^{}_{ip_0+iq_0,\bp}\,&&\cF^\alf_\bp\,&&\cG^{}_{ip_0,\bp}\,&&\cF^\delta_\bp\,&&\cG^{}_{ip_0,\bp}\,&&\cE^{\beta\gamma}_\bp\,&&\Big] \, .
\end{alignat}
When performing the matrix trace explicitly, the three Green's functions are of the same band in each term in the intraband contribution. In the interband contribution, the three Green's functions of the two bands mix. Note that the final result does involve the inverse quasiparticle effective mass $\cE^{\beta\gamma}_\bp$ instead of the inverse generalized effective mass $(\dM^{-1})^{\beta\gamma}_\bp$. Be aware that the steps from \eqref{eqn:PiHTriSpecial} and \eqref{eqn:PiHRecSpecial} to \eqref{eqn:PiHDecomp} crucially rely on the assumption of a momentum-independent gap. 

%
%

\subsection{Matsubara summation}
\label{sec:theory:Hall:matsubara}

In the following, we perform the Matsubara summation and the analytic continuation of the intra- and interband contributions in \eqref{eqn:PiHIntra}, \eqref{eqn:PiHInter} and \eqref{eqn:PiHInter2}. We omit the momentum dependence and introduce the reduced notation $\cG\equiv\cG_{ip_0}$ and $\cG_{\pm iq_0}\equiv \cG_{ip_0\pm iq_0}$ for simplicity in this section. The involved Matsubara summation over the (fermionic) Matsubara frequency $p_0$ is of the form
\begin{align}
 \label{eqn:MatsumH}
 I^H_{iq_0}\equiv T\sum_{p_0}\tr\Big[\big(\cG^{}_{iq_0}-\cG^{}_{-iq_0}\big)\,M^{}_1\,\cG\, M^{}_2 \,\cG \,M^{}_3\Big]
\end{align}
with three different particular choices of the matrices $M_i$. $T$ is the temperature. After performing the Matsubara summation and the analytic continuation of the external Matsubara frequency $iq_0\rightarrow \omega+i0^+$ as shown in Appendix~\ref{appendix:Matsubara}, we obtain  the general result
\begin{alignat}{3}
\label{eqn:IH}
 \lim_{\omega\rightarrow 0}\frac{I^H_\omega}{\omega}=
 2\hspace{-0.1cm}\int\hspace{-0.2cm}d\eps\,f_\eps\,\tr\big[&+\pi^2\,A^{}_\eps\,M^{}_1\,A'_\eps\,M^{}_2\,A^{}_\eps\,M^{}_3&&+\pi^2\,A^{}_\eps\,M^{}_1\,A^{}_\eps\,M^{}_2\,A'_\eps\,M^{}_3&&
 \\[-1mm]
 \label{eqn:IH2}&+P'_\eps\,M^{}_1\,P^{}_\eps\,M^{}_2\,A^{}_\eps\,M^{}_3&&+P'_\eps\,M^{}_1\,A^{}_\eps\,M^{}_2\,P^{}_\eps\,M^{}_3&&
 \\[1mm]
 \label{eqn:IH3}&-A^{}_\eps\,M^{}_1\,P'_\eps\,M^{}_2\,P^{}_\eps\,M^{}_3&&-A^{}_\eps\,M^{}_1\,P^{}_\eps\,M^{}_2\,P'_\eps\,M^{}_3\big] \, .
\end{alignat}
for the zero frequency limit. The six terms involve the Fermi function $f_\eps=(e^{\eps/T}+1)^{-1}$, the spectral function matrix $A_\eps=-(\cG^R_\eps-\cG^A_\eps)/2\pi i$ and the principle-value function matrix $P_\eps=(\cG^R_\eps+\cG^A_\eps)/2$. We denote the derivative with respect to the internal frequency $\eps$ as $A'_\eps=\partial_\eps A_\eps$ and $P'_\eps=\partial_\eps P_\eps$. Note that the involved matrices may not commute so that the order is crucial. Using the explicit form of the retarded and advanced Green's function, $G^{R/A}_\eps=[\eps-\cE\pm i\Gamma]^{-1}$, we have the identity $P'_\eps=2\pi^2 A^2_\eps-\pi\,A_\eps/\Gamma$, which have already been used in the derivation of the formula for the anomalous Hall conductivity. Plugging this into \eqref{eqn:IH2} and \eqref{eqn:IH3}, the second term, which is linear in $A_\eps$, drops. We get
\begin{alignat}{3}
 \lim_{\omega\rightarrow 0}\frac{I^H_\omega}{\omega}=
 2\pi^2\hspace{-0.1cm}\int\hspace{-0.2cm}d\eps\,f_\eps\,\tr\big[&+A^{}_\eps\,M^{}_1\,A'_\eps\,M^{}_2\,A^{}_\eps\,M^{}_3&&+A^{}_\eps\,M^{}_1\,A^{}_\eps\,M^{}_2\,A'_\eps\,M^{}_3&&
 \\[-1mm]
 \label{eqn:IHw2}&+2A^2_\eps\,M^{}_1\,P^{}_\eps\,M^{}_2\,A^{}_\eps\,M^{}_3&&+2A^2_\eps\,M^{}_1\,A^{}_\eps\,M^{}_2\,P^{}_\eps\,M^{}_3&&
 \\[1mm]
 \label{eqn:IHw3}&-2A^{}_\eps\,M^{}_1\,A^2_\eps\,M^{}_2\,P^{}_\eps\,M^{}_3&&-2A^{}_\eps\,M^{}_1\,P^{}_\eps\,M^{}_2\,A^2_\eps\,M^{}_3\big] \, .
\end{alignat}
In order to perform further simplifications, we consider the three terms in \eqref{eqn:PiHIntra}, \eqref{eqn:PiHInter} and \eqref{eqn:PiHInter2} explicitly. We use that $\cE^\nu,\cE^{\nu\mu} \propto \mathds{1}$ and $\cF^\nu,\dF^{\nu\mu}\propto \pauma_x$ for all combinations of indices $\nu,\mu=\alf,\beta,\gamma,\delta$. Thus, $\cE^\nu$ and $\cE^{\nu\mu}$ do commute with the diagonal matrices $A_\eps$, $A'_\eps$ and $P_\eps$. The off-diagonal matrices $\cF^\nu$ and $\dF^{\nu\mu}$ flip the diagonal elements of the diagonal matrices $A^{}_\eps$, $A'_\eps$ and $P^{}_\eps$ under commutation, for instance $A_\eps \cF^\nu=\cF^\nu \overline{A_\eps}$ with $\overline{A_\eps}$ being the (diagonal) matrix with  diagonal elements $A^+_\eps\leftrightarrow A^-_\eps$ exchanged. The intraband contribution in \eqref{eqn:PiHIntra} involves $M_1=\cE^\alf$, $M_2=\cE^\delta$ and $M_3=\cE^{\beta\gamma}$. The two lines in \eqref{eqn:IHw2} and \eqref{eqn:IHw3} cancel. We have
\begin{align}
 \label{eqn:IHintra}I^H_\text{intra}\equiv
 4\pi^2\hspace{-0.1cm}\int\hspace{-0.2cm}d\eps\,f_\eps\,\tr\big[\cE^\alf\,\cE^\delta\,\cE^{\beta\gamma}\,A^2_\eps\,A'_\eps\big]=
 -\frac{4\pi^2}{3}\hspace{-0.1cm}\int\hspace{-0.2cm}d\eps\,f'_\eps\,\tr\big[\cE^\alf\,\cE^\delta\,\cE^{\beta\gamma}\,A^3_\eps\big] \, ,
\end{align}
where we used the chain rule and performed partial integration in the internal frequency $\eps$. The first term of the interband contribution in \eqref{eqn:PiHInter} involves $M_1=\cF^\alf$, $M_2=\cE^\delta$ and $M_3=\dF^{\beta\gamma}$. We have
\begin{align}
 I^H_\text{inter,1}\equiv
 4\pi^2\hspace{-0.1cm}\int\hspace{-0.2cm}d\eps\,f_\eps\,\tr\big[\cF^\alf\,\cE^\delta\,\dF^{\beta\gamma}\,\big(A^{}_\eps\,\overline{A'_\eps\,A^{}_\eps}+2A^2_\eps\,\overline{P^{}_\eps\,A^{}_\eps}-2A^{}_\eps\,\overline{A^2_\eps\,P^{}_\eps}\big)\big] \, .
\end{align}
Note that $\overline{A'_\eps\,A^{}_\eps}=\overline{A'_\eps}\,\,\overline{A^{}_\eps}$, and for $\overline{P^{}_\eps\,A^{}_\eps}$ and $\overline{A^2_\eps\,P^{}_\eps}$ accordingly, since the involved matrices are diagonal. Using the explicit form of the spectral function, $A_\eps=\Gamma/\pi\,[(\eps-\cE)^2+\Gamma^2]^{-1}$, we can write its derivative as the product of the principle-value function $P_\eps=(\eps-\cE)[(\eps-\cE)^2+\Gamma^2]^{-1}$ and the spectral function itself, that is, $A'_\eps=-2P_\eps A_\eps $. Using this, we obtain
\begin{align}
 I^H_\text{inter,1}=4\pi^2\hspace{-0.1cm}\int\hspace{-0.2cm}d\eps\,f_\eps\,\tr\big[\cF^\alf\,\cE^\delta\,\dF^{\beta\gamma}\,\big(2A^2_\eps\,\overline{P^{}_\eps\,A^{}_\eps}-4A^{}_\eps\,\overline{A^2_\eps\,P^{}_\eps}\big)\big] \, .
\end{align}
We can relate the second term to the frequency derivative of $\overline{A^2_\eps}A^{}_\eps$ by using the identity $-4A^{}_\eps \overline{A^2_\eps\,P^{}_\eps}=\big(\overline{A^2_\eps}A^{}_\eps\big)'+2A^{}_\eps \overline{A^2_\eps}P_\eps$. This identity can be checked by using the explicit form of the spectral function. Thus, we obtain 
\begin{align}
 I^H_\text{inter,1}=4\pi^2\hspace{-0.1cm}\int\hspace{-0.2cm}d\eps\,f_\eps\,\tr\Big[\cF^\alf\,\cE^\delta\,\dF^{\beta\gamma}\,\Big(\big(\overline{A^2_\eps}\,A^{}_\eps\big)'+2A^{}_\eps\, \overline{A^{}_\eps}\,\big(\overline{A^{}_\eps}\,P^{}_\eps+A^{}_\eps\,\overline{P^{}_\eps}\big)\Big)\Big] \, .
\end{align}
In a final step, we can use the identity $\overline{A_\eps}P_\eps+A_\eps\overline{P_\eps}=A_\eps\,\cS+\overline{A_\eps\,\cS}$ with $\cS=1/(E^+-E^-)\,\pauma_z$ in order to express the full result only in combinations of the spectral function or its derivative. We obtain
\begingroup
 \allowdisplaybreaks[0]
\begin{align}
 \label{eqn:IHinter1}I^H_\text{inter,1}=&-4\pi^2\hspace{-0.1cm}\int\hspace{-0.2cm}d\eps\,f'_\eps\,\tr\big[\cF^\alf\,\cE^\delta\,\dF^{\beta\gamma}\,\overline{A^2_\eps}\,A^{}_\eps\big]\\&\label{eqn:IHinter1b}+8\pi^2\hspace{-0.1cm}\int\hspace{-0.2cm}d\eps\,f_\eps\,\tr\big[\cF^\alf\,\cE^\delta\,\dF^{\beta\gamma}\,\big(A^2_\eps\, \overline{A^{}_\eps}\cS+A^{}_\eps\,\overline{A^2_\eps\cS^{}}\big)\big] \, ,
\end{align}
\endgroup
where we performed partial integration in the internal frequency $\eps$ in the first term. We have, thus, obtained one contribution proportional to the derivative of the Fermi function and one contribution proportional to the Fermi function itself. The second term of the interband contribution in \eqref{eqn:PiHInter2} involves $M_1=\cF^\alf$, $M_2=\cF^\delta$ and $M_3=\cE^{\beta\gamma}$. We have
\begin{align}
 I_\text{inter,2}\equiv
 2\pi^2\hspace{-0.1cm}\int\hspace{-0.2cm}d\eps\,f_\eps\,\tr\big[\cF^\alf\,\cF^\delta\,\cE^{\beta\gamma}\big(A^{}_\eps\,\overline{A'_\eps}\,A^{}_\eps\hspace{-0.05cm}+\hspace{-0.05cm}A^{}_\eps\,\overline{A^{}_\eps}\,A'_\eps
 \hspace{-0.05cm}+\hspace{-0.05cm}2A^2_\eps\,\overline{A^{}_\eps}\,P^{}_\eps
 \hspace{-0.05cm}-\hspace{-0.05cm}2A^{}_\eps\,\overline{A^2_\eps}\,P^{}_\eps\big)\big] \, .
\end{align}
Expressing the derivative of the spectral function as $A'_\eps=-2P_\eps A_\eps$ and using the identity $\overline{A_\eps}P_\eps+A_\eps\overline{P_\eps}=A_\eps\,\cS+\overline{A_\eps\,\cS}$, we get
\begin{align}
 \label{eqn:IHinter2}I_\text{inter,2}=&-4\pi^2\hspace{-0.1cm}\int\hspace{-0.2cm}d\eps\,f_\eps\,\tr\big[\cF^\alf\,\cF^\delta\,\cE^{\beta\gamma}\,\big(A^2_\eps\,\overline{A^{}_\eps}\,\cS
 +A^{}_\eps\,\overline{A^2_\eps\,\cS}\big)\big] \, .
\end{align}
Note that \eqref{eqn:IHinter2} involves the same combination of spectral functions as \eqref{eqn:IHinter1b}. The polarization tensor $\lim_{\omega\rightarrow 0}\Pi^{H,\alf\beta\gamma\delta}_{iq_0\rightarrow \omega+i0^+}/\omega$ is obtained by combing \eqref{eqn:IHintra} with \eqref{eqn:PiHIntra}, \eqref{eqn:IHinter1} and \eqref{eqn:IHinter1b} with \eqref{eqn:PiHInter}, and \eqref{eqn:IHinter2} with \eqref{eqn:PiHInter2}.

%
%

\subsection{Formulas for the Hall conductivity tensor}
\label{sec:theory:Hall:formulas}

As the final step of the derivation, we combine all our results for the Hall conductivity under the assumption of a momentum-independent gap. We have shown that the tensor $\sigma^{\alf\beta\eta}_\text{\tiny EB}$ is purely antisymmetric in $\alf\leftrightarrow\beta$. Thus, we can identify it with the Hall conductivity tensor $\sigma^{\alf\beta\eta}_\text{H}=-\sigma^{\beta\alf\eta}_\text{H}$. The (Hall) conductivity tensor is related to the polarization tensor via \eqref{eqn:sH}. We write out the trace over the two bands explicitly. The zero-frequency (DC) Hall conductivity tensor $\sigma^{\alf\beta\eta}_\text{H}$ decomposes into four different contributions: 
\begin{align}
\label{eqn:DecompSigmaH}
 \sigma^{\alf\beta\eta}_\text{H}&=\sigma^{\alf\beta\eta}_{\text{H},\text{intra},+}+\sigma^{\alf\beta\eta}_{\text{H},\text{intra},-}+\sigma^{\alf\beta\eta}_{\text{H},\text{inter},+}+\sigma^{\alf\beta\eta}_{\text{H},\text{inter},-} \, .
\end{align}
Each contribution consists of three essential parts: i) the Fermi function $f(\epsilon)$ or its derivative $f'(\epsilon)$, ii) a spectral weighting factor involving a specific combination of the quasiparticle spectral functions $A^n_\bp(\epsilon)$ of band $n=\pm$, that is,
\begingroup
 \allowdisplaybreaks[0]
\begin{align}
 \label{eqn:WintraH}&w^{H,n}_{\bp,\text{intra}}(\epsilon)=\frac{2\pi^2}{3}\big(A^n_\bp(\epsilon)\big)^3 \, ,\\[2mm]
 \label{eqn:WinterH}&w^{H,n}_{\bp,\text{inter}}(\epsilon)=2\pi^2(E^+_\bp-E^-_\bp)^2 \big(A^n_\bp(\epsilon)\big)^2A^{-n}_\bp(\epsilon)\equiv w^{a,n}_{\bp,\text{inter}}(\eps)\, ,
\end{align}
\endgroup
with $-n$ denoting the opposite band, and iii) a momentum-dependent weighting factor involving the momentum derivatives of $\g_\bp=(\eps_{\bp,A}+\eps_{\bp,B})/2$ and $h_\bp=(\eps_{\bp,A}-\eps_{\bp,B})/2$. Note that the spectral weighting factor of the interband contribution in \eqref{eqn:WinterH} is equal to $w^{a,n}_{\bp,\text{inter}}$ in \eqref{eqn:Wainter}. For consistency of notation in this section, we replaced $r_\bp=(E^+_\bp-E^-_\bp)/2$. The spectral weighting factor of the intraband contribution in \eqref{eqn:WintraH} is very similar to $w^{n}_{\bp,\text{intra}}$ in \eqref{eqn:Wintra}, but involves the quasiparticle spectral function to the power of three instead to the power of two. The diagonal elements of the eigenenergy $\cE_\bp$, the quasiparticle velocity $\cE^\alf_\bp$ and the inverse quasiparticle effective mass $\cE^{\alf\beta}_\bp$ in terms of the simplified Bloch Hamiltonian in \eqref{eqn:lamSpecial} are given as 
\begin{alignat}{2}
 &E^{\pm}_\bp&&=\g_\bp\pm \sqrt{h^2_\bp+\Delta^2} \, ,\\[3mm]
 &E^{\pm,\alf}_\bp&&=\g^\alf_\bp\pm \frac{2h_\bp h^\alf_\bp}{E^+_\bp-E^-_\bp} \, ,\\[1mm]
 &E^{\pm,\alf\beta}_\bp&&=\g^{\alf\beta}_\bp\pm \bigg(\frac{2h_\bp h^{\alf\beta}_\bp}{E^+_\bp-E^-_\bp}+\bigg[1-\frac{h^2_\bp}{\Delta^2+h^2_\bp}\bigg]\frac{2h^\alf_\bp h^\beta_\bp}{E^+_\bp-E^-_\bp}\bigg) \, .
\end{alignat}
The off-diagonal elements of $\cF^\alf_\bp$ and $\dF^{\alf\beta}_\bp$ read
\begin{align}
 &F^\alf_\bp=\frac{2\Delta h^\alf_\bp}{E^+_\bp-E^-_\bp} \, , 
 &F^{\alf\beta}_\bp=\frac{2\Delta h^{\alf\beta}_\bp}{E^+_\bp-E^-_\bp}\, .
\end{align}
We write $\g^\nu_\bp=\partial_\nu \g_\bp$ , $\g^{\nu\mu}_\bp=\partial_\nu\partial_\mu \g_\bp$ and similar for $h_\bp$ with the momentum derivative in $\nu=x,y,z$ direction $\partial_\nu=\partial/\partial p^\nu$. We express the Hall conductivity in units of $e^3/\hbar^2$ by restoring $\hbar$, which is set to unity throughout this thesis. We perform the thermodynamic limit by replacing $L^{-1}\sum_\bp\rightarrow \int \frac{d^d\bp}{(2\pi)^d}$, where $d$ is the dimension of the system. We have  
\begingroup
 \allowdisplaybreaks[0]
\begin{alignat}{4}
 &\sigma^{\alf\beta\eta}_{\text{H},\text{intra},n}=&&-\hspace{-0.5mm}\sum_{\gamma,\delta}\frac{\eps^{\eta\gamma\delta}}{2}\,&&\frac{e^3}{\hbar^2}&&\int\hspace{-1.9mm}\frac{d^d\bp}{(2\pi)^d}\hspace{-1.7mm}\int\hspace{-1.5mm}d\epsilon \,f'(\epsilon)\, w^{H,n}_{\bp,\text{intra}}(\epsilon) \Big[E^{n,\alf}_\bp\,E^{n,\beta\gamma}_\bp\,E^{n,\delta}_\bp\hspace{-0.5mm}-\hspace{-0.5mm}(\alf\hspace{-0.5mm}\leftrightarrow\hspace{-0.5mm}\beta)\Big]\,, \label{eqn:SHintra}
 \\[3mm]
 &\sigma^{\alf\beta\eta}_{\text{H},\text{inter},n}=&&-\hspace{-0.5mm}\sum_{\gamma,\delta}\frac{\eps^{\eta\gamma\delta}}{2}\,&&\frac{e^3}{\hbar^2}&&\int\hspace{-1.9mm}\frac{d^d\bp}{(2\pi)^d}\hspace{-1.7mm}\int\hspace{-1.5mm}d\epsilon\,f'(\epsilon)w^{H,n}_{\bp,\text{inter}}(\epsilon) \bigg[\frac{F^\alf_\bp\,F^{\beta\gamma}_\bp\,E^{n,\delta}_\bp}{(E^+_\bp-E^-_\bp)^2}\hspace{-0.5mm}-\hspace{-0.5mm}(\alf\hspace{-0.5mm}\leftrightarrow\hspace{-0.5mm}\beta)\bigg] \nonumber
 \\[2mm]
 & &&+\hspace{-0.5mm}\sum_{\gamma,\delta}\eps^{\eta\gamma\delta}\,&&\frac{e^3}{\hbar^2}&&\int\hspace{-1.9mm}\frac{d^d\bp}{(2\pi)^d}\hspace{-1.7mm}\int\hspace{-1.5mm}d\epsilon\,f(\epsilon)\big(w^{H,+}_{\bp,\text{inter}}(\epsilon)-w^{H,-}_{\bp,\text{inter}}(\epsilon)\big) \nonumber \\
 & && && &&\hspace{1.5cm}\times\bigg[\frac{F^\alf_\bp\,F^{\beta\gamma}_\bp\, E^{n,\delta}_\bp-F^\alf_\bp\,E^{n,\beta\gamma}_\bp\,F^\delta_\bp}{(E^+_\bp-E^-_\bp)^3}\hspace{-0.5mm}-\hspace{-0.5mm}(\alf\hspace{-0.5mm}\leftrightarrow\hspace{-0.5mm}\beta)\bigg]
 \,. \label{eqn:SHinter}
\end{alignat}
\endgroup
If we restore SI units, the conductivity has units $[\Omega\,\text{m}^{d-2}\,\text{T}]^{-1}$ for dimension $d$. For a given Bloch Hamiltonian $\lam_\bp$ of the form \eqref{eqn:lamSpecial}, chemical potential $\mu$, temperature $T$ and relaxation rate $\Gamma$, the evaluation of \eqref{eqn:SHintra} and \eqref{eqn:SHinter} is straightforward. 

%
%

\subsection{Hall coefficient and Hall number}
\label{sec:theory:Hall:coefficient}

The Hall conductivity $\sigma^{\alf\beta\eta}_\text{H}$ is a tensor with indices $\alf$, $\beta$ and $\eta$, where the first index $\alf=x,y,z$ is the direction of the induced Hall current density $j^\alf_\text{H}$, the second index $\beta=x,y,z$ is the direction of the external electric field $E^\beta$ and the third index $\eta=x,y,z$ is the direction of the external magnetic field $B^\eta$. The Hall conductivity tensor is antisymmetric in the indices $\alf\leftrightarrow\beta$ and, thus, has nine independent elements. Therefore, we can write the induced current density in a vector form as
\begin{align}
 \bj_\text{H} = \bE\times (\bsigma_\text{H}\bB)
\end{align}
with the vectors $\bj_\text{H}=(j^\alf_\text{H})$, $\bE=(E^\beta)$ and $\bB=(B^\eta)$. The matrix $\bsigma_\text{H}$ is defined as
\begin{align}
\label{eqn:bsigmaH}
 \bsigma_\text{H}=\begin{pmatrix} \sigma^{yzx}_\text{H} & \sigma^{yzy}_\text{H} & \sigma^{yzz}_\text{H} \\ -\sigma^{xzx}_\text{H} & -\sigma^{xzy}_\text{H} & -\sigma^{xzz}_\text{H}\\ \sigma^{xyx}_\text{H} & \sigma^{xyy}_\text{H} & \sigma^{xyz}_\text{H} \end{pmatrix} \, ,
\end{align}
involving the nine independent elements of $\sigma^{\alf\beta\eta}_\text{H}$. We consider a two-dimensional system, which is assumed to lie in the $x$-$y$ plane without loss of generality. Then, the dispersion is independent of the $z$ direction, so that $\partial_z \eps_{\bp,A}=\partial_z \eps_{\bp,B}=0$.  By writing the components in \eqref{eqn:SHintra} and \eqref{eqn:SHinter} explicitly, we find that only the component $\sigma^{xyz}_\text{H}$ is nonzero as expected. This drastically simplifies \eqref{eqn:bsigmaH}. The only nonzero components read $j^x_\text{H}=\sigma^{xyz}_\text{H}E^yB^z$ and $j^y_\text{H}=-\sigma^{xyz}_\text{H}E^xB^z$. This is a transverse current with strength $|\bj_\text{H}|=\sigma^{xyz}_\text{H}B^z|\bE|$. 

The Hall conductivity tensor $\sigma^{\alf\beta\eta}_\text{H}$ for the simplified Bloch Hamiltonian in \eqref{eqn:lamSpecial} decomposes into four components in \eqref{eqn:DecompSigmaH}. These components were given in \eqref{eqn:SHintra} and \eqref{eqn:SHinter} for arbitrary indices. In the following, we give the Hall conductivity  
\begin{align}
 \sigma^{xyz}_\text{H}=\sigma^{xyz}_{\text{H},\text{intra},+}+\sigma^{xyz}_{\text{H},\text{intra},-}+\sigma^{xyz}_{\text{H},\text{inter},+}+\sigma^{xyz}_{\text{H},\text{inter},-}
\end{align}
explicitly for a two-dimensional system in the $x$-$y$ plane with a perpendicular external magnetic field in $z$ direction. The intraband contributions for the quasiparticle bands $n=\pm$ read
\begin{alignat}{1}
 &\sigma^{xyz}_{\text{H},\text{intra},n}=\frac{1}{2}\,\frac{e^3}{\hbar^2}\int\hspace{-1.9mm}\frac{d^2\bp}{(2\pi)^2}\hspace{-1.7mm}\int\hspace{-1.5mm}d\epsilon \,f'(\epsilon)\, w^{H,n}_{\bp,\text{intra}}(\epsilon) \Big[(E^{n,x}_\bp)^2\,E^{n,yy}_\bp-E^{n,x}_\bp\,E^{n,y}_\bp\,E^{n,xy}_\bp+(x\leftrightarrow y)\Big]\,. \label{eqn:SHintraXYZ}
 \end{alignat}
We write $\hbar$ explicitly here and in the rest of this section. We indicated the additions of the previous terms with exchanged $x$ and $y$ indices via $(x\leftrightarrow y)$. The interband contribution reads
\begingroup
 \allowdisplaybreaks[0]
\begin{alignat}{1}
 &\sigma^{xyz}_{\text{H},\text{inter},n}=-\frac{1}{2}\,\frac{e^3}{\hbar^2}\int\hspace{-1.9mm}\frac{d^2\bp}{(2\pi)^2}\hspace{-1.7mm}\int\hspace{-1.5mm}d\epsilon\,f'(\epsilon)\,w^{H,n}_{\bp,\text{inter}}(\epsilon) \bigg[\frac{F^x_\bp\,F^{yx}_\bp\,E^{n,y}_\bp-F^y_\bp\,F^{xx}_\bp\,E^{n,y}_\bp}{(E^+_\bp-E^-_\bp)^2}+(x\leftrightarrow y)\bigg] \nonumber
 \\[3mm]
 &\hspace{1.9cm} +\,\frac{e^3}{\hbar^2}\int\hspace{-1.9mm}\frac{d^2\bp}{(2\pi)^2}\hspace{-1.7mm}\int\hspace{-1.5mm}d\epsilon\,f(\epsilon)\big(w^{H,+}_{\bp,\text{inter}}(\epsilon)-w^{H,-}_{\bp,\text{inter}}(\epsilon)\big) \nonumber \\[2mm]
 & \hspace{1.9cm}\times\bigg[\frac{F^x_\bp\,F^{yx}_\bp\, E^{n,y}_\bp-F^y_\bp\,F^{xx}_\bp\, E^{n,y}_\bp-F^x_\bp\,E^{n,yx}_\bp\,F^y_\bp+F^y_\bp\,E^{n,xx}_\bp\,F^y_\bp}{(E^+_\bp-E^-_\bp)^3}+(x\leftrightarrow y)\bigg]
 \,. \label{eqn:SHinterXYZ}
\end{alignat}
\endgroup
Note that the exchange of $x$ and $y$ in \eqref{eqn:SHintraXYZ} and \eqref{eqn:SHinterXYZ} is not equivalent with the symmetry in $\alf\leftrightarrow\beta$ of the tensor $\sigma^{\alf\beta\eta}_\text{H}$ itself and, thus, not in contradiction with its antisymmetry, $\sigma^{xyz}_\text{H}=-\sigma^{yxz}_\text{H}$. The Hall conductivity $\sigma_\text{H}^{xyz}$ and the longitudinal conductivities determine the {\em Hall coefficient}
\begin{equation}
 \label{eqn:RH}
 R_\text{H} = \frac{\sigma_\text{H}^{xyz}}{\sigma^{xx} \sigma^{yy}} \, .
\end{equation}
where $\sigma^{xx}$ and $\sigma^{yy}$ are the longitudinal conductivities given in \eqref{eqn:DecompSigma}. We assumed that $\sigma^{xy}=\sigma^{yx}=0$. Unlike the longitudinal and the Hall conductivities, the Hall coefficient is finite in the limit of small relaxation rate $\Gamma$. In order to see this, note that the formulas in \eqref{eqn:SHintraXYZ} and \eqref{eqn:SHinterXYZ} hold for a relaxation rate $\Gamma$ of arbitrary size. The dependence on $\Gamma$ is captured by the spectral weighting factors $w^{H,n}_{\bp,\text{intra}}(\eps)$ and $w^{H,n}_{\bp,\text{inter}}(\eps)$, which were defined in \eqref{eqn:WintraH} and \eqref{eqn:WinterH}, respectively. We can perform the limit of small relaxation rate for these spectral weighting factors and obtain
\begin{align}
 &w^{H,n}_{\bp,\text{intra}}(\eps)\approx \frac{1}{(2\Gamma)^2}\delta(\eps+\mu-E^n_\bp)\sim \mathcal{O}(\Gamma^{-2})\, ,\\[2mm]
 &w^{H,n}_{\bp,\text{inter}}(\eps)\approx \delta(\eps+\mu-E^n_\bp)\sim \mathcal{O}(\Gamma^{0}) \, .
\end{align}
The limit applies under the same conditions as for the ordinary conductivity, which we discussed in detail in Sec.~\ref{sec:theory:conductivity:limits}: The momentum-dependent functions $E_\bp^{\pm,\alf}$, $F_\bp^\alf$, and so on, must be almost constant in the momentum range in which the variation of $E_\bp^\pm$ is of order $\Gamma$, and $\Gamma$ must be much smaller than the direct band gap $E_\bp^+ - E_\bp^-$. Here, for the Hall conductivity, the interband contributions are suppressed by a factor $\Gamma^2$ compared to the intraband contributions. Such a suppression was also found for the interband contribution compared to the longitudinal conductivity. Since the longitudinal conductivity scales as $\sigma^{xx}\sim\sigma^{yy}\sim\Gamma^{-1}$ and the Hall conductivity scales as $\sigma^{xyz}_\text{H}\sim \Gamma^{-2}$ in the clean limit, we see that the Hall coefficient in \eqref{eqn:RH} is independent of $\Gamma$ in the clean limit.

After having performed the limit of a small relaxation rate, we can integrate over the internal frequency $\eps$ and obtain
\begin{align} \label{sg_Hintra3}
 \sigma_{\text{H},\text{intra},n}^{xyz} \rightarrow\,\,&
 \frac{1}{2}\frac{e^3\tau^2}{\hbar^2} \int \frac{d^2\bp}{(2\pi)^2}  f'(E^n_\bp) \Big[ (E_\bp^{n,x})^2 E_\bp^{n,yy} - E_\bp^{n,x} E_\bp^{n,y} E_\bp^{n,xy}
 + (x \leftrightarrow y) \Big] \, ,
\end{align}
where we replaced the relaxation rate by the relaxation time $\tau=1/2\Gamma$. Using $f'(E^n_\bp) \, E_\bp^{n,\alf}= \partial_\alf f(E_\bp^n)$, and performing a partial integration in momentum, this can also be written as
\begin{align} \label{sg_Hintra4}
 \sigma_{\text{H},\text{intra},n}^{xyz} \rightarrow 
 - \frac{e^3 \tau^2}{\hbar^2} &\int \frac{d^2\bp}{(2\pi)^2}f(E_\bp^n) \Big[ E_\bp^{n,xx} E_\bp^{n,yy} - E_\bp^{n,xy} E_\bp^{n,yx} \Big] \, .
\end{align}
These results in \eqref{sg_Hintra3} and \eqref{sg_Hintra4} agree with the corresponding expressions derived by Voruganti {\it et al.} \cite{Voruganti1992} and are equivalent to the result obtained in Boltzmann transport theory in \eqref{eqn:HallBoltzmann} when replacing the bare dispersion with the dispersion of the quasiparticle band.

There are special cases where the Hall coefficient is determined by the charge density $\rho_c$ via the simple relation $R_\text{H} = \rho_c^{-1}$. For free electrons with a parabolic dispersion, this relation holds for any magnetic field, with $\rho_c = -e n_e$. Note that we use the convention $e>0$. For band electrons, it still holds in the high-field limit $\omega_c \tau \gg 1$, if the semiclassical electron orbits of all occupied (or all unoccupied) states are closed \cite{Ashcroft1976}. For Fermi surfaces enclosing unoccupied states, the relevant charge density is then $\rho_c = +e n_h$, where $n_h$ is the density of holes. If both electron and hole-like Fermi surfaces are present, one has $\rho_c = e(n_h - n_e)$ \cite{Ashcroft1976}. Results for the Hall conductivity are, thus, frequently represented in terms of the so-called {\em Hall number} $n_\text{H}$, defined via the relation
\begin{equation}
 \label{eqn:nH}
 R_\text{H} = \frac{1}{e \, n_\text{H}} \, .
\end{equation}
We use the convention that electron-like contributions are counted negatively and hole-like contributions are counted positively in the Hall number $n_\text{H}$ . However, $n_\text{H}$ is given by the electron and hole densities only in the special cases described above. We will see an example in Sec.~\ref{sec:application:spiral}.

We finally emphasize that our derivation is valid under the assumption of a momentum-independent gap $\Delta_\bp=-\Delta$ and a momentum-independent relaxation rate $\Gamma$.
A generalization to a momentum-dependent gap is not straightforward, since numerous additional terms appear and the simplifications in Sec.~\ref{sec:theory:Hall:simplifications} are not valid anymore.

\cleardoublepage



\rhead[\fancyplain{}
{\bfseries Applications}
]{\fancyplain{}{
}}
\lhead[\fancyplain{}{
}]{\fancyplain{}
{\bfseries Applications}
}

\chapter{Applications}
\label{sec:application}  %

The general two-band model that we introduced in Chapter~\ref{sec:theory} captures a broad variety of very different physical systems. We discussed the longitudinal conductivity as well as the anomalous and ordinary Hall conductivity in this general context and derived several formulas including interband contributions, which we analyzed in detail. In this chapter, we now apply our general theory to different physically motivated models in order to exemplify the general conclusions and to gain further insight into physical problems that were not possible to obtain before without formulas that include interband contributions. 

We split this chapter into two parts. In Sec.~\ref{sec:application:anomalousHallEffect}, we focus on the conductivity in the context of material with topological properties. The anomalous transport behavior is intrinsically linked to the interplay between the two bands. The generalization to a relaxation rate of arbitrary size as well as the unique decomposition allows us to clarify several aspects that remained unclear in earlier treatments. We go beyond the ``dissipationless'' limit of the quantum anomalous Hall effect and discuss the scaling behavior of the anomalous Hall conductivity with respect to the relaxation rate and with respect to the longitudinal conductivity in application to experimental results on ferromagnets. In Sec.~\ref{sec:application:spiral}, we analyze the phenomenology and the transport properties of a two-dimensional tight-binding model with onsite spiral magnetic order. This model breaks translational symmetry but has combined lattice translation and spin rotation symmetry and has, thus, several unconventional properties. We relate this model to recent experiments on cuprates and discuss the relevance of interband contributions in the analysis of the Hall number, which was studied experimentally in very strong magnetic fields.

%
%

\section{Anomalous Hall effect}
\label{sec:application:anomalousHallEffect}

In recent years, there is an increasing interest in the transport properties of systems with topological properties \cite{Culcer2020, Xu2020, Sun2020, Kubler2014, Manna2018, Noky2020, Destraz2020, Li2020, Nagaosa2020, Sharpe2019, McIver2020, Gianfrate2020, Cooper2019}. The established connection between the intrinsic anomalous Hall effect and the Berry curvature \cite{Thouless1982, Niu1985, Kohmoto1985, Onoda2002, Jungwirth2002} has become a powerful tool for combining theoretical predictions and experimental results. In the following two sections, we apply our general theory to two examples, a Chern insulator and a quasi-two-dimensional ferromagnetic multi-d-orbital model, which both involve a nonzero Berry curvature and, thus, show the anomalous Hall effect.

In Sec.~\ref{sec:application:anomalousHallEffect:Wilson}, we discuss the quantum anomalous Hall effect, the quantized version of the anomalous Hall effect, in a Chern insulator. The Berry curvature integrated over a full band is quantized to integer values \cite{Thouless1982}, where a zero or a nonzero integer is the defining character of a topologically trivial or non-trivial band, respectively. A Chern insulator may involve such topological non-trivial bands for certain parameter ranges. Due to the direct link between the Berry curvature and the intrinsic anomalous Hall effect, the anomalous Hall conductivity can, thus, be quantized in this range. We motivate a model of a Chern insulator via a tight-binding model presented by Nagaosa {\it et al.} \cite{Nagaosa2010} and discuss its transport properties. We identify a regime of quantized anomalous Hall conductivity, study the impact of the relaxation rate $\Gamma$ on the quantization, and discuss the different contributions to the conductivities. 

The formulas that we derived in our general theory are valid for a relaxation rate $\Gamma$ of arbitrary size. This allows us to study the scaling behavior of the conductivities with respect to the relaxation rate for both small and large $\Gamma$. As a consequence, we can apply those results in order to understand the scaling of the anomalous Hall conductivity with respect to the longitudinal conductivity, $\sigma^{xy}\propto (\sigma^{xx})^\nu$, which is often used in the analysis of experimental results (see, for instance, Ref.~\cite{Onoda2008}). In Sec.~\ref{sec:application:anomalousHallEffect:scaling}, we re-discuss the scaling behavior of a quasi-two-dimensional ferromagnetic multi-d-orbital model with spin-orbit coupling proposed by Kontani {\it et al.} \cite{Kontani2007} and show that a formerly proposed non-integer scaling behavior can be understood as a crossover regime in good agreement with experimental results.

%
%

\subsection{Quantum anomalous Hall effect in a Chern insulator}
\label{sec:application:anomalousHallEffect:Wilson}

We discuss the Wilson fermion model, a two-dimensional lattice model of a Chern insulator \cite{Grushin2018}. The main focus lies on the quantized anomalous Hall effect due to a finite Chern number of the fully occupied band in order to illustrate our discussion in Sec.~\ref{sec:theory:conductivity:anomalousHall}. The Wilson fermion model is motivated via a tight-binding model presented by Nagaosa {\it et al.} \cite{Nagaosa2010}, which we recapitulate in the following. We assume a two-dimensional square lattice with three orbitals $s$, $p_x$, $p_y$, and spin. The three orbitals are located at the same lattice site. We include the hopping between these sites and a simplified spin-orbit interaction between the $z$ component of the spin and the orbital moment. Furthermore, we assume to be in the ferromagnetic state with spin $\uparrow$ only. Due to the spin-orbit interaction, the $p$-orbitals are split into $p_x\pm ip_y$. We identify that the effective two-band low-energy model has an Hamiltonian of the form in \eqref{eqn:H} with the two subsystems $A=(s,\uparrow)$ and $B=(p_x-ip_y,\uparrow)$ and with $\brho_A=\brho_B=0$ and $\bQ_A=\bQ_B=0$. The Bloch Hamiltonian reads
\begin{align}
 \lam_\bp=\begin{pmatrix} \eps_s-2t_s\big(\cos p^x+\cos p^y\big) & \sqrt{2}\,t_{sp} \big(i \sin p^x+\sin p^y\big) \\[3mm] \sqrt{2}\,t^*_{sp}(-i\sin p^x+\sin p^y\big) & \eps_p+t_p\big(\cos p^x+\cos p^y\big) \end{pmatrix} \, ,
\end{align}
where $\eps_s$ and $\eps_p$ are the energy levels of the two orbitals. The real numbers $t_s$ and $t_p$ describe the hopping within one orbital and the complex number $t_{sp}$ describes the hopping between the two orbitals. We refer for a more detailed motivation to Nagaosa {\it et al.} \cite{Nagaosa2010}. In the following, we further reduce the number of parameters by setting $t_s=t$, $t_p/t=2$, $t_{sp}/t=1/\sqrt{2}$ and $\eps_s/t=-\eps_p/t=m$. We recover the two-dimensional Wilson fermion model \cite{Grushin2018} with only one free dimensionless parameter, which we labeled as $m$, and energy scale $t$. We discuss the conductivity of this model as a function of $m$ and the chemical potential $\mu$. 

We give some basic properties of the model. The quasiparticle dispersions are 
\begin{align}
E^\pm_\bp/t=\pm\sqrt{(m-2\cos p^x-2\cos p^y)^2+\sin^2 p^x+\sin^2 p^y} \,.
\end{align}
The gap closes in form of a Dirac point at $(p^x,p^y)=(\pm\pi,\pm\pi)$ for $m=-4$, at $(0,\pm\pi)$ and $(\pm \pi,0)$ for $m=0$, and at $(0,0)$ for $m=4$. For instance, the linearized Hamiltonian for $m=4$ near the gap reads $\lam_\bp/t=p^y\pauma_x-p^x\pauma_y$. The Chern number of the lower band calculated by its formula in \eqref{eqn:Chern} is $C_-=-1$ for $-4<m<0$, $C_-=1$ for $0<m<4$, and $C_-=0$ for $|m|>4$. As expected for the Chern number of the upper band, $C_+=-C_-$. The bandwidth is $W/t=4+|m|$.

\begin{figure}[t!]
\centering
\includegraphics[width=0.6\textwidth]{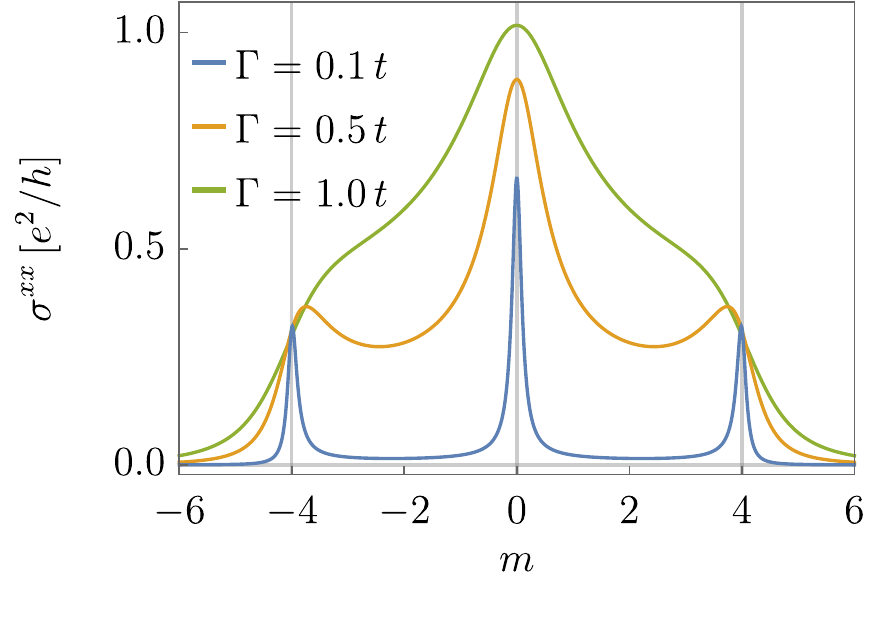}\\
\includegraphics[width=0.6\textwidth]{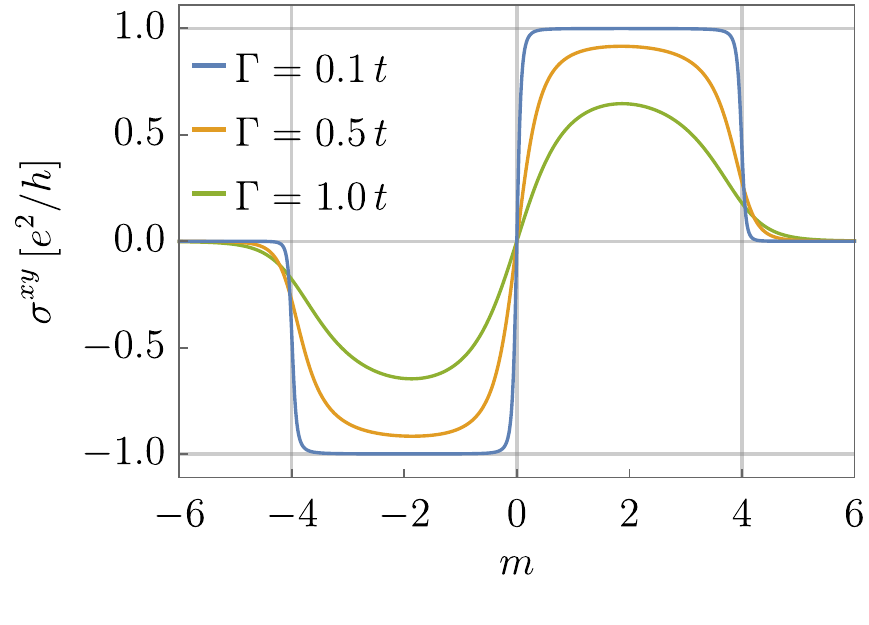}
\caption{The longitudinal conductivity $\sigma^{xx}$ and the anomalous Hall conductivity $\sigma^{xy}$ for different $\Gamma/t=0.1,\,0.5,\,1$ at $\mu=0$ and $T=0$. The vertical lines indicate the gap closings at $m=\pm4$ and $m=0$. \label{fig:3}}
\end{figure}

We calculate the diagonal conductivity $\sigma^{xx}$ and the off-diagonal conductivity $\sigma^{xy}$ by using \eqref{eqn:SintraN}-\eqref{eqn:SinterAN} in the zero temperature limit. The intraband and the symmetric interband contribution to the off-diagonal conductivity vanish after integrating over momenta, so that $\sigma^{xx}=\sigma^{yy}$ is the longitudinal conductivity and $\sigma^{xy}$ is the (antisymmetric) anomalous Hall conductivity. In Fig.~\ref{fig:3}, we plot $\sigma^{xx}=\sigma^{xx}_{\text{intra},+}+\sigma^{xx}_{\text{intra},-}+\sigma^{xx,s}_\text{inter}$ (upper figure) and $\sigma^{xy}=\sigma^{xy,a}_{\text{inter},+}+\sigma^{xy,a}_{\text{inter},-}$ (lower figure) as a function of the parameter $m$ at half filling, $\mu=0$. For a small relaxation rate $\Gamma=0.1\,t$, we find peaks of high longitudinal conductivity (blue) only when the gap closes at $m=\pm 4 $ and $m=0$, indicated by the vertical lines. For an increased relaxation rate $\Gamma=0.5\,t$ (orange), the peaks are broaden and the conductivity inside the gap is nonzero. For an even higher relaxation rate $\Gamma=1\,t$ (green), the peak structure eventually disappears and a broad range of finite conductivity is present. The anomalous Hall conductivity $\sigma^{xy}$ is quantized to $e^2/h$ due to a nonzero Chern number of the fully occupied lower band for low relaxation rate $\Gamma=0.1\,t$ (blue). At higher relaxation rates $\Gamma=0.5\,t$ (orange) and $\Gamma=1\,t$ (green), the quantization is no longer present most prominent for $m=\pm4$ and $m=0$, where the gap closes. 
\begin{figure}[t!]
\centering
\includegraphics[width=0.6\textwidth]{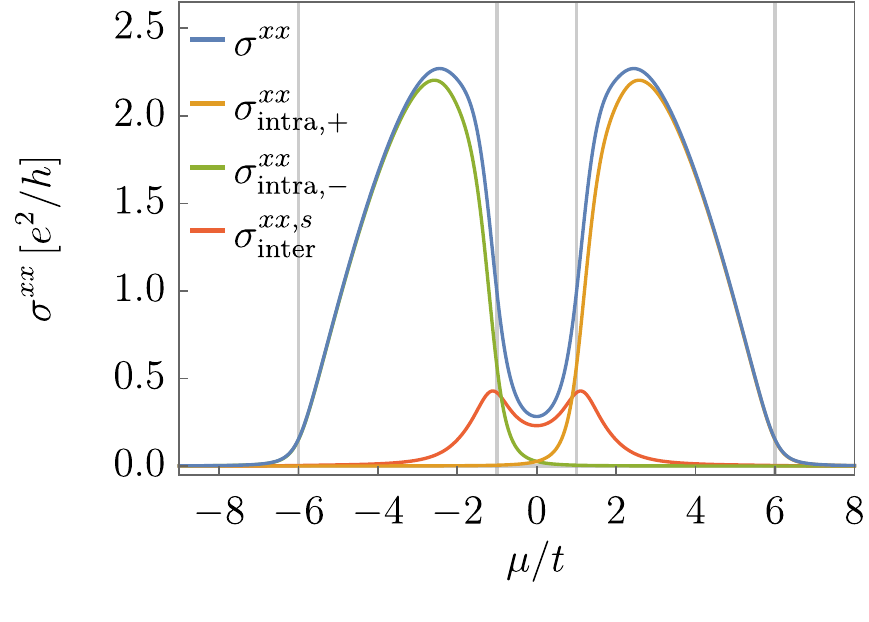}
\includegraphics[width=0.6\textwidth]{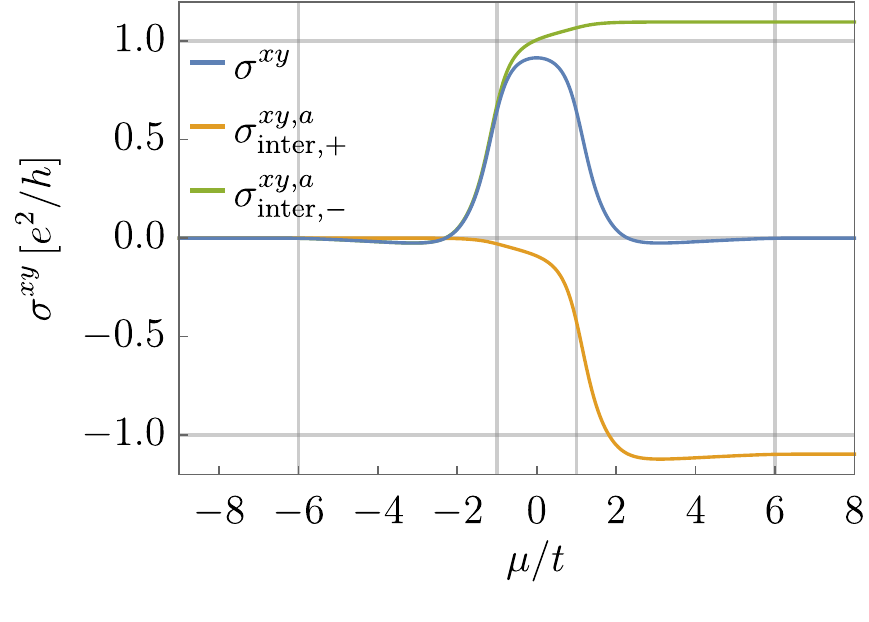}
\caption{The different contributions to $\sigma^{xx}$ and $\sigma^{xy}$ as a function of the chemical potential $\mu$ for $m=2$, $\Gamma=0.5\,t$ and $T=0$. The vertical lines indicate the upper and the lower end of the bands at $\mu/t=\pm6$, and the gap between $\mu/t=\pm1$.  \label{fig:4}}
\end{figure}

In Fig.~\ref{fig:4}, we show the different contributions to the longitudinal and the anomalous Hall conductivity as a function of the chemical potential $\mu$ for $m=2$ and $\Gamma=0.5\,t$. The lower and upper band end at $\mu/t=\pm 6$, respectively, and we have a gap of size $2\,t$ between $\mu/t=\pm 1$, both indicated by vertical lines. In the upper figure, we show the longitudinal conductivity $\sigma^{xx}$ (blue) and its three contributions, the intraband conductivity of the lower band $\sigma^{xx}_{\text{intra},-}$ (green), the intraband conductivity of the upper band $\sigma^{xx}_{\text{intra},+}$ (orange) and the symmetric interband conductivity $\sigma^{xx,s}_\text{inter}$ (red). We see that for $-6<\mu/t<-1$ the conductivity is dominated by the lower band, whereas it is dominated by the upper band for $1<\mu/t<6$. Inside the gap $-1<\mu/t<1$ the main contribution is due to the symmetric interband conductivity. We further see smearing effects at $\mu/t=\pm6$ and $\mu/t=\pm 1$. In the lower figure, we show the anomalous Hall conductivity $\sigma^{xy}$ (blue) as well as their two contributions, the antisymmetric interband conductivity of the lower band $\sigma^{xy,a}_{\text{inter},-}$ (green) and the upper band $\sigma^{xy,a}_{\text{inter},+}$ (orange). Both contributions are essentially zero for $\mu/t\lesssim-1$. The nonzero tail is caused by the finite relaxation rate $\Gamma$ and is suppressed for smaller values of $\Gamma$. Inside the gap $-1<\mu/t<1$, only the contribution of the lower band rises to approximately $e^2/h$, whereas the contribution of the upper band remains close to zero. Thus, we obtain a nonzero anomalous Hall conductivity. Above $\mu/t\gtrsim1$, the contribution of the upper band compensates the contribution of the lower band. The finite relaxation rate $\Gamma$ leads to a crossover regime with incomplete compensation. A large anomalous Hall effect is only present for a chemical potential inside the band gap. We see that a finite relaxation rate $\Gamma$ leads to a maximal value of the anomalous Hall conductivity of the two individual bands that is larger than $e^2/h$ as shown in Sec.~\ref{sec:theory:conductivity:limits}. Inside the gap the total anomalous Hall conductivity is reduced due to the nonzero contribution of the upper band. Around $\mu/t=\pm1$, we see smearing effects due to finite $\Gamma$, which we have described above for the individual contributions.

%
%

\subsection{Scaling behavior in Ferromagnets}
\label{sec:application:anomalousHallEffect:scaling}

We discuss a quasi-two-dimensional ferromagnetic multi-d-orbital model with spin-orbit coupling based on the work of Kontani {\it et al.} \cite{Kontani2007}. Similar to the previous example this model involves a nonzero Berry curvature and we expect a nonzero anomalous Hall conductivity, which is, by contrast, not quantized. We mainly focus on the scaling dependence with respect to the relaxation rate $\Gamma$ of the different contributions using our results of Sec.~\ref{sec:theory:conductivity:limits}. We comment on the consequences when analyzing experimental results in the dirty limit by determining the scaling behavior $\sigma^{xy}\propto (\sigma^{xx})^\nu$.

Following Kontani {\it et al.} \cite{Kontani2007}, we consider a square lattice tight-binding model with onsite $d_{xz}$ and $d_{yz}$ orbitals. We assume nearest-neighbor hopping $t$ between the $d_{xz}$ orbitals in $x$ direction and between the $d_{yz}$ orbitals in $y$ direction. Next-nearest-neighbor hopping $t'$ couples both types of orbitals. We assume a ferromagnetic material with magnetic moments in $z$ direction that is, completely spin-polarized in the spin $\downarrow$ direction. The Hamiltonian is of the form \eqref{eqn:H}, when we identify the two subsystems with quantum numbers $A=(d_{xz},\downarrow)$ and $B=(d_{yz},\downarrow)$. We have $\brho_A=\brho_B=0$ and $\bQ_A=\bQ_B=0$. The Bloch Hamiltonian reads
\begin{align}
 \lam_\bp=\begin{pmatrix} -2t\cos p^x & 4t'\sin p^x \sin p^y + i \lam \\[1mm] 4t'\sin p^x \sin p^y - i \lam & -2t\cos p^y \end{pmatrix} \, .
\end{align}
We included spin-orbit coupling $\lam$. Further details and physical motivations can be found in Kontani {\it et al.} \cite{Kontani2007}. We take the same set of parameters setting $t'/t=0.1$ and $\lam/t=0.2$ as in Ref.~\cite{Kontani2007}. We fix the particle number per unit cell to $n=0.4$ and adapt the chemical potential adequately. We consider temperature zero. 

\begin{figure}[t!]
\centering
\includegraphics[width=0.6\textwidth]{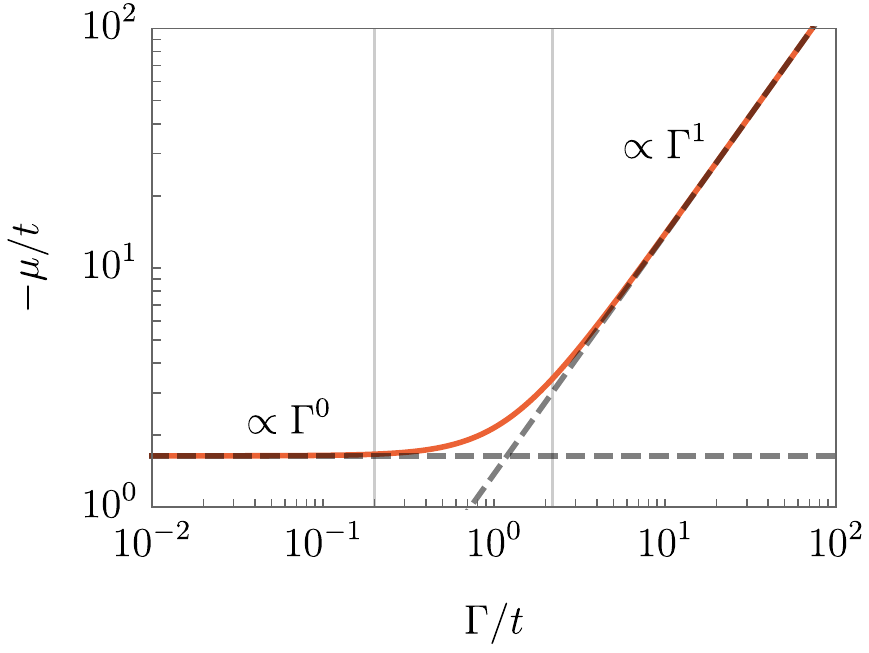}
\caption{The (negative) chemical potential $\mu$ as a function of the relaxation rate $\Gamma$ for $t'/t=0.1$ and $\lam/t=0.2$ at $n=0.4$. The chemical potential $\mu$ is $\Gamma$ independent below $\Gamma/t\ll0.2$ and scales linearly $\mu=\mu_\infty \Gamma$ above $\Gamma/t\gg2.2$ with $\mu_\infty=-1.376$ (dashed lines).  \label{fig:5}}
\end{figure}
The chemical potential $\mu$ becomes a function of the relaxation rate for fixed particle number (per unit cell) $n\equiv\rho_N$ according to \eqref{eqn:muGamma}. Whereas it is constant in the clean limit, the linear dependence on $\Gamma$ in the dirty limit is crucial and has to be taken into account carefully via a nonzero $\mu_\infty=-\tan(1-n)\pi/2\approx -1.376$ for $n=0.4$. The center of the two bands $c=(E^+_\text{max}+E^-_\text{min})/2=0$ drops out in \eqref{eqn:muGamma}. In Fig.~\ref{fig:5}, we plot the (negative) chemical potential $\mu/t$ as a function of the relaxation rate $\Gamma/t$, which was obtained by inverting $n(\mu,\Gamma)=0.4$ numerically for fixed $\Gamma$. We find the expected limiting behavior in the clean and the dirty limit indicated by dashed lines. The vertical lines are at those $\Gamma/t$, where $\Gamma/t$ is equal to the spin-orbit coupling $\lambda/t=0.2$, which is the minimal gap between the lower and the upper band $E^\pm_\bp$, and the band width $W/t=2.2$. Both scales give a rough estimate for the crossover region between the constant and the linear regime of the chemical potential. 

\begin{figure}[t!]
\centering
\includegraphics[width=0.6\textwidth]{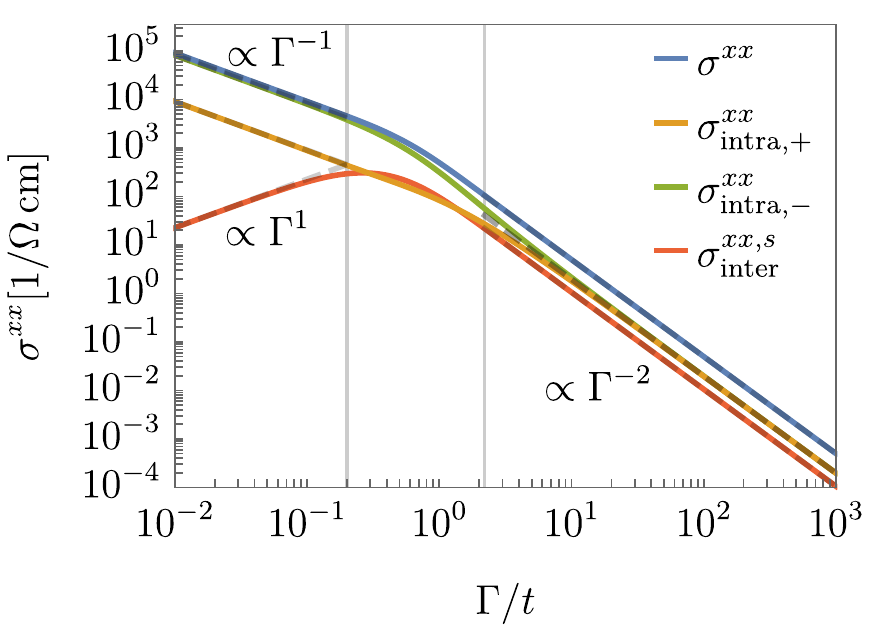}\\
\includegraphics[width=0.6\textwidth]{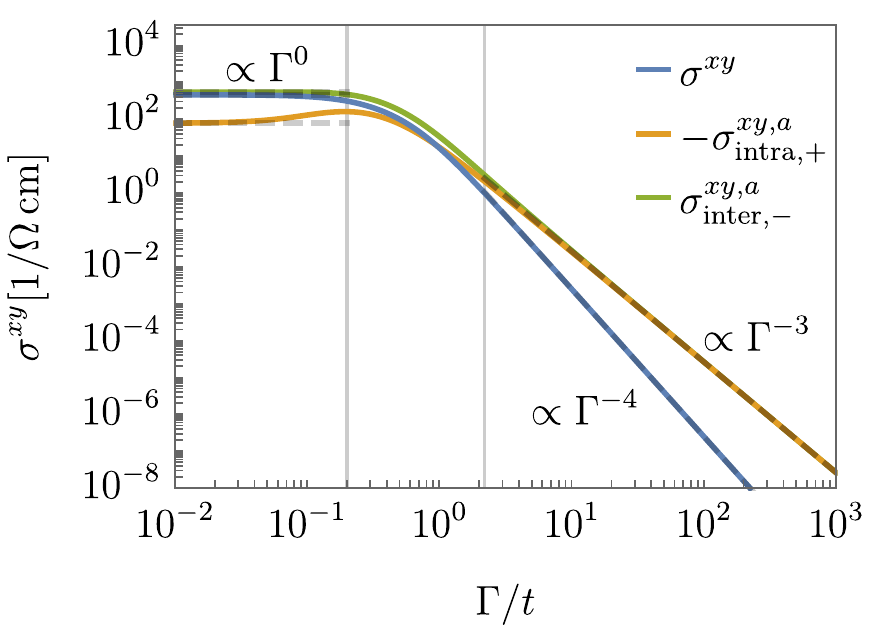}
\caption{The longitudinal (top) and anomalous Hall (bottom) conductivity and their nonzero contributions as a function of the relaxation rate $\Gamma/t$ for $t'/t=0.1$ and $\lam/t=0.2$ at $n=0.4$. For $\Gamma/t\ll 0.2$, we find the scaling of the clean limit given by \eqref{eqn:winG0}-\eqref{eqn:wanG0} (dashed lines). For $\Gamma/t\gg 2.2$, we find the scaling of the dirty limit given by \eqref{eqn:WnInfty}-\eqref{eqn:WaDiffInfty} with vanishing lowest order for $\sigma^{xy}$ (dashed lines). For $0.2<\Gamma/t<2.2$, we have a crossover regime. \label{fig:6}}
\end{figure}

We discuss the diagonal conductivity $\sigma^{xx}=\sigma^{yy}$ and the off-diagonal conductivity $\sigma^{xy}$ as a function of the relaxation rate $\Gamma/t$. The off-diagonal symmetric contributions $\sigma^{xy}_{\text{intra},\pm}$ and $\sigma^{xy,s}_\text{inter}$ vanish by integration over momenta. We calculate the longitudinal conductivity $\sigma^{xx}=\sigma^{xx}_{\text{intra},+}+\sigma^{xx}_{\text{intra},-}+\sigma^{xx,s}_\text{inter}$ and the (antisymmetric) anomalous Hall conductivity $\sigma^{xy}=\sigma^{xy,a}_{\text{inter},+}+\sigma^{xy,a}_{\text{inter},-}$ by using \eqref{eqn:SintraN}-\eqref{eqn:SinterAN} at zero temperature. In a stacked quasi-two-dimensional system, the conductivities are proportional to $e^2/ha$, where $a$ is the interlayer distance. When choosing $a\approx \SI{4}{\angstrom}$ \cite{Kontani2007, Onoda2008}, we have $e^2/ha\approx\SI[parse-numbers=false]{10^{3}}{\ohm^{-1}\cm^{-1}}$. In this chapter, we express the conductivities in SI units $1/\Omega\,\text{cm}$ for a simple comparison with experimental results on ferromagnets (see Ref.~\cite{Onoda2008} and references therein). 

In Fig.~\ref{fig:6}, we plot the longitudinal (top) and the anomalous Hall (bottom) conductivity (blue lines) and their nonzero contributions as a function of the relaxation rate $\Gamma/t$. In the clean limit, $\Gamma/t\ll0.2$, we obtain the expected scaling \eqref{eqn:winG0}-\eqref{eqn:wanG0} indicated by dashed lines. The intraband contributions (orange and green lines in the upper figure) scale as $1/\Gamma$, whereas the symmetric intraband contribution (red line) scales as $\Gamma$. The anomalous Hall conductivity becomes ``dissipationless'' \cite{Nagaosa2010} with $\Gamma^0$ in the clean limit. In absolute scales both the longitudinal and anomalous Hall conductivity are dominated by the lower band $E^-_\bp$ (green lines), consistent with a filling of $n=0.4$. In the dirty limit, $\Gamma/t\gg2.2$, the intraband and the symmetric interband contributions of the longitudinal conductivity scale as $\Gamma^{-2}$, which is the lowest order in the expansions in \eqref{eqn:WnInfty} and in \eqref{eqn:WsInfty}. The anomalous Hall conductivities $\sigma^{xy,a}_{\text{inter},\pm}$ scale as $\Gamma^{-3}$ in agreement with \eqref{eqn:WaInfty}. The lowest order $\Gamma^{-2}$ in \eqref{eqn:WaInfty} vanishes after integration over momenta. We have $\sigma^{xy,a}_{\text{inter},+}=-\sigma^{xy,a}_{\text{inter},-}$ that leads to a $\Gamma^{-4}$-dependence of the anomalous Hall conductivity summed over both bands, which is different than expected previously \cite{Kontani2007, Tanaka2008}. The dashed lines in the dirty limit are explicitly calculated via our results in Sec.~\ref{sec:theory:conductivity:limits}. In the intermediate range $0.2<\Gamma<2.2$, we find a crossover between the different scalings. We could only reproduce results consistent with those of Kontani {\it et al.} \cite{Kontani2007} by assuming a constant chemical potential that is fixed to its value in the clean limit, that is, if we neglect the $\Gamma$ dependence of the chemical potential in \eqref{eqn:muGamma} for fixed particle number $n=0.4$ within our calculation.

\begin{figure}[t!]
\centering
\includegraphics[width=0.6\textwidth]{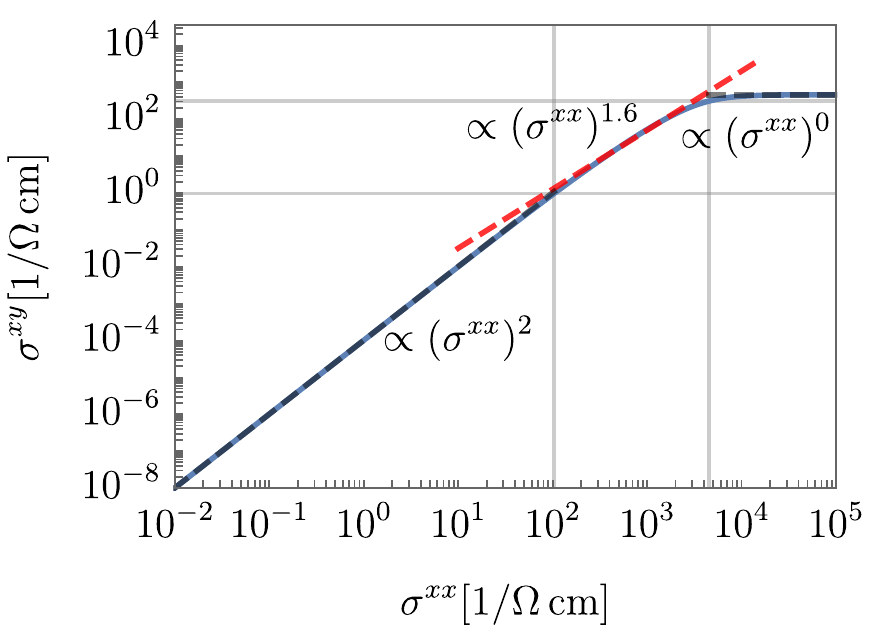}
\caption{The anomalous Hall conductivity $\sigma^{xy}$ as a function of the longitudinal conductivity $\sigma^{xx}$ for $t'/t=0.1$ and $\lam/t=0.2$ at $n=0.4$. The two vertical and the two horizontal lines indicate the values of $\sigma^{xx}$ and $\sigma^{xy}$ at $\Gamma/t=0.2$ and $\Gamma/t=2.2$, respectively. In the clean and the dirty limit, we find $\sigma^{xy}\propto (\sigma^{xx})^0$ and $\sigma^{xy}\propto (\sigma^{xx})^2$, respectively, in agreement with the individual scaling in $\Gamma$ (gray dashed lines). The crossover regime can be approximated by a scaling $\sigma^{xy}\propto (\sigma^{xx})^{1.6}$ (red dashed line).
\label{fig:scalingAH}}
\end{figure}
In Fig.~\ref{fig:scalingAH}, we plot the anomalous Hall conductivity as a function of the longitudinal conductivity. The representation is useful for comparison with experimental results, where the dependence on the relaxation rate is not known explicitly. The result is both qualitatively and quantitatively in good agreement with experimental results for ferromagnets (see Ref.~\cite{Onoda2008} and references therein). We find three regimes: In the clean regime, we get $\sigma^{xy}\propto (\sigma^{xx})^0$ since the anomalous Hall conductivity becomes $\Gamma$ independent. In the dirty regime, we have $\sigma^{xy}\propto (\sigma^{xx})^2$, which can be easily understood by the scaling behavior that shown in Fig.~\ref{fig:6}. The black dashed lines are calculated explicitly via \eqref{eqn:WnInfty}-\eqref{eqn:WaDiffInfty}. We indicated the regime boundaries by gray lines that correspond to the conductivities at $\Gamma/t=0.2$ and $\Gamma/t=2.2$. In the intermediate regime that corresponds to the crossover between the different scalings in Fig.~\ref{fig:6}, we get a good agreement with a scaling $\sigma^{xy}\propto (\sigma^{xx})^{1.6}$ (red dashed line). 

\begin{figure}[t!]
\centering
\includegraphics[width=0.6\textwidth]{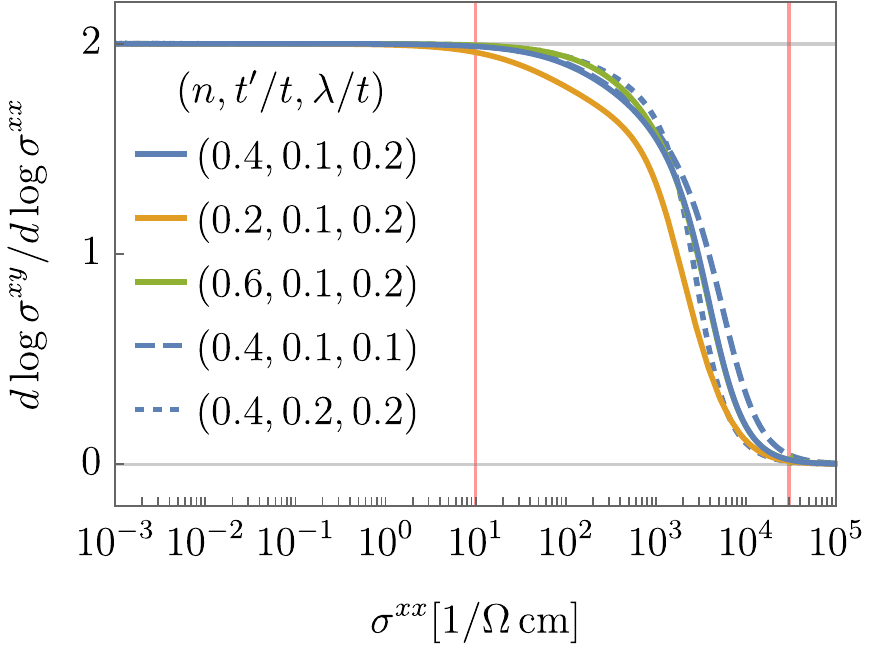}
\caption{The logarithmic derivative of the anomalous Hall conductivity $\sigma^{xy}$ as a function of the longitudinal conductivity $\sigma^{xx}$ for different particle numbers $n$, next-nearest neighbor hoppings $t'/t$, and spin-orbit couplings $\lam/t$. In between $\sigma^{xx}=10-3\times 10^4\,\si{(\ohm\,\cm)^{-1}}$ (red lines), we have a crossover regime between the scaling $\sigma^{xy}\propto(\sigma^{xx})^0$ in the clean limit and $\sigma^{xy}\propto(\sigma^{xx})^2$ in the dirty limit (gray lines). The range is insensitive to parameters over a broad range.
\label{fig:scalingAHLog}}
\end{figure}

The scaling behavior $\sigma^{xy}\propto (\sigma^{xx})^{1.6}$ is observed experimentally and discussed theoretically in various publications in the recent years (see \cite{Onoda2006, Miyasato2007, Onoda2008, Kovalev2009, Xiong2010, Lu2013, Zhang2015, Sheng2020} and references therein). Within our theory, we clearly identify the intermediate regime, $\sigma^{xx}\approx 100-5000\,(\Omega\,\text{cm})^{-1}$, as a crossover regime not related to a (proper) scaling behavior. This is most prominent when showing the logarithmic derivative of the anomalous Hall conductivity as a function of the longitudinal conductivity in Fig.~\ref{fig:scalingAHLog} for different particle numbers $n=0.2,0.4,0.6$, next-nearest neighbor hoppings $t'/t=0.1,0.2$, and spin-orbit couplings $\lam/t=0.1,0.2$. We see a clear crossover from $\sigma^{xy}\propto(\sigma^{xx})^0$ to $\sigma^{xy}\propto(\sigma^{xx})^2$ in a range of $\sigma^{xx}=10-30000\,(\Omega\,\text{cm})^{-1}$ (red vertical lines), which is even larger than estimated by the scales $\Gamma=\lam=0.2\,t$ and $\Gamma=W=2.2\,t$ indicated by the gray lines in Fig.~\ref{fig:scalingAH}. This crossover regime is insensitive to parameters over a broad range. Interestingly, various experimental results are found within the range $10-30000\,(\Omega\,\text{cm})^{-1}$ (see Fig.~12 in Ref.~\cite{Onoda2008} for a summary). We have checked that a smooth crossover similar to the presented curve in Fig.~\ref{fig:scalingAH} qualitatively agrees with these experimental results within their uncertainty. 

Following the seminal work of Onoda {\it et al.} \cite{Onoda2006,Onoda2008}, which treated intrinsic and extrinsic contributions on equal footing, the experimental and the theoretical investigation of the scaling that includes, for instance, vertex correction, electron localization and quantum corrections from Coulomb interaction is still ongoing research \cite{Kovalev2009, Xiong2010, Lu2013, Zhang2015, Sheng2020} and is beyond the scope of this thesis.

%
%

\section{Spiral magnetic order}
\label{sec:application:spiral}

Recent experimental results in very high magnetic fields \cite{Badoux2016, Laliberte2016, Collignon2017} shed new light on the non-superconducting ground state of the cuprate high-temperature superconductors, whose phase diagram is not yet fully understood \cite{Keimer2015}. A drop of the Hall number as a function of hole doping was found at a critical doping $p^*$ at the edge of the pseudogap regime, which indicates a Fermi surface reconstruction at $p^*$. Assuming spiral magnetic order for hole dopings below $p^*$ within a phenomenological model, Eberlein {\it et al.} \cite{Eberlein2016} showed that this order can cause a drop of the Hall number consistent with the experimental results. There are various other theoretical proposals that lead to similar conclusions \cite{Storey2016, Storey2017, Verret2017, Chatterjee2017, Caprara2017, Sharma2018,Maharaj2017, Qi2010, Chatterjee2016, Morice2017, Scheurer2018, Yang2006a}, which have already been discussed in the introduction of this thesis. However, it is hard to confirm or rule out any of these candidates experimentally since not many tools can be applied in the very high magnetic fields. In this thesis, we focus on the proposal of spiral magnetic order by Eberlein {\it et al.} \cite{Eberlein2016}. They calculated the Hall number by using the expressions of Voruganti {\it et al.} \cite{Voruganti1992}, which suggest to replace the bare dispersion by the dispersion after Fermi surface reconstruction in the semiclassical transport formulas in \eqref{eqn:condBoltzmann} and \eqref{eqn:HallBoltzmann}. However, the validity of those formulas, which do not include interband contributions, remained unclear. Although the experiments are performed at very high magnetic field, it was shown that the low-field limit $\omega_c\tau\ll 1$ is still valid \cite{Collignon2017}. Thus, a sizable relaxation rate $\Gamma=1/2\tau$ might cause that interband contributions are not negligible and relevant at the onset of the spiral magnetic order, where the magnetic gap and the relaxation rate are of similar size. After having derived a generalization for both the formulas of the longitudinal and the Hall conductivity including interband contributions within Chapter~\ref{sec:theory}, we are now able to provide an answer, whether or not interband contributions are relevant in the application of spiral magnetic order to cuprates and how this might affect the previous conclusions. 

This section is structured as follows: We give the definition and fundamental properties of the spiral magnetic order (Sec.~\ref{sec:application:spiral:definition}). The broken lattice-translational invariance of the model leads to a spectral function of single-particle excitations with Fermi-arc characteristics resembling those found in the pseudogap phase of cuprates (Sec.~\ref{sec:application:spiral:spectralweights}). We discuss spiral magnetic order in the two-dimensional Hubbard model and present results found by a Hartree-Fock approximation and by dynamical mean-field theory (Sec.~\ref{sec:application:spiral:hubbard}). In the subsequent chapters, we focus on transport properties. The spiral magnetic order reduces the lattice symmetry and, thus, may lead to an asymmetry of the longitudinal conductivities in $x$ and $y$ direction (Sec.~\ref{sec:application:spiral:coordinatesystem}). We discuss the impact of interband contributions on the longitudinal conductivity and the Hall number within a simplified phenomenological model for the doping dependence of the spiral magnetic order as well as using {\it ab initio} results of the Hubbard model obtained via dynamical mean-field theory. We relate them to the experimental results for cuprates (Sec.~\ref{sec:application:spiral:CondCuprates}).

%
%

\subsection{Definition of spiral magnetic order}
\label{sec:application:spiral:definition}

We assume a two-dimensional tight-binding model with spin. Using our notation in Sec.~\ref{sec:theory:twobandsystem}, the two subsystems are the spins $A=\,\,\uparrow$ and $B=\,\,\downarrow$ located at the lattice sites $\bR_i$ with $\brho_\uparrow=\brho_\downarrow=0$. Furthermore, we set $\bQ_\uparrow=\bQ$ and $\bQ_\downarrow=0$ and assume a Bloch Hamiltonian 
\begin{align}
\label{eqn:spiralH}
\lam_\bp=\begin{pmatrix} \epsilon_{\bp+\bQ} && -\Delta \\[2mm] -\Delta && \epsilon_\bp \end{pmatrix} \,,
\end{align}
where the dispersion reads
\begin{align}
 \label{eqn:dispersion}
 \eps_\bp=-2t(\cos p^x+\cos p^y)-4t'\cos p^x \cos p^y- 2t''(\cos 2p^x + \cos 2p^y) \, ,
\end{align}
which includes nearest-, next-nearest-, and next-next-nearest neighbor hopping $t$, $t'$, and $t''$, respectively, on a two-dimensional square lattice with lattice constant $a=1$. Different choices of $\bQ_\uparrow$ and $\bQ_\downarrow$ are equivalent by redefining the momentum summation in the Hamiltonian in \eqref{eqn:H} as long as $\bQ_\uparrow-\bQ_\downarrow=\pm\bQ$. We choose $t$ as our unit of energy. To make connection to experiments, hopping amplitudes in cuprates have been determined by downfolding {\it ab initio} band structures on effective single-band Hamiltonians \cite{Andersen1995, Pavarini2001}. We assume a real, positive, and momentum- and frequency-independent onsite coupling $\Delta$ between the states $|\bp+\bQ,\uparrow\rangle$ and $|\bp,\downarrow\rangle$. A finite momentum difference $\bQ=\bQ_\uparrow-\bQ_\downarrow$ between the two subsystems in the spinor \eqref{eqn:spinor} breaks the lattice-translation invariance of the Hamiltonian \eqref{eqn:H}. However, the Hamiltonian is still invariant under a combined translation in real space and rotation in spin space \cite{Sandratskii1998}.

The coupling $\Delta$ leads to a nonzero onsite magnetic moment of the form
\begin{align}
 \langle \bS_i\rangle=\frac{1}{2}\sum_{\nu,\nu'=\uparrow,\downarrow}\langle \cdag_{i,\nu}\,\bpauma^{}_{\nu\nu'}\,\c_{i,\nu'}\rangle=m\, \mathbf{n}_i 
\end{align}
with direction $\mathbf{n}_i$ and magnetization amplitude $m$. $\cdag_{i,\nu}$ and $\c_{i,\nu}$ are the fermionic creation and annihilation operators for site $i$ and spin $\nu$, respectively. The vector $\bpauma=(\pauma_x,\pauma_y,\pauma_z)$ is composed of the three Pauli matrices. The direction vector $\mathbf{n}_i$ lies in the $x$-$y$-plane and is given by
\begin{align}
 \label{eqn:magnDir}
 \mathbf{n}_i=\begin{pmatrix}\cos (\bQ\cdot\bR_i)\\[1mm]-\sin (\bQ\cdot \bR_i)\\[1mm]0 \end{pmatrix}\,.
\end{align}
The direction $\mathbf{n}_i$ between neighboring sites $i$ and $j$ differs by an angle $\bQ\cdot(\bR_i-\bR_j)$. The global phase, which rotates all $\mathbf{n}_i$ by the same angle, is captured by the complex phase of the onsite coupling and is, thus, fixed by choosing $\Delta$ to be real and positive.  The magnetization amplitude $m$ is uniform and controlled by the coupling via
\begin{align}
\label{eqn:magnAmp}
 m=-\frac{\Delta}{L} \sum_\bp\int d\eps f(\eps) \frac{A^+_\bp(\eps)-A^-_\bp(\eps)}{E^+_\bp-E^-_\bp}\,,
\end{align}
where $E^\nu_\bp$ are the two quasiparticle bands and $A^\nu_\bp(\eps)$ are the quasiparticle spectral functions. 

\begin{figure}[t!]
\centering
\includegraphics[width=0.75\textwidth]{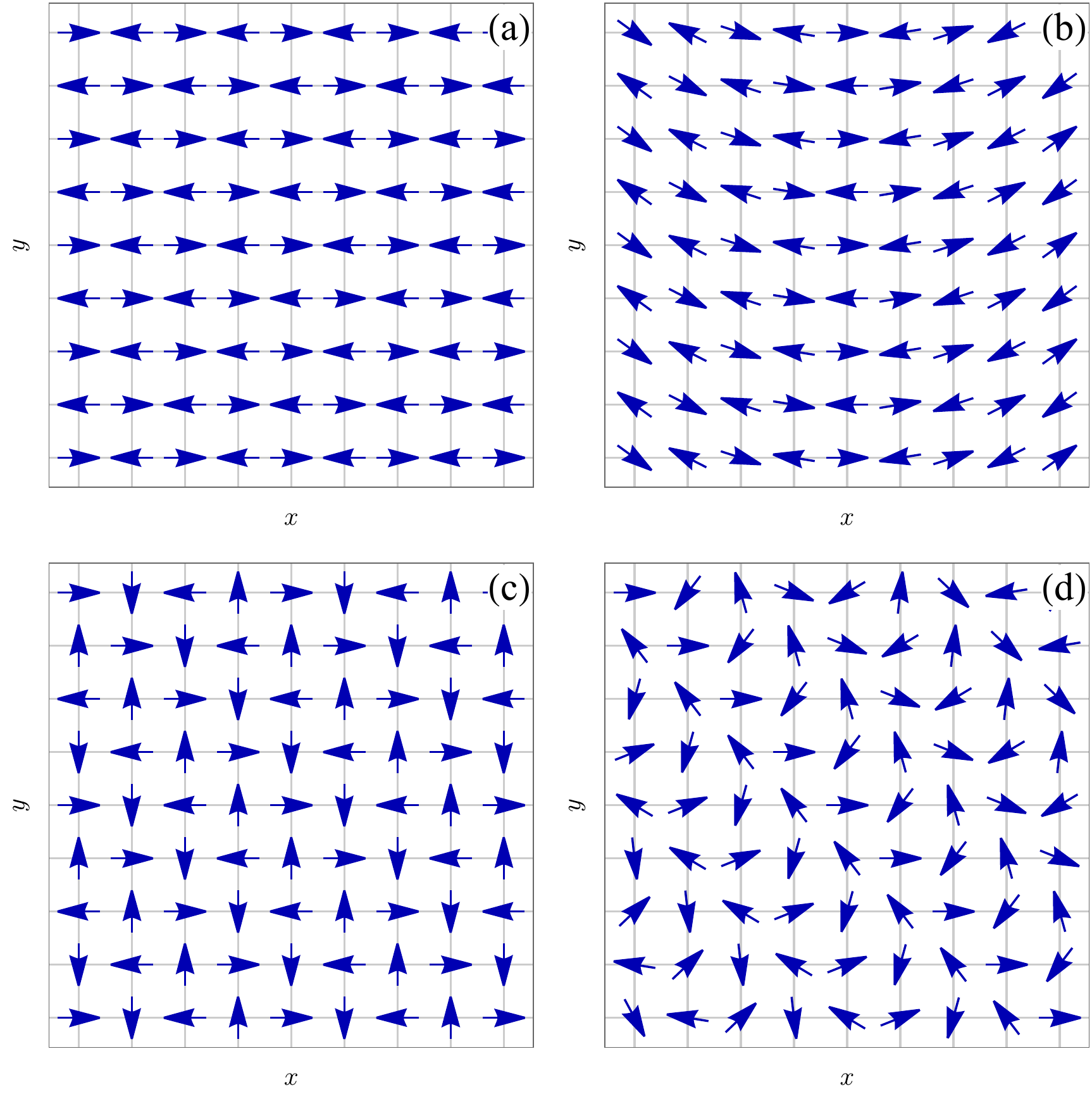}
\caption{The magnetization patterns $\langle \bS_i\rangle\propto\mathbf{n}_i$ for different ordering vectors (a) $\bQ=(\pi,\pi)$, (b) $\bQ=(0.95\pi,\pi)$, (c) $\bQ=(\pi/2,\pi/2)$ and (d) $\bQ=(\pi/\sqrt{2},\pi/\sqrt{2})$ on a square lattice.  We have N\'eel antiferromagnetic and spiral order in (a) and (b)-(d), respectively. The spiral order shown in (d) is incommensurate with the underlying lattice. \label{fig:magPattern}}
\end{figure}

The magnetic moment of the form $\langle \bS_i\rangle=m\,\mathbf{n}_i$ is the defining character of a spiral spin density wave in contrast to collinear spin density waves with magnetic moments of the form $\langle \bS_i\rangle=m_i\,\mathbf{n}$, where the direction remains constant but the length is modulated. Collinear spin density waves are not invariant under combined translation and spin-rotation. In Fig.~\ref{fig:magPattern}, we show magnetization patterns $\langle\bS_i\rangle\propto \mathbf{n}_i$ of spiral spin density waves for different wave vectors $\bQ$ on a square lattice. The two special cases $\bQ=(0,0)$ and $\bQ=(\pi,\pi)$ correspond to ferromagnetic and N\'eel-antiferromagnetic order, respectively. We show N\'eel-antiferromagnetic order in Fig.~\ref{fig:magPattern} (a). We refer to an order different than these two special cases as (purely) spiral. For instance, $\bQ=(\pi/2,\pi/2)$ describes a $90^\circ$ rotation per lattice site in both $x$ and $y$ direction as shown in Fig.~\ref{fig:magPattern} (c). Due to the invariance under combined translational and spin-rotation, this case can be described via \eqref{eqn:spiralH} without considering a four-times larger unit cell. The $2\times 2$ structure of the Hamiltonian also captures order wave vectors $\bQ$ that are incommensurate with the underlying lattice, when enlarging the unit cell to any size does not restore translation symmetry \cite{Sandratskii1998}. In Fig.~\ref{fig:magPattern} (d), we show such an incommensurate order with $\bQ=(\pi/\sqrt{2},\pi/\sqrt{2})$. Spiral order of the form $\bQ=(\pi-2\pi \eta,\pi)$ or symmetry related with $\eta>0$, where $\eta$ is the so-called incommensurability, is found in the two-dimensional $t-J$ model \cite{Shraiman1989, Kotov2004, Sushkov2004,Sushkov2006} and in the two-dimensional Hubbard model \cite{Schulz1990, Kato1990, Fresard1991, Chubukov1992, Chubukov1995, Raczkowski2006, Igoshev2010, Igoshev2015, Yamase2016, Bonetti2020} by various theoretical methods. We will discuss this in more detail in Sec.~\ref{sec:application:spiral:hubbard}. A visualization of the magnetization pattern for $\eta=0.025$ is shown in Fig.~\ref{fig:magPattern} (b).

%
%

\subsection{Quasiparticle and single-electron spectral functions}
\label{sec:application:spiral:spectralweights}

As we have just seen, the nonzero difference $\bQ=\bQ_\uparrow-\bQ_\downarrow$ is essential and the defining property of the spiral magnetic state. The following analysis is even valid beyond the specific case of spiral magnetic order, so that we will use again $\ic=A,B$ for arbitrary subsystems in this section. A key difference between a zero and a nonzero momentum difference $\bQ=\bQ_A-\bQ_B$ in the spinor \eqref{eqn:spinor} is the distinction between the spectral function for quasiparticles and the spectral function for single-electron excitations. This distinction is relevant, for instance, for spectroscopic measurements like angle-resolved photoemission spectroscopy (ARPES) but also in the context of our discussion later in this chapter when we will discuss the change of the Fermi surface topology due to the onset of spiral magnetic order and when we interpret the contribution of different momenta to the conductivities. 

The spectral functions of single-electron excitations $A_{\bp,\ic}(\eps)$ with $\ic=A,B$ involve the respective creation and annihilation operators in the form $\cdag_{\bp,\ic}\c_{\bp,\ic}$. In contrast, the quasiparticle spectral functions involve a mixture of the particle operators $\cdag_{\bp+\bQ_\ic,\ic}\c_{\bp+\bQ_\ic,\ic}$ of both subsystems, whose momenta are furthermore shifted by $\bQ_\ic$. We can relate both types of spectral functions by taking the diagonal elements of $\U_\bp \,A^{}_{\bp}(\eps)\, \Udag_\bp$ and shift the momenta of those diagonal elements by $\bQ_\ic$ \cite{Eberlein2016}. The diagonal matrix $A^{}_\bp(\eps)$ consists of the quasiparticle spectral functions $A^\pm_\bp(\eps)$ in \eqref{eqn:Apm}. The transformation matrix $\U_\bp=\big(|+_\bp\rangle \,\, |-_\bp\rangle\big)$ contains the eigenstates $|\pm_\bp\rangle$ in \eqref{eqn:+} and \eqref{eqn:-}. We get the spectral functions of single-electron excitations
\begin{align}
\label{eqn:spectralA}
 A_{\bp,A}(\eps)&=A^+_{\bp-\bQ_A}(\eps)\,\cos^2\frac{\theta_{\bp-\bQ_A}}{2}+A^-_{\bp-\bQ_A}(\eps)\,\sin^2\frac{\theta_{\bp-\bQ_A}}{2}\,,\\
 \label{eqn:spectralB}
 A_{\bp,B}(\eps)&=A^+_{\bp-\bQ_B}(\eps)\,\sin^2\frac{\theta_{\bp-\bQ_B}}{2}+A^-_{\bp-\bQ_B}(\eps)\,\cos^2\frac{\theta_{\bp-\bQ_B}}{2}\,.
\end{align}
We see that the quasiparticle spectral function are summed up with momentum-dependent weighting factors, which are controlled by the angle $\theta_\bp$ defined in \eqref{eqn:Theta}. 
If $\bQ_A-\bQ_B=\bQ=0$, that is, in a lattice-translational-invariant system, the weighting factors drop out in the total spectral function $A_{\bp,A}(\eps)+A_{\bp,B}(\eps)$ and the momentum shift $\bQ_A=\bQ_B$ is a simple shift of the full Brillouin zone, which is physically irrelevant. Thus, the sum of the quasiparticle spectral functions is equal to the sum of the spectral functions of single-electron excitations. However, for a system that is no longer translational invariant, but has an invariance under combined translation in real space and rotation in the subsystem space, that is, $\bQ_A\neq \bQ_B$ or $\bQ\neq0$, this is no longer the case. The two types of spectral functions are different even after summation. 
\begin{figure}[t!]
\centering
\includegraphics[width=0.45\textwidth]{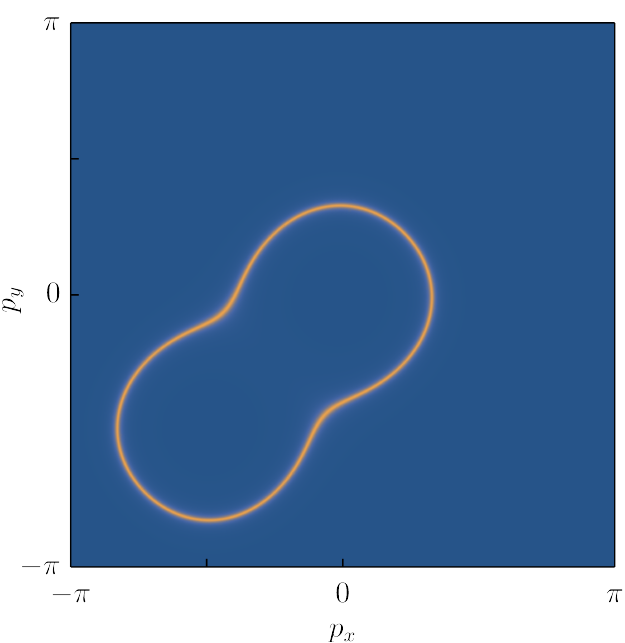}\hspace{0.02\textwidth}
\includegraphics[width=0.45\textwidth]{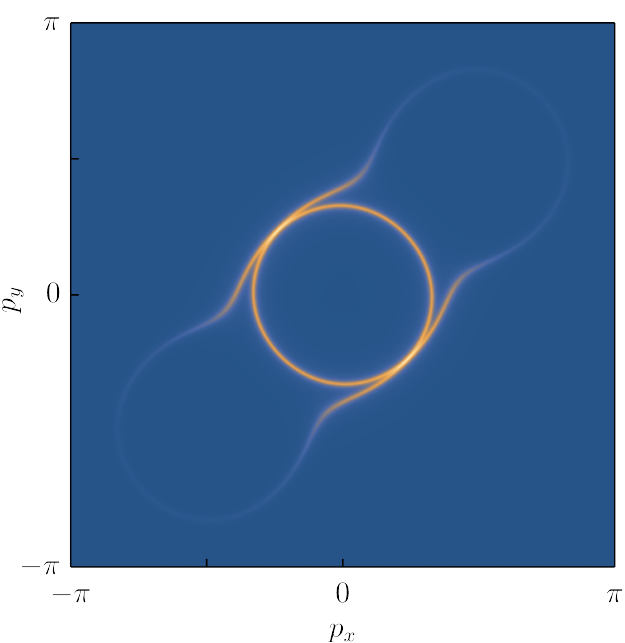}\\[0.02\textwidth]
\includegraphics[width=0.45\textwidth]{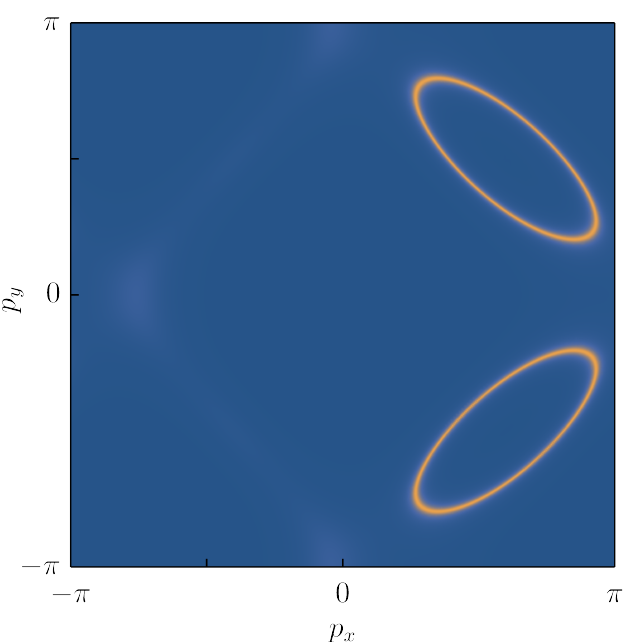}
\hspace{0.02\textwidth}
\includegraphics[width=0.45\textwidth]{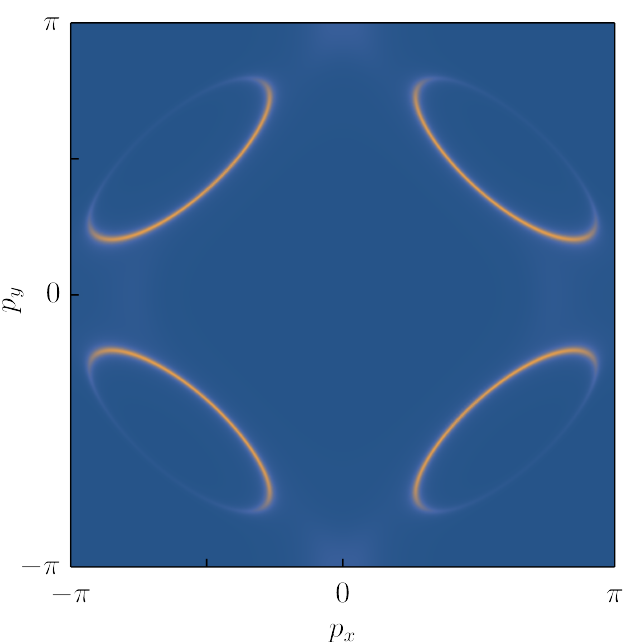}
\caption{The spectral function $A^+_\bp+A^-_\bp$ of the quasiparticles (left column) and the spectral function $A_{\bp,\uparrow}+A_{\bp,\downarrow}$ for single-electron excitations (right column) at zero frequency for $t'/t=0.1$, $t''/t=0$, $\Delta/t=1$, $\bQ=(\pi/2,\pi/2)$, $n=0.2$ (upper row) and $t'/t=-0.17$, $t''/t=0.05$, $\Delta/t=0.5$, $\bQ=(\pi-2\pi\eta,\pi)$ with $\eta=0.1$ and $n=0.9$ (lower row) with $\Gamma/t=0.05$.  \label{fig:spectral}}
\end{figure}

In Fig.~\ref{fig:spectral}, we plot the sum of the quasiparticle spectral functions $A^+_\bp+A^-_\bp$ (left column) and the sum of the single-electron spectral functions $A_{\bp,\uparrow}+A_{\bp,\downarrow}$ (right column) at zero frequency for two different sets of parameters (upper and lower row). In the upper row, we show the spectral functions for an ordering wave vector $\bQ=(\pi/2,\pi/2)$, whose magnetization pattern is shown in Fig.~\ref{fig:magPattern} (c). At particle number $n=0.2$ only the lower band contributes to the quasiparticle spectral function. We see explicitly that the spiral magnetic order lowers the symmetry. The single-electron spectral function consists of two copies of the quasiparticle spectral function of the lower band with one of them shifted by momentum $\bQ$. The momentum-dependent weighting factors in \eqref{eqn:spectralA} and \eqref{eqn:spectralB} cause a reduction of the spectral weights at specific parts of the quasiparticle Fermi surfaces. This reduction must not be confused with the reduction due to a momentum- or frequency dependent relaxation rate $\Gamma(\eps,\bp)$. Furthermore, the inversion symmetry is restored for $A_{\bp,A}(\eps)+A_{\bp,B}(\eps)$ for inversion-symmetric dispersion relations ($\eps_\bp = \eps_{-\bp}$), as for the dispersion in \eqref{eqn:dispersion}: The Fermi surface corresponding to peaks in $A_\bp(\eps) = A_{\bp,\uparrow}(\eps) + A_{\bp,\downarrow}(\eps)$ at $\eps=0$ is given by the points in momentum space obeying $E_\bp^\pm-\mu = 0$ or $E_{\bp - \bQ}^\pm-\mu= 0$. The latter equation is equivalent to $E_{-\bp}^\pm-\mu = 0$ for inversion symmetric $\eps_\bp$. We have $A_{\bp,\uparrow}(\eps) = A_{-\bp,\downarrow}(\eps)$. Thus, the quasiparticle dispersions $E_\bp^\pm$ and the quasiparticle Fermi surfaces are not inversion symmetric, while the total single-electron spectral function $A_{\bp,\uparrow}(\eps)+A_{\bp,\downarrow}(\eps)$ is. The spectral weight is maximal for momenta close to the bare Fermi surface, where $\eps_\bp-\mu=0$. 

In the lower row of Fig.~\ref{fig:spectral}, we show the total quasiparticle spectral function $A^+_\bp+A^-_\bp$ and the total single-electron spectral function $A_{\bp,\uparrow}+A_{\bp,\downarrow}$ at zero frequency for band parameters that are commonly used for the cuprate compound $\rm La_{2-x} Sr_x Cu O_4$ (LSCO). At small hole doping $p=1-n=0.1$ and $\bQ=(\pi-2\pi\eta)$ with $\eta=p$ the Fermi surface consists only of two hole pockets. The inversion symmetry is restored in the spectral function of single-electron excitations. Since the spectral weights are maximal close to the bare Fermi surface $\eps_\bp-\mu=0$, only the inside half of the four pockets has significant spectral weight, whereas the weight at the backside is strongly suppressed. The result resembles the Fermi arcs observed in underdoped cuprates in the pseudogap phase \cite{Damascelli2003}. It is consistent with theoretical calculations of the ARPES spectral functions for the hole pockets in the N\'eel antiferromagnetic state of $\rm YBa_2Cu_3O_y$ (YBCO) close to half filling \cite{Chen2011}. A shadow of the upper band can be seen in the lower left figure. For larger hole doping, electron-like pockets will eventually appear at those regions.

%
%

\subsection{Spiral magnetic order in the Hubbard model}
\label{sec:application:spiral:hubbard}

The competition between antiferromagnetism and superconductivity in cuprates seems to be well captured by the two-dimensional Hubbard model \cite{Scalapino2012}. The Hamiltonian of the Hubbard model reads
\begin{align}
\label{eqn:Hubbard}
 H=\sum_{\ic=\uparrow,\downarrow}\sum_\bp \eps^{}_\bp \cdag_{\bp,\ic}\c_{\bp,\ic}+\sum_i U n_{i,\uparrow}n_{i,\downarrow} \, ,
\end{align}
where $n_{i,\nu}=\cdag_{i,\nu}\c_{i,\nu}$ is the particle number operator for lattice site $i$ and spin $\nu=\uparrow,\downarrow$. $U$ is the (repulsive) on-site interaction for doubly-occupied lattice sites. The dispersion $\eps_\bp$ for a square lattice including nearest-, next-nearest- and next-next-nearest-neighbor hoppings is given in \eqref{eqn:dispersion}. For the two-dimensional Hubbard model, antiferromagnetic order with wave vectors $\bQ$ away from the N\'eel point $(\pi,\pi)$ was found in numerous mean-field calculations \cite{Schulz1990, Kato1990, Fresard1991, Raczkowski2006, Igoshev2010, Igoshev2015}, and also by expansions for small hole density, where fluctuations are taken into account \cite{Chubukov1992, Chubukov1995}. At weak coupling (small $U$), magnetic order with $\bQ \neq (\pi,\pi)$ was confirmed by functional renormalization group calculations \cite{Metzner2012, Yamase2016}, and at strong coupling (large $U$) by state-of-the-art numerical techniques \cite{Zheng2017}. Recent dynamical mean-field calculations with vertex corrections suggest that the Fermi-surface geometry determines the (generally incommensurate) ordering wave vector not only at weak coupling, but also at strong coupling \cite{Vilardi2018}. For the two-dimensional $t$-$J$ model, which is the strong-coupling limit of the Hubbard model, expansions for small hole density indicate that the N\'eel state is stable only at half filling, and is replaced by a spiral antiferromagnet upon doping \cite{Shraiman1989, Kotov2004, Sushkov2004, Sushkov2006, Luscher2007, Sushkov2009}. Long-range magnetic order is forbidden at finite temperature $T\neq 0$ in a two-dimensional system according to the Mermin-Wagner theorem \cite{Mermin1966, Hohenberg1967}. Nevertheless, it was shown for the t-J model that also dynamic antiferromagnetic fluctuations can cause a Fermi surface reconstruction \cite{Holt2012}. In the Hubbard model, the correlation length of the magnetic order can become sufficiently large at low temperature, so that the electrons experience a local environment of antiferromagnetic order, which has an impact on the low-lying states, for instance, by opening a pseudogap \cite{Vilk1997}.

There is a whole zoo of distinct magnetic states.  The most favorable, or at least the most popular, are planar spiral states and collinear states, combined with charge order to form spin-charge stripes. Stripe order has been observed in Lanthanum-based cuprates \cite{Tranquada1995}. Theoretically, commensurate stripe order was shown to minimize the ground-state energy of the strongly interacting Hubbard model with pure nearest-neighbor hopping at doping $1/8$ \cite{Zheng2017}. However, this is a very special choice of parameters, and stripe order is not ubiquitous in cuprates. Recently, it was shown that it is difficult to explain the recent high-field transport experiments in cuprates by collinear magnetic order \cite{Charlebois2017}. Generally, the energy difference between different magnetic states seems to be rather small.

We recapitulate how planar spiral antiferromagnetic order arises in a Hartree-Fock approximation of the Hubbard model. The Green's function of the noninteracting Hamiltonian with $U=0$ reads
\begin{align}
 \sG^0_{ip_0,\bp}=\begin{pmatrix} ip_0+\mu-\eps_{\bp+\bQ} & 0 \\ 0 & ip_0+\mu-\eps_\bp \end{pmatrix}^{-1} \, .
\end{align}
In order to prepare for the following calculation, we separated the parts of spin $\uparrow$ and $\downarrow$ and used the momentum summation to shift the momentum of the spin-$\uparrow$ component by an arbitrary vector $\bQ$, which will eventually be the ordering vector of the spiral state, which we discussed previously. We added a chemical potential $\mu$. We assume a full Green's function of the form
\begin{align}
\label{eqn:GreenHubbard}
 \sG_{ip_0,\bp}=\begin{pmatrix} ip_0+\mu-\eps^\uparrow_\bp & \Delta \\ \Delta & ip_0+\mu-\eps^\downarrow_\bp \end{pmatrix}^{-1} \, ,
\end{align}
where the bands $\eps^\uparrow_\bp$ and $\eps^\downarrow_\bp$ as well as the (real and constant) gap $\Delta$ has to be determined.  
Thus on the one hand, the self-energy takes the form
\begin{align}
\label{eqn:selfCond}
 \Sigma_{ip_0,\bp}=\big(\sG^0_{ip_0,\bp}\big)^{-1}-\big(\sG_{ip_0,\bp}\big)^{-1}=\begin{pmatrix} \eps^\uparrow_\bp-\eps_{\bp+\bQ} & -\Delta \\ -\Delta & \eps^\downarrow_\bp-\eps_\bp \end{pmatrix} \, .
\end{align}
On the other hand, we determine the self-energy by a diagrammatic expansion that includes the Hartree and the Fock contribution. The diagonal components of the self-energy are
\begin{align}
\label{eqn:selfCond1}
 \big(\Sigma_{iq_0,\bp}\big)_{\ic\ic}=U\,\frac{T}{L}\sum_{ip_0,\bp} \big(\sG_{ip_0,\bp}\big)_{\overline{\ic}\overline{\ic}}=U n^{}_{\overline{\ic}} 
\end{align}
with $\ic=\uparrow,\downarrow$, where $\overline{\ic}$ indicates the opposite spin. $n_\uparrow$ and $n_\downarrow$ are the occupation number of the respective spin. The off-diagonal components of the self-energy are  
\begin{align}
\label{eqn:selfCond2}
 \big(\Sigma_{iq_0,\bp}\big)_{\ic\overline{\ic}}=-U\,\frac{T}{L}\sum_{ip_0,\bp}\big(\sG_{ip_0,\bp}\big)_{\ic\overline{\ic}}=U\,\frac{\Delta}{L}\sum_\bp\frac{f^+_\bp-f^-_\bp}{E^+_\bp-E^-_\bp} \, .
\end{align}
We introduced the Fermi function $f^\pm_\bp=f(E^\pm_\bp-\mu)$ with $f(\omega)=(1+e^{\omega/T})^{-1}$ of the quasiparticle bands $E^\pm_\bp=\g_\bp\pm\sqrt{h^2_\bp+\Delta^2}$ with $\g_\bp=\frac{1}{2}(\eps^\uparrow_\bp+\eps^\downarrow_\bp)$ and $h_\bp=\frac{1}{2}(\eps^\uparrow_\bp-\eps^\downarrow_\bp)$, which are obtained by diagonalizing the full Green's function matrix in \eqref{eqn:GreenHubbard}.  
The Hartree term \eqref{eqn:selfCond1} combined with \eqref{eqn:selfCond} leads to the condition $\eps^\uparrow_\bp-\eps^{}_{\bp+\bQ}=U\,n^{}_\downarrow$ and $\eps^\downarrow_\bp-\eps^{}_\bp=U\,n^{}_\uparrow$, so that we can read off 
\begingroup
 \allowdisplaybreaks[0]
\begin{align}
 &\g_\bp=\frac{1}{2}(\eps_{\bp+\bQ}+\eps_\bp)+\frac{U n}{2} \, ,\\
 &h_\bp=\frac{1}{2}(\eps_{\bp+\bQ}-\eps_\bp)-Um_z \, ,
\end{align}
\endgroup
where we defined the total particle number $n=n_\uparrow+n_\downarrow$ and the relative particle number $m_z=(n_\uparrow-n_\downarrow)/2$, which corresponds to the out-of-plane magnetization. We expect that $m_z=0$ since the Hubbard Hamiltonian is symmetric in spin. A Zeeman term in the Hamiltonian would lead to a nonzero $m_z$. The Fock term \eqref{eqn:selfCond2} combined with \eqref{eqn:selfCond} leads to the gap equation
\begin{align}
\label{eqn:gapEq}
 \frac{1}{U}=-\frac{1}{L}\sum_\bp\frac{f^+_\bp-f^-_\bp}{E^+_\bp-E^-_\bp} \, .
\end{align}
Comparing with \eqref{eqn:magnAmp} we can read of the magnetization amplitude $m=\Delta/U$ for the onsite magnetic moment $\langle \bS_i\rangle=m\,\mathbf{n}_i$. The grand canonical potential is given by 
\begin{align}
\label{eqn:LuttingerOmega}
 \Omega=\Phi - \Tr\,(\Sigma G) -\Tr \ln (-G^{-1})\, ,
\end{align}
where $\Tr$ is understood as the trace over the Green's function and self-energy matrices $G$ and $\Sigma$ involving all degrees of freedoms \cite{Eder2019}. $\Phi$ is the Luttinger-Ward functional, of which we only consider the Hartree and the Fock contribution,
\begin{align}
 \Phi=\Phi_\text{Hartree}+\Phi_\text{Fock}+...\, .
\end{align}
The Hartree contribution reads
\begin{align}
 \Phi_\text{Hartree}=Un_\uparrow n_\downarrow = U\bigg(\frac{n^2}{4}-m_z^2\bigg) \, ,
\end{align}
where we expressed the occupation number of the individual spins by the total and the relative particle number $n$ and $m_z$. The Fock contribution reads
\begin{align}
 \Phi_\text{Fock}=-U\bigg(\frac{\Delta}{L}\sum_\bp\frac{f^+_\bp-f^-_\bp}{E^+_\bp-E^-_\bp}\bigg)^2=-U\,\bigg(\frac{\Delta}{U}\bigg)^2=-\frac{\Delta^2}{U}\,,
\end{align}
where we used the gap equation \eqref{eqn:gapEq} in the second step. After performing the trace over the spin and using \eqref{eqn:GreenHubbard} and \eqref{eqn:selfCond}, the second term of the grand canonical potential in \eqref{eqn:LuttingerOmega} reduces to 
\begin{align}
 \Tr\,(\Sigma G)=2\,\bigg(U\,n_\uparrow n_\downarrow-\,\frac{\Delta^2}{U}\bigg)=2\,\big(\Phi_\text{Hartree}+\Phi_\text{Fock}\big) \, .
\end{align}
Performing the Matsubara summation of the last term in \eqref{eqn:LuttingerOmega} leads to the grand canonical potential within the Hartree-Fock approximation 
\begin{align}
\Omega=-\frac{U n^2}{4}+U\, m_z^2+\frac{\Delta^2}{U}-\sum_{\nu=\pm}\frac{T}{L}\sum_\bp\ln\big(1+e^{-(E^\nu_\bp-\mu)/T}\big) \,.
\end{align}
A similar result can be obtained by a Hubbard-Stratonovich transformation of the fermionic interaction in the Hubbard model into bosonic fields with a subsequent saddle-point approximation \cite{Voruganti1992}. However, the decoupling is not unique, which is called Fierz ambiguity, and, thus, leads to different results under approximations \cite{Ayral2017}. 

The grand canonical potential is a function of the chemical potential $\mu$. In our application, we will treat the hole doping $p=1-n$ as the fixed variable. Thus, it is convenient to perform the Legendre transformation to the free energy $F=\Omega+\mu\, n$ by inverting the total particle number
\begin{align}
\label{eqn:nMF}
 n(\mu)=\frac{1}{L}\sum_\bp \big(f^+_\bp+f^-_\bp\big)
\end{align}
numerically and using the thermodynamic identity $\partial \Omega/\partial \mu=-n$. Besides the dependence on the model parameters $t$, $t'$, $t''$ and $U$, the free energy $F(n,m_z,\Delta,\bQ,T)$ is a function of the total particle number $n$, the relative particle number $m_z$, the three parameters $\Delta$ and $\bQ=(Q_x,Q_y)$ of the spiral magnetic order as well as the temperature $T$. We determine $m_z$, $\Delta$ and $\bQ$ by minimizing the free energy numerically. The stationary condition for $m_z$ reads
\begin{align}
\label{eqn:stationaryMz}
 \frac{\partial F}{\partial m_z}=\frac{\partial\Omega}{\partial m_z}=0\Leftrightarrow m_z=\frac{1}{L}\sum_\bp\frac{h_\bp}{E^+_\bp-E^-_\bp}\big(f^+_\bp-f^-_\bp\big) \, ,
\end{align}
which has to be determined self-consistently, since all quantities on the right-hand side also involve $m_z$. We see that $m_z=0$ is a solution to \eqref{eqn:stationaryMz} due to the inversion symmetry of the bare band $\eps_\bp=\eps_{-\bp}$ in \eqref{eqn:dispersion}. We find that minimizing the free energy always leads to $m_z=0$ in agreement with Ref.~\cite{Igoshev2010} and as expected by the symmetry in spin of the Hubbard model. We recover the vanishing $z$ component of the direction $\mathbf{n}_i$ of the onsite magnetic magnetic moment in \eqref{eqn:magnDir}. We set $m_z=0$ in the following. The stationary condition for the gap, $\partial F/\partial\Delta=\partial\Omega/\partial\Delta=0$, reproduces the gap equation in \eqref{eqn:gapEq}. 

\begin{figure}[t!]
\centering
\includegraphics[width=0.6\textwidth]{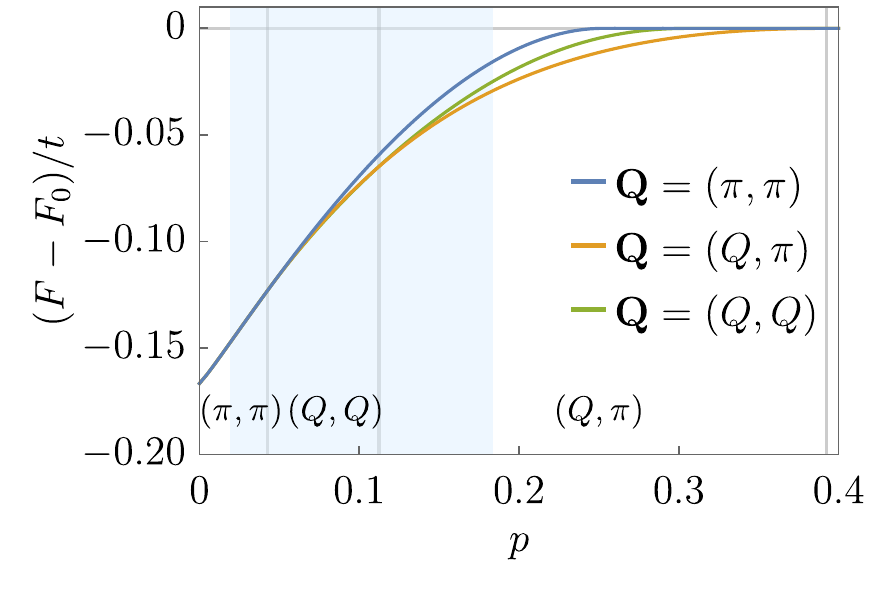}\\
\hspace*{0.02\textwidth}\includegraphics[width=0.58\textwidth]{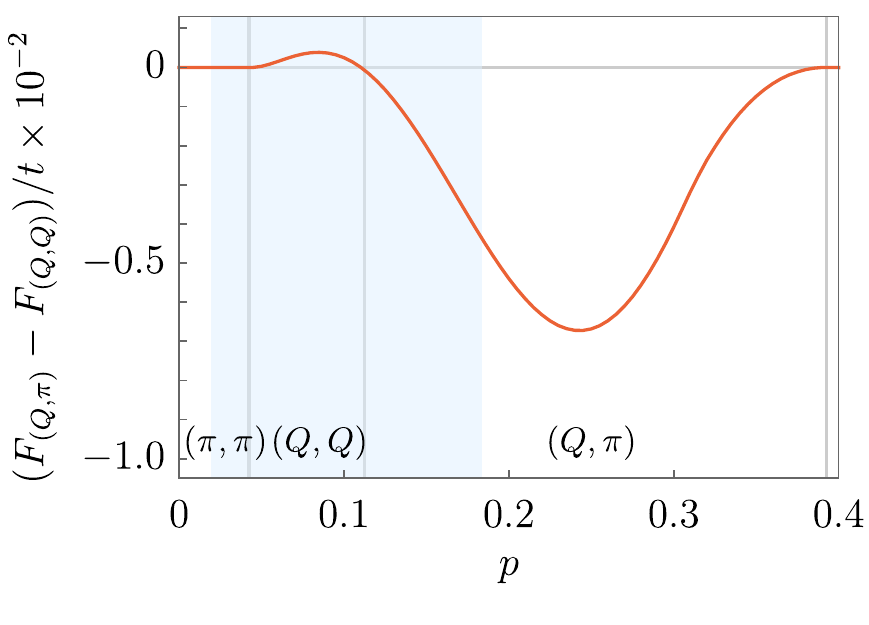}
\caption{The free energy after minimizing $\Delta$ and $\bQ$ for $t'/t=t''/t=0$ and $U/t=4$ at $T/t=0.1$ as a function of hole doping $p=1-n$. We compare the free energy $F$ for different wave vectors with respect to the free energy $F_0$ of the non-ordered state (top). We find three different regions (vertical lines), where the energy gain for one type of $\bQ$ is the largest. The energy difference between these types is small (bottom). We indicate a region of phase separation by a shaded blue area.
\label{fig:MF1}}
\end{figure}
\begin{figure}[t!]
\centering
\includegraphics[width=0.6\textwidth]{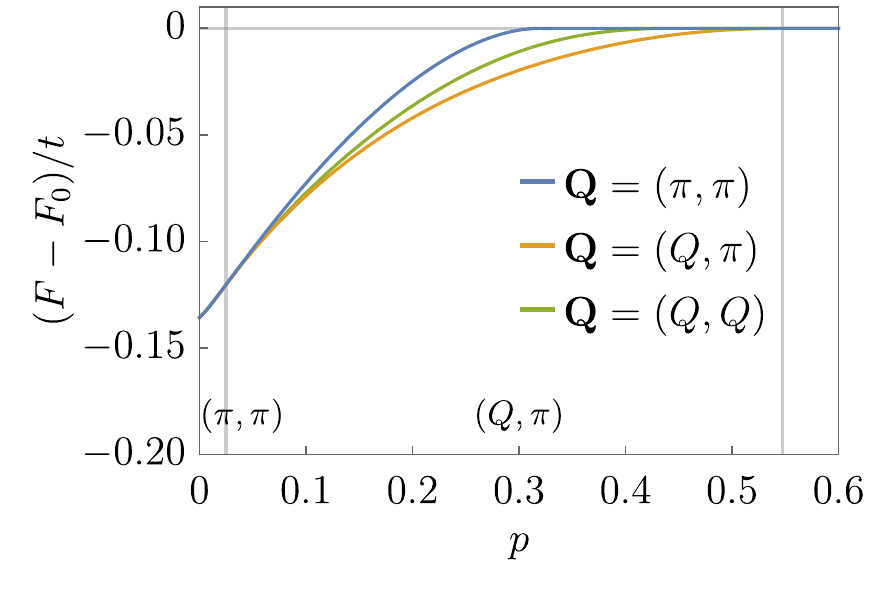}\\
\hspace*{0.02\textwidth}\includegraphics[width=0.58\textwidth]{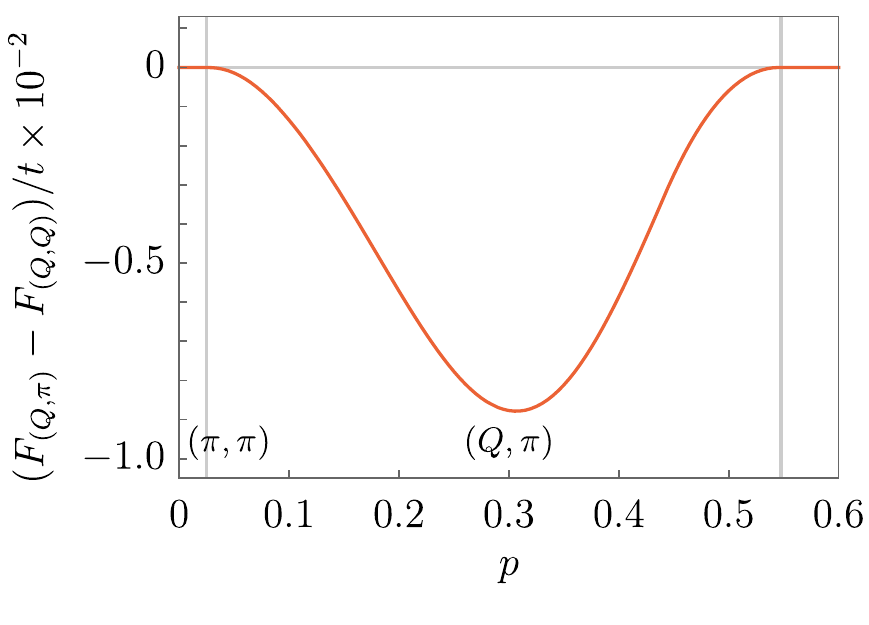}
\caption{The free energy after minimizing $\Delta$ and $\bQ$ for the same parameters as in Fig.~\ref{fig:MF1} but different dispersion $t'/t=-0.17$ and $t''/t=0.05$ as a function of hole doping $p=1-n$. We see that an ordering wave vector $\bQ=(Q,\pi)$ causes the largest energy gain over almost the entire doping range. There is no region of phase separation.
\label{fig:MF2}}
\end{figure}
In Fig.~\ref{fig:MF1}, we show the free energy after minimizing $\Delta$ and $\bQ$ for $t'/t=t''/t=0$ and $U/t=4$ at $T/t=0.1$ as a function of hole doping $p=1-n$. In the upper figure, we show the free energy $F$ relative to the bare free energy $F_0$ that is obtained by setting $\Delta=0$. We compare three different types of spiral magnetic order that are known to be relevant for this parameter range \cite{Voruganti1992, Igoshev2010}: N\'eel antiferromagnetic order with ordering wave vector $\bQ=(\pi,\pi)$ as blue curve and spiral magnetic order with $\bQ=(Q,\pi)$ as orange and $\bQ=(Q,Q)$ as green curve. The value $Q$ is obtained by the minimization of the free energy. We see that magnetic order reduces the free energy for a large doping range from $p=0$ up to $p\approx0.39$. For large hole doping beyond $p\approx 0.11$, the largest energy gain is obtained for $\bQ=(Q,\pi)$. For small hole doping up to $p\approx0.04$, we find N\'eel antiferromagnetism. In the intermediate range, magnetic order with $\bQ=(Q,Q)$ is favored as can be seen by comparing the difference between the free energies of ordering vector $(Q,\pi)$ and $(Q,Q)$ in the lower figure. Overall, we see explicitly that the energy difference between the different magnetic states are very small. The finite region of N\'eel antiferromagnetism at low doping is a finite temperature effect \cite{Voruganti1992}. The chemical potential as a function of doping, which is obtained by inverting Eq.~\eqref{eqn:nMF}, violates the thermodynamic stability criterion $\partial \mu/\partial n=-\partial\mu/\partial p>0$ over some doping range, which indicates a region of phase separation. We obtain the region of phase separation by a Maxwell construction \cite{Igoshev2010}. The region (blue shaded area) ranges form $p\approx0.02$ up to $p\approx0.18$. The range is already slightly reduced due to finite temperature. We see that spiral magnetic order with $\bQ=(Q,Q)$ is completely covered by the area of phase separation and is, thus, not physical. Our result and conclusions are consistent with the zero temperature results of Igoshev {\it et al.} \cite{Igoshev2010} and the finite temperature results of Voruganti {\it et al.} \cite{Voruganti1992}.

In Fig.~\ref{fig:MF2}, we show the energy gain for a different dispersion with $t'/t=-0.17$ and $t''/t=0.05$, which is commonly used to describe the cuprate LSCO \cite{Verret2017}. The ordering wave vector $\bQ=(Q,\pi)$ has the lowest energy over the full doping range from $p\approx 0.025$ to $p\approx 0.55$ except of very close to half filling. The finite region of N\'eel antiferromagnetism close to half filling is again a finite temperature effect.  Next-nearest- and next-next-nearest hopping $t'$ and $t''$ generally reduces the region of phase separation \cite{Igoshev2010}. For our set of parameter we do not find any phase separation. 

\begin{figure}[t!]
\centering
\includegraphics[width=0.6\textwidth]{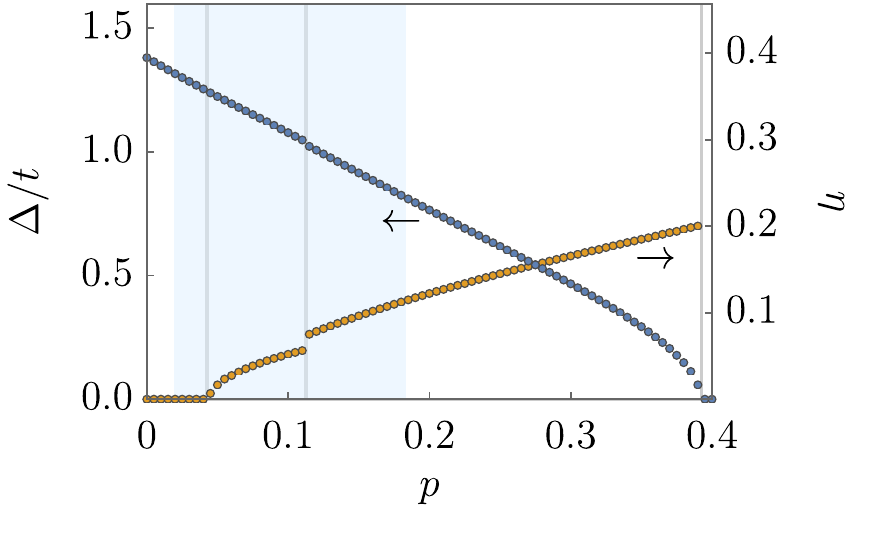}\\
\includegraphics[width=0.6\textwidth]{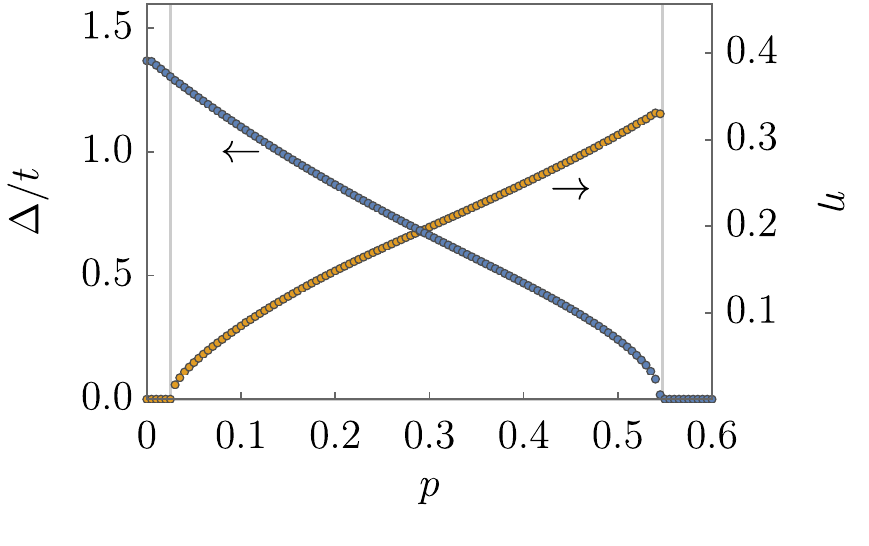}
\caption{The gap $\Delta$ (blue) and the incommensurability $\eta$ (orange) with $Q=\pi-2\pi\eta$ as a function of hole doping $p$. We used the two sets of parameters in Fig.~\ref{fig:MF1} and Fig.~\ref{fig:MF2} for the upper and lower figure, respectively. We find a nearly linear dependence over a large doping range. A second-order phase transition is present between the paramagnetic and the spiral magnetic state and the spiral magnetic and N\'eel antiferromagnetic state. There is a first-order transition between the states with $\bQ=(Q,Q)$ and $\bQ=(Q,\pi)$, which is hidden beneath the region of phase separation (blue shaded area).
\label{fig:MF3}}
\end{figure}

So far, we have focused on the energy gain due to spiral magnetic order but not on the precise doping dependence of the magnetic gap and the ordering wave vector. In Fig.~\ref{fig:MF3}, we show the gap $\Delta(p)$ (blue) and the incommensurability $\eta(p)$ (orange) with $Q=\pi-2\pi\eta$ as a function of hole doping $p$ for the two sets of band parameters that we discussed so far. The gap $\Delta(p)$ decreases from its largest value at half filling and eventually vanishes at a critical doping $p^*$. It is nearly linear at low and intermediate doping. Close to the (second-order) phase transition at $p^*$, it takes the form $\Delta(p)\propto \sqrt{p^*-p}$ as expected for a mean-field-like calculation. The incommensurability $\eta(p)$ is zero in the N\'eel antiferromagnetic state at low doping and increases within the spiral magnetic state. It is nearly linear for large $p$ and vanishes continuously at the second-order phase transition into the N\'eel state. For $t'/t=t''/t=0$ in the upper figure, we have a first-order transition between $\bQ=(Q,Q)$ and $(Q,\pi)$, which is hidden beneath the region of phase separation and, thus, unphysical \cite{Igoshev2010}.

The Hartree-Fock result generally overestimates order since fluctuations of the order parameter are neglected within this approach. In Bonetti {\it et al.} \cite{Bonetti2020,Bonetti2020Authors}, the Hubbard model \eqref{eqn:Hubbard} was studied by using the dynamical mean-field theory (DMFT) \cite{Metzner1989,Georges1992} in the strong-coupling regime. Spiral magnetic order in a DMFT solution of the Hubbard model has been analyzed previously for the square lattice by Fleck {\it et al.} \cite{Fleck1999} and for the triangular lattice by Goto {\it et al.} \cite{Goto2016}. Similar to the Hartree-Fock approach, the non-superconducting ground state is stable within DMFT even without an external magnetic field since there is no pairing instability for $U>0$ within DMFT. In the experiments on cuprates, high magnetic fields are required in order to suppress superconductivity. We expect a reduction of the spiral magnetic gap due to the inclusion of local fluctuations. Nonlocal fluctuation of the magnetic order parameter orientation are not included, so that the presence of magnetic long-range order is expected also at low finite temperature irrespective of the low dimensionality of the system. We study the order parameter for two sets of parameters relevant for the cuprates LSCO and YBCO. In the following and when we study the conductivities based on the following DMFT results, we assume a dispersion with $t'/t=-0.17$ and $t''/t=0.05$ and an onsite repulsion $U/t=8$ for LSCO. We assume a dispersion with $t'/t=-0.3$, $t''/t=0.15$ and an onsite repulsion $U/t=10$ for YBCO. We incorporate the bilayer structure of YBCO, which is known to be crucial for this cuprate compound \cite{Sushkov2009}, by modifying the dispersion via
\begin{align}
 \eps_{\bp,p^z}=\eps_\bp-t^\perp_\bp\cos p^z
\end{align}
with two value $0$ and $\pi$ for $p^z$, which correspond to the bonding and antibonding band, respectively. The interband hopping $t^\perp_\bp=t^\perp\big(\cos p^x-\cos p^y\big)^2$ is set to $t^\perp/t=0.15$. We express all results in units of $t$. For comparison, $t=0.35\,\text{eV}$. The DMFT calculations are performed at $T/t=0.027$ and $T/t=0.04$ for LSCO and $T/t=0.04$ for YBCO. We do not find any phase separation. The gap $\Delta$ is obtained by a zero-frequency extrapolation of the off-diagonal self-energy, which is in general frequency dependent in DMFT. The incommensurability $\eta$ is determined by minimizing the DMFT free energy. A more detailed description of the methodology can be found in Bonetti {\it et al.} \cite{Bonetti2020}. 

\begin{figure}[t!]
\centering
\includegraphics[width=0.6\textwidth]{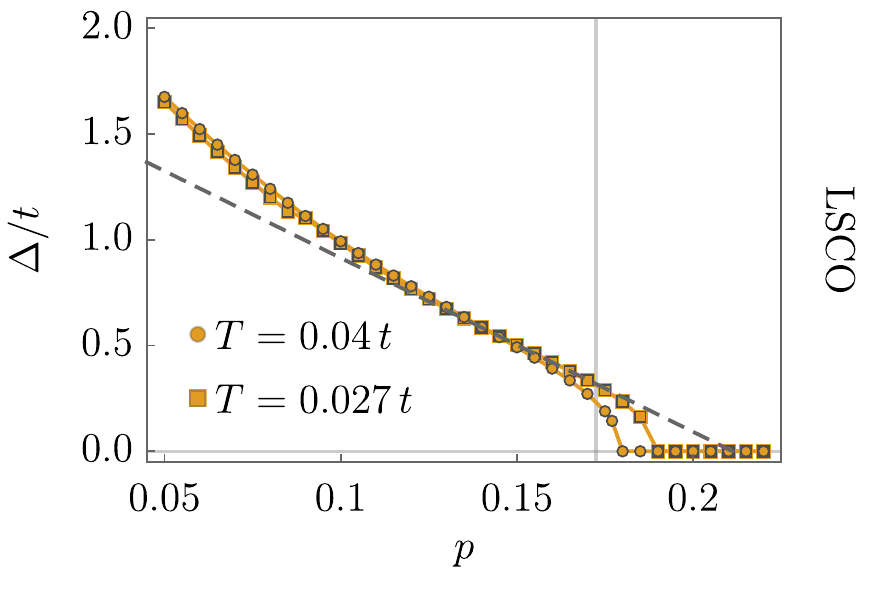}\\
\includegraphics[width=0.6\textwidth]{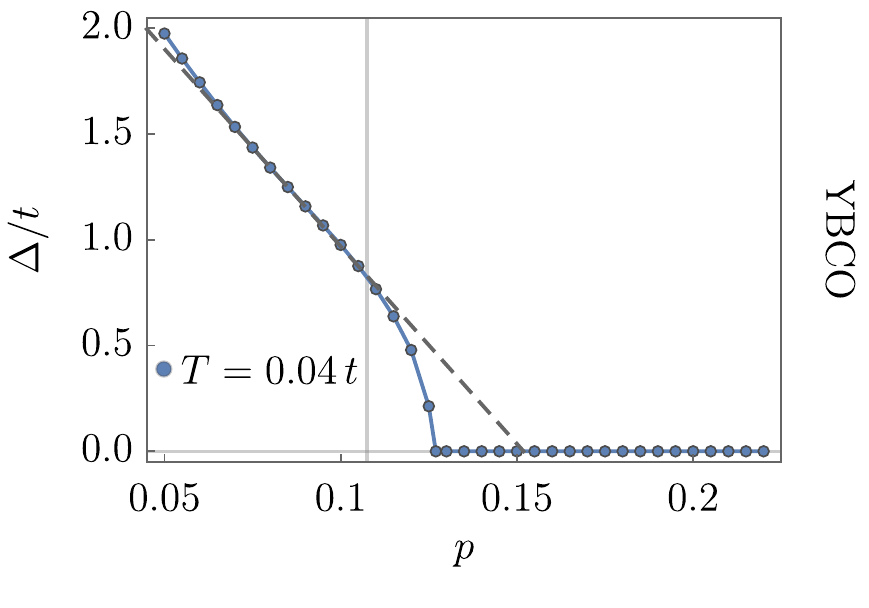}
\caption{The gap $\Delta$ as a function of hole doping $p$ obtained by the DMFT calculation for LSCO (top) and YBCO (bottom). The vertical line indicates the presence of electron pockets beyond that doping for $T/t=0.04$. A linear extrapolation yields an estimate for a gap at $T=0$ (gray dashed lines).
\label{fig:DeltaDMFT}}
\end{figure}

In Fig.~\ref{fig:DeltaDMFT}, we show the gap $\Delta$ obtained by the DMFT calculation as a function of hole doping for LSCO (top) and YBCO (bottom). The results are qualitatively very similar to those that were obtained within the Hartree-Fock approach, which is shown in Fig.~\ref{fig:MF3}. The onset of order is drastically reduced to lower doping as expected by including order parameter fluctuations. The onset of order for LSCO is at $p^*\approx 0.18$ at $T/t=0.04$ and $p^*\approx 0.19$ at $T/t=0.027$, whereas it is smaller for YBCO at $p^*\approx 0.13$ at $T/t=0.04$ due to larger in-plane hoppings. The inter-plane hopping of the bilayer YBCO slightly increases the onset. In the vicinity of $p^*$, both electron and hole pockets are present (vertical lines for $T/t=0.04$). For smaller doping, there are only hole-like pockets. We will discuss the evolution of the Fermi surface topology in more detail in Sec.~\ref{sec:application:spiral:CondCuprates}. We see that a lower temperature only slightly modifies the linear regime but increases $p^*$. We can give an estimate of the gap in the zero temperature limit by a linear approximation
\begin{align}
 \Delta(p)=D\cdot(p^*-p)
\end{align}
for $p<p^*$ with a prefactor $D$ to be determined. We obtain $p^*=0.21$ with $D/t=8.2$ for LSCO and $p^*= 0.15$ with $D/t=18.7$ for YBCO. The onset of order for LSCO parameters is remarkably close to the critical value that was found in recent experiments \cite{Frachet2020}. 

\begin{figure}[t!]
\centering
\includegraphics[width=0.6\textwidth]{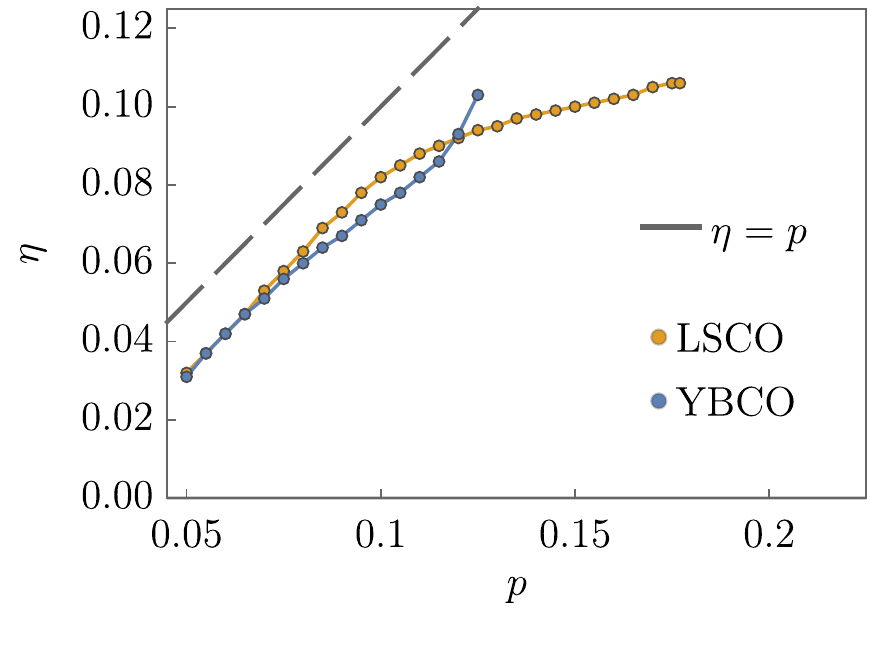}
\caption{The incommensurability $\eta$ as a function of hole doping $p$ obtained by the DMFT calculation of LSCO (orange) and YBCO (blue). We show $\eta=p$ for comparison.
\label{fig:etaDMFT}}
\end{figure}

Comparing different ordering wave vectors within the DMFT calculation showed that spiral magnetic order with $\bQ=(Q,\pi)$ is  favored. In Fig.~\ref{fig:etaDMFT}, we show the incommensurability $\eta$ with $Q=\pi-2\pi\eta$ as a function of doping obtained by minimizing the DMFT free energy. The incommensurability increases linear for both materials. For LSCO, we find two linear regimes. At higher doping, the slope of the second linear regime is reduced. We show $\eta(p)=p$ for comparison. Experimentally, a linear doping dependence $\eta(p)=p$ is approximately valid for LSCO in the doping range $0.06<p<0.12$. It saturates at a value $\eta\approx1/8$ beyond that doping range \cite{Yamada1998}. In YBCO, the incommensurability $\eta(p)$ is found to be significantly smaller than $p$ \cite{Haug2010}. The precise value of $\eta$ is not only doping but also strongly temperature dependent. The free energy depends only very weakly on $\eta$ so that an optimal choice of $\eta$ crucially depends on details. We will clarify in the following discussions on the conductivities that the Hall coefficient only slightly depends on $\eta$, whereas it plays an important role for the anisotropy in the $x$ and $y$ direction of the conductivities.

%
%

\subsection{Symmetric off-diagonal conductivity}
\label{sec:application:spiral:coordinatesystem}

After the discussion on the theoretical evidence of spiral magnetic order in the Hubbard model, we continue with the transport properties of a spiral magnetic state. The real and constant coupling $\Delta$ in the Hamiltonian \eqref{eqn:spiralH} leads to the angle $\varphi_\bp=-\pi$ in the spherical representation \eqref{eqn:lamPolar}, which we considered within our general theory in Sec.~\ref{sec:theory}. This angle, which describes the negative phase of the interband coupling, is momentum independent for a spiral magnetic state. As a consequence, the Berry curvature \eqref{eqn:Omega} and, thus, the antisymmetric interband contributions \eqref{eqn:SinterAN} are identically zero. We express the diagonal and the (symmetric) off-diagonal conductivity 
\begin{align}
 \label{eqn:longCondSpiral}
 \sigma^{\alf\beta}=\sigma^{\alf\beta}_{\text{intra},+}+\sigma^{\alf\beta}_{\text{intra},-}+\sigma^{\alf\beta,s}_\text{inter}
\end{align}
with $\alf,\beta=x,y$ in an orthogonal basis $\mathbf{e}_x$ and $\mathbf{e}_y$ aligned with the underlying square lattice (see Fig.~\ref{fig:magPattern}). We calculate the different contributions via \eqref{eqn:SintraN} and \eqref{eqn:SinterS} at zero temperature.

We have three independent quantities that are described by $\sigma^{\alf\beta}$ for the spiral magnetic state in the two-dimensional plane: the two diagonal conductivities $\sigma^{xx}$ and $\sigma^{yy}$ and the off-diagonal conductivity $\sigma^{xy}=\sigma^{yx}$. We have a closer look at the condition, under which the off-diagonal conductivity $\sigma^{xy}$ vanishes. The off-diagonal intraband conductivities $\sigma^{xy}_{\text{intra},\pm}$ of the two bands $\pm$ involve the product of the two quasiparticle velocities $E^{\pm,x}_\bp E^{\pm,y}_\bp$ in $x$ and $y$ direction. Besides the trivial case of a constant quasiparticle band, we expect a nonzero product for almost all momenta. Thus, in general, $\sigma^{xy}_{\text{intra},\pm}$ only vanishes by integration over momenta. Let us consider the special cases $\bQ=(Q,0)$ and $\bQ=(Q,\pi)$, where we fixed the $y$ component to $0$ or $\pi$. The $x$ component is arbitrary. The following arguments also holds for fixed $x$ and arbitrary $y$ component. Those two special cases include ferromagnetic $(0,0)$, N\'eel antiferromagnetic $(\pi,\pi)$ and the order $(\pi-2\pi \eta,\pi)$ found in the Hubbard model. For those $\bQ$, the two quasiparticle bands $E^\pm_\bp\equiv E^\pm(\bp)$ are symmetric under reflection on the $x$ axis, that is, $E^\pm(p^x,-p^y)=E^\pm(p^x,p^y)$. Thus, the momentum components of the off-diagonal conductivity are antisymmetric, $\sigma^{xy}(p^x,-p^y)=-\sigma^{xy}(p^x,p^y)$, which leads to a zero off-diagonal conductivity when integrating over momenta.  
\begin{figure}[t!]
\centering
\includegraphics[width=0.6\textwidth]{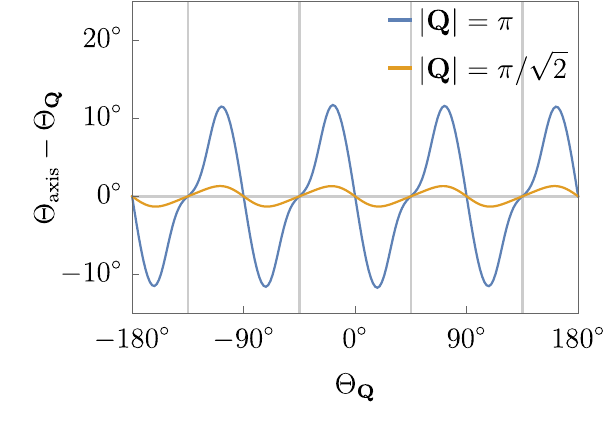}
\caption{The relative angle between the principle axis and the ordering vector $\bQ\propto (\cos\Theta_\bQ,\sin \Theta_\bQ)$ as a function of $\Theta_\bQ$ for $t'/t=0.1$, $t''/t=0$, $\Delta/t=1$, $\Gamma/t=0.05$, $n=0.2$ and different lengths $|\bQ|$. Both axes are aligned for $0^\circ$, $\pm 90^\circ$ and $\pm 180^\circ$ since $\sigma^{xy}$ vanishes as well as for $\pm 45^\circ$ and $\pm 135^\circ$ since $\sigma^{xx}=\sigma^{yy}$ are equal (vertical lines). 
\label{fig:10}}
\end{figure}

As discussed in Sec.~\ref{sec:theory:conductivity:anomalousHall}, a non-diagonal symmetric conductivity matrix $\sigma=(\sigma^{\alf\beta})$ due to nonzero off-diagonal conductivities $\sigma^{xy}=\sigma^{yx}$ can be diagonalized by a rotation of the coordinate system. So far, we have expressed all quantities in the basis vectors $\mathbf{e}_x$ and $\mathbf{e}_y$ that are aligned with the underlying square lattice (see Fig.~\ref{fig:magPattern}). In our two-dimensional case, we can describe the rotation of the basis by a single angle $\Theta$. In Fig.~\ref{fig:10}, we plot the difference between the rotation angle $\Theta_\text{axis}$ that diagonalizes the conductivity matrix $\sigma=(\sigma^{\alf\beta})$ and the direction of the ordering wave vector $\bQ\propto (\cos\Theta_\bQ,\sin\Theta_\bQ)$ as a function of $\Theta_\bQ$ for $t'/t=0.1$, $t''/t=0$, $\Delta/t=1$, $\Gamma/t=0.05$ and $n=0.2$ at different lengths $|\bQ|$. The chemical potential is adapted adequately. We see that both angles are close to each other but not necessarily equal with a maximal deviation of a few degrees. The angles $\Theta_\bQ=0^\circ,\pm 90^\circ,\,\pm 180^\circ$ corresponds to the case of vanishing $\sigma^{xy}$ that we have discussed above, so that the rotated basis axes are parallel to the original $\mathbf{e}_x$ and $\mathbf{e}_y$ axes. At the angles $\Theta_\bQ=\pm 45^\circ,\pm 135^\circ$, the ordering vector $\bQ$ is of the form $(Q,Q)$. Thus, the $x$ and $y$ direction are equivalent, which results in equal diagonal conductivities $\sigma^{xx}=\sigma^{yy}$. A $2\times 2$ conductivity matrix $\sigma=(\sigma^{\alf\beta})$ with equal diagonal elements is diagonalized by rotations with angles $\Theta_\text{axis}=\pm 45^\circ,\pm 135^\circ$ independent of the precise value of the entries and, thus, independent on the length of $\bQ$. These angles are indicated by vertical lines. 
\begin{figure}[t!]
\centering
\includegraphics[width=0.6\textwidth]{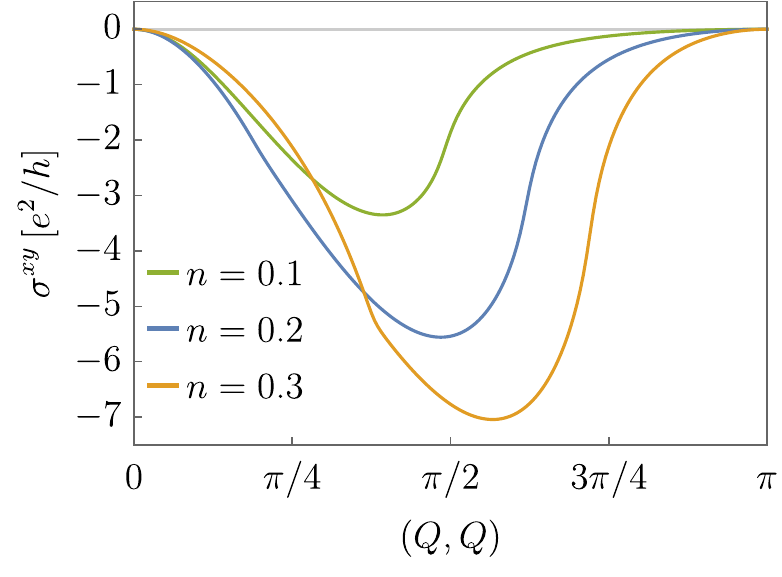}
\caption{The off-diagonal conductivity $\sigma^{xy}$ as a function of $\bQ=(Q,Q)$ for the same parameters as in Fig.~\ref{fig:10} at different particle numbers $n=0.1,\,0.2,\,0.3$.  
\label{fig:11}}
\end{figure}

In the following, we focus on the special case of an ordering vector $\bQ=(Q,Q)$. The conductivity matrix is diagonal within the basis $(\mathbf{e}_x\pm \mathbf{e}_y)/\sqrt{2}$, which corresponds to both diagonal directions in Fig.~\ref{fig:magPattern}. The longitudinal conductivities are $\sigma^{xx}\pm \sigma^{xy}$ with $\sigma^{xx}=\sigma^{yy}$. Thus, the presence of spiral magnetic order results in an anisotropy (or ''nematicity``) of the longitudinal conductivity. The ''strength`` of the anisotropy is given by $2\sigma^{xy}$ for $\bQ=(Q,Q)$. In Fig~\ref{fig:11}, we show $\sigma^{xy}$ as a function of $\bQ=(Q,Q)$ for the same parameters as in Fig.~\ref{fig:10} at different particle numbers $n=0.1,\,0.2,\,0.3$. The chemical potential is adapted adequately. The values $|(\pi/\sqrt{2},\pi/\sqrt{2})|=\pi$ and $|(\pi/2,\pi/2)|=\pi/\sqrt{2}$ correspond to the cases presented in Fig.~\ref{fig:10}. We see that the anisotropy vanishes for ferromagnetic $(0,0)$ and N\'eel-antiferromagnetic $(\pi,\pi)$ order as expected. The largest anisotropy for the presented set of parameters is close to $(\pi/2,\pi/2)$. In Fig.~\ref{fig:magPattern} (a), (c) and (d), we show the corresponding magnetization patterns.

\begin{figure}[t!]
\centering
\includegraphics[width=0.6\textwidth]{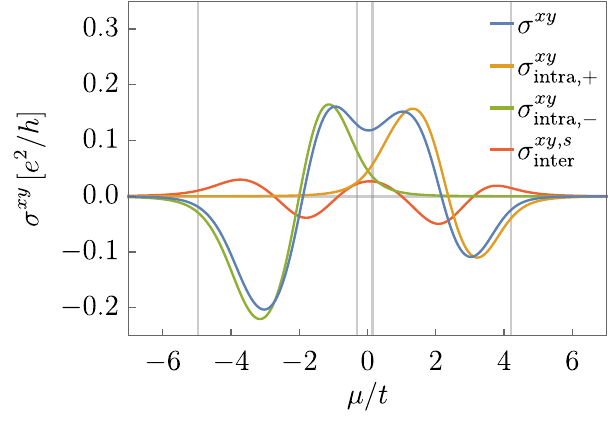}
\caption{The off-diagonal conductivity $\sigma^{xy}$ and its nonzero contributions as a function of the chemical potential $\mu/t$ for $t'/t=0.1$, $t''/t=0$, $\Delta/t=2$, $\bQ=(\pi/\sqrt{2},\pi/\sqrt{2})$ and $\Gamma/t=1$. The vertical lines indicate the bandwidth and the band gap. 
\label{fig:12}}
\end{figure}

In Fig.~\ref{fig:12}, we show the off-diagonal conductivity, that is, the anisotropy, and its three different contributions as a function of the chemical potential $\mu/t$ for $t'/t=0.1$, $t''/t=0$, $\Delta/t=2$, $\bQ=(\pi/\sqrt{2},\pi/\sqrt{2})$ and $\Gamma/t=1$. The value of the conductivity is reduced compared to the previous examples by approximately one order of magnitude as expected by the scaling $\sigma^{xy}\propto 1/\Gamma$. As we vary the chemical potential, we get nonzero conductivity within the bandwidth given by approximately $-4.9\,t$ to $4.2\,t$. Both the off-diagonal conductivity and its different contributions given in \eqref{eqn:longCondSpiral} take positive and negative values in contrast to the diagonal conductivities $\sigma^{xx}$ and $\sigma^{yy}$, which are always positive. For $\Delta/t=2$, we have a band gap between $-0.3\,t$ and $0.1\,t$ with nonzero conductivities due to the large value of $\Gamma$. We see that for negative and positive chemical potential outside the gap, $\sigma^{xy}$ is mainly given by the contribution of the lower band $\sigma^{xy}_{\text{intra},-}$ or the upper band $\sigma^{xy}_{\text{intra},+}$, respectively. Inside the gap, we have both contributions of the two bands due to smearing effects and the symmetric interband contribution $\sigma^{xy,s}_\text{inter}$, which are all comparable in size. 
\begin{figure}[t!]
\centering
\includegraphics[width=0.6\textwidth]{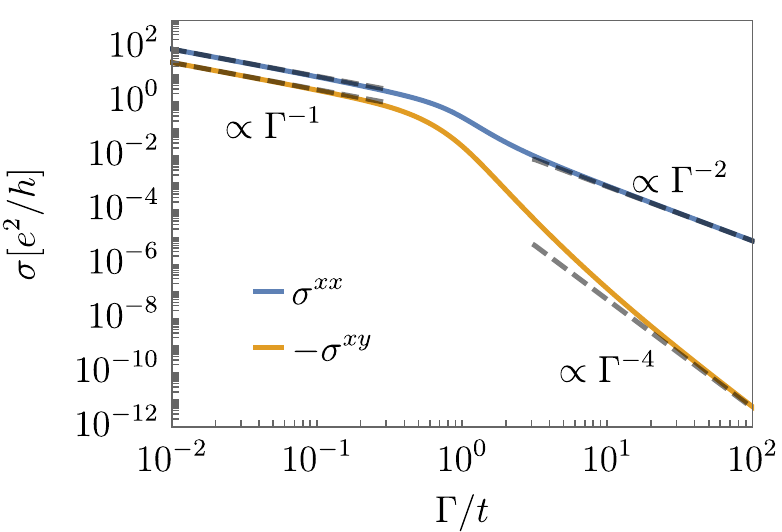}
\caption{The diagonal (blue) and off-diagonal (orange) conductivity as a function of $\Gamma/t$ for $t'/t=0.1$, $t''/t=0$, $\Delta/t=1$ and $\bQ=(\pi/2,\pi/2)$ at $n=0.2$. The calculated limiting behaviors in the clean and the dirty limit are indicated by dashed lines. 
\label{fig:13}}
\end{figure}

In Fig.~\ref{fig:13}, we show the diagonal (blue) and off-diagonal (orange) conductivity as a function of the relaxation rate $\Gamma/t$ for $t'/t=0.1$, $t''/t=0$, $\Delta/t=1$ and $\bQ=(\pi/2,\pi/2)$ at $n=0.2$. We fixed the particle number by calculating the chemical potential at each $\Gamma$. In the clean limit (low $\Gamma$), both $\sigma^{xx}$ and $\sigma^{xy}$ scale as $1/\Gamma$ as expected for the limiting behavior of the intraband contributions in \eqref{eqn:winG0} (dashed lines). In Sec.~\ref{sec:theory:conductivity:limits}, we showed that, in the dirty limit (large $\Gamma$), both the diagonal and the off-diagonal conductivities scale as $\Gamma^{-2}$ in first order due to their intraband character. However for the considered parameters, the diagonal conductivity $\sigma^{xx}$ scales as $\Gamma^{-2}$, whereas the off-diagonal conductivity $\sigma^{xy}$ scales as $\Gamma^{-4}$. The dashed lines are calculated via \eqref{eqn:WnInfty} for the respective order. The off-diagonal conductivity eventually scales as $\Gamma^{-2}$ for $\Gamma$ far beyond the numerically accessible range due to very small prefactors in the expansion. 
We see explicitly that the analysis of the individual prefactors of the expansion in the dirty limit as discussed in Sec.~\ref{sec:theory:conductivity:limits} is useful in order to understand this or similar unexpected scaling behaviors.

%
%

\subsection{Longitudinal conductivity and Hall number in cuprates} 
\label{sec:application:spiral:CondCuprates}

We continue with the transport properties of a spiral magnetic state. The onset of spiral magnetic order can explain the drop in the longitudinal conductivity and the Hall number \cite{Eberlein2016} seen first experimentally by Badoux {\it et al.} \cite{Badoux2016}. However, the range of validity of the used formulas that relate the model of spiral magnetic order with its transport properties remained unclear. As we have already discussed in the introduction of this thesis and recapitulated at the beginning of Sec.~\ref{sec:application:spiral}, a general argument suggests that interband contributions, which were neglected in the previous formulas, may be important at the onset of order. After having provided our general theory of the longitudinal and the Hall conductivity including interband contributions in Chapter~\ref{sec:theory}, we are now able to discuss the recent charge transport experiments in cuprates in more detail. 

In Sec.~\ref{sec:application:spiral:hubbard}, we have obtained the doping-dependence of the magnetic gap $\Delta(p)$ and the incommensurability $\eta(p)$, which describes the ordering wave vector of the form $\bQ=(\pi-2\pi \eta,\pi)$, for the spiral magnetic order via a DMFT calculation. We will use these results to calculate the corresponding conductivities and the Hall number as a function of doping for LSCO and YBCO. Furthermore, we will calculate the conductivities by using a simplified phenomenological model for the doping dependences. In particular, we will use this simplified ansatz to study the importance of interband contributions systematically. 

\subsubsection*{Phenomenological model of the spiral magnetic order in cuprates} 

We give a more detailed description of the phenomenological model that we will use in the following. Some details have already been mentioned throughout this thesis when discussing the model that was used by Eberlein {\it et al.} \cite{Eberlein2016}. The phenomenological model is in close analogy to previous theoretical studies \cite{Storey2016, Eberlein2016, Chatterjee2017}. Theoretical results for spiral states in the two-dimensional $t$-$J$ model \cite{Sushkov2004} suggest a linear dependence of the magnetic gap on the hole doping $p$ of the form
\begin{align}
 \label{eqn:doping}
 \Delta(p) = D\,(p^* - p)\,\Theta(p^* - p) \, ,
\end{align}
where $D$ is a prefactor and $p^*$ is the critical doping, at which the magnetic order vanishes. $\Theta(x)$ is the Heaviside step function, so that the gap is zero for $p>p^*$. We have seen explicitly in Sec.~\ref{sec:application:spiral:hubbard} that such a form is reasonable for an approximation of the magnetic gap $\Delta(p)$ that was obtained for the Hubbard model by a Hartree-Fock approximation and via DMFT. Both $D$ and $p^*$ are material-dependent parameters of the phenomenological model and, thus, need to be fitted to experimental data. For comparison, we have obtained $p^*=0.21$ with $D/t=8.2$ for LSCO and $p^*= 0.15$ with $D/t=18.7$ for YBCO in the DMFT calculation. A linear doping dependence of the gap for $p < p^*$ is also found in resonating valence bond mean-field theory for the $t$-$J$ model \cite{Yang2006a}. 
In Ref.~\cite{Eberlein2016}, a phenomenological quadratic doping dependence of $\Delta(p)$ was considered, too.

The wave vector of the incommensurate magnetic states obtained in the theoretical literature \cite{Schulz1990, Kato1990, Fresard1991, Raczkowski2006, Igoshev2010, Chubukov1992, Chubukov1995, Metzner2012,Igoshev2015, Yamase2016, Zheng2017, Vilardi2018, Shraiman1989, Kotov2004, Sushkov2004, Sushkov2006, Sushkov2009, Luscher2007} has the form $\bQ = (\pi-2\pi\eta,\pi)$ or a form that is symmetry related to $\bQ$, that is, $(-\pi+2\pi\eta,\pi)$, $(\pi,\pi-2\pi\eta)$, and $(\pi,-\pi+2\pi\eta)$. The incommensurability $\eta > 0$ measures the deviation from the N\'eel wave vector $(\pi,\pi)$. Peaks in the magnetic structure factors seen in neutron-scattering experiments are also situated at such wave vectors \cite{Yamada1998, Fujita2002, Haug2009, Haug2010}. The incommensurability $\eta$ is a monotonically increasing function of doping. In Ref.~\cite{Eberlein2016}, the doping dependence of $\eta$ was determined by minimizing the mean-field free energy, resulting in $\eta \approx p$, which is roughly consistent with experimental observations in LSCO \cite{Yamada1998}. A linear doping dependence of $\eta$ was also found in the $t$-$J$ model in the underdoped regime close to half filling \cite{Luscher2007, Sushkov2009}. In YBCO, $\eta$ values below $p$ are observed \cite{Haug2010}, and functional renormalization group calculations for the Hubbard model also yield $\eta < p$ \cite{Yamase2016}.
Some further aspects of the doping dependence of $\eta$ for the Hubbard model have already been discussed in Sec.~\ref{sec:application:spiral:hubbard}. Our results supported the approximately linear doping dependence of $\eta$. In the following, we choose $\eta = p$ in the phenomenological model for simplicity and discuss consequence of a different doping dependence if required. We have already sketched the reconstruction of the Fermi surface and the corresponding magnetization pattern of the phenomenological model in Fig.~\ref{fig:modelSpiral} in the introduction of this thesis. We will discuss the change in the Fermi surface topology throughout this section in detail.

\subsubsection*{Relaxation rate for cuprates} 

Due to the specific form of the interband coupling $\Delta$, there are no interband contributions to the conductivity based on the Berry curvature. Thus for small relaxation rates $\Gamma$, the interband contributions of the longitudinal and the Hall conductivity in the spiral state are suppressed by a factor $\Gamma^2$ compared to the intraband contributions (see Sec.~\ref{sec:theory:conductivity:limits} and \ref{sec:theory:Hall:coefficient}). Nevertheless, it has still to be clarified whether interband contributions are relevant for certain parameter choices. To get a feeling for the typical size of $\Gamma$ in the recent high-field experiments, we estimate $\Gamma$ from the experimental result $\omega_c \tau = 0.075$ reported for $\rm La_{1.6-x} Nd_{0.4} Sr_x Cu O_4$ (Nd-LSCO) samples at zero temperature by Collignon {\it et al.} \cite{Collignon2017}. The cyclotron frequency can be written as $\omega_c = e|\bB|/m_c$, which defines the cyclotron mass $m_c$. For free electrons, $m_c$ is just the bare electron mass $m_e$. Inserting the applied magnetic field of 37.5 Tesla and assuming $m_c = m_e$, one obtains $\Gamma = (2\tau)^{-1} \approx 0.03 \, \text{eV}$. With the typical value $t \approx 0.3 \, \text{eV}$ for the nearest-neighbor hopping amplitude in cuprates, one thus gets $\Gamma/t \approx 0.1$. The cyclotron mass in cuprates is actually larger than the bare electron mass. Mass ratios $m_c/m_e$ equal to 3 or even larger have been observed \cite{Stanescu2008}. Hence, $\Gamma/t = 0.1$ is just an upper bound; the actual value can be expected to be even smaller. Indeed, an estimate from the observed residual resistivity in Nd-LSCO yields $\Gamma \approx 0.008 \, \text{eV}$ \cite{Taillefer2018}.

The relaxation rates in cuprate superconductors are actually momentum dependent. However, we do not expect the momentum dependence to affect the order of magnitude of interband contributions. Concerning the doping dependence of $\Gamma$, we are using experimental input. Magnetoresistance data suggest that the electron mobility does not change significantly in the doping range where the Hall number drop is observed \cite{Collignon2017}. Since the mobility is directly proportional to the inverse relaxation rate, it is a reasonable assumption to consider a relaxation rate $\Gamma$ that is independent of doping.

\subsubsection*{Interband contributions for the longitudinal conductivity in cuprates} 

We have raised the question of the relevance of interband contributions to the conductivities, which were often neglected in earlier calculations for N\'eel and spiral magnetic states. In our theory, the interband effects are captured by the symmetric interband contribution of the longitudinal conductivities $\sigma^{xx,s}_\text{inter}$ and $\sigma^{yy,s}_\text{inter}$ in \eqref{eqn:SinterS} and the interband contributions of the Hall conductivity $\sigma^{xyz}_{\text{H},\text{inter},\pm}$ in \eqref{eqn:SHinterXYZ}. In order to discuss its relevance in the context of recent transport measurements \cite{Badoux2016, Laliberte2016, Collignon2017}, we have a closer look at the size of those interband contributions using the phenomenological model in \eqref{eqn:doping}. We first discuss the interband contributions for the longitudinal conductivity and come back to the interband contributions to the Hall number later on. Interband contributions have been taken into account in a calculation of the optical conductivity in a $d$-density wave state \cite{Aristov2005}, and in a very recent evaluation of the longitudinal DC conductivity in the spiral state \cite{Chatterjee2017}. 

\begin{figure}[t!]
\centering
\includegraphics[width=0.53\textwidth]{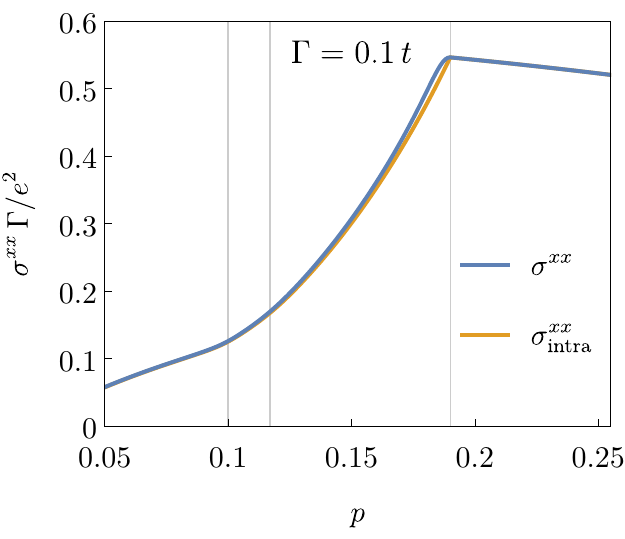}\\
\includegraphics[width=0.53\textwidth]{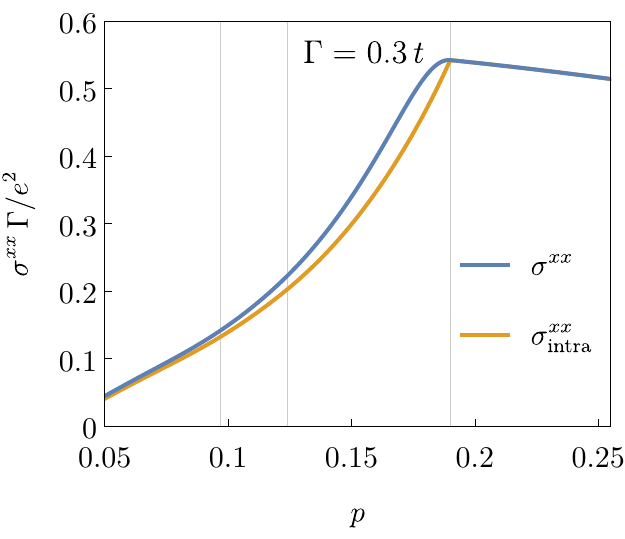}
\caption{Longitudinal conductivity $\sigma^{xx}$ at zero temperature as a function of doping $p$ for a doping-dependent magnetic order parameter $\Delta(p) = 12t (p^* - p) \Theta(p^* - p)$ with $p^* = 0.19$. The intraband contribution $\sigma_\text{intra}^{xx}$ is also shown for comparison. We use the hopping parameters $t'/t = -0.3$ and $t''/t = 0.2$ and the relaxation rates $\Gamma/t = 0.1$ (top) and $\Gamma/t = 0.3$ (bottom). The vertical lines indicate changes of the Fermi-surface topology at the three doping values $p_e^*$, $p_h^*$, and $p^*$ (from left to right).}
\label{fig:cond}
\end{figure}

In Fig.~\ref{fig:cond}, we show results for the longitudinal conductivity $\sigma^{xx}$ of the phenomenological model in \eqref{eqn:doping} with parameters $D/t=12$, $p^*=0.19$, and $\eta=p$. The conductivity is obtained from Eq.~\eqref{eqn:longCondSpiral} at zero temperature for two values of the relaxation rate $\Gamma$. We chose hopping parameters that are used for YBCO in the literature (see, for instance, \cite{Verret2017}). We neglected the double-layer structure of YBCO for simplicity. The critical doping $p^* = 0.19$ is the onset doping for the Hall number drop observed in the experiments on YBCO by Badoux {\it et al.} \cite{Badoux2016}. The total conductivity $\sigma^{xx}$, which is calculated via \eqref{eqn:longCondSpiral}, is compared to the intraband contribution $\sigma_\text{intra}^{xx}=\sigma_{\text{intra},+}^{xx}+\sigma_{\text{intra},-}^{xx}$, where the interband contribution is not taken into account. Note that $\hbar=1$, so that $\sigma^{\alf\beta}/e^2$ is dimensionless. The conductivities can be written in units of the conductance quantum $e^2/h$ simply by multiplying with $2\pi$. Here, we rescaled the longitudinal conductivity by the leading order in $\Gamma$ so that the plots for different $\Gamma$ are easier to compare. One can see a pronounced drop of the conductivity for $p < p^*$, as expected from the drop of charge-carrier density in the spiral state. For $\Gamma/t = 0.1$, the interband contributions are practically negligible, while they are already sizable for $\Gamma/t = 0.3$. In particular, the interband contributions shift the drop of $\sigma^{xx}$ induced by the spiral order towards smaller values of $p$, and they smooth the sharp kink exhibited by $\sigma^{xx}$ at $p^*$ for $\Gamma \to 0$. We see that the interband contributions are of particular importance at the onset of order, where the gap and the relaxation rate are comparable in size, in agreement with the general argument that we gave in the introduction. In other words, there is no general theoretical argument that holds so that those interband contributions are negligible for all dopings. However, the interband contributions are indeed also negligible close to the onset of order due to their small numerical value for relaxation rates $\Gamma/t\lesssim 0.1$, which is a reasonable estimate of the size of the relaxation rate for the recent experiments on cuprates.

Chatterjee {\it et al.} \cite{Chatterjee2017} have derived expressions for the electrical and the heat conductivities in the spiral state, for a momentum-independent relaxation rate, and showed that the two quantities are related by the Wiedemann-Franz law. While their formulas for the conductivities have a different form than ours, we have checked that the numerical results are consistent.

\subsubsection*{Fermi surface topology} 

The presence and the doping dependence of the spiral magnetic order has a big impact on the Fermi surface topology. The change from a large Fermi surface with volume of size $1+p$ to (several) pockets with total volume of size $p$ led to the naive explanation of the observed drop in the Hall number. Although this explanation may not be correct in general for the regime $\omega_c\tau\ll 1$, which is relevant for those experiments, but only for $\omega_c\tau\gg 1$ or the particular case of parabolic dispersions, the evolution of the Fermi surface still gives useful insights to understand the different regimes of the longitudinal conductivity, which we indicated by vertical lines in Fig.~\ref{fig:cond}, and of the Hall number, which we will discuss later on. 

\begin{figure}[t!]
\centering 
\includegraphics[width=0.45\textwidth]{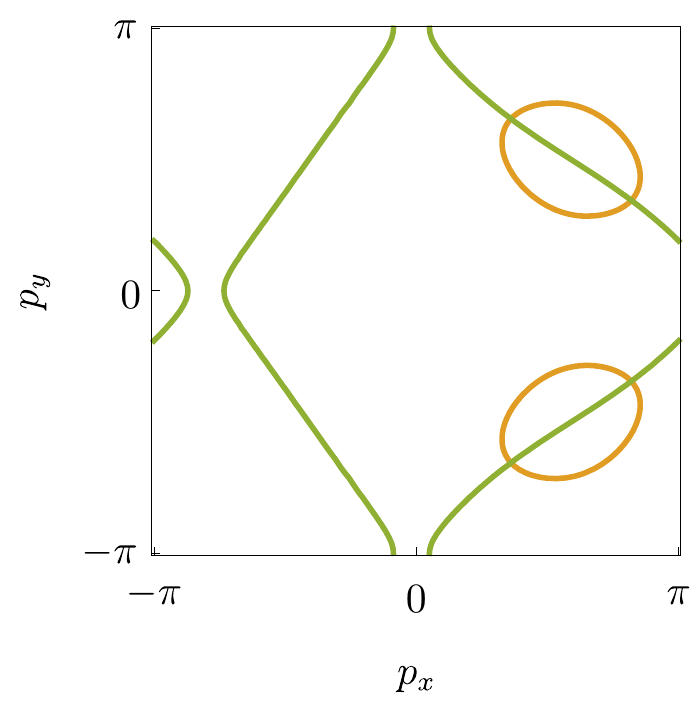}\\[2mm]
\includegraphics[width=0.45\textwidth]{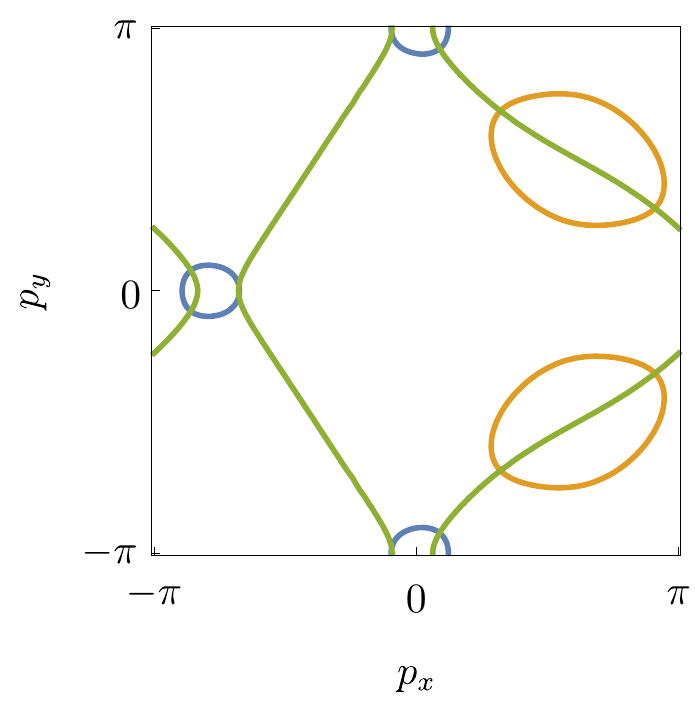}
\hspace{0.04\textwidth}
\includegraphics[width=0.45\textwidth]{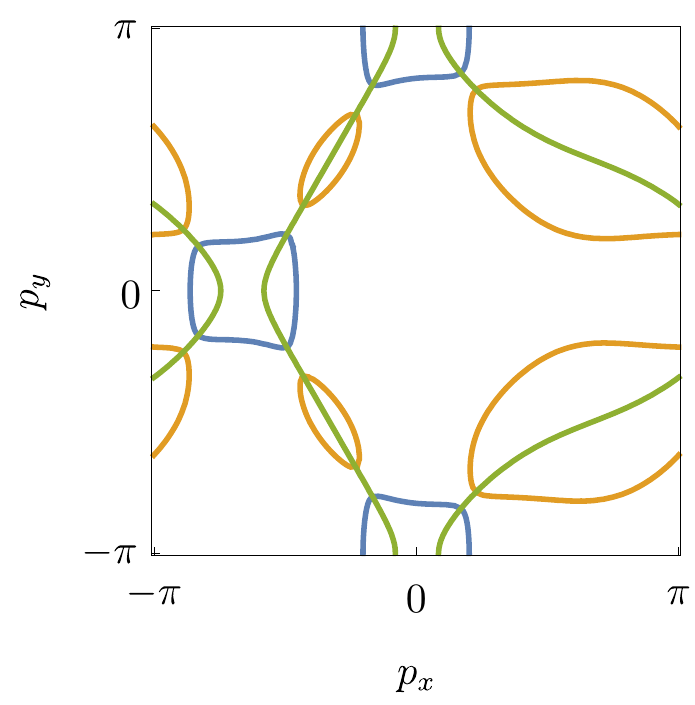}
\caption{Quasiparticle Fermi surfaces for $p = 0.09$ (top), $p = 0.115$ (bottom left), and $p = 0.17$ (bottom right). Fermi-surface sheets surrounding hole (orange) and electron (blue) pockets correspond to zeros of $E_\bp^--\mu$ and $E_\bp^+-\mu$, respectively. The green ``nesting line'' indicates momenta $\bp$ satisfying the condition $\eps_{\bp} = \eps_{\bp+\bQ}$. The band and gap parameters are the same as in Fig.~\ref{fig:cond}, with $\Gamma/t = 0.1$.}
\label{fig:fs}
\end{figure}

In Fig.~\ref{fig:fs}, we plot the Fermi surfaces of the phenomenological model for different dopings $p=0.09$, $0.115$, $0.17$. We use the the same parameters as in Fig.~\ref{fig:cond}. We identify different doping regimes with different Fermi surface topologies. For $p < p_e^*$, the quasiparticle Fermi surface consists exclusively of hole pockets (orange), while for $p_e^* < p < p^*$ also electron pockets (blue) are present. Note that $p_e^*$ depends (slightly) on the relaxation rate $\Gamma$ since the relation between the chemical potential $\mu$ and the hole doping depends on $\Gamma$. For $p < p_h^*$, there are only two hole pockets, while a second (smaller) pair of hole pockets appears for $p_h^* < p < p^*$. The dopings in Fig.~\ref{fig:fs} are chosen to give an example for those three regimes with $p < p_e^*$ (top), $p_e^* < p < p_h^*$ (bottom left), and $p_h^* < p < p^*$ (bottom right). At $p = p^*$, electron and hole pockets merge, and, for $p > p^*$, there is only a single large Fermi-surface sheet, which is closed around the unoccupied (hole) states. In Fig.~\ref{fig:cond}, the doping dependence of the conductivity changes its slope at $p_e^*$, while there is no pronounced feature at $p_h^*$. However, choosing a smaller relaxation rate $\Gamma/t \ll 0.1$, a change of the slope of $\sigma^{xx}$ is visible also at $p_h^*$, while no pronounced feature in $\sigma^{yy}$ is visible. The sequence of Fermi-surface topologies as a function of doping depends on the doping dependence of the incommensurability $\eta$. The above results were obtained for $\eta = p$. Choosing, for example, a smaller $\eta(p)$, one may have four (not just two) hole pockets at low doping.
The green lines in Fig.~\ref{fig:fs} indicate momenta $\bp$ that satisfy the condition $\eps_\bp=\eps_{\bp+\bQ}$. At those momenta, the band gap $E^+_\bp-E^-_\bp=2\Delta$ is minimal. For N\'eel antiferromagnetic order with $\bQ=(\pi,\pi)$, this line corresponds to the antiferromagnetic Brillouin zone boundary independent of the band parameters.

Note that the spiral order with ordering vector component $Q_x=\pi-2\pi\eta$ and nonzero $\eta$ breaks the mirror symmetry in $x$ direction, which is  reflected by the asymmetry in the Fermi surfaces in Fig.~\ref{fig:fs}. Inversion symmetry is restored in the spectral function of single-electron excitations as discussed in Sec.~\ref{sec:application:spiral:spectralweights}. In the lower row of Fig.~\ref{fig:spectral}, we plotted the spectral function of the quasiparticle and the single-electron excitations for the regime, where only two hole-pockets are present. The shown spectral functions corresponds to a Fermi surface similar to the one shown in Fig.~\ref{fig:fs} (top). Besides the Fermi surface, which is clearly visible in the spectral functions in Fig.~\ref{fig:spectral}, there are also precursors of the electron pockets due to the finite relaxation rate, which will eventually appear at higher doping and are shown in Fig.~\ref{fig:fs} (bottom left).

\subsubsection*{Momentum-resolved longitudinal conductivity}

It is instructive to see which quasiparticle states yield the dominant contributions to the conductivity. In two dimensions, the conductivity formulas in \eqref{eqn:SintraN} and \eqref{eqn:SinterS} are given by a momentum integral of the form $\sigma^{\alf\beta} = \int \frac{d^2\bp}{(2\pi)^2} \, \sigma^{\alf\beta}(\bp)$. The Fermi function derivative $f'(\eps)$ restricts the energies $\eps$ up to values of order $T$. For a temperature $T=0$, one has $f'(\eps) = - \delta(\eps)$. For a small relaxation rate $\Gamma$, the quasiparticle spectral functions $A_\bp^{\pm}(\eps)$ are peaked at the quasiparticle energies. Hence, for low $T$ and small or moderate $\Gamma$, the dominant contributions to the conductivity come from momenta where either $|E_\bp^+-\mu|$ or $|E_\bp^--\mu|$ is small, that is, in particular from momenta near the quasiparticle Fermi surfaces.
\begin{figure}[t!]
\centering
\includegraphics[width=0.48\textwidth]{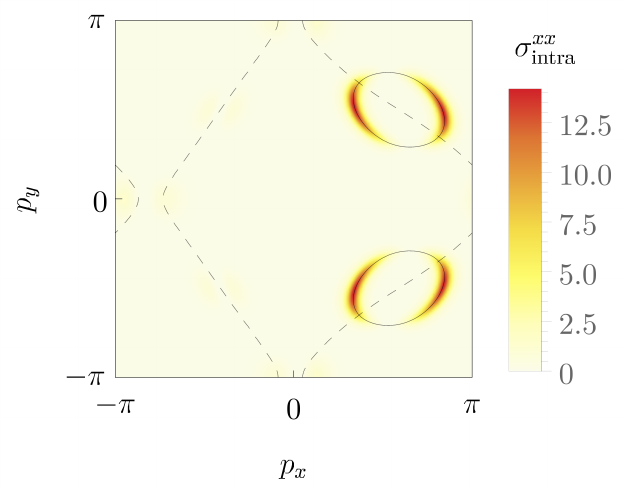}
\includegraphics[width=0.48\textwidth]{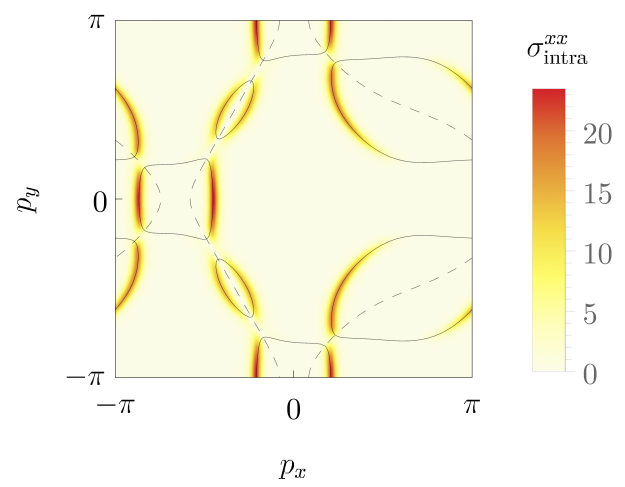} \\
\includegraphics[width=0.48\textwidth]{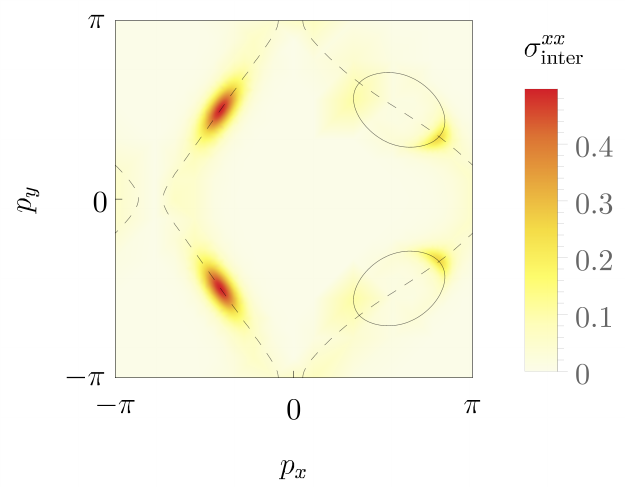}
\includegraphics[width=0.48\textwidth]{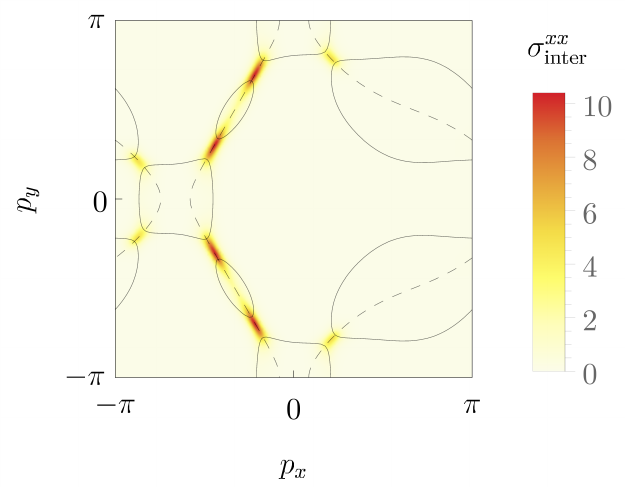}
\caption{Top: Color plot of the momentum resolved intraband contribution to the longitudinal conductivity $\sigma_\text{intra}^{xx}(\bp)$ for $p = 0.09$ (left) and $p = 0.17$ (right). Bottom: Interband contribution $\sigma_\text{inter}^{xx}(\bp)$ for the same choices of $p$. The band and gap parameters are the same as in Fig.~\ref{fig:cond}, and the relaxation rate is $\Gamma/t = 0.3$. The Fermi surfaces and the nesting line (cf.\,\,Fig.~\ref{fig:fs}) are plotted as thin black lines.}
\label{fig:sgp}
\end{figure}

In Fig.~\ref{fig:sgp}, we show color plots of $\sigma_\text{intra}^{xx}(\bp)$ and $\sigma_\text{inter}^{xx}(\bp)$ in the Brillouin zone. Although a sizable $\Gamma/t = 0.3$ has been chosen, the intraband contributions are clearly restricted to the vicinity of the quasiparticle Fermi surface. Variations of the size of intraband contributions along the Fermi surfaces are due to the momentum dependence of the quasiparticle velocities $E_\bp^{\pm,x} = \partial E_\bp^\pm/\partial p^x$. The interband contributions are particularly large near the ``nesting line'' defined by $\eps_{\bp+\bQ} = \eps_{\bp}$, where the direct band gap between the quasiparticle energies $E_\bp^+$ and $E_\bp^-$ assumes the minimal value $2\Delta$. For $p=0.09$, the largest interband contributions come from regions on the nesting line remote from the Fermi surfaces. Note, however, that they are much smaller than the intraband contributions, and $|E_\bp^--\mu|$ has a local minimum in these regions. For $p=0.17$, the interband contributions are generally larger, and they are concentrated in regions between neighboring electron and hole pockets.

\subsubsection*{Longitudinal conductivity obtained by using the DMFT results}

So far, we have discussed the longitudinal conductivity for the phenomenological model. In Sec.~\ref{sec:application:spiral:hubbard}, we have presented  results for the Hubbard model within a dynamical mean-field theory (DMFT) approach. In principle, one could compute charge transport properties from linear response theory within the DMFT approximation \cite{Georges1996}. However, this involves a rather delicate analytic continuation from Matsubara to real frequencies. Moreover, the relaxation rates obtained from the DMFT cannot be expected to provide a good approximation in two dimensions. Hence, we compute only the magnetic gap, the incommensurability, and the so-called $Z$-factor from the DMFT, while we take the relaxation rates from estimates obtained from experiments. By this approach, the doping dependence of the quantities as well as their size are no longer parameters that have to be adjusted but that have been obtained {\it ab initio} from the Hubbard model itself. 

In Bonetti {\it et al.} \cite{Bonetti2020, Bonetti2020Authors}, the gap $\Delta$ was extracted via the zero-frequency limit of the off-diagonal term of the self-energy. The $Z$-factor captures one of the main effects of the normal (diagonal) term $\Sigma(ip_0)$ of the self-energy at low energies and low temperature besides the renormalization of the quasiparticle energies, which can be incorporated into a modified chemical potential. The self-energy is momentum independent within DMFT but, in general, frequency dependent, where $ip_0$ is the fermionic Matsubara frequency. It reduces the quasiparticle weight by a factor 
\begin{align}
 Z=\left[1-\left.\frac{\partial\, \im \Sigma(ip_0)}{\partial p_0}\right|_{p_0=0}\,\right]^{-1}\,.
\end{align}
At finite temperatures, the differential quotient may be approximated by the quotient $\im\Sigma(i\pi T)/(\pi T)$, where $\pi T$ is the lowest positive Matsubara frequency. The $Z$-factor reduces the bare single-particle excitation energy and the gap to $Z \eps_\bp$ and $Z \Delta$, respectively, and, thus, also the quasi-particle energies to $Z E_\bp^\pm$. In other words, it narrows the overall bandwidth by rescaling the unit of energy $t$. Moreover, it reduces the quasi-particle contributions in the spectral functions in  Eqs.~\eqref{eqn:spectralA} and \eqref{eqn:spectralB} by a global factor $Z$. The missing spectral weight is shifted to incoherent contributions at higher energies.

\begin{figure}[t!]
\centering
\includegraphics[width=0.58\textwidth]{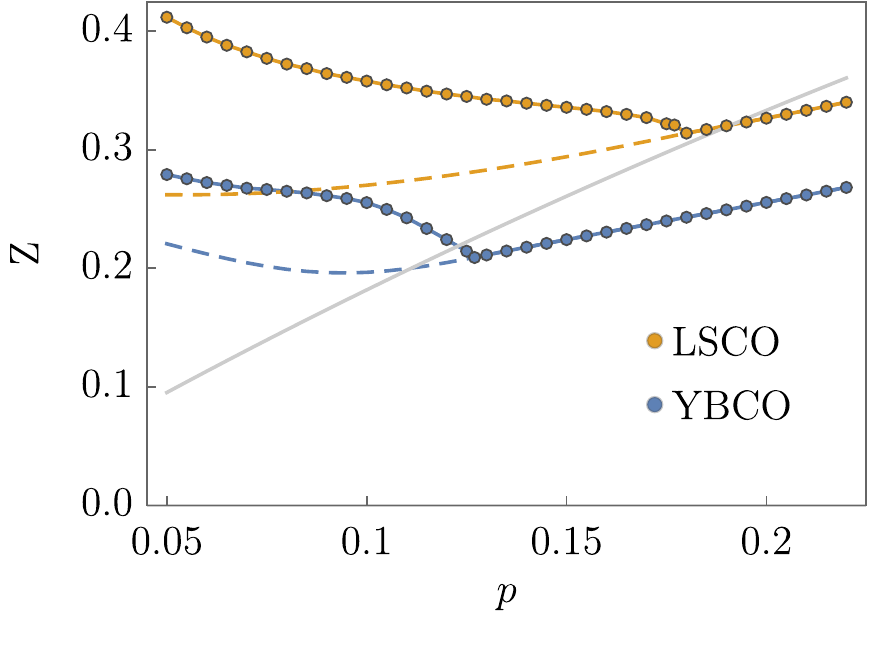}
\caption{ $Z$-factor as a function of doping $p$ for LSCO and YBCO at $T/t=0.04$ obtained by the DMFT calculation (cf.\,\,Sec.~\ref{sec:application:spiral:hubbard}). The $Z$-factor obtained from the unstable (below $p^*$) paramagnetic solution is also shown for comparison (dashed lines). We also show the Gutzwiller factor $2p/(1+p)$ \cite{Yang2006a} for comparison (gray line).
\label{fig:ZDMFT}}
\end{figure}

In Fig.~\ref{fig:ZDMFT}, we show the $Z$-factor as obtained from the DMFT calculation \cite{Bonetti2020,Bonetti2020Authors}, which was presented in Sec.~\ref{sec:application:spiral:hubbard}, as a function of hole doping $p$. For $p< p^*$, we also show the $Z$-factor that is found in the unstable paramagnetic solution. One can see that the magnetic order enhances $Z$ compared to the paramagnetic phase. The $Z$-factor exhibits only a moderate doping dependence and assumes material-dependent values between $0.2$ and $0.4$. The strongest renormalization is found for YBCO. Note that the paramagnetic $Z$-factors do not vanish for $p \to 0$, because the paramagnetic DMFT solution at half-filling is still on the metallic side of the Mott transition for the choice of parameters in the DMFT calculation. The overall scale is similar to the one that is obtained by a Gutzwiller factor, which captures phenomenologically the loss of metallicity in the doped Mott insulator. Such a factor is used in the Yang-Rice-Zhang (YRZ) ansatz for the pseudogap phase \cite{Yang2006a}.
\begin{figure}[t!]
\centering
\includegraphics[width=0.58\textwidth]{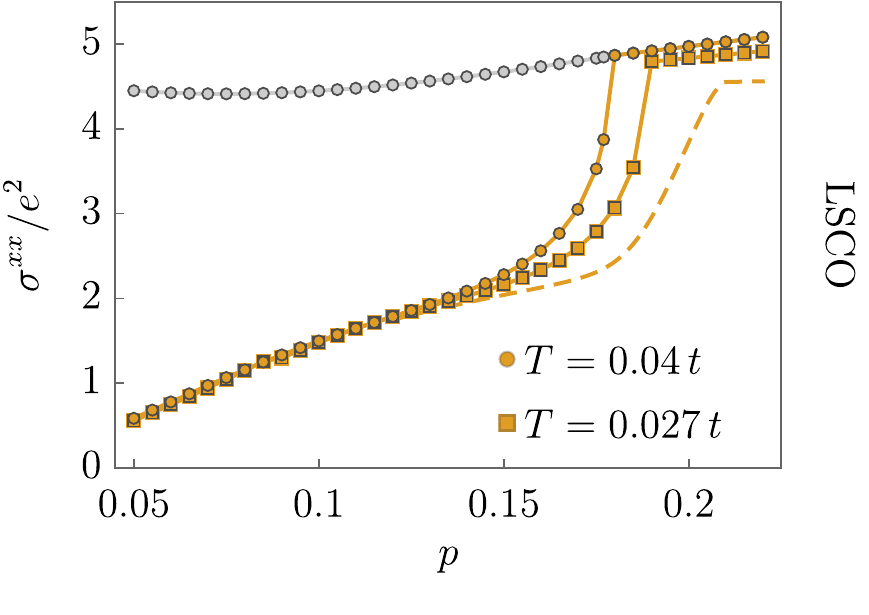}\\
\includegraphics[width=0.58\textwidth]{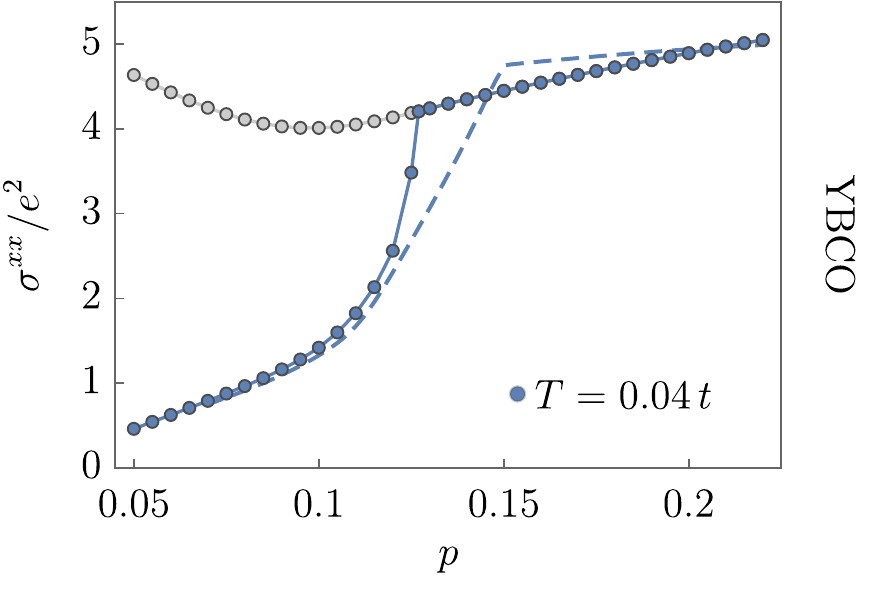}
\caption{Longitudinal conductivity as a function of doping for LSCO at $T/t=0.027$ (squares) and $T/t=0.04$ (circles) and for YBCO at $T/t=0.04$ (circles), together with an extrapolation to zero temperature (dashed lines). The conductivity in the unstable paramagnetic phase is also shown for comparison at $T/t=0.04$ (gray lines). The parameters $\Delta(p)$, $\eta(p)$ and $Z(p)$ were extracted from the DMFT calculation (cf. Sec.~\ref{sec:application:spiral:hubbard}).
\label{fig:sxxDMFT}}
\end{figure}

To take the renormalization of the quasi-particle energies into account, we replace the dispersion and the gap by $Z\eps_\bp$ and $Z\Delta$, respectively, in the conductivity formulas in \eqref{eqn:SintraN} and \eqref{eqn:SinterS}. Note that the reduction of the spectral weight of single-particle excitations by the $Z$-factor does not apply to the conductivities. The reduction of the quasi-particle contribution to the Green's functions by $Z$ is canceled by vertex corrections to the conductivities \cite{Nozieres1964}. For the bilayer compound YBCO, we modify the momentum integration to $\frac{1}{2} \sum_{k_z=0,\pi} \int \frac{d^2\mathbf{k}}{(2\pi)^2}$. We assume the doping independent value $\Gamma/t = 0.025$ for the relaxation rate, which corresponds to the estimate for $\rm La_{1.6-x}Nd_{0.4}Sr_xCuO_4$ (Nd-LSCO) at low temperatures, which we discussed above. We re-calculate the chemical potential with the new parameters for a consistent description at fixed doping. The magnetic gap in the zero temperature results is based on a linear extrapolation of $\Delta(p)$ as shown in Fig.~\ref{fig:DeltaDMFT}. The zero temperature limit of $\eta(p)$ and $Z(p)$ was obtained by a linear temperature extrapolation at each doping, and a subsequent linear fit in $p$ up to the zero temperature extrapolation of $p^*$. The linear extrapolation in temperature is based on data for $T/t=0.027$ and $T/t=0.04$ for LSCO, and data for $T/t=0.04$ and $T/t=0.05$ for YBCO.

In Fig.~\ref{fig:sxxDMFT}, we show the longitudinal conductivity $\sigma^{xx}$ as a function of doping for LSCO parameters at $T=0.027t$ and $T=0.04t$, and for YBCO parameters at $T=0.04t$, together with an extrapolation to zero temperature. The conductivities were calculated for the respective set of parameters and the corresponding temperature. We used the complete conductivity formulas including the interband contribution in \eqref{eqn:longCondSpiral}. Note that $\sigma^{xx}/e^2$ is a dimensionless quantity since we use natural units where $\hbar=1$. Our results for the two-dimensional conductivity correspond to three-dimensional resistivities of the order $100\,\mu\hspace{0.2mm}\Omega\hspace{0.2mm}\text{cm}$, in agreement with experimental values.
In order to see this, note that $h/e^2$ is the von Klitzing constant $R_K\approx 25813\,\Omega$. The two-dimensional conductivity of a CuO-layer in SI units is, thus, obtained by multiplying our dimensionless result by $2\pi/R_K$. To obtain the conductivity of the three-dimensional sample, one has to divide by the average distance between the layers. The expected drop below $p^*$ is clearly visible. It is particularly steep at $T > 0$, which is due to the square root type onset of the order parameter at finite temperature, see Fig.~\ref{fig:DeltaDMFT}. Since the relaxation rate is fixed in our calculations, the drop of $\sigma^{xx}$ is exclusively due to a drop of the charge carrier concentration related to the Fermi surface reconstruction by the magnetic gap. The results are consistent with the results that were obtained by the phenomenological model and are shown in Fig.~\ref{fig:cond}. Note the rescaling by the leading order in $\Gamma$ when comparing the overall scale.

\subsubsection*{Anisotropy or ``nematicity'' of the conductivity in cuprates} 

\begin{figure}[t!]
\centering
\includegraphics[width=0.56\textwidth]{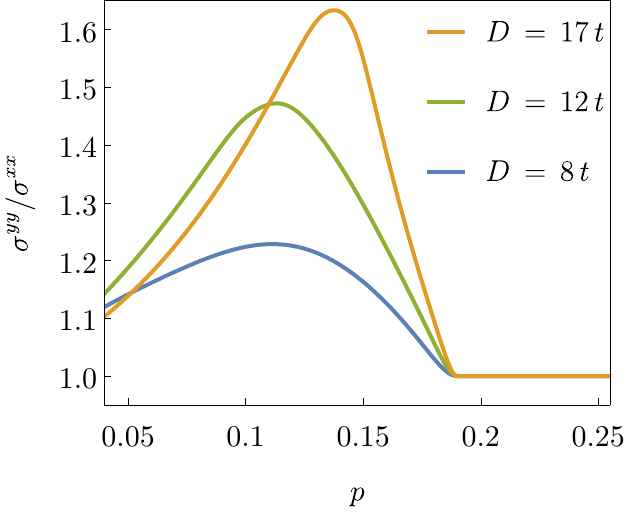}
\caption{Anisotropy ratio of the longitudinal conductivity $\sigma^{yy}/\sigma^{xx}$ at zero temperature as a function of doping $p$ for three choices of $D$. The band parameters are the same as in Fig.~\ref{fig:cond}, and the relaxation rate is $\Gamma/t = 0.1$.
}
\label{fig:anisotropy}
\end{figure}
\begin{figure}[t!]
\centering
\includegraphics[width=0.58\textwidth]{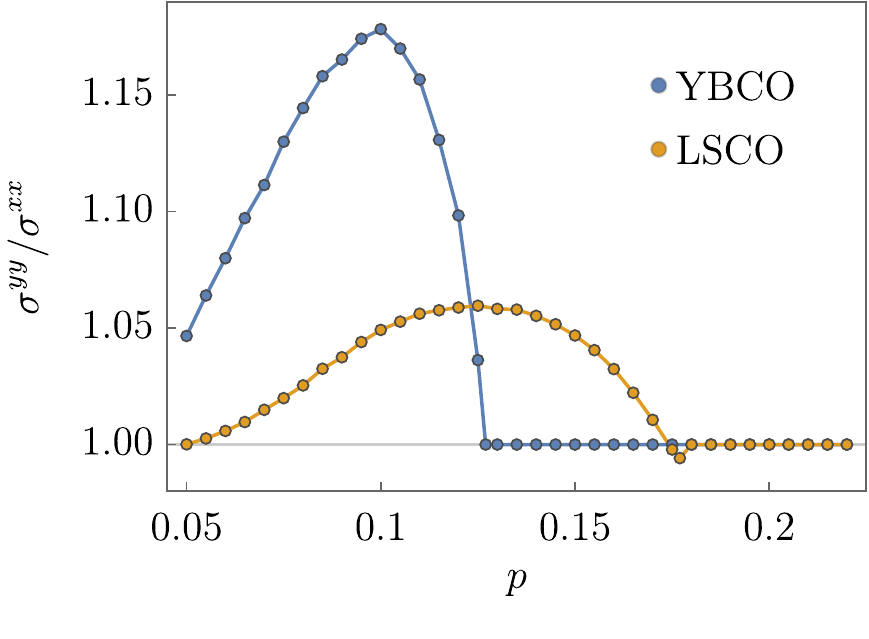}
\caption{Ratio $\sigma^{yy}/\sigma^{xx}$ as a function of doping for LSCO (orange) and YBCO (blue) at $T/t=0.04$. The parameters $\Delta(p)$, $\eta(p)$ and $Z(p)$ were extracted from the DMFT calculation (cf.\,\,Sec.~\ref{sec:application:spiral:hubbard}).
}
\label{fig:anisotropyDMFT}
\end{figure}

The breaking of the tetragonal symmetry of the square lattice by a spiral order with $\eta > 0$ naturally leads to an anisotropy (or ``nematicity'') in the longitudinal conductivity, as we have already discussed in Sec.~\ref{sec:application:spiral:coordinatesystem}. In Fig.~\ref{fig:anisotropy}, we show the ratio $\sigma^{yy}/\sigma^{xx}$ as a function of doping for the phenomenological model with the same band parameters as in Fig.~\ref{fig:cond}. We plot the anisotropy for several choices of $D$, which is the slope of the linear doping dependence of the gap, at $\Gamma/t = 0.1$. For an ordering wave vector $\bQ = (\pi - 2\pi\eta,\pi)$ with an incommensurability in the $x$ direction, the conductivity in the $y$ direction is larger than in the $x$ direction. The anisotropy increases smoothly upon lowering the doping from the critical point $p^*$, and it decreases upon approaching half filling, where $\eta$ vanishes such that the square lattice symmetry is restored. In Fig.~\ref{fig:anisotropyDMFT}, we show the ratio $\sigma^{yy}/\sigma^{xx}$ as a function of doping for LSCO and YBCO at $T = 0.04t$ using our DMFT results. We clearly see the same general behavior. Note that a smaller anisotropy is expected since the gap $\Delta(p)$ is reduced by the $Z$-factor. A pronounced temperature and doping-dependent in-plane anisotropy of the longitudinal conductivities with conductivity ratios up to 2.5 has been observed in YBCO by Ando {\it et al.} \cite{Ando2002}. The observed anisotropy is, thus, much larger than those obtained in our calculation. There is no further contribution to the anisotropy due to a rotation of the coordinate system within our model since the off-diagonal contribution $\sigma^{xy}=\sigma^{yx}$ vanishes for spiral states with a wave vector of the form $\bQ = (\pi - 2\pi\eta,\pi)$ as discussed in Sec.~\ref{sec:application:spiral:coordinatesystem}.

\subsubsection*{Interband contributions for the Hall number in cuprates} 

\begin{figure}[t!]
\centering
\includegraphics[width=0.53\textwidth]{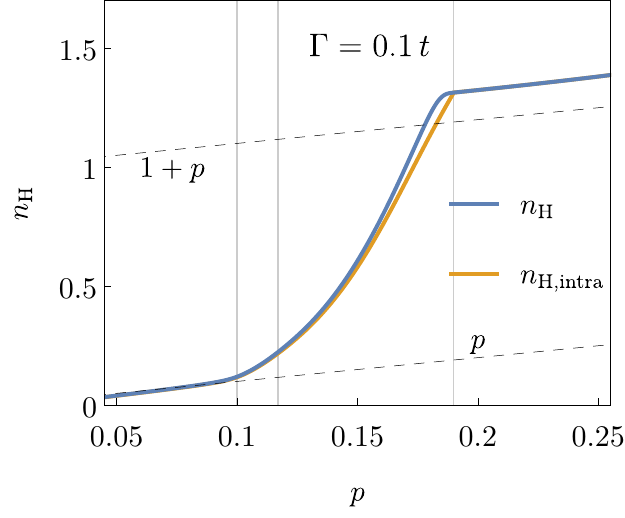}\\
\includegraphics[width=0.53\textwidth]{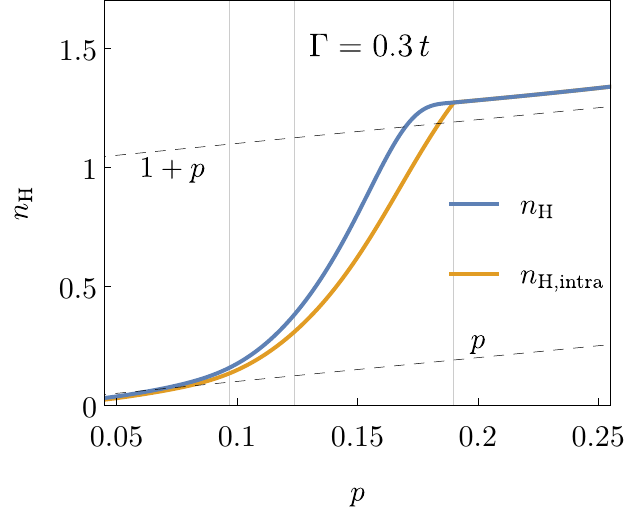}
\caption{Hall number $n_\text{H}$ as a function of doping $p$ for a doping dependent magnetic order parameter $\Delta(p) = 12t (p^* - p) \Theta(p^* - p)$ with $p^* = 0.19$. The intraband contribution $n_{\text{H},\text{intra}}$ is also shown for comparison. The straight dashed lines correspond to the naive expectation for large and reconstructed Fermi surfaces, $n_\text{H} = 1+p$ and $n_\text{H} = p$, respectively. We use the hopping parameters $t'/t = -0.3$ and $t''/t = 0.2$ and the relaxation rates $\Gamma/t = 0.1$ (top) and $\Gamma/t = 0.3$ (bottom). The vertical lines indicate the three special doping values $p_e^*$, $p_h^*$, and $p^*$ (from left to right). }
\label{fig:hall}
\end{figure}

So far, we have focused on the longitudinal conductivity.  We continue by calculating and discussing both the Hall number both for the phenomenological model and the Hall number that we calculate by using our DMFT results in Sec.~\ref{sec:application:spiral:hubbard}. In both approaches, we again use parameters that are common for cuprates in order to make connection to the recent transport measurements \cite{Badoux2016, Laliberte2016, Collignon2017}. We calculate the Hall conductivity 
\begin{align}
 \label{eqn:HallCondCuprates}
 \sigma^{xyz}_\text{H}=\sigma^{xyz}_{\text{H},\text{intra},+}+\sigma^{xyz}_{\text{H},\text{intra},-}+\sigma^{xyz}_{\text{H},\text{inter},+}+\sigma^{xyz}_{\text{H},\text{inter},-}
\end{align}
via the formulas presented in \eqref{eqn:SHintraXYZ} and \eqref{eqn:SHinterXYZ}. The Hall number is obtained by the ratio between the product of the longitudinal conductivities and the Hall conductivity via \eqref{eqn:nH}. We use the convention that hole-like contributions count positively to the Hall number, whereas electron-like contributions count negatively. In Fig.~\ref{fig:hall}, we show the Hall number $n_\text{H}$ for the phenomenological model in \eqref{eqn:doping} with $D/t=12$ and $p^*=0.19$ at zero temperature. It is calculated by including all contributions of the Hall conductivity in \eqref{eqn:HallCondCuprates} and all contributions of the longitudinal conductivity in \eqref{eqn:longCondSpiral}.  Again, we used $\eta=p$. The used gap parameters as well as the hopping parameters are the same as in Fig.~\ref{fig:cond}. We plot the Hall number $n_{\text{H},\text{intra}}$ that is calculated by neglecting all interband contributions of the conductivities in \eqref{eqn:longCondSpiral} and \eqref{eqn:HallCondCuprates} for comparison.

For $p \geq p^*$, where $\Delta = 0$, the Hall number is slightly above the value $1+p$ corresponding to the density of holes enclosed by the (large) Fermi surface. This is also seen in experiment in YBCO \cite{Badoux2016} and Nd-LSCO  \cite{Collignon2017}. Note that $n_\text{H}$ is not expected to be equal to $1+p$ for $\omega_c \tau \ll 1$, since the dispersion $\eps_\bp$ is not parabolic. For $p < p^*$ the Hall number drops drastically. For $\Gamma/t = 0.1$, the interband contributions are again quite small, as already observed for the longitudinal conductivity in Fig.~\ref{fig:cond}, and the Hall number gradually approaches the value $p$ upon lowering $p$. Hence, the naive expectation that the Hall number is given by the density of holes in the hole pockets turns out to be correct for sufficiently small $p$. Visible deviations from $n_\text{H} = p$ set in for $p > p_e^*$, where the electron pockets emerge. For $\Gamma/t = 0.3$, interband contributions are sizable. They shift the onset of the drop of $n_\text{H}$ to smaller doping. We can make the same conclusion on the relevance of interband contributions to the Hall number that we have obtained for the longitudinal conductivity: Whereas a general argument for negligible interband contributions does not hold for dopings sufficiently close to the onset of the order, the interband contributions are practically negligible due to their small numerical value for a relaxation rate $\Gamma/t\lesssim 0.1$, which is relevant for the recent experiments.

\subsubsection*{Momentum-resolved Hall conductivity} 

\begin{figure}[t!]
\centering
\includegraphics[width=0.48\textwidth]{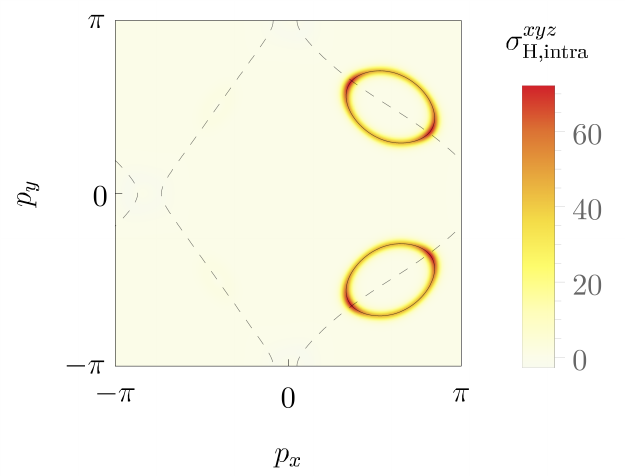}
\includegraphics[width=0.48\textwidth]{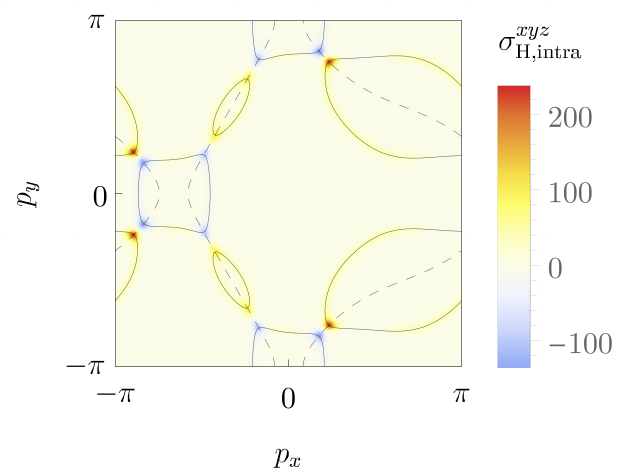} \\
\includegraphics[width=0.48\textwidth]{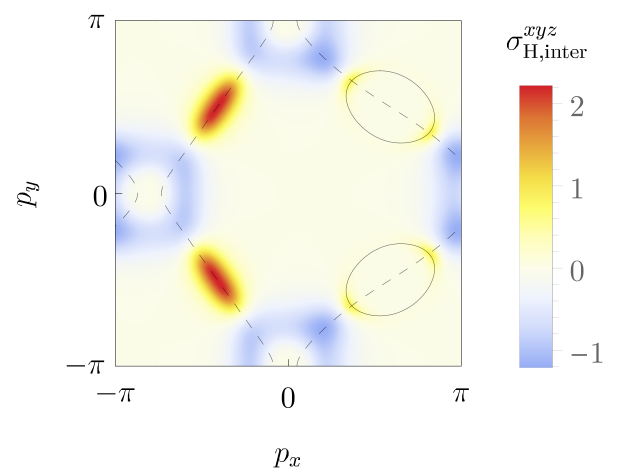}
\includegraphics[width=0.48\textwidth]{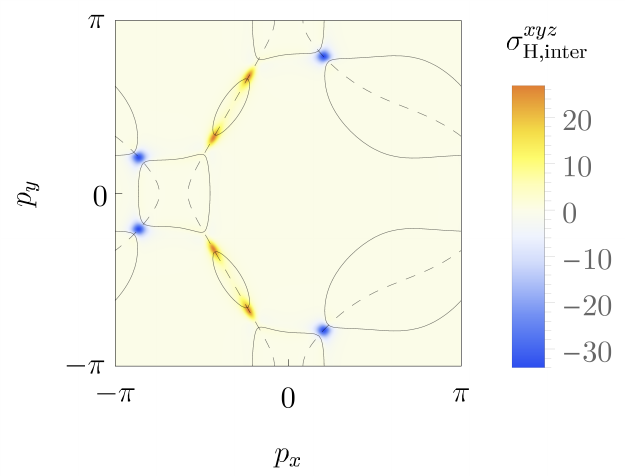}
\caption{{\it Top:} Color plot of the momentum resolved intraband contribution to the Hall conductivity $\sigma_{\text{H},\text{intra}}^{xyz}(\bp)$ for $p = 0.09$ {\it (left)} and $p = 0.17$ {\it (right)}. {\it Bottom:} Interband contribution $\sigma_{\text{H},\text{inter}}^{xyz}(\bp)$ for the same choices of $p$. The band and gap parameters are the same as in Fig.~\ref{fig:hall}, and the relaxation rate is $\Gamma/t = 0.3$. The Fermi surfaces and the nesting line (cf.\,\ Fig.~\ref{fig:fs}) are plotted as thin black lines.}
\label{fig:sgHp}
\end{figure}

We have already identified the momenta that contribute to the longitudinal conductivity in Fig.~\ref{fig:sgp} and continue by the same analysis for the Hall conductivity. The Hall conductivity in \eqref{eqn:SHintraXYZ} and \eqref{eqn:SHinterXYZ} is given by a momentum integral of the form  $\sigma_\text{H}^{xyz} = \int \frac{d^2\bp}{(2\pi)^2} \sigma_\text{H}^{xyz}(\bp)$. To see which momenta, that is, which quasiparticle states, contribute most significantly to the Hall conductivity, we show color plots of $\sigma_{\text{H},\text{intra}}^{xyz}(\bp)$ and $\sigma_{\text{H},\text{inter}}^{xyz}(\bp)$ for two choices of hole doping $p$, which represent a different Fermi surface topology, in Fig.~\ref{fig:sgHp}. The Fermi surface only consists of hole pockets at $p=0.09$, whereas a second set of hole pockets and electron pockets are present at $p=0.17$. The intraband contributions are concentrated near the quasiparticle Fermi surfaces, due to the peaks in $f'(\eps)$ and in the spectral functions, as for the longitudinal conductivity. Contributions from hole pockets count positively, and those from electron pockets negatively, as expected. The intraband contributions are particularly large near crossing points of the Fermi surfaces with the nesting line defined as $\eps_\bp=\eps_{\bp+\bQ}$, where the Fermi surfaces have a large curvature. The interband contributions lie mostly near the nesting line, not necessarily close to Fermi surfaces. For $p=0.17$, they are concentrated in small regions between electron and hole pockets.

\subsubsection*{Hall number obtained by using the DMFT results}

\begin{figure}[t!]
\centering
\includegraphics[width=0.58\textwidth]{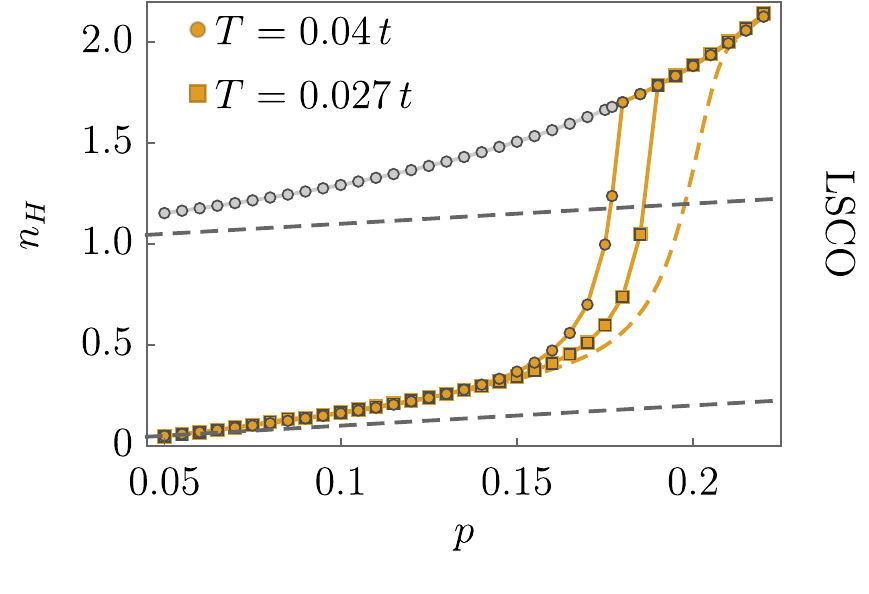}\\
\includegraphics[width=0.58\textwidth]{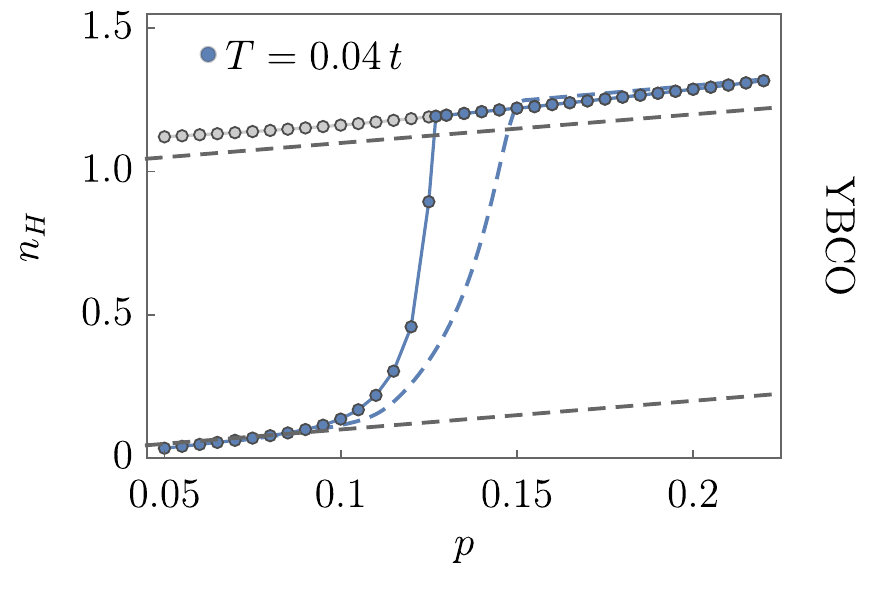}
\caption{Hall number as a function of doping for LSCO at $T/t=0.027$ (square) and $T/t=0.04$ (circle) and for YBCO at $T/t=0.04$ (circle), together with an extrapolation to zero temperature (dashed lines). The Hall number in the unstable paramagnetic phase is also shown for comparison at $T/t=0.04$ (gray lines). The black dashed lines correspond to the naive expectations $n_\text{H}=p$ for hole pockets and $n_\text{H}=1+p$ for a large hole-like Fermi surface. The parameters $\Delta(p)$, $\eta(p)$ and $Z(p)$ were extracted from the DMFT calculation (cf.\,\,Sec.~\ref{sec:application:spiral:hubbard}).
\label{fig:HallDMFT}}
\end{figure}

After having discussed the Hall number within the phenomenological approach, we continue by calculating the Hall number by using our results obtained by the DMFT calculations. We do this in the same fashion as we did previously for the longitudinal conductivity. We consider all contributions of the longitudinal and the Hall conductivity in \eqref{eqn:longCondSpiral} and \eqref{eqn:HallCondCuprates} including the interband contributions. The calculation was performed at the corresponding finite temperature. The Hall numbers as a function of doping are shown in Fig.~\ref{fig:HallDMFT}, again for LSCO at $T/t=0.027$ and $T/t=0.04$, and for YBCO parameters at $T/t=0.04$, together with an extrapolation to zero temperature. The extrapolation was obtained in the same way as discussed for the longitudinal conductivity. A pronounced drop is seen for doping values below $p^*$, indicating once again a drop of the charge carrier concentration. In the high-field limit $\omega_c \tau \gg 1$, the Hall number would be exactly equal to the charge carrier density enclosed by the Fermi lines, that is, $1+p$ in the paramagnetic phase and $p$ in the magnetically ordered phase, even in the presence of electron pockets \cite{Ashcroft1976}. However, the experiments, which motivated our analysis are in the {\em low-field}\/ limit $\omega_c \tau \ll 1$, since $\tau$ is rather small, and our expression for the Hall conductivity has been derived in this limit. In the low field limit, the Hall number is equal to the carrier density only for a parabolic dispersion. For low doping, the Hall number $n_\text{H}(p)$ shown in Fig.~\ref{fig:HallDMFT} indeed approaches the value $p$, which indicates a near-parabolic dispersion of the holes in the hole pockets for small $p$. For large doping, in the paramagnetic phase, the Hall number is only slightly above the naively expected value $1+p$ in YBCO, while it shoots up to significantly higher values in LSCO, indicating that the dispersion of charge carriers near the Fermi surface cannot be approximated by a parabolic form. The increase of $n_\text{H}(p)$ way above $1+p$ is a precursor of a divergence at the doping $p = 0.33$, well above the van Hove point at $p=0.23$, which is due to a cancellation of positive (hole-like) and negative (electron-like) contributions to the Hall coefficient $R_\text{H}$.

\subsubsection*{Fitting of the phenomenological model to the experimental results} 

The phenomenological model of the gap $\Delta(p)$ in \eqref{eqn:doping} with its free parameters $D$ and $p^*$ allows us to make closer connection to the experimental results of Badoux {\it et al.} \cite{Badoux2016}. To get a better estimate for the required set of parameters, we have fitted the parameters $p^*$ and $D$ such that we obtain quantitative agreement with the observed data points for YBCO. For the phenomenological model, we used so far the onset $p^* = 0.19$ extracted from the experimental data by Badoux {\it et al.} \cite{Badoux2016} and $D/t = 12$, which was used previously by Verret {\it et al.} \cite{Verret2017}. The Hall number in Fig.~\ref{fig:hall} was calculated with those parameters.

\begin{figure}[t!]
\centering
\includegraphics[width=0.57\textwidth]{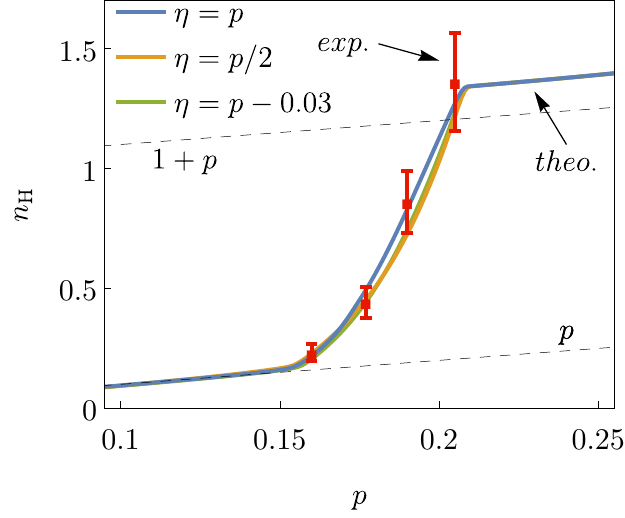} 
\caption{Fit of the Hall number as a function of doping to the experimental data (red) from Badoux {\it et al.} \cite{Badoux2016}. For the relaxation rate chosen as $\Gamma/t = 0.05$, best agreement for $\eta = p$ is obtained for $p^* = 0.21$ and $D/t = 16.5$. Also shown are results obtained with the same parameters but $\eta(p) = p/2$ and $\eta(p) = p - 0.03$.}
\label{fig_fit}
\end{figure}

The fit, obtained for a fixed $\eta=p$ and $\Gamma/t = 0.05$, and shown in Fig.~\ref{fig_fit}, is optimal for $p^* = 0.21$ and $D/t = 16.5$. Here we also compare to results obtained with the same values of $p^*$ and $D$, but with a different choice of the incommensurability $\eta(p)$, namely $\eta = p/2$ and $\eta = p - 0.03$. These alternative functions are closer to the incommensurabilities observed for YBCO \cite{Haug2010}. While the doping dependence of the Fermi-surface topologies depends on the choice of $\eta(p)$, one can see that the doping dependence of the Hall number is only weakly affected for hopping parameters that are relevant for YBCO.

The value of $D$ is unreasonably large. For a hopping amplitude $t \approx 0.3 \, \text{eV}$, the magnetic gap $\Delta(p) = D (p^* - p)$ would rise to a value $\Delta \approx 0.5 \, \text{eV}$ at $p = 0.1$. Large values for $D$ were also assumed in previous studies of the Hall effect in N\'eel and spiral antiferromagnetic states, to obtain a sufficiently steep decrease of the Hall number \cite{Storey2016, Eberlein2016, Verret2017}.
The required size of $D$ can be substantially reduced, if the bare hopping $t$ is replaced by a smaller effective hopping
\begin{equation}
 t_{\rm eff} = \frac{2p}{1+p} \, t \, ,
\end{equation}
where the doping-dependent prefactor of $t$, the Gutzwiller factor, on the right-hand side captures phenomenologically the loss of metallicity in the doped Mott insulator. Such a factor is used in the Yang-Rice-Zhang (YRZ) ansatz for the pseudogap phase \cite{Yang2006a}.
Replacing $t$ by $t_{\rm eff}$ with $t = 0.3 \, \text{eV}$, a prefactor $D = 1.5 \, \text{eV}$ of the gap $\Delta$ is sufficient to obtain the best fit for $n_\text{H}$, leading to $\Delta \approx 0.15 \, \text{eV}$ at $p=0.1$. This value is similar to the magnetic energy scale $J$ in cuprates. The value of $D$ obtained by fitting to the experimental result is comparable to the one that was obtained within the DMFT calculation in the zero temperature limit, namely $D/t=18.7$, although the onset of order is strongly reduced for the DMFT result. The effect of the effectively reduced hopping due to the Gutzwiller factor is consistent with the renormalization of the dispersion and the gap by the $Z$-factor. The Gutzwiller factor and the $Z$-factor are shown in Fig.~\ref{fig:ZDMFT} for comparison.

All our results of the phenomenological model including the fitting has been computed by evaluating the conductivity formulas with a Fermi function at zero temperature. We have checked that the temperature dependence from the Fermi function is negligible at the temperatures at which the recent transport experiments in cuprates \cite{Badoux2016, Laliberte2016, Collignon2017} have been carried out. The temperature effects are however important in the evaluation of the conductivity formulas for the temperatures, at which the DMFT calculations were performed. Thus, we performed the calculations for the DMFT results at the corresponding finite temperature and provided the zero temperature limit for comparison. 

\subsubsection*{Comparison of the transport properties of the spiral magnetic state to the experiments in cuprates} 

We have already discussed several aspects of our theory in comparison with the experimental results. We now summarize the most important conclusions. Qualitatively, the pronounced drop of the longitudinal conductivity and the Hall number observed in experiment is captured by our theory, both by the phenomenological ansatz proposed by Eberlein {\it et al.} \cite{Eberlein2016} and by the {\it ab initio} results obtained by DMFT calculations by Bonetti {\it et al.} \cite{Bonetti2020, Bonetti2020Authors}. In the DMFT calculation, the onset of the drop at $p^* = 0.21$ in the zero temperature extrapolation for LSCO (see Fig.~\ref{fig:DeltaDMFT}) is slightly above the experimental value $0.18$ for LSCO \cite{Laliberte2016}, but below the value $0.23$ observed for Nd-LSCO \cite{Collignon2017}. Why the observed $p^*$ differs so much between LSCO and Nd-LSCO is unclear. For YBCO, we obtain $p^* = 0.15$ (see Fig.~\ref{fig:DeltaDMFT}), while the charge carrier drop seen in experiment starts at $p^* = 0.19$ \cite{Badoux2016}. Cluster extensions of the DMFT \cite{Maier2005}, which go beyond the assumption of a momentum-independent self-energy, yield critical doping strengths for the onset of pseudogap behavior, which are also below the experimental value \cite{Wu2018, Reymbaut2019}. Hence, for a better quantitative agreement one probably needs to go beyond the single-band Hubbard model. The effect of interband contributions to both the longitudinal and the Hall conductivity are most prominent at the onset of order. They smooth the kink that was seen in the conductivities, when interband contributions were neglected, and they shift the onset of the drop to lower doping than the onset of the order itself. For an estimated relaxation rate $\Gamma\lesssim 0.1$ for cuprates, the interband contributions are small in numerical size and, thus, do not have any consequence on previous conclusions of Eberlein {\it et al.} \cite{Eberlein2016}, which neglected interband contributions completely.

The relatively narrow doping range of a few percents, in which the steep charge carrier drop occurs, also agrees between theory and experiment. The analysis of the phenomenological model shows that the doping range of the drop is mainly controlled by the prefactor $D$ of the phenomenological model. In order to get a good agreement with the experimental results, an unreasonably large prefactor $D$ is needed, which could be suppressed to reasonable values by introducing a Gutzwiller factor. The DMFT results for YBCO yield a gap of similar size, where the $Z$-factor plays the role of the Gutzwiller factor. A square-root onset further narrows the range, which is shown by comparing the finite and zero temperature results of the DMFT calculations. The Hall number obtained from our calculations reaches the value $n_\text{H}(p) = p$ only at much lower dopings than in the experiments. The convergence is particularly slow for LSCO (see Fig.~\ref{fig:HallDMFT}) and can be attributed to the non-parabolic dispersion of the charge carriers in the hole-pockets. In the experiments, the behavior $n_\text{H} \approx p$ is observed over an {\em extended}\/ doping range only at low doping far below $p^*$, too. At larger doping, a few percent below $p^*$, the Hall number becomes equal to $p$ only at a {\em single}\/ doping value. Upon further reducing the doping, it drops below $p$ and even becomes negative, presumably due to charge density wave order \cite{Leboeuf2007}. To obtain the steep drop of the Hall number down to $n_\text{H} = p$ and below in a theoretical description, one therefore needs to take the charge order into account. Charge order on top of spiral order was discussed by Eberlein {\it et al.} \cite{Eberlein2016}, but the corresponding transport properties were not yet computed.

For dopings $p \geq p^*$, the Hall number for YBCO is close to $1+p$ as naively expected. More precisely, it is slightly larger in agreement with the experimental observations \cite{Badoux2016}. By contrast, for LSCO parameters $n_\text{H}$ is much larger than $1+p$ for $p \geq p^*$ with an increasing deviation for larger doping. From a theoretical point of view this behavior is not surprising, since there is no reason why $n_\text{H}$ should be close to the charge carrier density for the strongly non-parabolic dispersion near the van Hove filling. In the experiments, $p^*$ practically coincides with the van Hove filling in Nd-LSCO, and $n_\text{H}$ is nevertheless only slightly above $1+p$ for $p$ near $p^*$ \cite{Collignon2017}.

A drop by a factor $p/(1+p)$ in a narrow doping range below $p^*$ has also been observed for the longitudinal conductivity in Nd-LSCO \cite{Collignon2017}. This drop corresponds to the expectation based on a Drude formula for the conductivity only if the relaxation rate and the effective electron mass remain constant, while the charge carrier concentration drops from $1+p$ to $p$. The drop of $\sigma^{xx}$ below $p^*$ obtained from our calculation for LSCO parameters is less pronounced (see Fig.~\ref{fig:sxxDMFT}). Since we assumed a doping-independent relaxation rate, this reduction of the drop compared to the carrier concentration ratio must be due to an increase of the average Fermi velocities below $p^*$, that is, a decrease of the effective electron mass $m_\text{eff}$ in a Drude picture. A priori, there is no reason why these quantities should remain constant when the Fermi surface gets fractionalized into pockets. Of course, it can be reconciled with the theory by assuming a suitable doping-dependent enhancement of the relaxation rate below $p^*$ \cite{Storey2017,Chatterjee2017}, so that the mobility $\mu\propto 1/\Gamma m_\text{eff}$ remains constant as seen in experiment \cite{Collignon2017}.

\cleardoublepage



\rhead[\fancyplain{}{\bfseries                                              
Conclusion}]{\fancyplain{}{
}}
\lhead[\fancyplain{}{
}]{\fancyplain{}{\bfseries            
Conclusion}}
\chapter*{Conclusions}
\addcontentsline{toc}{chapter}{Conclusion}    

The electrical conductivity is one of the fundamental properties of solids and, as such, of ongoing interest in physics for both experiment and theory. Recent experimental results revealed the need of reconsidering the electrical conductivity in systems with more than one valence band \cite{Proust2019,Nagaosa2020,Culcer2020}. In this thesis, we developed and analyzed a theory of the longitudinal conductivity, the intrinsic anomalous Hall conductivity and the ordinary Hall conductivity, which we used to explore several electrical transport phenomena that are directly linked to the presence of multiple bands. In the following, we summarize and conclude the main results that were presented throughout this thesis. We combine it with a short outlook on potential expansions beyond the scope of this thesis.

External electric and magnetic fields can induce a current in a solid. The induced current density up to first order in the electric and the magnetic field is captured by the electrical conductivity tensors $\sigma^{\alf\beta}$ and $\sigma^{\alf\beta\eta}$, where the indices $\alf$, $\beta$, and $\eta$ indicate the directions of the current, the electric field, and the magnetic field, respectively. From a theoretical point of view, the derivation of formulas for these two conductivity tensors is a crucial step in order to gain better insights in the physical mechanisms behind the related transport phenomena. The very successful band theory for solids is a key concept for a theoretical description of the conductivity, which is apparent, for instance, in the well-known semiclassical Boltzmann transport theory, where the conductivity is related to the momentum derivatives of the band dispersion \cite{Mahan2000}. However, the original approach of Boltzmann does not incorporate the effects that might arise due to the presence and interplay of several valence bands and, thus, a different approach is required. Microscopic approaches beyond the semiclassical approximations are indeed able to provide a systematic handling of interband effects \cite{Voruganti1992, Nagaosa2010}, so that it is possible to go beyond a one-band model or the simple summation of independent band contributions in a multiband model.

In Chapter~\ref{sec:theory}, we presented a microscopic derivation of the conductivity formulas at finite temperature for a general momentum-block-diagonal two-band model combined with detailed discussions of several aspects that arose during the derivation. In Chapter~\ref{sec:application}, we applied those formulas and results to models of recent interest. Our two-band model, which we specified in Sec.~\ref{sec:theory:twobandsystem:model}, acts as a minimal model that involves more than one band and, thus, potentially hosts interband effects. We were able to describe a broad variety of physically very different systems by the considered model ranging from systems with magnetic order like N\'eel antiferromagnetism and spiral spin density waves to systems that involve topological properties like non-zero Chern numbers. Despite the simplicity of our model, we could, thus, gain a broader and more general insight independent of details that might be particular for a more specific model. We derived the formulas for the longitudinal, the ordinary Hall and the intrinsic anomalous Hall conductivity. The longitudinal conductivity captures the induced (parallel) current due to an external electric field. The ordinary Hall conductivity describes a transverse current in the presence of a transverse external magnetic field. The intrinsic anomalous Hall conductivity describes a transverse current that is present without any external magnetic field and that is not caused by (skew) scattering but by the properties of the eigenstate manifold itself. Our main formulas of the longitudinal and the anomalous Hall conductivity are summarized in Sec.~\ref{sec:theory:conductivity:formulas}. The formulas of the ordinary Hall conductivity are summarized in Sec.~\ref{sec:theory:Hall:formulas}. We will go through several key results of this thesis in the following. 

\subsection*{Criteria for a unique and physically motivated decomposition}

On the one hand, microscopic approaches may have a tendency not to be very transparent and easy to interpret physically, which is seen as a disadvantage in comparison to semiclassical approaches, but, on the other hand, they can provide a more systematic treatment, which can be useful for identifying further relevant phenomena \cite{Nagaosa2010}. In Chapter~\ref{sec:theory}, we combined the derivation of the conductivity formulas with a systematic analysis of the underlying structure of the involved quantities, which led to the identification of two fundamental criteria for a unique and physically motivated decomposition of those formulas. {\it Intraband} and {\it interband} contributions are defined by the involved quasiparticle spectral functions of one or both bands, respectively. {\it Symmetric} and {\it antisymmetric} contributions are defined by the symmetry under the exchange of the current and the electric field directions. 

We identified the symmetry under the exchange of the current and the electric field directions as a powerful criterion for disentangling conductivity contributions of different physical origin. We found that this symmetry is strongly intertwined with a second symmetry that is fundamentally based on the presence of multiple bands. The two-band structure of our considered model is manifested by $2\times 2$ Green's function and vertex matrices that capture the occupation and the coupling of the external fields and the current to the conductivity, respectively. The trace over the product of those matrices eventually provides the conductivity formulas. The matrix structure that arises in contrast to an one-band system and its property to be, in general, non-abelian is crucial. We identified an one-to-one correspondence between the symmetry of these matrix quantities under transposition and the symmetry under the exchange of the current and the electric field directions. 

The intraband contribution of the conductivity tensor $\sigma^{\alf\beta}$, where $\alf$ and $\beta$ are the direction of the current and the external electric field, respectively, is purely symmetric in $\alf\leftrightarrow\beta$. The interband contribution of $\sigma^{\alf\beta}$ decomposes into a symmetric and an antisymmetric part. The symmetric interband contribution is shown to be caused by the quantum metric. The antisymmetric interband contribution is related to the Berry curvature and describes the intrinsic anomalous Hall effect. The conductivity tensor $\sigma^{\alf\beta\eta}$, where $\alf$, $\beta$ and $\eta$ are the directions of the current, the external electric and magnetic fields, respectively, describe both the ordinary Hall effect and linear magnetoresistance. The former one is captured by the antisymmetric part of $\sigma^{\alf\beta\eta}$ with respect to $\alf\leftrightarrow\beta$, which we denoted as $\sigma^{\alf\beta\eta}_\text{H}$, whereas the latter one is described by the symmetric part. In this thesis, we focused on the ordinary Hall conductivity and its interband contribution.

\subsection*{Momentum-relaxation rate of arbitrary size}

In order to obtain a non-diverging conductivity, a momentum relaxation process is required. We incorporated this in our theory by considering a simplified phenomenological momentum-relaxation rate $\Gamma$. It is assumed to be constant in frequency and momentum and both diagonal and equal for the two bands. We did not assume any restriction on the size of $\Gamma$, which allowed us to study the scaling behavior of the conductivity contributions in the clean (small $\Gamma$) and the dirty (large $\Gamma$) limit. We showed that so-called spectral weighting factors, which involve the product of quasiparticle spectral functions of one or both bands only, entirely capture the dependence on the relaxation rate $\Gamma$. 

In the clean limit, we obtained the expected $1/\Gamma$ scaling of the longitudinal conductivity \cite{Mahan2000} and the constant (or ''dissipationless`` \cite{Nagaosa2010}) limit of the intrinsic anomalous Hall conductivity. The widely used formula of the anomalous Hall effect by Onoda {\it et al.} \cite{Onoda2006} and Jungwirth {\it et al.} \cite{Jungwirth2002} are obtained by our formula of the intrinsic anomalous Hall effect in the clean limit. We have, thus, generalized those formulas to finite $\Gamma$. The symmetric interband contribution of $\sigma^{\alf\beta}$ scales as $\Gamma$ in the clean limit and is, thus, suppressed. We derived the ordinary Hall conductivity for a momentum-independent coupling between the two subsystems of the two-band model. We found that the interband contribution to the Hall conductivity tensor $\sigma^{\alf\beta\eta}_\text{H}$ obeys the expected $1/\Gamma^2$ scaling in the clean limit \cite{Voruganti1992}. The interband contributions are suppressed by a factor of the order $\Gamma^2$. The relevant energy scale for comparison with $\Gamma$ is the direct band gap between the lower and upper band. Therefore, interband contributions are of particular importance for small gaps, for instance, at the onset of order. 

The validity of the derived formulas for a relaxation rate of arbitrary size allowed us to perform a systematic scaling analysis of the different contributions in the dirty (large $\Gamma$) limit. We provided the precise prefactors of the expansion in powers of $1/\Gamma$, which were shown to be helpful in our applications that we will specify below.

\subsection*{Relation to quantum geometry}

The coupling of the current to the external electromagnetic fields are described by the vertex matrices. We showed that those vertex matrices, which are essentially given by the momentum derivative of the Bloch Hamiltonian, split into a diagonal and an off-diagonal part when expressing them in the eigenbasis of the Hamiltonian. The off-diagonal parts eventually led to the interband contributions. Since the eigenbasis is, in general, momentum dependent, the change of the basis results in terms that involve the momentum derivatives of the eigenbasis, that is, the Berry connection. This suggested a deeper connection to concepts of quantum geometry. Indeed, we found that the interband contributions of the conductivity tensor $\sigma^{\alf\beta}$ are controlled by the quantum geometric tensor $\cT^{\alf\beta,n}_\bp$ of the underlying eigenbasis manifold of the lower and upper bands $n=\pm$. Its real (symmetric) part is the quantum metric and its imaginary (antisymmetric) part is the Berry curvature. The symmetry in $\alf\leftrightarrow\beta$ shows again its usefulness in disentangling contributions of different physical origin. We saw that the intraband contribution of the Hall conductivity tensor $\sigma^{\alf\beta\eta}_\text{H}$ involves the quasiparticle effective mass since the quantum metric correction was shown to drop out. This was already recognized by Voruganti {\it et al.} \cite{Voruganti1992} without noticing the connection to quantum geometry.

\subsection*{Intrinsic anomalous Hall effect}

We identified the antisymmetric interband contribution of $\sigma^{\alf\beta}$ to describe the intrinsic anomalous Hall effect, which is caused by the Berry curvature and not by (skew) scattering. Our formulas of $\sigma^{\alf\beta}$ are closely related to the formulas derived by Bastin {\it et al.} \cite{Bastin1971, Crepieux2001} and St\v{r}eda \cite{Streda1982, Crepieux2001}. We gave a new derivation following Bastin {\it et al.} \cite{Bastin1971} in our notation and discussed the precise relation to our formulas. 

In Sec.~\ref{sec:theory:conductivity:BastinStreda}, we proposed a different definition of the so-called Fermi-sea and Fermi-surface contributions of the conductivity than previously proposed by St\v{r}eda \cite{Streda1982}. Historically, the definition is based on the involved Fermi function or its derivative so that the relevant energy states are restricted to the Fermi sea or the Fermi surface, respectively. However, this definition is not unique due to the possibility of partial integration in the integration over the internal frequency. In contrast, we have based our definition on the symmetry under exchange of the current and the electric field directions. We found that the symmetric contributions and the antisymmetric contribution of the conductivity tensor $\sigma^{\alf\beta}$ involve the derivative of the Fermi function and the Fermi function, respectively, when we expressed the contributions entirely in terms of quasiparticle spectral functions. The same decomposition naturally arises when decomposing the Bastin formula \cite{Bastin1971} into its symmetric and antisymmetric part. Therefore, we argued that the symmetry in $\alf\leftrightarrow\beta$ is the fundamental property to split contributions of different physical origin, whereas the distinction in Fermi-sea and Fermi-surface contributions is merely a change of view.

In Chapter~\ref{sec:application}, we applied our general theory to different models of recent interest. By this, we exemplified the broad and easy applicability, and highlighted different aspects of the general insights. In the first part of the applications in Sec.~\ref{sec:application:anomalousHallEffect}, we focused on the anomalous Hall effect. The anomalous Hall conductivity is connected to the (integer) Chern number, which is a topological invariant, 
via their dependence on the Berry curvature. This can lead to a quantized anomalous Hall conductivity under certain conditions, which we discussed in our general theory in Sec.~\ref{sec:theory:conductivity:anomalousHall}. Our result is consistent with previous work \cite{Thouless1982, Niu1985, Kohmoto1985, Onoda2002, Jungwirth2002}, which we have obtained in the clean (low $\Gamma$) limit of our formulas. In the first example in Sec.~\ref{sec:application:anomalousHallEffect:Wilson}, we discussed the quantized anomalous Hall conductivity for a Chern insulator at finite $\Gamma$. We saw that the quantization is violated for larger $\Gamma$ in the parameter range, where the band gap closes and the Chern number changes its value. We understood this effect by a partial occupation of the former unoccupied (upper) band due to finite $\Gamma$. This reduces the conductivity since the Berry curvature of the upper band is equal to the Berry curvature of the lower band except of the overall sign. 

In a second example in Sec.~\ref{sec:application:anomalousHallEffect:scaling}, we used our conductivity formulas for a relaxation rate $\Gamma$ of arbitrary size and their scaling behavior for small and large $\Gamma$, which we derived and summarized in Sec.~\ref{sec:theory:conductivity:limits}. We analyzed the scaling behavior of the anomalous Hall conductivity with respect to the longitudinal conductivity for a ferromagnetic multi-d-orbital model, which was proposed by Kontani {\it et al.} \cite{Kontani2007}. Our results are qualitatively and quantitatively in good agreement with experimental findings (see Ref.~\cite{Onoda2008} for an overview). Whereas there is a proper scaling of the anomalous Hall conductivity with $\sigma^{xy}\propto (\sigma^{xx})^0$ and $\sigma^{xy}\propto (\sigma^{xx})^2$ in the clean and dirty limit, respectively, we identified a crossover regime without any proper scaling behavior for intermediate conductivities $\sigma^{xx}=10-30000\,(\Omega\,\text{cm})^{-1}$. The conductivity of various ferromagnets were found in this range \cite{Onoda2008}. The treatment of intrinsic and extrinsic contributions on equal footing as well as the experimental and theoretical investigation of the scaling behavior for systems and models that involve, for instance, vertex corrections, electron localization effects and quantum corrections from Coulomb interaction is still ongoing research \cite{Onoda2006, Onoda2008, Kovalev2009, Xiong2010, Lu2013, Zhang2015, Sheng2020} and beyond the scope of this thesis.

\subsection*{Spiral antiferromagnetic order in cuprates}

We allowed for a relative momentum shift $\bQ$ in the spinor of our two-band system. The system is no longer lattice-translational invariant for a nonzero $\bQ$, but has a combined symmetry in lattice-translation and rotation inside the spinor space, for instance, in spin space \cite{Sandratskii1998}. In Sec.~\ref{sec:application:spiral}, we discussed spiral spin density waves as an example of such systems. A key property of spiral spin density waves, which are described by two parameters, the magnetic gap $\Delta$ and the ordering wave vector $\bQ$, is that the order needs not to be commensurate with the underlying lattice to be describable by the two-band model. The ordering wave vector $\bQ$ is precisely the relative momentum shift in the spinor. The magnetic gap $\Delta$ in the spiral magnetic state is momentum independent and, thus, consistent with our assumption that we used in the calculation of the Hall conductivity. 

We were interested to relate spiral magnetic order with recent Hall conductivity measurements on cuprates in very strong magnetic fields, so that the superconducting state is substantially suppressed even at low temperature \cite{Badoux2016, Laliberte2016, Collignon2017, Proust2019}. Following the proposal by Eberlein {\it et al.} \cite{Eberlein2016}, the observed drop in the Hall number, which is naively related to the volume of the Fermi surface, can be explained by a Fermi surface reconstruction due to the onset of spiral magnetic order at a critical doping $p^*$. This critical doping $p^*$ was experimentally found close to the onset of the pseudogap regime of cuprates, whose origin is still debated. The longitudinal and Hall conductivity for a spiral magnetic state was already derived by Voruganti {\it et al.} \cite{Voruganti1992}. However, interband contributions were neglected so far. In this thesis, we provided the conductivity formulas for the spiral magnetic state including interband contributions as a special case of our results in Sec.~\ref{sec:theory:Hall:coefficient}. 

In Sec.~\ref{sec:application:spiral:hubbard}, we discussed spiral magnetic order in the two-dimensional repulsive Hubbard model, which is known to describe the valence electrons in the $\rm CuO_2$-planes of the cuprate high-temperature superconductors \cite{Scalapino2012}. Spiral magnetic order was already found in previous theoretical studies \cite{Schulz1990, Kato1990, Fresard1991, Chubukov1992, Chubukov1995, Raczkowski2006, Igoshev2010, Igoshev2015, Yamase2016}. We computed the magnetic gap $\Delta(p)$ and the ordering wave vector $\bQ(p)$ as a function of doping $p$ for cuprate band parameters in a Hartree-Fock approach and by dynamical mean-field theory (DMFT) \cite{Bonetti2020Authors}. Whereas order parameter fluctuations are neglected in the Hartree-Fock approach, local fluctuations are included in DMFT. The wave vector has the form $\mathbf{Q} = (\pi-2\pi\eta,\pi)$, where the incommensurability $\eta$ increases with doping. The magnetic gap $\Delta$ decreases monotonically as a function of doping and vanishes at a critical doping $p^*$. An extrapolation of the DMFT results (obtained at low finite $T$) to zero temperature yields an approximately linear doping dependence $\Delta(p) \propto p^*-p$ in a broad doping range below $p^*$. The magnetic order leads to a Fermi surface reconstruction with electron and hole pockets. Electron pockets exist only in a restricted doping range near $p^*$. The spectral function for single-particle excitations, which can, for instance, be seen in angular resolved photoemission spectroscopy (ARPES), seems to exhibit Fermi arcs instead of hole-pockets due to a strong momentum dependence of the spectral weight along the reconstructed Fermi surface \cite{Eberlein2016}. This is a consequence of the broken translational invariance of the spiral magnetic state as pointed out in Sec.~\ref{sec:application:spiral:spectralweights}.

Whereas the onset of order in the Hartree-Fock approach is far off the experimentally observed $p^*$, the zero temperature extrapolation of our DMFT results for $\Delta(p)$ yields $p^* = 0.21$ for $\rm La_{2-x} Sr_x Cu O_4$ (LSCO) and $p^* = 0.15$ for bilayer $\rm YBa_2Cu_3O_y$ (YBCO) parameters. These values are close to those found experimentally \cite{Badoux2016,Collignon2017}, but we are obviously not in a position to provide accurate predictions for the experimentally observed critical doping $p^*$. For a better agreement one probably needs a material dependent modeling beyond the single-band Hubbard model.

In Sec.~\ref{sec:application:spiral:CondCuprates}, we calculated the longitudinal and the Hall conductivity for the phenomenological model proposed by Eberlein {\it et al.} \cite{Eberlein2016} and for the {\it ab initio} results obtained by DMFT. We used the phenomenological model in order to study the importance of interband contributions. A general comparison of energy scales suggests that interband contributions are of particular importance at the onset of order, where the magnetic gap $\Delta$ is of the order of a (doping-independent) relaxation rate $\Gamma$. A numerical evaluation of the conductivities for band parameters as in YBCO and various choices of the relaxation rate $\Gamma$ showed that interband contributions start playing a significant role only for $\Gamma/t > 0.1$, where $t$ is the nearest-neighbor hopping amplitude. Relaxation rates observed in recent high-field transport experiments for cuprates are smaller \cite{Collignon2017, Taillefer2018}, so that we concluded that the interband contributions are not important not due to a general argument comparing energy scales but due to the small numerical value at the relevant parameters for YBCO. The application of the formulas derived by Voruganti {\it et al.} \cite{Voruganti1992} in previous studies was, thus, justified.

Using our DMFT results, the longitudinal and Hall conductivities were computed by inserting the magnetic gap, the magnetic wave vector, and the quasi-particle renormalization $Z$ as obtained from the DMFT into transport equations for spin-density wave states with a phenomenological scattering rate. A pronounced drop of the longitudinal conductivity and the Hall number in a narrow doping range of few percent below $p^*$ is obtained in agreement with the corresponding high-field experiments. Note that our expansion of the current up to first order in the magnetic field is still sufficient for those high-field experiments since $\omega_c\tau\ll 1$ still holds despite the high magnetic field \cite{Collignon2017}. $\omega_c$ is the cyclotron frequency and $\tau=1/2\Gamma$ the relaxation time. The doping range in which electron pockets exist matches approximately with the range of the steepest Hall number drop, but there is no simple theoretical relation between these two features. For $p>p^*$ the calculated Hall number $n_H(p)$ is close to the naively expected value $1+p$ for YBCO parameters, but significantly higher for LSCO. From a theoretical point of view, this is not surprising since the band structure near the Fermi surface of LSCO cannot be approximated by a parabolic band in a broad doping range around $p^*$. For $p<p^*$ the Hall number approaches the value $p$ from above, but converges to this limiting value only for dopings well below the point where the electron pockets disappear. $n_H(p) \approx p$ is obtained only in a regime where the hole pockets are sufficiently small so that the quasi-particle dispersion in the pockets is approximately parabolic. In the cuprates, $n_H(p)$ does not {\em converge}\/ to $p$ but rather {\em crosses}\/ the line $n_H(p) = p$ at a doping value a few percent below $p^*$, and becomes negative at lower doping, presumably due to electron pockets associated with charge density wave order  \cite{Leboeuf2007}. Computing charge transport properties in the presence of charge density wave order on top of magnetic order could thus be an interesting extension of this work.

To conclude, spiral magnetic order is consistent with transport experiments in cuprates, where superconductivity is suppressed by high magnetic fields. We finally note that fluctuating instead of static magnetic order should yield similar transport properties, as long as pronounced magnetic correlations are present.

\subsection*{Outlook and closing remarks}

Both the theoretical study in Chapter~\ref{sec:theory} and the applications in Chapter~\ref{sec:application} provided insights that raise further questions beyond the scope of this thesis. In the following, we shortly present several paths for extensions.

The conductivity tensor $\sigma^{\alf\beta\eta}$ does not only capture the Hall conductivity but also the effect of linear magnetoresistance. We could relate the formula of the linear magnetoresistance to the symmetric contribution of $\sigma^{\alf\beta\eta}$ with respect to the exchange of the current and electric field directions, that is, the exchange of $\alf\leftrightarrow\beta$. The further derivation of a real-frequency formula for the linear magnetoresistance was not the focus of this thesis. Nevertheless, the provided general formulas in Sec.~\ref{sec:theory:Hall:symmetryindirections} may act as a starting point for this derivation. We see already at this stage that linear magnetoresistance is a pure interband effect since the intraband contribution of the symmetric contribution of $\sigma^{\alf\beta\eta}$ vanishes identically.

In this thesis, we considered a general two-band model as a minimal model for a system with multiple bands. A treatment of a model with more than two bands is both interesting in order to broaden the range of applicability and to potentially gain deeper structural insights. Some of our derivations in Chapter~\ref{sec:theory} were already performed for an arbitrary number of bands, which gave evidence that our developed concepts might be applicable even for more than two bands and that further phenomena might arise. An obvious key difference for a $n$-band system is the presence of $n\times n$ matrices instead of the $2\times 2$ matrices in this thesis. During our derivation, the conductivity involved the matrix trace over the two subsystems of the two-band model. In general, the evaluation of this matrix trace may lead to numerous terms and, thus, may make an analytical treatment tedious or even impossible for multiple bands. We presented the analysis of the involved matrices with respect to their behavior under transposition as a useful strategy to reduce this computational effort. This strategy may also be useful for an analytical treatment of multiband systems. Furthermore, it could be used for higher order expansions in electric and magnetic fields. 

We could relate the different interband contributions to concepts of quantum geometry. A nonzero quantum metric and Berry curvature are responsible for the symmetric and antisymmetric interband contribution of $\sigma^{\alf\beta}$, respectively. It might be an interesting question how those or other concepts of quantum geometry can be connected to transport phenomena beyond the scope of this thesis. Our microscopic derivation suggests that the precise way in which those concepts have to be included in other transport quantities is nontrivial. We presented the gauge invariance with respect to the $U(1)$ gauge in momentum space as a potential guide for identifying further physical (gauge-independent) quantities beyond the quantum metric and the Berry curvature.

We considered a phenomenological relaxation rate $\Gamma$ that is constant in momentum and frequency as well as diagonal and equal for both bands. Abandoning any of those assumptions needs a re-analysis of the presented derivation. We expect that several concepts, which we found throughout this thesis under those assumptions on $\Gamma$, will still be useful for these generalizations. A microscopically derived relaxation rate $\Gamma$, for instance, due to interaction or impurity scattering depends on details of the models, which we did not further specify in our general two-band model. A microscopic derivation can, for instance, be performed within a Born approximation \cite{Rickayzen1980}, which then can be used to concretize the range of validity. For example in the context of the anomalous Hall effect, such a microscopically derived $\Gamma$ is important to treat intrinsic and extrinsic effects on equal footing as we have already pointed out above.

To conclude, we presented a theory of conductivity that includes interband contributions. In our analysis of the conductivity for a very general two-band model, we could identify several concepts that helped us to gain deeper insights, which were not only useful for the analysis and applications in this thesis but may also be useful for further research beyond the presented scope. The application to recent experiments showed that the consideration of interband phenomena is not only important to gain a solid assessment of the applicability of formulas but also for a better understanding of fascinating phenomena that are intrinsically linked to the presence and interplay of multiple bands.

\newpage\leavevmode\thispagestyle{empty}\newpage



\rhead[\fancyplain{}{\bfseries                                              
Appendix \thechapter}]{\fancyplain{}{
}}                               
\lhead[\fancyplain{}{
}]{\fancyplain{}{\bfseries            
Appendix \thechapter}} 
\begin{appendices}
\appendixpage
\noappendicestocpagenum
\addappheadtotoc
\addtocontents{toc}{\protect\setcounter{tocdepth}{0}}

%
%

\chapter{Conductivity in the low-field limit}
\label{appendix:lowfield}

For a given dispersion $\eps_\bp$, the conductivity tensor $\sigma^{\alf\beta}[\bB]$ that includes the effect of the external magnetic field $\bB(\br,t)$ is shown to be 
\begin{align}
\label{eqn:sigmaB}
 (\sigma^{\alf\beta})[\bB]=-e^2\tau\int\hspace{-2mm}\frac{d^d\bp}{(2\pi)^d} \,f'(\eps_\bp-\mu)\,\bv_\bp \overline{\bv}_\bp \, .
\end{align}
where $\overline{\bv}_\bp\equiv \overline{\bv}(\bp)$ is the average velocity over the history of the electron orbit \cite{Ashcroft1976}. For simplicity, we assumed the lifetime $\tau$ to be constant. The weighted average of the velocity is given by
\begin{align}
 \overline{\bv}_\bp=\int_{-\infty}^0\frac{dt}{\tau}\,e^{t/\tau}\,\bv_{\bp(t)} \,,
\end{align}
where $\bv_{\bp(t)}$ is the solution of the semiclassical equation of motion in a uniform magnetic field,
\begin{align}
 &\partial_t\, \br(t) = \bv_{\bp(t)}\equiv \left.\nabla_\bp \eps_{\bp}\right|_{\bp=\bp(t)}\,,\\[1mm]
 \label{eqn:DGLp}
 &\partial_t\, \bp(t) = -e\,\bv_{\bp(t)}\times\bB\,,
\end{align}
with initial condition $\bp=\bp(t=0)$ and momentum derivative $\nabla_\bp=(\partial_{p^x},\partial_{p^y},\partial_{p^z})$. Solving \eqref{eqn:DGLp} up to linear order in $\bB$, we obtain the velocity
\begin{align}
 \bv^{}_{\bp(t)}\approx \bv^{}_{\bp-e\,\bv_\bp\times \bB\, t}\approx \bv_\bp-e\,\dM^{-1}_\bp\big(\bv^{}_\bp\times \bB\big)\, t \, ,
\end{align}
where we introduced the effective mass tensor
\begin{align}
 \big(\dM^{-1}_\bp\big)^{\alf\beta}\equiv (\nabla^{}_\bp v^\alf_\bp)^\beta=\frac{\partial^2\eps_\bp}{\partial p^\alf\partial p^\beta} \, .
\end{align}
After performing the remaining integration over time $t$, we obtain the weighted average of the velocity up to linear order
\begin{align}
 \label{eqn:expv}
 \overline{\bv}^{}_\bp=\bv^{}_\bp+e\tau\,\dM^{-1}_\bp\big(\bv^{}_\bp\times \bB\big) \, .
\end{align}
For a quadratic dispersion $\eps_\bp=\bp^2/2m$, we have $e\tau|\bB|\dM^{-1}=\omega_c\tau\,\mathds{1}$ with cyclotron frequency $\omega_c=e|\bB|/m$. The first term in \eqref{eqn:expv} gives \eqref{eqn:condBoltzmann}. The $y$ component of \eqref{eqn:expv} for a magnetic field in $z$ direction $\bB=B\,\mathbf{e}_z$ reads
\begin{align}
 \overline{v}^y_\bp=v^y_\bp+e\tau B\bigg(\frac{\partial^2 \eps_\bp}{\partial p^x \partial p^y}\frac{\partial \eps_\bp}{\partial p^y}-\frac{\partial^2 \eps_\bp}{\partial p^y \partial p^y}\frac{\partial \eps_\bp}{\partial p^x}\bigg)\,.
\end{align}
By inserting this result in $\sigma^{xy}[\bB]/B$ in \eqref{eqn:sigmaB} and performing partial integration in momentum by using the chain rule $f'(\eps_\bp-\mu)\,\partial \eps_\bp/\partial p^\alf=\partial f(\eps_\bp-\mu)/\partial p^\alf$, we obtain $\sigma^{xyz}_\text{H}$ in \eqref{eqn:HallBoltzmann}.

%
%

\chapter{Peierls substitution}
\label{appendix:peierls}

%
%

\section{Hopping in real space}

The Peierls substitution adds a phase factor to the hoppings in real space. Thus in order to apply \eqref{eqn:Peierls}, we Fourier transform the diagonal elements $\epsilon_{\bp,\ic}$ of the two subsystems $\ic=A,B$ and the coupling between these two systems $\Delta_\bp$ of Hamiltonian \eqref{eqn:H} to real space. The Fourier transformation of the creation operator $\c_{i,\ic}$ and $c_{\bp,\ic}$ were defined in \eqref{eqn:FourierC} and \eqref{eqn:FourierCInv}. The intraband hopping $t_{jj',\ic}\equiv t_{jj',\ic\ic}$ of one subsystem, which is defined by
\begin{align}
 \sum_\bp \cdag_{\bp+\bQ_\ic,\ic}\epsilon^{}_{\bp,\ic} \c_{\bp+\bQ_\ic,\ic}=\sum_{j,j'}\cdag_{j,\ic}t^{}_{jj',\ic}\c_{j',\ic} \, ,
\end{align}
is given by 
\begin{align}
\label{eqn:tijsigma}
 t_{jj',\ic}=\left(\frac{1}{L}\sum_\bp\eps_{\bp,\ic}\,e^{i\br_{jj'}\cdot\bp}\right)\,e^{i\br_{jj'}\cdot \bQ_\ic}\, .
\end{align}
We see that the intraband hopping is only a function of the difference between the unit cells, $\br_{jj'}=\bR_j-\bR_{j'}$. The fixed offset $\bQ_\ic$ leads to a phase shift. The interband hopping $t_{jj',AB}$ between the two subsystems, which is defined by
\begin{align}
 \sum_\bp \cdag_{\bp+\bQ_A,A}\Delta^{}_\bp \c_{\bp+\bQ_B,B}=\sum_{j,j'}\cdag_{j,A}t^{}_{jj',AB}\c_{j',B} \, ,
\end{align}
is given by
\begin{align}
\label{eqn:tijAB}
 t_{jj',AB}=&\left(\frac{1}{L}\sum_\bp\Delta_\bp\,e^{i\bp\cdot(\br_{jj'}+\brho_A-\brho_B)}\right)\,e^{i\br_{jj'}\cdot(\bQ_A+\bQ_B)/2}\nonumber\\
 &\times e^{i\bR_{jj'}\cdot(\bQ_A-\bQ_B)}\,e^{i(\brho_A\cdot\bQ_A-\brho_B\cdot\bQ_B)} \, .
\end{align}
We see that it is both a function of $\br_{jj'}$ and the mean position between the unit cells $\bR_{jj'}=(\bR_j+\bR_{j'})/2$, which breaks translational invariance and is linked to nonequal $\bQ_A\neq\bQ_B$. Similar to \eqref{eqn:tijsigma}, we have different phase shifts due to $\brho_\ic$ and $\bQ_\ic$. Those phases are necessary to obtain a consistent result in the following derivations.

%
%

\section{Derivation of the electromagnetic vertex $\sV_{\bp\bp'}$}

We derive the Hamiltonian $H[\bA]$ given in \eqref{eqn:HA} after Peierls substitution. We omit the time dependence of the vector potential $\bA[\br]\equiv \bA(\br,t)$ for a shorter notation in this section. The Peierls substitution in \eqref{eqn:Peierls} of the diagonal and off-diagonal elements of $\lam_{jj'}$ defined in \eqref{eqn:FourierH} and calculated in \eqref{eqn:tijsigma} and \eqref{eqn:tijAB} in the long-wavelength regime read 
\begin{align}
\label{eqn:Peierls1}
 &t_{jj',\ic}\rightarrow t^\bA_{jj',\ic}\equiv t_{jj',\ic}\,e^{-ie\bA[\bR_{jj'}+\brho_\ic]\cdot \br_{jj'}} \, ,\\
 &t_{jj',AB}\rightarrow t^\bA_{jj',AB}\equiv t_{jj',AB}\,e^{-ie\bA[\bR_{jj'}+\frac{\brho_A+\brho_B}{2}]\cdot\big(\br_{jj'}+\brho_A-\brho_B\big)} \, .
\end{align}
In a first step, we consider the diagonal elements. We expand the exponential of the hopping $t^\bA_{jj',\ic}$ after Peierls substitution given in \eqref{eqn:Peierls1} and Fourier transform the product of the vector potentials $\big(\bA[\bR_{jj'}+\brho_\ic]\cdot \br_{jj'}\big)^n$ via \eqref{eqn:FourierAq}. We get 
\begin{align}
 t^\bA_{jj',\ic}=&\sum_n\frac{(-i e)^n}{n!}\sum_{\bq_1...\bq_n}t_{jj',\ic}\,e^{i\sum_m\bq_m\cdot(\bR_{jj'}+\brho_\ic)}\prod^n_m\br_{jj'}\cdot\bA_{\bq_m}\,.
\end{align}
After insertion of the hopping \eqref{eqn:tijsigma}, we Fourier transform $t^\bA_{jj',\ic}$ back to momentum space defining $\epsilon^\bA_{\bp\bp',\ic}$ via
\begin{align}
 \sum_{jj'}\cdag_{j,\ic}t^\bA_{jj',\ic}\c_{j',\ic}=\sum_{\bp\bp'} \cdag_{\bp+\bQ_\ic,\ic}\epsilon^\bA_{\bp\bp',\ic} \c_{\bp'+\bQ_\ic,\ic} \, .
\end{align}
The summation over $\bR_{jj'}$ leads to momentum conservation. The phase factor proportional to the position $\brho_\ic$ inside the unit cell cancels. During the calculation, we can identify 
\begin{align}
&-\frac{i}{L}\sum_\bp\sum_{\br_{jj'}}\epsilon_{\bp,\ic}\, e^{i\br_{jj'}\cdot(\bp-\bp_0)}\,\big(\br_{jj'}\cdot \bA_\bq\big)=\sum_{\alf=x,y,z}\left.\frac{\partial\epsilon_{\bp,\ic}}{\partial p^\alf}\right|_{\bp=\bp_0}A^\alf_\bq
\end{align}
as the derivative of the band $\epsilon_{\bp,\ic}$ at $\bp_0=(\bp+\bp')/2$. We continue with the off-diagonal element. The derivation of the interband coupling after Peierls substitution, which we label as $\Delta^\bA_{\bp\bp'}$, is analogue to the derivation above. The phase factors in \eqref{eqn:tijAB} assure that we can identify the derivative of the interband coupling $\Delta_\bp$ via
\begin{align}
&-\frac{i}{L}\sum_\bp\sum_{\br_{jj'}}\Delta_\bp\, e^{i(\br_{jj'}+\brho_A-\brho_B)\cdot(\bp-\bp_0)}\big(\br_{jj'}+\brho_A-\brho_B\big)\cdot \bA_\bq
 =\sum_{\alf=x,y,z}\left.\frac{\partial\Delta_\bp}{\partial p^\alf}\right|_{\bp=\bp_0}A^\alf_\bq \, .
\end{align}
As in the diagonal case, the summation over $\bR_{jj'}$ leads to momentum conservation and additional phase factors drop. Finally, we write the result in matrix form and separate the zeroth element of the exponential expansion. We end up with the Hamiltonian of the form \eqref{eqn:HA} and the electromagnetic vertex $\sV_{\bp\bp'}$ given in \eqref{eqn:Vpp'}.

\newpage\leavevmode\thispagestyle{empty}\newpage

%
%

\chapter{Current}
\label{appendix:current}

%
%

\section{Grand canonical potential}
\label{appendix:current:grandcanonicalpotential}

Since the action $S[\Psi,\Psi^*]$ in \eqref{eqn:S} is quadratic in the Grassmann fields $\Psi$ and $\Psi^*$, the Gaussian path integral leads to the partition function $Z=\det\big(\sG^{-1}-\sV\big)$, where the Green's function $\sG$ and the electromagnetic vertex $\sV$ are understood as matrices of both Matsubara frequencies and momenta. The grand canonical potential $\Omega$ is related to the partition function via $\Omega=-T\ln Z$ with temperature $T$ and $k_B=1$. We factor out the part that is independent of the vector potential, that is $\Omega_0=-T\,\Tr \ln \sG^{-1}$, and expand the logarithm $\ln(1-x)=-\sum_n x^n/n$ of the remaining part. We obtain
\begin{align}
 \label{eqn:OmegaExpansion}
 \Omega[\bA]=\Omega_0+T\sum_{n=1}^\infty \frac{1}{n}\Tr\big(\sG\sV\big)^n \, .
\end{align}
Using the definition of the Green's function and the vertex in \eqref{eqn:Vpp'} and \eqref{eqn:Green}, one can check explicitly that $\Omega[\bA]$ is real at every order in $n$. 

%
%

\section{Expansion of the current in the vector potential}
\label{appendix:current:expansion}

The current density $j^\alf_q$ in direction $\alf=x,y,z$ and Matsubara frequency and momentum $q=(iq_0,\bq)$ is given as functional derivative of the grand canonical potential with respect to the vector potential, $j^\alf_q=-1/L\, \delta\Omega[\bA]/\delta A^\alf_{-q}$. The Green's function $\sG$ in \eqref{eqn:Green} has no dependence on the vector potential. We denote the derivative of the electromagnetic vertex, the current vertex, as $\dot \sV^\alf_q = -1/L\,\delta \sV/\delta A^\alf_{-q}$ and expand $\Omega[\bA]$ in \eqref{eqn:OmegaExpansion} up to third order. We obtain  
\begin{align}
\label{eqn:jexp}
 j^\alf_q=T\,\Tr\big(\sG\dot \sV^\alf_q\big)+T\,\Tr\big(\sG\dot \sV^\alf_q \sG\sV\big)+T\,\Tr\big(\sG\dot \sV^\alf_q \sG\sV\sG\sV\big)+... \, ,
\end{align}
where we used the invariance of the trace under cyclic permutation to recombine the terms of the same order. Both the electromagnetic vertex $\sV$ and the current vertex $\dot \sV^\alf_q$ are a series of the vector potential. In the following, we expand the current up to second order in the vector potential. The expansion of the electromagnetic vertex $\sV$ is given in \eqref{eqn:Vpp'}. The expansion of the current vertex reads
\begin{align}
 \dot \sV^\alf_{q,pp'}=&-\frac{e}{L}\sum^\infty_{n=0}\frac{e^n}{n!}\sum_{\substack{q_1...q_n \\ \alf_1...\alf_n}}
 \lam^{\alf\alf_1...\alf_n}_{\frac{p+p'}{2}}\,\,A^{\alf_1}_{q_1}...A^{\alf_n}_{q_n}\,\,\delta_{\sum_m q_m,p-p'+q} \,.
\end{align}
Note that the current vertex $\dot \sV^\alf_{q,pp'}$ has a zeroth order, which is independent of the vector potential, whereas the electromagnetic vertex $\sV_{pp'}$ is at least linear. 

We expand the three terms in \eqref{eqn:jexp} in the following. The first term in \eqref{eqn:jexp} involves
\begin{align}
 \Tr\big(\sG\dot \sV^\alf_q\big)
 =\label{eqn:j11}&-e\frac{T}{L}\sum_p \tr\big[\sG^{}_p\lam^\alf_p\big]\delta^{}_{q,0}
 \\[1mm]
 &\label{eqn:j12}-e^2\sum_\beta \frac{T}{L}\sum_p \tr\big[\sG^{}_p\lam^{\alf\beta}_p \big]A^\beta_q 
 \\[1mm]
 &\label{eqn:j13}-\frac{1}{2}\sum_{\beta\gamma}\frac{T}{L}\sum_{p,q'}\tr\big[\sG^{}_p\lam^{\alf\beta\gamma}_p\big]A^\beta_{q'}A^\gamma_{q-q'}
 \\[1mm]
 &+\mathcal{O}(A^3)\, .
\end{align}
The first term \eqref{eqn:j11} corresponds to a current without any external fields. The second term \eqref{eqn:j12} is known as {\it diamagnetic} contribution of the linear conductivity. We expand the second term in \eqref{eqn:jexp} up to second order in the vector potential and obtain
\begin{align}
 \Tr\big(\sG\dot \sV^\alf_q \sG\sV\big)
 =\label{eqn:j21}&-e^2\sum_\beta\frac{T}{L}\sum_p \tr\big[\sG^{}_p\lam^\alf_{p+\frac{q}{2}}\sG^{}_{p+q}\lam^\beta_{p+\frac{q}{2}}\big]A^\beta_q
 \\[1mm]
 \label{eqn:j22}&-\frac{1}{2}\sum_{\beta\gamma}\frac{T}{L}\sum_{p,q'}\tr\big[\sG^{}_p\lam^\alf_{p+\frac{q}{2}}\sG^{}_{p+q}\lam^{\beta\gamma}_{p+\frac{q}{2}}\big]A^\beta_{q'}A^\gamma_{q-q'} 
 \\[1mm]
 \label{eqn:j23}&-\frac{1}{2}\sum_{\beta\gamma}\frac{T}{L}\sum_{p,q'}\tr\big[\sG^{}_p\lam^{\alf\gamma}_{p+\frac{q'}{2}}\sG^{}_{p+q'}\lam^\beta_{p+\frac{q'}{2}}\big]A^\beta_{q'}A^\gamma_{q-q'}
 \\[1mm]
 \label{eqn:j24}&-\frac{1}{2}\sum_{\beta\gamma}\frac{T}{L}\sum_{p,q'}\tr\big[\sG^{}_{p+q}\lam^\gamma_{p+\frac{q'}{2}+\frac{q}{2}}\sG^{}_{p+q'}\lam^{\alf\beta}_{p+\frac{q'}{2}+\frac{q}{2}}\big]A^\beta_{q'}A^\gamma_{q-q'}
 \\[1mm]
 &+\mathcal{O}(A^3)\, .
\end{align}
The first term \eqref{eqn:j21} is known as {\it paramagnetic} contribution of the linear conductivity. The two contributions \eqref{eqn:j23} and \eqref{eqn:j24} are equal by shifting and renaming the summations. The third term in \eqref{eqn:jexp} up to second order in the vector potential involves
\begin{align}
 \Tr&\big(\sG\dot\sV^\alf_q\sG\sV\sG\sV\big)\nonumber
 \\[1mm]=&\label{eqn:j31}-\frac{1}{2}\sum_{\beta\gamma}\frac{T}{L}\sum_{p,q'}\tr\big[\sG^{}_p\lam^\alf_{p+\frac{q}{2}}\sG^{}_{p+q}\lam^\gamma_{p+\frac{q'}{2}+\frac{q}{2}}\sG^{}_{p+q'}\lam^\beta_{p+\frac{q'}{2}}\big]A^\beta_{q'}A^\gamma_{q-q'}
 \\[1mm]
 &\label{eqn:j32}-\frac{1}{2}\sum_{\beta\gamma}\frac{T}{L}\sum_{p,q'}\tr\big[\sG^{}_{p+q}\lam^\gamma_{p+\frac{q'}{2}+\frac{q}{2}}\sG^{}_{p+q'}\lam^\alf_{p+\frac{q}{2}+q'}\sG^{}_{p+q+q'}\lam^\beta_{p+q+\frac{q'}{2}}\big]A^\beta_{q'}A^\gamma_{q-q'}
 \\[1mm]
 &+\mathcal{O}(A^3)\, .
\end{align}
The two terms \eqref{eqn:j31} and \eqref{eqn:j32} are equal by shifting and renaming the summations. By collecting the zeroth-, first- and second-order contributions we can identify $j^\alf_0$, $\Pi^{\alf\beta}_q$ and $\Pi^{\alf\beta\gamma}_{q,q'}$ defined in \eqref{eqn:jexpDef}, respectively. Note that the involved Green's function and vertex matrices in the expressions above do not commute in general. Our derivation above respects this issue.

%
%

\section{Current without any external fields}
\label{appendix:current:paramagneticcurrent}

The current without any external fields, that is the first term of \eqref{eqn:jexpDef}, is independent of the vector potential and is given by
\begin{align}
 j^\alf_0=-e\frac{T}{L}\sum_p \tr\big[\sG^{}_p\lam^\alf_p\big]\delta^{}_{q,0} \, .
\end{align}
We perform the Matsubara summation and diagonalize the Bloch Hamiltonian $\lam_\bp$. The current without any external fields reads
\begin{align}
 j^\alf_0=-\frac{e}{L}\sum_\bp\int d\eps\,f(\eps) \sum_{n=\pm} A^n_\bp(\eps) E^{n,\alf}_\bp \, ,
\end{align}
involving the Fermi function $f(\eps)$, the quasiparticle velocities $E^{n,\alf}_\bp=\partial_\alf E^n_\bp$ of the two quasiparticle bands $E^\pm_\bp=\frac{1}{2}(\eps_{\bp,A}+\eps_{\bp,B})\pm\sqrt{\frac{1}{4}(\eps_{\bp,A}-\eps_{\bp,B})^2+|\Delta_\bp|^2}$ and the spectral functions $A^\pm_\bp(\eps)=\Gamma/\pi\big[(\eps-E^\pm_\bp)^2+\Gamma^2\big]^{-1}$. In general, the contributions at momentum $\bp$ are nonzero. If the quasiparticle bands fulfill $E^\pm(\bp)=E^\pm(-\bp-\bp^\pm)$ for a fixed momentum $\bp^\pm$, we have $E^{\pm,\alf}(\bp)=-E^{\pm,\alf}(-\bp-\bp^\pm)$. Thus, the current $j^\alf_0$ vanishes by integrating over momenta \cite{Voruganti1992}.  

%
%

\section{Linear electrical conductivity} 
\label{appendix:current:piE}

We combine the linear terms \eqref{eqn:j12} and \eqref{eqn:j21} in order to identify the polarization tensor $\Pi^{\alf\beta}_q$ in \eqref{eqn:jexpDef}. We write the Matsubara frequencies and the momenta explicitly, shift the momentum summation and obtain
\begin{align}
\label{eqn:LinPi}
 \Pi^{\alf\beta}_{iq_0,\bq}=e^2\frac{T}{L}\sum_{ip_0,\bp}\tr\big[\sG^{}_{ip_0,\bp-\bq/2}\,\,\lam^\alf_\bp\,\,\sG^{}_{ip_0+iq_0,\bp+\bq/2}\,\,\lam^\beta_\bp+\sG^{}_{ip_0,\bp}\,\,\lam^{\alf\beta}_\bp\big] \, .
\end{align}
Note that the second term is the $\big((iq_0,\bq)=0\big)$ contribution of the first term: We use the definition $\lam^{\alf\beta}_\bp=\partial^{}_\alf \lam^\beta_\bp$ in \eqref{eqn:DlamDef} and perform partial integration in the momentum integration over $p^\alf$. The derivative of the Green's function is $\partial^{}_\alf \sG^{}_{ip_0,\bp}=\sG^{}_{ip_0,\bp}\,\lam^\alf_\bp \,\sG^{}_{ip_0,\bp}$, which follows by \eqref{eqn:Green}. 

We assume a uniform electric field $\bE(t)$, which is entirely described by a uniform vector potential $\bA^E(t)$. Thus, the corresponding vector potential after Wick rotation and Fourier transformation yields $A^\beta_q=A^{E,\beta}_{iq_0}\delta^{}_{\bq,0}$. We drop the momentum dependence of the current and set $\bq=0$ in \eqref{eqn:LinPi}. After using the invariance of the matrix trace under cyclic permutation, we obtain \eqref{eqn:PiE}.

\section{Hall conductivity}
\label{appendix:current:piH}

We assume a uniform electric field $\bE(t)$ and a static magnetic field $\bB(\br)$, which can be described by the sum of a uniform and a static part of the vector potential $\bA(t,\br)=\bA^E(t)+\bA^B(\br)$. After Wick rotation and Fourier transformation, we have $A^\beta_{iq_0,\bq}=A^{E,\beta}_{iq_0}\delta^{}_{\bq,0}+A^{B,\beta}_\bq\delta^{}_{iq_0,0}$. The product of two spectral functions with this form yields
\begin{align}
\label{eqn:DecompAA}
 A^\beta_{q'}A^\gamma_{q-q'}=A^{E,\beta}_{iq_0'}A^{B,\gamma}_{\bq-\bq'}\,\,\delta^{}_{\bq',0}\delta^{}_{iq_0',iq_0}+A^{B,\beta}_{\bq'}A^{E,\gamma}_{iq_0-iq_0'}\delta^{}_{iq_0',0}\,\,\delta^{}_{\bq',\bq}+\cdots \, ,
\end{align}
where we do not consider the terms that will be eventually quadratic in the electric and in the magnetic field. We combine the six second-order terms \eqref{eqn:j13}, \eqref{eqn:j22}, \eqref{eqn:j23}, \eqref{eqn:j24}, \eqref{eqn:j31} and \eqref{eqn:j32} to $\Pi^{\alf\beta\gamma}_{q,q'}$. We use \eqref{eqn:DecompAA}, which defines the polarization tensor of the Hall conductivity via  $j^\alf_{\text{\tiny EB},q}=-\sum_{\beta\gamma}\Pi^{\alf\beta\gamma}_{\text{\tiny EB},q}A^{E,\beta}_{iq_0}A^{B,\gamma}_\bq$ with $q=(iq_0,\bq)$. We obtain after shifting the momentum integration
\begin{alignat}{8}
 \Pi^{\alf\beta\gamma}_{\text{\tiny EB},q}=
 \label{eqn:PiH1} e^3\frac{T}{L}\sum_{ip_0,\bp}\tr\big[& &&\sG^{}_{ip_0,\bp}&&\lam^{\alf\beta\gamma}_\bp&& && && && &&
 \\[-4mm]
 \label{eqn:PiH2}&+\,&&\sG^{}_{ip_0,\bp}&&\lam^{\alf\gamma}_{\bp}&&\sG^{}_{ip_0+iq_0,\bp}&&\lam^\beta_{\bp}&& && &&
 \\
 \label{eqn:PiH3}&+\,&&\sG^{}_{ip_0,\bp-\frac{\bq}{2}}&&\lam^{\alf\beta}_{\bp}&&\sG^{}_{ip_0,\bp+\frac{\bq}{2}}&&\lam^\gamma_{\bp}&& && &&
 \\
 \label{eqn:PiH4}&+\,&&\sG^{}_{ip_0,\bp-\frac{\bq}{2}}\,&&\lam^\alf_{\bp}&&\sG^{}_{ip_0+iq_0,\bp+\frac{\bq}{2}}\,&&\lam^{\beta\gamma}_{\bp}&& && &&
 \\
 &+\,&&\sG^{}_{ip_0,\bp-\frac{\bq}{2}}&&\lam^\alf_{\bp}&&\sG^{}_{ip_0+iq_0,\bp+\frac{\bq}{2}}&&\lam^\gamma_{\bp}&&\sG^{}_{ip_0+iq_0,\bp-\frac{\bq}{2}}\,&&\lam^\beta_{\bp-\frac{\bq}{2}}\,&&
 \\
 \label{eqn:PiH6}&+\,&&\sG^{}_{ip_0,\bp-\frac{\bq}{2}}&&\lam^\alf_{\bp}&&\sG^{}_{ip_0+iq_0,\bp+\frac{\bq}{2}}&&\lam^\beta_{\bp+\frac{\bq}{2}}&&\sG^{}_{ip_0,\bp+\frac{\bq}{2}}&&\lam^\gamma_{\bp}&&\big]
 \, .
\end{alignat}
By partial integration in momentum $p^\gamma$, we find that \eqref{eqn:PiH1} and \eqref{eqn:PiH2} are equal up to a negative sign to \eqref{eqn:PiH3} and to \eqref{eqn:PiH4}-\eqref{eqn:PiH6} at zero $\bq$, respectively. We will indicate this subtraction by the notation $(\bq=0)$ in the following. We get
\begin{alignat}{7}
 \Pi^{\alf\beta\gamma}_{\text{\tiny EB},q}=
 \label{eqn:PiH1b} e^3\frac{T}{L}\sum_{ip_0,\bp}\tr\big[& &&\sG^{}_{ip_0,\bp-\frac{\bq}{2}}\,&&\lam^{\alf\beta}_{\bp}\,&&\sG^{}_{ip_0,\bp+\frac{\bq}{2}}\,&&\lam^\gamma_{\bp} && &&
 \\[-4mm]
 &+\,&&\sG^{}_{ip_0,\bp-\frac{\bq}{2}}&&\lam^\alf_{\bp}&&\sG^{}_{ip_0+iq_0,\bp+\frac{\bq}{2}}&&\lam^{\beta\gamma}_{\bp} && &&
\\
 &+\,&&\sG^{}_{ip_0,\bp-\frac{\bq}{2}}&&\lam^\alf_{\bp}&&\sG^{}_{ip_0+iq_0,\bp+\frac{\bq}{2}}&&\lam^\beta_{\bp+\frac{\bq}{2}}\,&&\sG^{}_{ip_0,\bp+\frac{\bq}{2}}\,&&\lam^\gamma_{\bp}
 \\
 &+\,&&\sG^{}_{ip_0,\bp-\frac{\bq}{2}}&&\lam^\alf_{\bp}&&\sG^{}_{ip_0+iq_0,\bp+\frac{\bq}{2}}\,&&\lam^\gamma_{\bp}&&\sG^{}_{ip_0+iq_0,\bp-\frac{\bq}{2}}\,&&\lam^\beta_{\bp-\frac{\bq}{2}}
 \\&-&&(\bq=0)\big]
 \, .
\end{alignat}
After partial integration in momentum $p^\beta$, we see that \eqref{eqn:PiH1b} is equal up to a negative sign to the three other contributions at zero $iq_0$. We denote this subtraction by $(iq_0=0)$ in the following and get
\begin{alignat}{7} 
 \Pi^{\alf\beta\gamma}_{\text{\tiny EB},q}=
 e^3\frac{T}{L}\sum_{ip_0,\bp}\tr\big[& &&\sG^{}_{ip_0+iq_0,\bp+\frac{\bq}{2}}&&\lam^{\beta\gamma}_{\bp}&&\sG^{}_{ip_0,\bp-\frac{\bq}{2}}&&\lam^\alf_{\bp}
 \\[-4mm]
 &+\,&&\sG^{}_{ip_0+iq_0,\bp+\frac{\bq}{2}}\,&&\lam^\beta_{\bp+\frac{\bq}{2}}\,&&\sG^{}_{ip_0,\bp+\frac{\bq}{2}}\,&&\lam^\gamma_{\bp}&&\sG^{}_{ip_0,\bp-\frac{\bq}{2}}&&\lam^\alf_{\bp}
 \\
 &+\,&&\sG^{}_{ip_0-iq_0,\bp-\frac{\bq}{2}}&&\lam^\alf_{\bp}&&\sG^{}_{ip_0,\bp+\frac{\bq}{2}}&&\lam^\gamma_{\bp}\,&&\sG^{}_{ip_0,\bp-\frac{\bq}{2}}\,&&\lam^\beta_{\bp-\frac{\bq}{2}}
 \\
 & && && && && && &&\hspace{-7cm}-\,(\bq=0)-(iq_0=0) \big] 
 \, .
\end{alignat}
We continue by an expansion in $\bq$. The constant zeroth order drops.  The linear order $\Pi^{\alf\beta\gamma\delta}_{\text{\tiny EB},iq_0}=\partial/\partial q^\delta\,\Pi^{\alf\beta\gamma}_{\text{\tiny EB},q}|_{\bq=0}$ with $q=(iq_0,\bq)$ and $q^\delta$ the $\delta$ component of $\bq$ reads
\begin{alignat}{9}
 \Pi^{\alf\beta\gamma\delta}_{\text{\tiny EB},iq_0}=\label{eqn:PiHPartial}\frac{1}{2}\frac{T}{L}\sum_{ip_0,\bp}\tr\big[& &&\sG^{}_{ip_0+iq_0,\bp}&&\lam^{\beta\delta}_{\bp}\,&&\sG^{}_{ip_0,\bp}&&\lam^\gamma_{\bp}\,&&\sG^{}_{ip_0,\bp\,}&&\lam^\alf_{\bp}&& &&
 \\[-4mm]
 &+\,&&\sG^{}_{ip_0-iq_0,\bp}\,&&\lam^\alf_{\bp}&&\sG^{}_{ip_0,\bp}&&\lam^\delta_\bp&&\sG^{}_{ip_0,\bp}&&\lam^{\beta\gamma}_{\bp} && &&
 \\
 &+&&\sG^{}_{ip_0+iq_0,\bp}&&\lam^\beta_{\bp}&&\sG^{}_{ip_0,\bp}&&\lam^\delta_\bp&&\sG^{}_{ip_0,\bp}&&\lam^\gamma_{\bp}&&\sG^{}_{ip_0,\bp}&&\lam^\alf_{\bp}
  \\
 &+&&\sG^{}_{ip_0-iq_0,\bp}&&\lam^\alf_{\bp}&&\sG^{}_{ip_0,\bp}&&\lam^\delta_\bp&&\sG^{}_{ip_0,\bp}&&\lam^\gamma_{\bp}&&\sG^{}_{ip_0,\bp}&&\lam^\beta_{\bp}
 \\
 &+&&\sG^{}_{ip_0+iq_0,\bp}&&\lam^\delta_\bp&&\sG^{}_{ip_0+iq_0,\bp}\,&&\lam^\beta_{\bp}&&\sG^{}_{ip_0,\bp}&&\lam^\gamma_{\bp}&&\sG^{}_{ip_0,\bp}\,&&\lam^\alf_{\bp}
 \\
 & && && && && && && && &&\hspace{-7.55cm}-(iq_0=0)-(\gamma\leftrightarrow \delta)\big]
 \, ,
\end{alignat}
where we indicated the subtraction of the same terms with exchanged indices by $(\gamma\leftrightarrow\delta)$. We perform partial integration in momentum $p^\delta$ of the term \eqref{eqn:PiHPartial} and obtain
\begin{alignat}{12}
 \label{eqn:PiHPartial1}&\Pi^{\alf\beta\gamma\delta}_{\text{\tiny EB},iq_0} &&=\,&&\frac{1}{2}\frac{T}{L}\sum_{ip_0,\bp}\tr\big[&& &&\sG^{}_{ip_0+iq_0,\bp}&&\lam^{\beta}_{\bp}&&\sG^{}_{ip_0,\bp}&&\lam^\delta_{\bp}&&\sG^{}_{ip_0,\bp}&&\lam^{\alf\gamma}_{\bp}&& &&
 \\[-4mm]
 \label{eqn:PiHPartial2}& && && &&+\,&&\sG^{}_{ip_0-iq_0,\bp}&&\lam^\alf_{\bp}&&\sG^{}_{ip_0,\bp}&&\lam^\delta_\bp&&\sG^{}_{ip_0,\bp}&&\lam^{\beta\gamma}_{\bp}&& &&
 \\
 & && && &&+\,&&\sG^{}_{ip_0+iq_0,\bp}&&\lam^{\beta}_{\bp}&&\sG^{}_{ip_0,\bp}&&\lam^\delta_{\bp}&&\sG^{}_{ip_0,\bp}&&\lam^\gamma_\bp&&\sG^{}_{ip_0,\bp}&&\lam^\alf_{\bp}
 \\
 & && && &&+\,&&\sG^{}_{ip_0-iq_0,\bp}\,&&\lam^\alf_{\bp}\,&&\sG^{}_{ip_0,\bp}\,&&\lam^\delta_\bp\,&&\sG^{}_{ip_0,\bp}\,&&\lam^\gamma_{\bp}\,&&\sG^{}_{ip_0,\bp}\,&&\lam^\beta_{\bp}
 \\
 & && && && && && && && && && && &&\hspace{-6.85cm}-(iq_0=0)-(\gamma\leftrightarrow \delta)\big] \, .
\end{alignat}
In a last step we split \eqref{eqn:PiHPartial1} and \eqref{eqn:PiHPartial2} into two equal contributions and perform partial integration in $p^\gamma$. We separate the four contributions that involve three vertices and the two contributions that involve four vertices and end up with \eqref{eqn:PiHTri} and \eqref{eqn:PiHRec}.


%
%

\chapter{Second-order vertex of a multiband system}
\label{appendix:Mass}

We calculate the second-order vertex $\hat\lam^{\alf\beta}_\bp=\partial^{}_\alf\partial^{}_\beta \hat \lam^{}_\bp$ of a general multiband (and not necessarily two-band) Bloch Hamiltonian $\hat \lam_\bp$ in its orthonormal and complete eigenbasis $|n_\bp\rangle$ with eigenvalues $E^n_\bp$. By considering the momentum dependence of the eigenbasis and the product rule, we get
\begin{align}
 \langle n^{}_\bp|\big(\partial^{}_\alf\partial^{}_\beta\hat\lam^{}_\bp\big)|m^{}_\bp\rangle
 &=
 \label{eqn:M1}\frac{1}{2}\partial^{}_\alf\partial^{}_\beta\Big(\langle n^{}_\bp|\hat\lam^{}_\bp|m^{}_\bp\rangle\Big)
 \\
 &\label{eqn:M2}-\frac{1}{2}\Big(\langle \partial^{}_\alf\partial^{}_\beta n^{}_\bp|\hat \lam^{}_\bp|m^{}_\bp\rangle
 +\langle n^{}_\bp|\hat\lam^{}_\bp|\partial^{}_\alf\partial^{}_\beta m^{}_\bp\rangle\Big)
 \\[0.5mm]
 &\label{eqn:M3}-\Big(\langle\partial^{}_\alf n^{}_\bp|\lam^\beta_\bp|m^{}_\bp\rangle
 + \langle n^{}_\bp|\lam^\alf_\bp|\partial^{}_\beta m^{}_\bp\rangle\Big)
 \\[1mm]
 &\label{eqn:M4}- \langle \partial_\alf n^{}_\bp|\lam^{}_\bp|\partial^{}_\beta m^{}_\bp\rangle 
 \\[2mm]
 &+ (\alf\leftrightarrow \beta) \, .
\end{align}
The expression is symmetric in $\alf\leftrightarrow\beta$. We indicate the addition of all previous terms with exchanged indices as $(\alf\leftrightarrow\beta)$. We use  $\hat \lam_\bp|n_\bp\rangle=E^n_\bp|n_\bp\rangle$ and $\langle n_\bp|\hat\lam_\bp=E^n_\bp\langle n_\bp|$ in the terms \eqref{eqn:M1}, \eqref{eqn:M2} and \eqref{eqn:M4}. The first term \eqref{eqn:M1} is the inverse quasiparticle effective mass $E^{n,\alf\beta}_\bp=\partial^{}_\alf\partial^{}_\beta E^n_\bp$. In order to calculate the first terms in \eqref{eqn:M2}, we use the identity
\begin{align}
 \langle \partial_\alf\partial_\beta n_\bp|m_\bp\rangle=\frac{1}{2}\Big[\partial_\alf\Big(\langle \partial_\beta n_\bp|m_\bp\rangle\Big)-\langle \partial_\beta n_\bp|\partial_\alf n_\bp\rangle+\partial_\beta\Big(\langle \partial_\alf n_\bp|m_\bp\rangle\Big)-\langle \partial_\alf n_\bp|\partial_\beta n_\bp\rangle\Big] \, ,
\end{align}
which can be checked by performing the derivatives on the right-hand side, explicitly. We immediately identify the definition of $\cA^{\alf,nm}_\bp=i\langle n_\bp|\partial_\alf m_\bp\rangle$ in the first and third term and after inserting the identity $\mathds{1}=\sum_l |l_\bp\rangle\langle l_\bp|$ in the second and the forth term. After similar steps for the second term in \eqref{eqn:M2}, we obtain
\begin{alignat}{2}
 \langle \partial^{}_\alf\partial^{}_\beta n^{}_\bp|m^{}_\bp\rangle=& &&\frac{i}{2}\big(\partial^{}_\alf A^{\beta,nm}_\bp+\partial^{}_\beta A^{\alf,nm}_\bp\big)-\frac{1}{2}\sum_l\big(A^{\beta,nl}_\bp A^{\alf,lm}_\bp + A^{\alf,nl}_\bp A^{\beta,lm}_\bp\big)\, , \\
 \langle n^{}_\bp|\partial^{}_\alf\partial^{}_\beta m^{}_\bp\rangle=& -&&\frac{i}{2}\big(\partial^{}_\alf A^{\beta,nm}_\bp+\partial^{}_\beta A^{\alf,nm}_\bp\big)-\frac{1}{2}\sum_l\big(A^{\beta,nl}_\bp A^{\alf,lm}_\bp + A^{\alf,nl}_\bp A^{\beta,lm}_\bp\big) \, .
\end{alignat}
In order to further simplify \eqref{eqn:M3}, we insert the identity $\mathds{1}=\sum_l |l_\bp\rangle\langle l_\bp|$ and use $\langle l^{}_\bp|\hat \lam^\beta_\bp|m^{}_\bp\rangle=\delta^{}_{lm}\,E^{m,\beta}_\bp+i\big(E^l_\bp-E^m_\bp\big)A^{\beta,lm}_\bp$, which is given in \eqref{eqn:DlamGeneral}. We get
\begin{alignat}{3}
 \langle \partial^{}_\alf n^{}_\bp|\hat \lam^\beta_\bp|m^{}_\bp\rangle=& &&iA^{\alf,nm}_\bp E^{m,\beta}_\bp&&-\sum_l\big(E^l_\bp-E^m_\bp\big) A^{\alf,nl}_\bp A^{\beta,lm}_\bp\, , \\ 
 \langle n^{}_\bp|\hat \lam^\alf_\bp|\partial^{}_\beta m^{}_\bp\rangle=&-&&iE^{n,\alf}_\bp A^{\beta,nm}_\bp&&-\sum_l\big(E^l_\bp-E^n_\bp\big) A^{\alf,nl}_\bp A^{\beta,lm}_\bp \, .
\end{alignat}
We re-express the last term \eqref{eqn:M4} by inserting the identity $\mathds{1}=\sum_l |l_\bp\rangle\langle l_\bp|$ and using that $\hat \lam^{}_\bp|l_\bp\rangle=E^l_\bp|l^{}_\bp\rangle$. We obtain $\langle \partial^{}_\alf n^{}_\bp|\hat \lam^{}_\bp|\partial^{}_\beta m^{}_\bp\rangle = \sum_l E^l_\bp A^{\alf,nl}_\bp A^{\beta,lm}_\bp$. After combining all terms, we end up with the generalized effective mass in the orthonormal eigenbasis of the Bloch Hamiltonian
\begin{align}
 \langle n_\bp|\big(\partial_\alf\partial_\beta&\hat\lam_\bp\big)|m_\bp\rangle\\&=\frac{1}{2}\delta_{nm}\,E^{n,\alf\beta}_\bp
 \\
 &+i\big(E^{n,\alf}_\bp-E^{m,\alf}_\bp\big)A^{\beta,nm}_\bp
 \\
 &+\frac{i}{2}\big(E^n_\bp-E^m_\bp\big)\big(\partial^{}_\alf A^{\beta,nm}_\bp\big)+\sum_l\big[E^l_\bp-\frac{1}{2}\big(E^n_\bp+E^m_\bp\big)\big]A^{\alf,nl}_\bp A^{\beta,lm}_\bp
 \\[-2mm]
 &+(\alf\leftrightarrow\beta) \, .
\end{align}
Note that the expression only involves the eigenenergies or its derivatives as well as the Berry connection or its derivative. The full expression is symmetric in $\alf\leftrightarrow\beta$. Performing the gauge transformation $\cA^{\alf,nm}_\bp\rightarrow \cA^{\alf,nm}_\bp e^{-i(\phi^n_\bp-\phi^m_\bp)}-\delta^{}_{nm}\,\phi^{n,\alf}_\bp$ in the result above explicitly shows that each line individually transforms with a phase factor $e^{-i(\phi^n_\bp-\phi^m_\bp)}$. The momentum derivatives of $\phi^n_\bp$ and $\phi^m_\bp$ drop. The diagonal component for $n=m$ is explicitly given in \eqref{eqn:Mn}. The off-diagonal component for $n\neq m$ reads \eqref{eqn:Mnm}-\eqref{eqn:Mnm4}.


%
%

\chapter{Simplifying the triangular and the rectangular contributions}
\label{appendix:DecompRecomp}

We present the detailed derivation from  Eqs.~\eqref{eqn:PiHTriSpecial} and \eqref{eqn:PiHRecSpecial} to Eq.~\eqref{eqn:PiHDecomp}. We start by expressing all involved matrices of the triangular and the rectangular contributions in the eigenbasis of the Bloch Hamiltonian $\lam_\bp$. We decompose the first-order and second-order vertices in their diagonal and off-diagonal components via $\Udag_\bp\lam^\alf_\bp\U_\bp=\cE^\alf_\bp+\cF^\alf_\bp$ and $\Udag_\bp\lam^{\alf\beta}_\bp\U_\bp=(\dM^{-1})^{\alf\beta}_\bp+\dF^{\alf\beta}_\bp$, which were defined in \eqref{eqn:UdagLamU} and \eqref{eqn:DecompLamAB}, respectively. Using that only an even number of off-diagonal matrices give a nonzero matrix trace, the triangular contribution in \eqref{eqn:PiHTriSpecial} decomposes into four contributions, which we label as
\begin{alignat}{6}
 (\text{tri},\,\text{I})&=\frac{1}{4}\wideTrH\Big[
 \cG^{}_{ip_0+iq_0,\bp}\,&&\cE^{\alf}_{\bp}\,&&\cG^{}_{ip_0,\bp}\,&&\cE^\gamma_\bp\,&&\cG^{}_{ip_0,\bp}\,&&(\dM^{-1})^{\beta\delta}_\bp\Big]\, ,\\
 (\text{tri},\,\text{II})&=\frac{1}{4}\wideTrH\Big[
 \cG^{}_{ip_0+iq_0,\bp}\,&&\cF^{\alf}_{\bp}\,&&\cG^{}_{ip_0,\bp}\,&&\cE^\gamma_\bp\,&&\cG^{}_{ip_0,\bp}\,&&\dF^{\beta\delta}_\bp\Big]\, ,\\
 (\text{tri},\,\text{III})&=\frac{1}{4}\wideTrH\Big[
 \cG^{}_{ip_0+iq_0,\bp}\,&&\cE^\alf_\bp\,&&\cG^{}_{ip_0,\bp}\,&&\cF^\gamma_\bp\,&&\cG^{}_{ip_0,\bp}\,&&\dF^{\beta\delta}_\bp\Big]\, , \\
 (\text{tri},\,\text{IV})&=\frac{1}{4}\wideTrH\Big[
 \cG^{}_{ip_0+iq_0,\bp}\,&&\cF^\alf_\bp\,&&\cG^{}_{ip_0,\bp}\,&&\cF^\gamma_\bp\,&&\cG^{}_{ip_0,\bp}\,&&(\dM^{-1})^{\beta\delta}_\bp\Big] \, .
\end{alignat}
We use our results in \eqref{eqn:ESpecial}, \eqref{eqn:cFSpecial} and \eqref{eqn:dFSpecial} for the considered special case of a momentum-independent gap. The inverse generalized effective mass is given by $(\dM^{-1})^{\alf\beta}_\bp=\cE^{\alf\beta}_\bp-2\cS^{}_\bp\cF^\alf_\bp\cF^\beta_\bp$, where we defined $\cS_\bp=1/\big(E^+_\bp-E^-_\bp\big)\pauma_z$. The part of the fourth contribution involving $2\cS_\bp\cF^\beta_\bp\cF^\delta_\bp$ vanishes by the antisymmetry in $\delta\leftrightarrow\gamma$, so that
\begin{align}
 (\text{tri},\,\text{IV})=\frac{1}{4}\wideTrH\Big[
 \cG^{}_{ip_0+iq_0,\bp}\,\cF^\alf_\bp\,\cG^{}_{ip_0,\bp}\,\cF^\gamma_\bp\,\cG^{}_{ip_0,\bp}\,\cE^{\beta\delta}_\bp\Big] \, .
\end{align}
Performing the same decomposition and dropping zero contributions due to the matrix trace, the rectangular contribution in \eqref{eqn:PiHRecSpecial} decomposes into eight contributions, which we label as
\begin{alignat}{9}
 (\text{rec},\,\text{I})&=\frac{1}{4}\wideTrH\Big[\cG^{}_{ip_0+iq_0,\bp}\,&&\cE^\alf_\bp\,&&\cG^{}_{ip_0,\bp}\,&&\cE^\gamma_\bp\,&&\cG^{}_{ip_0,\bp}\,&&\cE^\delta_\bp\,&&\cG^{}_{ip_0,\bp}\,&&\cE^\beta_\bp\,&&\Big] \, ,\\[1mm]
 (\text{rec},\,\text{II})&=\frac{1}{4}\wideTrH\Big[\cG^{}_{ip_0+iq_0,\bp}\,&&\cE^\alf_\bp\,&&\cG^{}_{ip_0,\bp}\,&&\cE^\gamma_\bp\,&&\cG^{}_{ip_0,\bp}\,&&\cF^\delta_\bp\,&&\cG^{}_{ip_0,\bp}\,&&\cF^\beta_\bp\,&&\Big] \, ,\\[1mm]
 (\text{rec},\,\text{III})&=\frac{1}{4}\wideTrH\Big[\cG^{}_{ip_0+iq_0,\bp}\,&&\cE^\alf_\bp\,&&\cG^{}_{ip_0,\bp}\,&&\cF^\gamma_\bp\,&&\cG^{}_{ip_0,\bp}\,&&\cE^\delta_\bp\,&&\cG^{}_{ip_0,\bp}\,&&\cF^\beta_\bp\,&&\Big] \, ,\\[1mm]
 (\text{rec},\,\text{IV})&=\frac{1}{4}\wideTrH\Big[\cG^{}_{ip_0+iq_0,\bp}\,&&\cE^\alf_\bp\,&&\cG^{}_{ip_0,\bp}\,&&\cF^\gamma_\bp\,&&\cG^{}_{ip_0,\bp}\,&&\cF^\delta_\bp\,&&\cG^{}_{ip_0,\bp}\,&&\cE^\beta_\bp\,&&\Big] \, ,\\[1mm]
 (\text{rec},\,\text{V})&=\frac{1}{4}\wideTrH\Big[\cG^{}_{ip_0+iq_0,\bp}\,&&\cF^\alf_\bp\,&&\cG^{}_{ip_0,\bp}\,&&\cE^\gamma_\bp\,&&\cG^{}_{ip_0,\bp}\,&&\cE^\delta_\bp\,&&\cG^{}_{ip_0,\bp}\,&&\cF^\beta_\bp\,&&\Big] \, ,\\[1mm]
 (\text{rec},\,\text{VI})&=\frac{1}{4}\wideTrH\Big[\cG^{}_{ip_0+iq_0,\bp}\,&&\cF^\alf_\bp\,&&\cG^{}_{ip_0,\bp}\,&&\cE^\gamma_\bp\,&&\cG^{}_{ip_0,\bp}\,&&\cF^\delta_\bp\,&&\cG^{}_{ip_0,\bp}\,&&\cE^\beta_\bp\,&&\Big] \, ,\\[1mm]
 (\text{rec},\,\text{VII})&=\frac{1}{4}\wideTrH\Big[\cG^{}_{ip_0+iq_0,\bp}\,&&\cF^\alf_\bp\,&&\cG^{}_{ip_0,\bp}\,&&\cF^\gamma_\bp\,&&\cG^{}_{ip_0,\bp}\,&&\cE^\delta_\bp\,&&\cG^{}_{ip_0,\bp}\,&&\cE^\beta_\bp\,&&\Big] \, , \\[1mm]
 (\text{rec},\,\text{VIII})&=\frac{1}{4}\wideTrH\Big[\cG^{}_{ip_0+iq_0,\bp}\,&&\cF^\alf_\bp\,&&\cG^{}_{ip_0,\bp}\,&&\cF^\gamma_\bp\,&&\cG^{}_{ip_0,\bp}\,&&\cF^\delta_\bp\,&&\cG^{}_{ip_0,\bp}\,&&\cF^\beta_\bp\,&&\Big] \, .
\end{alignat}
The four contributions
\begin{align}
 (\text{rec},\text{I}) =
 (\text{rec},\,\text{IV}) =
 (\text{rec},\,\text{V}) =
 (\text{rec},\,\text{VIII}) = 0  
 \label{eqn:RecVanish}
\end{align}
vanish due to their antisymmetric counterpart in $(\gamma\leftrightarrow\delta)$, which follows from $\cE^\gamma_\bp\cG^{}_{ip_0,\bp}\cE^\delta_\bp=\cE^\delta_\bp\cG^{}_{ip_0,\bp}\cE^\gamma_\bp$ and $\cF^\gamma_\bp\cG^{}_{ip_0,\bp}\cF^\delta_\bp=\cF^\delta_\bp\cG^{}_{ip_0,\bp}\cF^\gamma_\bp$. Note that \eqref{eqn:RecVanish} is only valid under the assumptions and resulting simplifications of the momentum-independent gap and the specific gauge choice, such that $\cF^\nu_\bp\propto \pauma_x$. Furthermore, we have $(\text{rec},\,\text{II})=(\text{rec},\,\text{VII})$ and $(\text{rec},\,\text{III})=(\text{rec},\,\text{VI})$. In order to see this, note that all terms are symmetric under matrix transposition and do not change sign under simultaneous change of both indices $\alf\leftrightarrow\beta$ and $\gamma\leftrightarrow\delta$. We continue by applying the identity \eqref{eqn:RedGreenFunc} to the remaining two contributions. We get
\begin{alignat}{9}
 \big(\text{rec},\text{II}\big)+\big(\text{rec},\text{VII}\big)=&+\frac{1}{2}\wideTrH\Big[\cG^{}_{ip_0+iq_0,\bp}\,&&\cE^\alf_\bp\,&&\cG^{}_{ip_0,\bp}\,&&\cE^\gamma_\bp\,&&\cF^\delta_\bp\,&&\cG^{}_{ip_0,\bp}\,&&\cS^{}_\bp\,&&\cF^\beta_\bp\,&&\Big] \label{rec1}
 \\[1mm]
 &+\frac{1}{2}\wideTrH\Big[\cG^{}_{ip_0+iq_0,\bp}\,&&\cE^\alf_\bp\,&&\cG^{}_{ip_0,\bp}\,&&\cE^\gamma_\bp\,&&\cS^{}_\bp\,&&\cG^{}_{ip_0,\bp}\,&&\cF^\delta_\bp\,&&\cF^\beta_\bp\,&&\Big] \, ,\label{rec2}
\end{alignat}
\begin{alignat}{9}
 \big(\text{rec},\text{VI}\big)+\big(\text{rec},\text{III}\big)=&+\frac{1}{2}\wideTrH\Big[\cG^{}_{ip_0+iq_0,\bp}\,&&\cF^\alf_\bp\,&&\cG^{}_{ip_0,\bp}\,&&\cE^\gamma_\bp\,&&\cF^\delta_\bp\,&&\cG^{}_{ip_0,\bp}\,&&\cS^{}_\bp\,&&\cE^\beta_\bp\,&&\Big] \label{rec3}
 \\[1mm]
 &+\frac{1}{2}\wideTrH\Big[\cG^{}_{ip_0+iq_0,\bp}\,&&\cF^\alf_\bp\,&&\cG^{}_{ip_0,\bp}\,&&\cE^\gamma_\bp\,&&\cS^{}_\bp\,&&\cG^{}_{ip_0,\bp}\,&&\cF^\delta_\bp\,&&\cE^\beta_\bp\,&&\Big] \label{rec4}\, .
\end{alignat}
Let us first combine the two lines \eqref{rec1} and \eqref{rec3}. 
Similar to \eqref{eqn:RedGreenFunc}, we have the algebraic relation $G^+_{ip_0,\bp} G^-_{ip_0,\bp} =
\left(G^+_{ip_0,\bp} - G^-_{ip_0,\bp} \right)/\left(E^+_\bp-E^-_\bp\right)$ with $G_{ip_0,\bp}^\pm = [ip_0 - E_\bp^\pm + i\Gamma\text{sign}(p_0)]^{-1}$. Thus, it immediately follows that 
\begin{align}
 &T\sum_{p_0}G^+_{ip_0,\bp}G^-_{ip_0,\bp}\big(G^+_{ip_0+iq_0,\bp}-G^+_{\bp,ip_0-iq_0}\big)\\=\,&T\sum_{p_0}G^+_{ip_0,\bp}G^-_{ip_0,\bp}\big(G^-_{ip_0+iq_0,\bp}-G^-_{\bp,ip_0-iq_0}\big) \, .
\end{align}
Thus, summing up the two lines \eqref{rec1} and \eqref{rec3} and performing the matrix trace, explicitly, leads to
\begin{align} \label{rec1+rec3}
 &\eqref{rec1}+\eqref{rec3}
 =\frac{1}{2L}\sum_{\bp} \left[\frac{F^\alf_\bp F^\delta_\bp}{E^+_\bp-E^-_\bp}(E^{+,\gamma}_\bp+E^{-,\gamma}_{\bp}) (E^{+,\beta}_{\bp}-E^{-,\beta}_\bp)\hspace{-0.5mm}-\hspace{-0.5mm}(\alf\hspace{-0.5mm}\leftrightarrow\hspace{-0.5mm}\beta)\hspace{-0.5mm}-\hspace{-0.5mm}(\delta\hspace{-0.5mm}\leftrightarrow\hspace{-0.5mm}\gamma)\right] \nonumber
 \\&\hspace{5cm}\times T \sum_{p_0} G^+_{ip_0,\bp} G^-_{ip_0,\bp} (G^+_{ip_0+iq_0,\bp}-G^+_{\bp,ip_0-iq_0})  \, .
\end{align}
It involves the eigenenergies $E^\pm_\bp$, their derivatives as well as the off-diagonal component $F^\nu_\bp$ of the first-order vertex $\cF^\nu_\bp$. After using the identity \eqref{eqn:changingindices}, the bracket $\left[\cdots\right]$ in \eqref{rec1+rec3} vanishes by antisymmetry in the indices $\alf\leftrightarrow\beta$, so that 
\begin{align}
 \eqref{rec1}+\eqref{rec3}=0\,.
\end{align}
We continue with the term in line \eqref{rec2}. We commute the two diagonal matrices $\cS^{}_\bp$ and $\cG_{ip_0,\bp}$. We identify $2\cS^{}_\bp\cF^\beta_\bp\cF^\delta_\bp=\cE^{\beta\delta}_\bp-(\dM^{-1})^{\beta\delta}_\bp$. Thus, we get
\begin{align}
 \big(\text{tri},\text{II}\big)+\eqref{rec2}=\frac{1}{4}\wideTrH\left[\cG^{}_{ip_0+iq_0,\bp}\,\cE^\alf_\bp\,\cG^{}_{ip_0,\bp}\,\cE^\gamma_\bp\,\cG^{}_{ip_0,\bp}\,\cE^{\beta\delta}_\bp\right] \, .
\end{align}
We consider the remaining term in \eqref{rec4}. We split it into two equal parts. In the first part, we reintroduce the derivative with respect to $p^\gamma$ of the Green's function after commuting the diagonal matrices $\cS_\bp$ and $\cG_{ip_0,\bp}$:
\begin{align} \label{rec4part1}
 \frac{1}{2}\times\eqref{rec4}=\frac{1}{4}\wideTrH\Big[\cG^{}_{ip_0+iq_0,\bp}\,\cF^\alf_\bp\,\big(\partial^{}_\gamma\cG^{}_{ip_0,\bp}\big)\,\cS^{}_\bp\,\cF^\delta_\bp\,\cE^\beta_\bp\Big] \, .
\end{align}
In the second part, we first shift the Matsubara summation $ip_0\rightarrow -ip_0$ and change the overall sign via its corresponding contribution in $(ip_0\rightarrow -ip_0)$. After reversing the matrix order under the trace, commuting $\cS_\bp$ with $\cG_{ip_0+iq_0,\bp}$, and reintroducing a derivative with respect to $p^\gamma$ of a Green's function, we get
\begin{align} \label{rec4part2}
 \frac{1}{2}\times\eqref{rec4}=-\frac{1}{4}\wideTrH\Big[\big(\partial^{}_\gamma\cG^{}_{ip_0+iq_0,\bp}\big)\,\cF^\alf_\bp\,\cG^{}_{ip_0,\bp}\,\cE^\beta_\bp\,\cF^\delta_\bp\,\cS^{}_\bp\Big] \, .
\end{align}
We use the identity $\cE^\beta_\bp\cF^\delta_\bp\cS^{}_\bp = \cS^{}_\bp\cF^\beta_\bp\cE^\delta_\bp+\cE^\delta_\bp\cF^\beta_\bp\cS^{}_\bp-\cS^{}_\bp\cF^\delta_\bp\cE^\beta_\bp$,
which immediately follows from \eqref{eqn:changingindices}. Only the last term is nonzero. The first two terms vanish by the antisymmetry in $\alf\leftrightarrow\beta$. We sum up \eqref{rec4part1} and \eqref{rec4part2} and perform a partial integration in $p^\gamma$ in \eqref{rec4part1}. 
The term with a derivative acting on the Green's function cancels \eqref{rec4part2}, and we obtain four remaining contributions:
\begin{align} 
 \eqref{rec4}= &-\frac{1}{4}\wideTrH\Big[\cG^{}_{ip_0+iq_0,\bp}\,\big(\partial^{}_\gamma\cF^\alf_\bp\big)\,\cG^{}_{ip_0,\bp}\,\cS^{}_\bp\,\cF^\delta_\bp\,\cE^\beta_\bp\Big] \label{D51+D52part1} \\[1mm]
 &-\frac{1}{4}\wideTrH\Big[\cG^{}_{ip_0+iq_0,\bp}\,\cF^\alf_\bp\,\cG^{}_{ip_0,\bp}\,\big(\partial^{}_\gamma\cS^{}_\bp\big)\,\cF^\delta_\bp\,\cE^\beta_\bp\Big] \label{D51+D52part2} \\[1mm]
 &-\frac{1}{4}\wideTrH\Big[\cG^{}_{ip_0+iq_0,\bp}\,\cF^\alf_\bp\,\cG^{}_{ip_0,\bp}\,\cS^{}_\bp\,\big(\partial^{}_\gamma\cF^\delta_\bp\big)\,\cE^\beta_\bp\Big] \label{D51+D52part3}\\[1mm]
 &-\frac{1}{4}\wideTrH\Big[\cG^{}_{ip_0+iq_0,\bp}\,\cF^\alf_\bp\,\cG^{}_{ip_0,\bp}\,\cS^{}_\bp\,\cF^\delta_\bp\,\big(\partial^{}_\gamma\cE^\beta_\bp\big)\Big] \label{D51+D52part4}\, .
\end{align}
We go through the four terms: 
(1) The derivative in \eqref{D51+D52part4} is by definition $\partial^{}_\gamma\cE^\beta_\bp=\cE^{\beta\gamma}_\bp$. (2) The term in \eqref{D51+D52part2} containing $\partial_\gamma\cS_\bp$ cancels by the corresponding part in $(\gamma\leftrightarrow\delta)$, since $(\partial^{}_\gamma\cS^{}_\bp)\cF^\delta_\bp=\partial^{}_\gamma\big(\frac{1}{E^+_\bp-E^-_\bp}\big)F^\delta_\bp\,\pauma_z\pauma_x=-\frac{1}{(E^+_\bp-E^-_\bp)^2}\frac{2h_\bp}{\Delta}F^\gamma_\bp F^\delta_\bp\,\pauma_z\pauma_x$, where we used the explicit form of $\cF^\gamma_\bp$ in \eqref{eqn:cFSpecial}. (3) In order to see the cancellation of \eqref{D51+D52part3} containing $\partial^{}_\gamma\cF^\delta_\bp$ we use $\cF^\delta_\bp=F^\delta_\bp\,\pauma_x=\frac{2\Delta}{E^+_\bp-E^-_\bp}h^\delta_\bp\,\pauma_x$ given in \eqref{eqn:cFSpecial}. The derivatives of $1/(E^+_\bp-E^-_\bp)$ and $h^\delta_\bp$ cancel by the corresponding part in $(\gamma\leftrightarrow\delta)$. (4) For \eqref{D51+D52part1} containing $\partial^{}_\gamma \cF^\alf_\bp$, we use the explicit form of $\cF^\alf_\bp$ in \eqref{eqn:cFSpecial}. Whereas the derivative 
of $1/(E^+_\bp-E^-_\bp)$ cancels due to the corresponding part in $(\gamma\leftrightarrow\delta)$, the derivative of $h^\alf_\bp$
now produces the off-diagonal matrix of the second-order vertex $\dF^{\alf\gamma}_\bp= \frac{2\Delta}{E^+_\bp-E^-_\bp}h^{\alf\gamma}_\bp\,\pauma_x$ given in \eqref{eqn:dFSpecial}. Thus, the four terms finally reduce to
\begin{align}
 \eqref{rec4}=&-\frac{1}{4}\wideTrH\Big[\cG^{}_{ip_0+iq_0,\bp}\,\dF^{\alf\gamma}_\bp\,\cS^{}_\bp\,\cG^{}_{ip_0,\bp}\,\cF^\delta_\bp\,\cE^\beta_\bp\Big]\\[1mm]&-\frac{1}{4}\wideTrH\Big[\cG^{}_{ip_0+iq_0,\bp}\,\cF^\alf_\bp\,\cS^{}_\bp\,\cG^{}_{ip_0,\bp}\,\cF^\delta_\bp\,\cE^{\beta\gamma}_\bp\Big] \, .
\end{align}
We commuted $\cS_\bp$ and $\cG^{}_{ip_0,\bp}$. We reinstall three Green's functions in both terms by using the identity $\cS^{}_\bp\cG^{}_{ip_0,\bp}\cF^\delta_\bp=\cG^{}_{ip_0,\bp}\cF^\delta_\bp\cG^{}_{ip_0,\bp}-\cF^\delta_\bp\cG^{}_{ip_0,\bp}\cS^{}_\bp$. The terms containing $\cS^{}_\bp$ cancel by the corresponding term in $(iq_0\leftrightarrow -iq_0)$ when shifting the Matsubara summation and commuting the matrices. We end up with identifying 
\begin{align} 
 \eqref{rec4}=-(\text{tri},\text{III})+(\text{tri},\text{IV}) \, .
\end{align}
Combining all contributions leads to the final result in \eqref{eqn:PiHDecomp}.


\newpage\leavevmode\thispagestyle{empty}\newpage

%
%

\chapter{Matsubara summation} 
\label{appendix:Matsubara}

We perform the summation over the internal Matsubara frequency $ip_0$ and the subsequent analytic continuation of the external Matsubara frequency $iq_0\rightarrow \omega+i0^+$ of the three relevant quantities $I^s_{iq_0}$, $I^a_{iq_0}$ and $I^H_{iq_0}$ in \eqref{eqn:Isq0}, \eqref{eqn:Iaq0} and \eqref{eqn:MatsumH}, respectively. In this section, we omit the momentum dependence for shorter notation. We can represent any Matsubara Green's function matrix $G_{ip_0}$ in the spectral representation as
\begin{align}
\label{eqn:Gip0}
 G_{ip_0}=\int\hspace{-1.5mm}d\epsilon \,\frac{A(\epsilon)}{ip_0-\epsilon} 
\end{align}
with corresponding spectral function matrix $A(\epsilon)\equiv A_\epsilon$. The retarded and the advanced Green's function matrices are 
\begin{align}
 \label{eqn:GRbasic}&G^R_\epsilon=\int\hspace{-1.5mm}d\epsilon'\,\frac{A(\epsilon')}{\epsilon-\epsilon'+i0^+}\, ,\\[0.5mm]
 \label{eqn:GAbasic}&G^A_\epsilon=\int\hspace{-1.5mm}d\epsilon'\,\frac{A(\epsilon')}{\epsilon-\epsilon'-i0^+} \, .
\end{align}
We define the principle-value matrix $P(\epsilon)\equiv P_\epsilon$ via
\begin{align}
 \label{eqn:Pbasic}
 P(\epsilon)=P.V.\hspace{-1mm}\int\hspace{-1.5mm}d\epsilon'\,\frac{A(\epsilon')}{\epsilon-\epsilon'} \, ,
\end{align}
where $P.V.$ denotes the principle value of the integral. Using the integral identity $\frac{1}{\epsilon-\epsilon'\pm i0^+}=P.V. \frac{1}{\epsilon-\epsilon'}\mp i\pi \,\delta(\epsilon-\epsilon')$, we have
\begin{align}
 \label{eqn:ImGR}&A_\epsilon=-\frac{1}{\pi}\im\,G^R_\eps\equiv-\frac{1}{2\pi i}\big(G^R_\epsilon-G^A_\epsilon\big) \, , \\[1mm]
 \label{eqn:ReGR}&P_\epsilon=\re\,G^R_\eps\equiv\frac{1}{2}\big(G^R_\epsilon+G^A_\epsilon\big)  \, .
\end{align}
Note that $A_\epsilon$ and $P_\epsilon$ are hermitian matrices.
The functions to be continued analytically have the following structure
\begin{equation}
 \label{eqn:Ibasic}
 I_{iq_0}^{m,n} \equiv T \sum_{p_0} {\rm tr} \bigg[
 \big(G_{ip_0+iq_0} \,M_1 \dots G_{ip_0+iq_0} \,M_m\big) \big(G_{ip_0} \,N_1 \dots G_{ip_0} \,N_n\big)
 \bigg] \, ,
\end{equation}
where $M_1,\dots,M_m$ and $N_1,\dots,N_n$ are frequency-independent $2 \times 2$ matrices. The first $m$ Green's function matrices involve the bosonic external Matsubara frequency $q_0$. The last $n$ Green's function only involve the internal fermionic Matsubara frequency $p_0$. We grouped the corresponding matrices by brackets.
We insert the spectral representation \eqref{eqn:Gip0} for each Green's function and perform the Matsubara frequency sum over the resulting product of energy denominators. Using $\text{Res}_{ip_0}[f(\eps)] = -T$, where $f(\eps)=(e^{\eps/T}+1)^{-1}$ is the Fermi function and $k_B=1$, we apply the residue theorem to replace the Matsubara frequency sum by a contour integral encircling the fermionic Matsubara frequencies counterclockwise. We then change the contour such that only the poles from the energy denominators are encircled. Applying the residue theorem again yields
\begin{eqnarray}
 && T \sum_{p_0} \bigg(\frac{1}{ip_0 + iq_0 - \eps_1} \dots \frac{1}{ip_0 + iq_0 - \eps_m}\bigg)
 \, \bigg(\frac{1}{ip_0 - \eps'_1} \dots \frac{1}{ip_0 - \eps'_n}\bigg)
 \nonumber \\[1mm]
 && = f(\eps_1) \,\bigg(\frac{1}{\eps_1 - \eps_2} \dots \frac{1}{\eps_1 - \eps_m}\bigg) \,\bigg(
 \frac{1}{-iq_0 + \eps_1 - \eps'_1} \dots \frac{1}{-iq_0 + \eps_1 - \eps'_n}\bigg) + \dots
 \nonumber \\[1mm]
 && + \, f(\eps_m) \bigg(\frac{1}{\eps_m - \eps_1} \dots \frac{1}{\eps_m - \eps_{m-1}}\bigg) \, \bigg(
 \frac{1}{-iq_0 + \eps_m - \eps'_1} \dots \frac{1}{-iq_0 + \eps_m - \eps'_n} \bigg)
 \nonumber \\[1mm]
 && + \, f(\eps'_1) \,\bigg(\frac{1}{iq_0 + \eps'_1 - \eps_1} \dots 
 \frac{1}{iq_0 + \eps'_1 - \eps_m}\bigg) \, \bigg(
 \frac{1}{\eps'_1 - \eps'_2} \dots \frac{1}{\eps'_1 - \eps'_n}\bigg) + \dots
 \nonumber \\[1mm]
 && + \, f(\eps'_n) \,\bigg(\frac{1}{iq_0 + \eps'_n - \eps_1} \dots 
 \frac{1}{iq_0 + \eps'_n - \eps_m}\bigg) \, \bigg(
 \frac{1}{\eps'_n - \eps'_1} \dots \frac{1}{\eps'_n - \eps'_{n-1}}\bigg)\, ,
\end{eqnarray}
where we grouped the product of $m$ and $n$ energy denominators for a more transparent representation. This expression can be analytically continued to real frequencies, easily, by replacing $iq_0$ with $\omega + i0^+$. Performing the integrals over $\eps_1,\dots,\eps_m$ and $\eps'_1,\dots,\eps'_n$ then yields
\begingroup
 \allowdisplaybreaks[0]
\begin{align} \label{eqn:Imn}
 I_\omega^{m,n} = &\int d\eps f_\eps \, {\rm tr} \bigg[
 \Big( A_\eps\,M_1\,P_\eps\,M_2\,\dots\,P_\eps\,M_m \,+\dots
 + \, P_\eps\,M_1\,\dots\,P_\eps\,M_{m-1}\,A_\eps \,M_m \Big) \nonumber
 \\
 &\hspace{2.5cm}\times G_{\eps-\omega}^A\, N_1 \,\dots \,G_{\eps-\omega}^A\, N_n \bigg]
 \nonumber \\
 + &\int d\eps f_\eps \, {\rm tr} \bigg[
 G_{\eps+\omega}^R \,M_1 \,\dots\,G_{\eps+\omega}^R \,M_m \nonumber
 \\
 &\hspace{2.5cm}\times\Big(
  A_\eps\,N_1\,P_\eps\,N_2\,\dots\,P_\eps\,N_n\,+ \dots
 + \, P_\eps\,N_1\,\dots\,P_\eps\,N_{n-1}\,A_\eps \,N_n \Big) \bigg] \, ,
\end{align}
\endgroup
where we identified the retarded Green's function matrix $G^R_{\eps+\omega}$ in \eqref{eqn:GRbasic} at frequency $\eps+\omega$, the advanced Green's function matrix $G^A_{\eps-\omega}$ in \eqref{eqn:GAbasic} at frequency $\eps-\omega$ and the principle-value matrix $P_\eps$ in \eqref{eqn:Pbasic}. We can understand the sum in the brackets as all combinations to place one spectral function $A_\eps$ at all positions before the matrices $M_i$ and $N_i$. Note that the involved matrices do not commute in general. The special case of no external Matsubara frequency $iq_0$ reads
\begin{align}
  I_\omega^{0,n} =  &\int d\eps f_\eps \, {\rm tr} \Big[
  A_\eps\,N_1\,P_\eps\,N_2\,\dots\,P_\eps\,N_n\,+ \dots
 + \, P_\eps\,N_1\,\dots\,P_\eps\,N_{n-1}\,A_\eps \,N_n \Big] \, .
\end{align}
Evaluating $I^{m,n}_{-iq_0}$, that is \eqref{eqn:Ibasic} with bosonic Matsubara frequencies at opposite sign, results in \eqref{eqn:Imn} with exchanged $G^R_{\eps+\omega}\leftrightarrow G^A_{\eps-\omega}$.

%
%

\section{Matsubara summation of $I^s_{iq_0}$ and $I^a_{iq_0}$}

We continue by performing the Matsubara summation and the analytic continuation of $I^s_{iq_0}$ and $I^a_{iq_0}$ in \eqref{eqn:Isq0} and \eqref{eqn:Iaq0}, respectively. They consist of three distinct cases. We use our general result in \eqref{eqn:Imn}. The first case involves the Green's function matrix $G_{ip_0+iq_0}$ leading to
\begin{align}
 T&\sum_{p_0}\left.\tr\big[G_{ip_0+iq_0}M_1G_{ip_0}M_2\big]\right|_{iq_0\rightarrow \omega+i0^+} = \int\hspace{-1.5mm}d\epsilon \,f_\epsilon\,\tr\big[A^{}_\epsilon M^{}_1G^A_{\epsilon-\omega}M^{}_2+G^R_{\epsilon+\omega}M^{}_1A^{}_\epsilon M^{}_2\big] \, .
\end{align}
The second case involves the Green's function matrix $G_{ip_0-iq_0}$ leading to
\begin{align}
 T&\sum_{p_0}\left.\tr\big[G_{ip_0-iq_0}M_1G_{ip_0}M_2\big]\right|_{iq_0\rightarrow \omega+i0^+}= \int\hspace{-1.5mm} d\epsilon\, f_\epsilon\,\tr\big[A^{}_\epsilon M^{}_1G^R_{\epsilon+\omega}M^{}_2+G^A_{\epsilon-\omega}M^{}_1A^{}_\epsilon M^{}_2\big] \, .
\end{align}
The third case involves no bosonic Matsubara frequency $iq_0$ and is given by
\begin{align}
 T&\sum_{p_0}\left.\tr\big[G_{ip_0}M_1G_{ip_0}M_2\big]\right|_{iq_0\rightarrow \omega+i0^+}= \int\hspace{-1.5mm} d\epsilon\, f_\epsilon\,\tr\big[A_\epsilon M_1P_\epsilon M_2+P_\epsilon M_1A_\epsilon M_2\big] \, .
\end{align}
We can rewrite these three cases by using
\begin{align}
 &G^R_\epsilon=P_\epsilon-i\pi A_\epsilon \, ,\\
 &G^A_\epsilon=P_\epsilon+i\pi A_\epsilon \, , \, 
\end{align}
which follows by \eqref{eqn:ImGR} and \eqref{eqn:ReGR}, in order to express all results only by the hermitian matrices $A_\epsilon$ and $P_\epsilon$. The Matsubara summation of \eqref{eqn:Isq0} after analytic continuation reads
\begin{align}
 I^s_\omega=\frac{1}{2}\int\hspace{-1.5mm} d\epsilon\, f_\epsilon 
 \,&\tr\Big[A_\epsilon M_1 \Big((P_{\epsilon+\omega}-P_\epsilon)
 +(P_{\epsilon-\omega}-P_\epsilon)\Big)M_2\nonumber\\[1mm]
 &+\Big((P_{\epsilon+\omega}-P_\epsilon)+(P_{\epsilon-\omega}-P_\epsilon)\Big) M_1 A_\epsilon M_2 
 \nonumber
 \\[1mm]
 &-i\pi A_\epsilon M_1 \Big((A_{\epsilon+\omega}-A_\epsilon)-(A_{\epsilon-\omega}-A_\epsilon)\Big)M_2\nonumber\\[1mm]
 &-i\pi \Big((A_{\epsilon+\omega}-A_\epsilon)-(A_{\epsilon-\omega}-A_\epsilon)\Big) M_1 A_\epsilon M_2\Big] \, .
\end{align}
We divide by $i\omega$ and perform the zero frequency limit leading to the frequency derivatives $\lim_{\omega\rightarrow 0}(P_{\epsilon\pm\omega}-P_\epsilon)/\omega=\pm P'_\epsilon$ and $\lim_{\omega\rightarrow 0}(A_{\epsilon\pm\omega}-A_\epsilon)/\omega=\pm A'_\epsilon$, which we denote by $(\cdot)'$. The first and the second line of the sum vanish. We get 
\begin{align}
 \lim_{\omega\rightarrow 0}\frac{I^s_\omega}{i\omega}&=-\pi\int\hspace{-1.5mm}d\epsilon\, f_\epsilon\,\tr\big[A_\epsilon M_1 A'_\epsilon M_2+A'_\epsilon M_1 A_\epsilon M_2\big] .
\end{align}
We can apply the product rule and partial integration in $\epsilon$ and end up with \eqref{eqn:Isw}. The Matsubara summation of \eqref{eqn:Iaq0} after analytic continuation is
\begin{align}
 I^a_\omega=\frac{1}{2}\int\hspace{-1.5mm}d\epsilon\, f_\epsilon\,\tr\Big[&-A_\epsilon M_1 \Big((P_{\epsilon+\omega}-P_\epsilon)-(P_{\epsilon-\omega}-P_\epsilon)\Big)M_2
 \nonumber\\[1mm]
 &+\Big((P_{\epsilon+\omega}-P_\epsilon)-(P_{\epsilon-\omega}-P_\epsilon)\Big) M_1 A_\epsilon M_2 \nonumber\\[1mm]
 &+i\pi A_\epsilon M_1 \Big((A_{\epsilon+\omega}-A_\epsilon)+(A_{\epsilon-\omega}-A_\epsilon)\Big)M_2\nonumber\\[1mm]
 &-i\pi \Big((A_{\epsilon+\omega}-A_\epsilon)+(A_{\epsilon-\omega}-A_\epsilon)\Big) M_1 A_\epsilon M_2\Big]\, .
\end{align}
We divide by $i\omega$ and perform the zero frequency limit. The two last lines of the summation drop. We end up with \eqref{eqn:Iaw}.

%
%

\section{Matsubara summation of $I^H_{iq_0}$}

We continue by performing the Matsubara summation and analytic continuation of $I^H_{iq_0}$ in \eqref{eqn:MatsumH}. We use our general result in \eqref{eqn:Imn}. We have
\begin{align}
 T&\sum_{p_0}\left.\tr\big[G_{ip_0+iq_0}M_1G_{ip_0}M_2G_{ip_0}M_3\big]\right|_{iq_0\rightarrow \omega+i0^+} \nonumber\\
 &= \int\hspace{-1.5mm}d\eps \,f_\eps\,\tr\big[A^{}_\eps M^{}_1G^A_{\eps-\omega}M^{}_2G^A_{\eps-\omega}M^{}_3+G^R_{\eps+\omega}M^{}_1A^{}_\eps M^{}_2P^{}_\eps M^{}_3+G^R_{\eps+\omega}M^{}_1P^{}_\eps M^{}_2 A^{}_\eps M^{}_3\big] 
\end{align}
and subtract
\begin{align}
 T&\sum_{p_0}\left.\tr\big[G_{ip_0-iq_0}M_1G_{ip_0}M_2G_{ip_0}M_3\big]\right|_{iq_0\rightarrow \omega+i0^+} \nonumber\\
 &= \int\hspace{-1.5mm}d\eps \,f_\eps\,\tr\big[A^{}_\eps M^{}_1G^R_{\eps+\omega}M^{}_2G^R_{\eps+\omega}M^{}_3+G^A_{\eps-\omega}M^{}_1A^{}_\eps M^{}_2P^{}_\eps M^{}_3+G^A_{\eps-\omega}M^{}_1P^{}_\eps M^{}_2 A^{}_\eps M^{}_3\big] \, .
\end{align}
We use $G^R_\eps=P_\eps-i\pi A_\eps$ and $G^A_\eps=P_\eps+i\pi A_\eps$ in order to express $I^H_{iq_0}$ only by the hermitian matrices $A_\eps$ and $P_\eps$. We obtain the lengthy expression
\begin{align}
 I^H_\omega=\int\hspace{-1.5mm}d\epsilon\, f_\epsilon\,\tr\Big[
 &-A_\eps M_1 \Big((P_{\eps+\omega}M_2P_{\eps+\omega}\hspace{-0.8mm}-\hspace{-0.8mm}P_\eps M_2 P_\eps)-(P_{\eps-\omega}M_2 P_{\eps-\omega}\hspace{-0.8mm}-\hspace{-0.8mm}P_\eps M_2 P_\eps) \Big)M_3
 \nonumber \\[-1mm]
 &+\pi^2 A_\eps M_1 \Big((A_{\eps+\omega}M_2A_{\eps+\omega}\hspace{-0.8mm}-\hspace{-0.8mm}A_\eps M_2 A_\eps)-(A_{\eps-\omega}M_2 A_{\eps-\omega}\hspace{-0.8mm}-\hspace{-0.8mm}A_\eps M_2 A_\eps) \Big)M_3
 \nonumber \\
 &+\Big((P_{\eps+\omega}\hspace{-0.8mm}-\hspace{-0.8mm}P_\eps)-(P_{\eps-\omega}\hspace{-0.8mm}-\hspace{-0.8mm}P_\eps)\Big)M_1 P_\eps M_2 A_\eps M_3
 \nonumber\\
 &+\Big((P_{\eps+\omega}\hspace{-0.8mm}-\hspace{-0.8mm}P_\eps)-(P_{\eps-\omega}\hspace{-0.8mm}-\hspace{-0.8mm}P_\eps)\Big)M_1 A_\eps M_2 P_\eps M_3
 \nonumber\\
 &+i\pi A_\eps M_1 \Big((P_{\eps+\omega}M_2A_{\eps+\omega}\hspace{-0.8mm}-\hspace{-0.8mm}P_\eps M_2 A_\eps)+(P_{\eps-\omega}M_2 A_{\eps-\omega}\hspace{-0.8mm}-\hspace{-0.8mm}P_\eps M_2 A_\eps) \Big)M_3
 \nonumber\\
 &+i\pi A_\eps M_1 \Big((A_{\eps+\omega}M_2P_{\eps+\omega}\hspace{-0.8mm}-\hspace{-0.8mm}A_\eps M_2 P_\eps)+(A_{\eps-\omega}M_2 P_{\eps-\omega}\hspace{-0.8mm}-\hspace{-0.8mm}A_\eps M_2 P_\eps) \Big)M_3
 \nonumber\\
 &-i\pi\Big((A_{\eps+\omega}\hspace{-0.8mm}-\hspace{-0.8mm}A_\eps)+(A_{\eps-\omega}\hspace{-0.8mm}-\hspace{-0.8mm}A_\eps)\Big)M_1 P_\eps M_2 A_\eps M_3
 \nonumber\\
&-i\pi\Big((A_{\eps+\omega}\hspace{-0.8mm}-\hspace{-0.8mm}A_\eps)+(A_{\eps-\omega}\hspace{-0.8mm}-\hspace{-0.8mm}A_\eps)\Big)M_1 A_\eps M_2 P_\eps M_3
 \Big] \, .
\end{align}
We divide by $\omega$ and perform the zero frequency limit leading to the frequency derivatives $\lim_{\omega\rightarrow 0}(P_{\eps\pm \omega}-P_\eps)/\omega=\pm P'_\eps$ and $\lim_{\omega\rightarrow 0}(P_{\eps\pm \omega}M_2 P_{\eps\pm\omega}-P_\eps M_2 P_\eps)/\omega=\pm (P_\eps M_2 P_\eps)'$ as well as the respective combinations with one or two $P_\eps$ replaced by $A_\eps$. The last four lines drop. We get
\begin{align}
 \lim_{\omega\rightarrow 0}\frac{I^H_\omega}{\omega}=2\int\hspace{-1.5mm}d\epsilon\, f_\epsilon\,\tr\Big[
 &-A_\eps M_1 \big(P_\eps M_2 P_\eps\big)'M_3
 +\pi^2 A_\eps M_1 \big(A_\eps M_2 A_\eps\big)'M_3 \nonumber
 \\
 &+P'_\eps M_1 P_\eps M_2 A_\eps M_3
 +P'_\eps M_1 A_\eps M_2 P_\eps M_3
 \Big] \,.
\end{align}
We perform the product rule and end up with \eqref{eqn:IH}-\eqref{eqn:IH3}.

\cleardoublepage

%
%
%
\end{appendices}






\rhead[\fancyplain{}{\bfseries                                             
Acknowledgments}]{\fancyplain{}{
}}                          
\lhead[\fancyplain{}{
}]{\fancyplain{}{\bfseries            
Acknowledgments}}
\thispagestyle{empty}
\chapter*{Acknowledgments}

My PhD thesis would not have been possible without the support of many people. It is a great pleasure to thank all those who helped me with their professional, scientific, technical, administrative and personal support. It was an extraordinary opportunity to study and work at the Max Planck Institute for Solid State Research in Stuttgart, Germany. I have experienced the institute as an inspiring, open, and friendly place. I have had the chance to meet, discuss and make friends with so many people I do not want to miss in my life. 

First and foremost, I want to thank my supervisor Prof. Dr. Walter Metzner. He gave me all the freedom and support to explore the opportunities that the institute provided me. His way of guiding me through the different stages of the PhD from entering a new research field, structuring my daily work, combining and writing up our results to developing my own creativity is outstanding. In particular, I deeply appreciate that I could always speak openly and ask him for his advice on everything. Dear Walter, thanks for having me in your Quantum Many-body theory department.

I cannot say how much I have acknowledged the daily exchange as well as the extraordinary PhD trips, retreats, schools, conferences, and private activities with my colleagues, in particular, with Jachym Sykora, Moritz Hirschmann, Andreas Leonhardt, Lukas Schwarz, Oleksii Maistrenko, Demetrio Vilardi and Pietro Maria Bonetti. The possibility to have common or related projects and to talk to them at any moment inspired me and contributed a lot to the results of this thesis.  I want to give a special thanks to my room mate Jachym Sykora, with whom I could have endless deep discussions about scientific and non-scientific topics. I also want to give a special thanks to my former senior colleagues Darshan Joshi and Eslam Khalaf, who helped me entering new fields by allowing me to ask all my fundamental questions, and to Andreas Eberlein, whose work inspired my project. 


Our group benefits a lot from a large group of outstanding senior scientist who are permanent at the institute or are visiting us regularly. Their different research interests as well as their different personalities are responsible for a very diverse pool of experiences from which I could benefit in many perspectives. A great thanks go to Andreas Schnyder, Dirk Manske, Pavel Ostrovsky, Hiroyuki Yamase, Elio K\"onig, Peter Horsch, Roland Zeyher and Andr\'es Greco. A special thanks go to the senior scientists in the other departments of the institute and to my external PhD advisor Alexander Yaresko. 

I am grateful to Prof. Dr. Maria Daghofer, Prof. Dr. Oleg Sushkov, and Prof. Dr. Tilman Pfau for being on my thesis committee.

During my PhD, I had the opportunity to meet and discuss with various scientists around the world. Communication is a key to science, which I could experience in numerous schools, workshops and conferences and in personal correspondence. In particular, it is a great pleasure to thank Andr\'e-Marie Tremblay for inviting me to his group at the University of Sherbrooke, Canada, in 2019 and for introducing me to scientists and students of his institute. For me, it was an extremely stimulating week of discussions and new impressions.

I have benefited a lot from the offers that the Graduate Academy of the University of Stuttgart (GRADUS) and the International Max Planck Research School for Condensed Matter Science (IMPRS-CMS) provided during my PhD. Besides numerous soft-skills seminars, which I could attend, the IMPRS-CMS organized annual internal conferences, where we as PhD students could practice our communication skills, reflect on the scientific practice and build a personal network inside the Max Planck institute. Furthermore, the IMPRS-CMS organized various schools and workshops. I want to thank the IMPRS coordinators Hans-Georg Libuda, Eva Benckiser and Kirsten Eppard for their personal support and their outstanding work.

The exchange and support between the students of different departments of the Max Planck institute is extraordinary. Due to the open and communicative environment, I could develop a much broader scientific background and learned from their different experiences. Representing all those that I had the opportunity to talk to, I want to thank, in particular, Katrin F\"ursich, Luzia Germann, Maximilian Krautloher, Daniel Putzky and Alexander P\"utz.

Besides the scientific support, the institute provides tremendous administrative help. I want to thank our secretary Jeanette Sch\"uller-Knapp and our senior executive manager Michael Eppard representatively for the full administration. A special thanks go to Regine Noack, Michaela Asen-Palmer and Anette Schleehauf.

The institute was much more than a working place for me. I really enjoyed being part of the PhD representatives 2018 and 
for organizing various events and our PhD trip together. A very special highlight was the opportunity of joining the organizing team for the TEDx MPIStuttgart event in 2019, the first TEDx event of a Max Planck institute in Germany. I want to express my deepest gratitude to Shai Mangel who proposed his idea of hosting a TEDx event at the institute at a very early stage to me. 
A special thanks go to Hrag Karakachian for bringing me on board of the institute band, the Band Gap.

Finally, I want to thank my family for their unconditional love and support. Especially, I want to thank my mother Birgit and my father J\"org. Dear Stefanie, thank you very much for your love, patience and support. This thesis would not have been possible without you.

\addcontentsline{toc}{chapter}{Acknowledgments}

\newpage\leavevmode\thispagestyle{empty}\newpage



\rhead[\fancyplain{}{\bfseries                                             
List of publications}]{\fancyplain{}{
}}                          
\lhead[\fancyplain{}{
}]{\fancyplain{}{\bfseries            
List of publications}}
\thispagestyle{empty}
\chapter*{List of publications}
\addcontentsline{toc}{chapter}{List of publications}

\begin{etaremune}
 \item \underline{\bf J.~Mitscherling}, \\ Longitudinal and anomalous Hall conductivity of a general two-band model, \\ \href{https://doi.org/10.1103/PhysRevB.102.165151}{Phys. Rev. B {\bf 102}, 165151 (2020)}. \\ {\it Selected as \href{https://twitter.com/PhysRevB/status/1325798050124673025}{\#PRBTopDownload by @PhysRevB on Twitter (9. Sept. 2020)}}
 \item P.~M.~Bonetti${}^*$, \underline{\bf J.~Mitscherling}${}^*$, D.~Vilardi, and W.~Metzner, \\ Charge carrier drop at the onset of pseudogap behavior in the two-dimensional Hubbard model, \\ \href{https://doi.org/10.1103/PhysRevB.101.165142}{Phys. Rev. B {\bf 101}, 165142 (2020)}.
 \item \underline{\bf J.~Mitscherling} and W.~Metzner, \\ Longitudinal conductivity and Hall coefficient in two-dimensional metals with spiral magnetic order, \\ \href{https://doi.org/10.1103/PhysRevB.98.195126}{Phys. Rev. B {\bf 98}, 195126 (2018)}. \\ {\it Selected as {\bf Editors' suggestion}}
 \item C.~Texier and \underline{\bf J.~Mitscherling}, \\ Nonlinear conductance in weakly disordered mesoscopic wires: Interaction and magnetic field asymmetry, \\ \href{https://doi.org/10.1103/PhysRevB.97.075306}{Phys. Rev. B {\bf 97}, 075306 (2018)}.
\end{etaremune}

*: equal contribution

\newpage\leavevmode\thispagestyle{empty}\newpage











\end{document}